\documentclass[a4paper,11pt]{article}
\pdfoutput=1 
\date{}
\usepackage{jheppub}
\usepackage{amsmath,amssymb,amsfonts,amsxtra, mathrsfs,graphics,graphicx,amsthm,epsfig, youngtab,bm,longtable,float,tikz,empheq}
\usepackage[margin=0.5cm]{caption}
\allowdisplaybreaks[1]
\usetikzlibrary{positioning}
\usetikzlibrary{automata}
\usetikzlibrary{arrows}
\usetikzlibrary{calc}
\usetikzlibrary{decorations.markings}
\usetikzlibrary{decorations.pathreplacing}
\usetikzlibrary{intersections}
\usetikzlibrary{positioning}
\usetikzlibrary{topaths}
\usetikzlibrary{shapes.geometric}
\usetikzlibrary{shapes.misc}
\tikzset{cf-group/.style = {
    shape = rounded rectangle, minimum size=1.0cm,
    rotate=90,
    rounded rectangle right arc = none,
    draw}}
\tikzset{cross/.style={path picture={ 
  \draw[black]
(path picture bounding box.south east) -- (path picture bounding box.north west) (path picture bounding box.south west) -- (path picture bounding box.north east);
}}}

\newcommand*\widefbox[1]{\fbox{\hspace{2em}#1\hspace{2em}}}

\newcommand{\be}{\begin{equation}}
\newcommand{\ee}{\end{equation}}
\newcommand{\ba}{\begin{array}}
\newcommand{\ea}{\end{array}} 
\newcommand{\bi}{\begin{itemize}}
\newcommand{\ei}{\end{itemize}}
\def\vec#1{\bm{#1}}
\def\bea#1\eea{\allowdisplaybreaMs \begin{align}#1\end{align}}
 \newcommand{\ben}{\begin{enumerate}}
\newcommand{\een}{\end{enumerate}}
\newcommand{\bean}{\begin{eqnarray*}}
\newcommand{\eean}{\end{eqnarray*}}
\newcommand{\eref}[1]{(\ref{#1})}

\newcommand{\nn}{\nonumber}

\newcommand{\tr}{\mathrm{Tr}}

\newcommand{\tq}{\widetilde{q}}

\newcommand{\bz}{\overline{z}}
\newcommand{\bQ}{\overline{Q}}

\newcommand{\BC}{\mathbb{C}}
\newcommand{\BR}{\mathbb{R}}
\newcommand{\BP}{\mathbb{P}}

\newcommand{\BZ}{\mathbb{Z}}

\newcommand{\BU}{\mathbf{1}}

\newcommand{\comment}[1]{}
\newcommand{\CF}{{\cal F}}

\newcommand{\CS}{{\cal S}}

\newcommand{\CT}{{\cal T}}
\newcommand{\CD}{{\cal D}}

\newcommand{\CO}{{\cal O}}

\newcommand{\CN}{{\cal N}}

\newcommand{\CP}{{\cal P}}

\newcommand{\CH}{{\cal H}}
\newcommand{\CZ}{{\cal Z}}
\newcommand{\CR}{{\cal R}}
\newcommand{\CI}{{\cal I}}
\newcommand{\CU}{{\cal U}}
\newcommand{\CA}{{\cal A}}

\newcommand{\de}{\mathrm{d}}

\newcommand{\frgl}{\mathfrak{gl}}

\newcommand{\frsu}{\mathfrak{su}}
\newcommand{\fru}{\mathfrak{u}}

\newcommand{\frg}{\mathfrak{g}}

\newcommand{\frm}{\mathfrak{m}}

\newcommand{\hk}{hyperk\"ahler }
\newcommand{\wt}{\widetilde}
\newcommand{\wh}{\widehat}

\newcommand{\sh}{\sinh \pi}
\newcommand{\ch}{\cosh \pi}
\newcommand{\s}{\sigma}

\newcommand{\Secref}[1]{Section~\ref{#1}}

\newcommand{\Appref}[1]{Appendix~\ref{#1}}
\newcommand{\appref}[1]{App.~\ref{#1}}

\newcommand{\figref}[1]{Fig.~\ref{#1}}
\renewcommand{\eqref}[1]{(\ref{#1})}

\title{Line Defects in Three Dimensional Mirror Symmetry beyond Linear Quivers}
\author{Anindya Dey}
\affiliation{Department of Physics and Astronomy, Johns Hopkins University, 3400 North Charles Street,
Baltimore, MD 21218, USA}
\emailAdd{anindya.hepth@gmail.com}

\abstract{The map of half-BPS line defects under mirror symmetry has previously been worked out for 3d $\CN=4$ linear quivers with unitary gauge groups, 
where these defects have a clear realization in terms of a brane picture in Type IIB String Theory. In this work, we initiate the study of line defects and the 
associated mirror maps for more general 3d $\CN=4$ quiver gauge theories from a QFT approach, using the $S$-type operations introduced in \cite{Dey:2020hfe}. 
In particular, our construction does not rely on any String Theory realization of the quiver gauge theories and the defects. 
After discussing the general framework for the construction of these line defects and their mirror maps, we focus on quiver gauge theories of the $D$-type and the affine $D$-type with unitary gauge groups, as a concrete set of examples. Some of the line defects we study admit a Hanany-Witten description and we show that the associated mirror maps predicted by the Type IIB construction in these cases agree with the QFT computation. In addition, we study an example involving defects in an affine $D$-type theory, for which the dual theory is not directly realized by the Type IIB description. In a companion paper, we will discuss defects in infinite families of non-ADE quivers using the general construction developed in this paper.}

\begin{document}

\maketitle

\section{Introduction and summary of results}

\subsection{Background and the basic idea of the paper}
The existence of UV/IR dualities is a ubiquitous feature of Quantum Field Theories in various space-time dimensions.
Existence of such a duality implies that a set of theories, having distinctly different descriptions (for example, 
theories with different Lagrangians) at a given energy scale, describe the same physics at another energy scale. 
The discovery and analysis of these UV/IR dualities, particularly for QFTs with supersymmetry, 
have relied heavily on String Theory constructions involving branes \cite{Hanany:1996ie}.
In the recent past, localization techniques (see \cite{Pestun:2016zxk} for a recent review) have provided an 
avenue for working out extremely non-trivial checks for dualities in QFTs with sufficient supersymmetry. 
Given this recent progress on the QFT side, a natural question to ask is : can one use the tools of localization 
to construct a systematic field theory prescription for generating new dualities, starting from a well-defined set of 
basic dualities?

In the context of a class of IR dualities in three dimensions, such a construction was presented in \cite{Dey:2020hfe}, 
which we will briefly summarize. Consider a class of 3d CFTs with a weakly coupled description 
that has a manifest global symmetry subgroup $G^{\rm sub}_{\rm global}=\prod_\gamma U(M_\gamma)$. 
We will refer to this class of 3d CFTs as class $\CU$. Given a UV theory $X$ in class $\CU$, one can define 
a map (see \Secref{SOps-def} for notation and details) which acts on $X$ to give a generically new theory $X'$, i.e.
\be \label{Smap}
\CO^\alpha_{\vec \CP}: X[\vec{\wh{A}}] \mapsto X'[\vec{\wh{B}}],
\ee
where $\wh{\vec{A}}$ and $\wh{\vec B}$ denote various background fields associated with global symmetries 
of the theories $X$ and $X'$ respectively\footnote{In the most general case, an $S$-operation can also turn on 
defects in the theory, but we will ignore this for the time being to keep the discussion simple. More details on this 
can be found in \Secref{SOps-def} and later in the paper.}.
In the trivial case, where $G^{\rm sub}_{\rm global}=U(1)$, and $\CO^\alpha_{\vec \CP}$ 
is a gauging operation of the said $U(1)$, the above map coincides with the $S$ generator of 
Witten's $SL(2,\BZ)$ action \cite{Witten:2003ya} on a 3d CFT. In analogy, we refer to the above map as an 
``elementary $S$-type operation". 

Suppose the theory $X[\wh{\vec A}]$ is IR dual to the theory $Y[\wh{\vec A}]$, where both theories 
have a weakly coupled description. Given an elementary $S$-type operation $\CO^\alpha_{\vec \CP}$ on $X[\wh{\vec A}]$, one can 
define a dual operation $\wt{\CO}^\alpha_{\vec \CP}$ on $Y[\wh{\vec A}]$, i.e.
\be
\wt{\CO}^\alpha_{\vec \CP}: Y[\wh{\vec A}] \mapsto Y'[\wh{\vec B}],
\ee
such that the pair of theories $(X'[\wh{\vec B}],Y'[\wh{\vec B}])$ are again IR dual. The four theories $X[\wh{\vec A}], Y[\wh{\vec A}]$,
$X'[\wh{\vec B}]$, and $Y'[\wh{\vec B}]$, are therefore related as shown in the following figure.

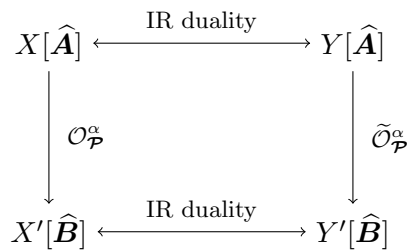
\begin{figure}[htbp]
\begin{center}
\begin{tikzpicture}
  \node (X) at (2,1.5) {$X[\wh{\vec A}]$};
  \node (Y) at (6,1.5) {$Y[\wh{\vec A}]$};
  \node (D) at (2,-1) {$X'[\wh{\vec B}]$};
  \node (E) at (6,-1) {$Y'[\wh{\vec B}]$};
   \draw[->] (X) -- (D) node [midway,  right=+3pt] {\footnotesize $\CO^\alpha_{\vec \CP}$};
  \draw[<->] (X) -- (Y)node [midway,above  ] {\footnotesize IR duality};
    \draw[->] (Y) -- (E) node [midway,  right=+3pt] {\footnotesize $\wt{\CO}^\alpha_{\vec \CP}$};
        \draw[<->] (D) -- (E)node  [midway,above  ] {\footnotesize IR duality};
\end{tikzpicture}
\end{center}
\caption{\footnotesize{Generating new dual pairs using an elementary $S$-type operation.} }
\label{S-OP-duality}
\end{figure}
The final step in the construction is to determine the dual operation $\wt{\CO}^\alpha_{\vec \CP}$, given the pair $(X[\wh{\vec A}],Y[\wh{\vec A}])$ 
and the map $\CO^\alpha_{\vec \CP}$, which should also allow one to read off the weakly coupled description of the theory $Y'[\wh{\vec B}]$, 
if it exists. In this paper, we will focus on 3d $\CN=4$ theories, and a specific IR duality -- the three dimensional mirror symmetry \cite{Intriligator:1996ex,deBoer:1996ck}. For such theories and the particular IR duality in question, one can solve the problem of determining the dual operation 
$\wt{\CO}^\alpha_{\vec \CP}$ explicitly, making use of RG-invariant observables computed using localization. The procedure, introduced in \cite{Dey:2020hfe}, is reviewed in \Secref{SOps-def}.

Given a recipe for finding the dual of the $S$-type operation, there exists a straightforward strategy for generating new dualities. 
For a given class of IR dualities, one can define a convenient subset of dual theories, which are well understood from the String Theory
and/or the QFT perspective, and refer to it as the set of ``basic dualities" for the given IR duality. Starting with a pair $(X,Y)$ in this set of 
basic dualities, one can then implement the prescription of \figref{S-OP-duality} sequentially to generate new dual pairs. For the specific case of 3d mirror symmetry, we pick the set of basic dualities to be the set of linear 
quiver (A-type) gauge theories with unitary gauge groups, obeying the additional constraint that they should be good theories in the 
Gaiotto-Witten sense \cite{Gaiotto:2008ak}. Mirror symmetry for this class of theories is well understood from a Type IIB brane construction 
\cite{Hanany:1996ie} as well as QFT considerations. Using the construction outlined above, one can then try to construct mirror pairs 
involving quivers of arbitrary shapes. Several infinite families of non-ADE type quiver gauge theories and their Lagrangian mirror duals
were constructed in this fashion \cite{Dey:2020hfe}.\\

In the present paper, we will extend the construction, summarized above in \figref{S-OP-duality}, to incorporate half-BPS 
vortex and Wilson defects in 3d $\CN=4$ theories. Consider a pair of dual theories decorated by line defects -- 
$X[\CD, \vec{\wh{A}}]$ and $Y[\CD^\vee, \wh{\vec A}]$, where $\CD$ and $\CD^\vee$ are line defects of different types 
(vortex and Wilson respectively or vice-versa) that are mapped to each other under mirror symmetry. 
We additionally require that $(X,Y)$ are good Lagrangian theories, with $X$ being in class $\CU$. 
Given a theory $X[\CD, \vec{\wh{A}}]$, an elementary $S$-type operation acts on $X$ to give a generically new theory $X'$ 
decorated by a defect $\CD'$, i.e.
\be \label{Smap-def}
\CO^\alpha_{\vec \CP}: X[\CD, \vec{\wh{A}}] \mapsto X'[\CD', \vec{\wh{B}}],
\ee
where $\CD$ and $\CD'$ are line defects of the same type (Wilson or vortex). 
As before, given the dual pair and the $S$-operation, one can define a dual operation, i.e.
\be
\wt{\CO}^\alpha_{\vec \CP}: Y[\CD^\vee, \wh{\vec A}] \mapsto Y'[\CD'^\vee, \wh{\vec B}],
\ee
such that the theories $X'[\CD', \vec{\wh{B}}], Y'[\CD'^\vee, \wh{\vec B}]$ are IR dual. This implies that 
the half-BPS defect $\CD'$ is mapped to the half-BPS defect $\CD'^\vee$ under mirror symmetry, and this 
relation between the two is referred to as the ``mirror map" of the line defects in question. 
Schematically, the construction can be summarized in \figref{S-OP-duality-def}. 
\begin{figure}[htbp]
\begin{center}
\begin{tikzpicture}
  \node (X) at (2,1.5) {$X[\CD, \wh{\vec A}]$};
  \node (Y) at (6,1.5) {$Y[\CD^\vee, \wh{\vec A}]$};
  \node (D) at (2,-1) {$X'[\CD',\wh{\vec B}]$};
  \node (E) at (6,-1) {$Y'[\CD'^\vee,\wh{\vec B}]$};
   \draw[->] (X) -- (D) node [midway,  right=+3pt] {\footnotesize $\CO^\alpha_{\vec \CP}$};
  \draw[<->] (X) -- (Y)node [midway,above  ] {\footnotesize IR duality};
    \draw[->] (Y) -- (E) node [midway,  right=+3pt] {\footnotesize $\wt{\CO}^\alpha_{\vec \CP}$};
        \draw[<->] (D) -- (E)node  [midway,above  ] {\footnotesize IR duality};
\end{tikzpicture}
\end{center}
\caption{\footnotesize{Generating new dual pairs with defects using an elementary $S$-type operation.} }
\label{S-OP-duality-def}
\end{figure}
The final step, as before, is to determine the dual operation $\wt{\CO}^\alpha_{\CP}$ given the pair $(X[\CD,\wh{\vec A}],Y[\CD^\vee, \wh{\vec A}])$ 
and the map $\CO^\alpha_{\vec \CP}$, which should allow one to read off the weakly coupled description of the theory $Y'$ (if it exists) as well as the 
defect $\CD'^\vee$. Note that a recipe for finding the dual operation therefore automatically leads to the mirror map for the line defects $\CD'$ and 
$\CD'^\vee$. One of the main results of this paper is to present a general recipe for finding the dual operation, using RG-invariant observables computed via localization. Given this recipe, one can again deploy the strategy outlined above to generate new dual pairs of 
quiver gauge theories of arbitrary shapes decorated by line defects, starting from a pair of $A$-type quivers with defects. 
After working out the details of this general construction, we use it to study the mirror maps of line defects in flavored 
$D$-type and affine $D$-type quiver gauge theories with unitary gauge groups. We also discuss how some of these line 
defects and the associated mirror maps can be realized in a Type IIB construction of the Hanany-Witten type, although the main 
focus of the paper is to work out these mirror maps without any reference to the String Theory constructions. We also study an example of mirror symmetry which does not have a known Type IIB realization. In a companion paper \cite{Dey:2020xyz}, we will apply our construction to study mirror maps of line defects in theories involving non-$ADE$ quiver gauge theories.\\

We would like to emphasize that finding the mirror map of line defects is in general a difficult problem for non-Abelian gauge theories. 
For $A$-type linear quivers with unitary gauge groups, the problem was solved relatively recently using a Type IIB brane construction 
\cite{Assel:2015oxa}, while line defects in $D$-type and $E$-type theories have not been addressed in this fashion in the literature, 
as far as we know. Our construction gives a systematic field theoretic procedure to generate line defects and the associated mirror 
maps in these quiver gauge theories and beyond, starting from the well understood dual linear quivers with defects.

\subsection{Outline and Summary}
The outline of the paper is as follows. In \Secref{Review}, we present a brief review of Wilson and vortex defects in 3d $\CN=4$ theories, 
and set up the notations for the rest of the paper. 
We review the Type IIB realization of these defects in an $A$-type quiver gauge theory, and the map of such defects under 
mirror symmetry. In addition, we discuss the localization computation for the expectation values of these defects on a round three-sphere,
which will be our principal tool.\\

In \Secref{SOps-def}, we concretely realize the construction, summarized in \figref{S-OP-duality-def}, in terms of the partition function on a 
round three-sphere. In particular, we discuss how to construct vortex defects as 3d-1d coupled quivers in a 
non- A-type 3d quiver gauge theory and their mirror maps. 
Analogous construction for the Wilson defects is also discussed.
In \Secref{Rev-SOps}, we review the $S$-type operations and their realization in terms of the $S^3$ partition function. 
Generically, a non-Abelian $S$-type operation (i.e. one that involves gauging a non-Abelian flavor symmetry) has a 
substantially more involved than the Abelian ones. However, at the level of the $S^3$ partition function, 
a non-Abelian $S$-type operation can be reduced to a set of Abelian $S$-type operations with 
certain Wilson defects. This ``abelianization'' procedure\footnote{This procedure should be understood simply as a convenient 
way of writing the matrix model, and not as a QFT operation.}, 
which is an important tool for constructing generic quiver gauge theories, is discussed 
in \Secref{NonAb}. In \Secref{SOps-defects}, we discuss the action of $S$-type operations on 3d quivers with defects, and therefore 
realize the map \eref{Smap-def} explicitly. 
The recipe for reading off the dual of the $S$-type operations for both cases is discussed in \Secref{SOps-defects-d}. Finally, we end 
the section with a simple example illustrating the general construction of \figref{S-OP-duality-def}, as given in the sections \ref{SOps-defects} -
\ref{SOps-defects-d}.\\

In \Secref{1-D}, we apply the machinery of \Secref{SOps-def} to engineer line defects in $D$-type/affine $D$-type quivers and find their duals, starting 
from a dual pair of linear quivers with defects. In particular, gauge vortex defects in $D$-type theories are realized as 3d-1d coupled quivers. 
Under mirror symmetry, these are shown to map to gauge/flavor Wilson defects, which we explicitly determine. Analogous construction for Wilson defects in 
$D$-type theories are also presented. The vortex defects dual to Wilson defects in $D$-type theories are realized by 3d-1d coupled systems  
involving symplectic and/or unitary 3d gauge groups.
A given vortex defect can be realized as multiple 3d-1d coupled quivers -- a feature that was called 
``hopping duality" in \cite{Assel:2015oxa}. We discuss the pattern of hopping dualities for vortex defects in the $D$-type and the affine 
$D$-type quiver gauge theories we study, as well as their mirror duals.

In \Secref{SwD-D4}, we begin by discussing an example of a $D_4$ quiver gauge theory dual to an $SU(2)$ gauge theory with four flavors. In \Secref{1-D-Sp(2)}, we extend our analysis to a $D_{N_f}$ quiver dual to an $Sp(N_c)$ theory with $N_f$ flavors. The former pair can be constructed by an Abelian $S$-type operation, while the latter pair requires a non-Abelian operation. In \Secref{SwD-F} and \Secref{SwD-Dhat}, we extend our discussion to more general $D$-type quivers and to affine $D$-type quivers respectively, using more involved $S$-type operations. In \Secref{SwD-D}, we extend our discussion to more general defects in a $D$-type quiver. 
The computation of the defect partition function is straightforward (although tedious in some cases) once the general expressions of  \ref{SOps-defects} -
\ref{SOps-defects-d} are given. In \Appref{app:SU(2)}, we present a sample case to familiarize the reader with the details of the computation for 
Abelian $S$-type operations. In \Appref{app: NAG-Ex}, we work out the details of the non-Abelian operation relevant for the example in \Secref{1-D-Sp(2)}. \\

In \Secref{LQ-TypeIIB}, we take a short detour to discuss how an important class of line defects in the $D$-type quivers and the associated mirror maps may be 
realized by a Type IIB brane construction. A vortex (Wilson) defect in a $D$-type quiver is realized by a D3-D5-NS5-D1(F1) system in the presence of an orbifold 5-plane. The dual Wilson (vortex) defect is realized by a D3-D5-NS5-F1(D1) system in the presence of an O5$^0$-plane, which is 
D5-brane coincident with an O5$^{-}$-plane. The S-duality that relates the two configurations is explicitly discussed. We observe that the 
Type IIB answer in this case matches the QFT answer precisely. A detailed study of the Type IIB brane construction for more general line defects in 
$D$-type quiver gauge theories will be the subject of a future paper.\\

Finally, in \Secref{3d-bad}, we discuss defects in an affine $D_4$ quiver, for which the mirror dual is not directly realized by a Type IIB construction.
The naive mirror dual, read off from S-dualizing the original brane configuration, gives a theory which is ``bad" in the Gaiotto-Witten sense 
\cite{Gaiotto:2008ak}. This implies, among other things, that the $S^3$ partition function of the mirror theory diverges. 
The procedure of \figref{S-OP-duality}, however, allows one to construct a good mirror dual for the $\wh{D}_4$ theory. 
In addition, it turns out that, for the aforementioned affine $D_4$ quiver decorated with a line defect, the procedure of \figref{S-OP-duality-def} 
allows one to construct the mirror map of such a line defect. These mirror maps can be worked out explicitly following the construction of 
sections \ref{SOps-defects} - \ref{SOps-defects-d}.

\section{Wilson loop and Vortex loop operators in three dimensions}\label{Review}
In this section, we review a few basic concepts and tools associated with half-BPS Wilson and vortex 
defects in 3d $\CN=4$ quiver gauge theories, from a field theory perspective as well as using a Type IIB 
String Theory construction. Readers familiar with these basic materials can skip this section.

\subsection{Generalities of 3d $\CN=4$ QFTs and half-BPS defects}
\subsubsection{Quiver notation for 3d $\CN=4$ QFTs} 
The 3d $\CN=4$ QFTs have dimensionful coupling constants - such theories 
are asymptotically free in the UV and flow to a 3d $\CN=4$ SCFT in the IR. The theories relevant to this paper are quiver gauge theories, 
the Lagrangian description for which involves a vector multiplet corresponding to a gauge group $G$, 
and hypermultiplets in a quaternionic representation of the gauge group $G$. The UV theory has an $SU(2)_H \times SU(2)_C$ 
R-symmetry, where the subscripts indicate that they act as isometries on the Higgs and the Coulomb branch of the moduli space 
respectively. Both branches are \hk cones and the corresponding $SU(2)$ R-symmetry rotates the three complex structure as a triplet. In addition, 
the theory has a global symmetry $G_H \times G_C$, which commutes with the R-symmetry and the Poincare supercharges. These 
are realized as tri-holomorphic isometries on the Higgs and Coulomb branches of the moduli space respectively.\\

The 3d $\CN=4$ gauge theories have two possible types of deformations which preserve the full $\CN=4$ supersymmetry -- 
hypermultiplet masses and FI parameters. The masses transform as triplets of $SU(2)_C$ and constitute the scalar components 
of a background vector multiplet in the Cartan subalgebra $\frg_H$. Generically, these parameters lift the Higgs branch and 
deform the Coulomb branch. 
The FI parameters transform as triplets of $SU(2)_H$ and constitute 
the scalar components of a background twisted vector multiplet in the Cartan subalgebra $\frg_C$. Generically, they will lift the 
Coulomb branch and deform the Higgs branch.\\

Finally, the quiver notation of a gauge theory that we will use in this paper is summarized in the following figure.
 
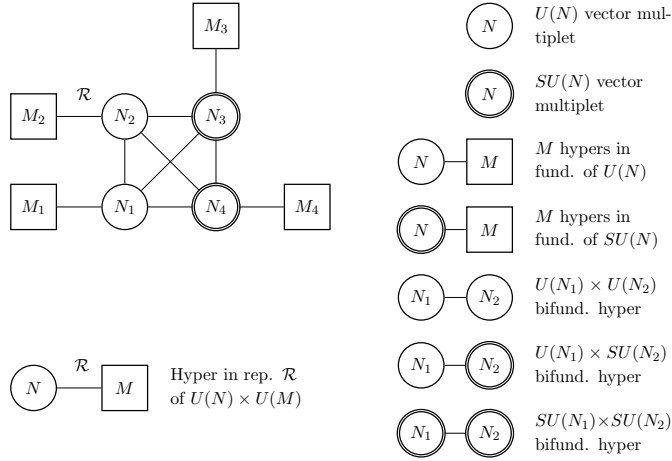
\begin{center}
\scalebox{0.6}{\begin{tikzpicture}[node distance=2cm,
cnode/.style={circle,draw,thick, minimum size=1.0cm},snode/.style={rectangle,draw,thick,minimum size=1cm}, pnode/.style={circle,double,draw,thick, minimum size=1.0cm}]
\node[cnode] (1) at (-2,0) {$N_1$};
\node[cnode] (2) at (-2,2) {$N_2$};
\node[pnode] (3) at (0,2) {$N_3$};
\node[pnode] (4) at (0,0) {$N_4$};
\node[snode] (5) at (-4,0) {$M_1$};
\node[snode] (6) at (-4,2) {$M_2$};
\node[snode] (7) at (0,4) {$M_3$};
\node[snode] (8) at (2,0) {$M_4$};
\draw[-] (1) -- (2);
\draw[-] (2) -- (3);
\draw[-] (3) -- (4);
\draw[-] (1) -- (4);
\draw[-] (1) -- (5);
\draw[-] (2) -- (6);
\draw[-] (3) -- (7);
\draw[-] (4) -- (8);
\draw[-] (2) -- (4);
\draw[-] (1) -- (3);
\node[cnode] (9) at (6,4) {$N$};
\node[text width=3cm](10) at (8.5, 4){$U(N)$ vector multiplet};
\node[pnode] (11) at (6,2.5) {$N$};
\node[text width=3cm](12) at (8.5, 2.5){$SU(N)$ vector multiplet};
\node[snode] (13) at (6,1) {$M$};
\node[cnode] (14) at (4.5,1) {$N$};
\draw[-] (13)--(14);
\node[text width=3cm](15) at (8.5, 1){$M$ hypers in fund. of $U(N)$};
\node[snode] (16) at (6, -0.5) {$M$};
\node[pnode] (17) at (4.5, -0.5) {$N$};
\draw[-] (16)--(17);
\node[text width=3cm](18) at (8.5, - 0.5){$M$ hypers in fund. of $SU(N)$};
\node[cnode] (19) at (6, -2) {$N_2$};
\node[cnode] (20) at (4.5, -2) {$N_1$};
\draw[-] (19)--(20);
\node[text width=3cm](21) at (8.5, -2){$U(N_1) \times U(N_2)$ bifund. hyper};
\node[pnode] (22) at (6, -3.5) {$N_2$};
\node[cnode] (23) at (4.5, -3.5) {$N_1$};
\draw[-] (22)--(23);
\node[text width=3cm](24) at (8.5, -3.5){$U(N_1) \times SU(N_2)$ bifund. hyper};
\node[pnode] (25) at (6, -5) {$N_2$};
\node[pnode] (26) at (4.5, -5) {$N_1$};
\draw[-] (25)--(26);
\node[text width=3cm](27) at (8.5, -5){$SU(N_1) \times SU(N_2)$ bifund. hyper};
\node[cnode] (31) at (-4,-4) {$N$};
\node[snode] (32) at (-2,-4) {$M$};
\draw[-] (31)--(32);
\node[text width=0.2cm](33) at (-3, -3.5){$\CR$};
\node[text width=3cm](34) at (0.5, -4){Hyper in rep. $\CR$ of $U(N) \times U(M)$};
\node[text width=0.1cm](34) at (-3,2.5){$\CR$};
\end{tikzpicture}}
\end{center}

The quiver diagram on the LHS represents the field content of a 3d $\CN=4$ theory with gauge group $G=U(N_1) \times U(N_2) \times SU(N_3) \times SU(N_4)$, 
and hypermultiplets in various representations. The conventions for the reading off the representations in which the hypermultiplets transform 
are listed on the RHS. In a quiver diagram, we will refer to the circles as gauge nodes and the 
boxes as flavor nodes. For more general gauge groups, we will use the notation:
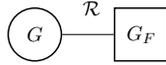
\begin{center}
\scalebox{0.7}{\begin{tikzpicture}[node distance=2cm,
cnode/.style={circle,draw,thick, minimum size=1.0cm},snode/.style={rectangle,draw,thick,minimum size=1cm}, pnode/.style={circle,double,draw,thick, minimum size=1.0cm}]
\node[cnode] (31) at (-4,-4) {$G$};
\node[snode] (32) at (-2,-4) {$G_F$};
\draw[-] (31)--(32);
\node[text width=0.2cm](33) at (-3, -3.5){$\CR$};
\end{tikzpicture}}
\end{center}
where $\CR$ is a representation of $G \times G_F$. 
For example, an $Sp(N)$ gauge theory with $N_f$ fundamental hypermultiplets (i.e. $2N_f$ half-hypers) will be represented by the above quiver diagram with $G=Sp(N)$, $G_F=SO(2N_f)$, and $\CR$ being the bifundamental representation of $Sp(N) \times SO(2N_f)$. The line connecting the gauge and the flavor node denotes a half-hypermultiplet in this case.\\

\subsubsection{Supersymmetry algebra and protected sectors}
 
The 3d $\CN=4$ supersymmetry algebra has a pair of inequivalent protected 1d subalgebras. 
Following the presentation of \cite{Dimofte:2019zzj}, we discuss these subalgebras on $\BR^3$.

The 3d $\CN=4$ Poincare supercharges $Q_{\alpha}^{ aa'}$ are doublets of $Spin(3)\cong SU(2)_E$ (indexed by $\alpha=1,2$) and 
transform as $(2,2)$ under the R-symmetry group, $SU(2)_H \times SU(2)_C$ (indexed by $a=1,2$ and $a'=1,2$ respectively). The complex supercharges 
on $\BR^{3}$ generate the following supersymmetry algebra:
\be
\{ Q_{\alpha}^{ a a'}, Q_{\beta}^{b b'}\}=   P_\mu \s^\mu_{\alpha \beta} \epsilon^{ab} \epsilon^{a'b'} -
 i \epsilon_{\alpha \beta} (\epsilon^{ab} m^{(a' b')} + \epsilon^{a'b'} t^{(ab)}),
\ee
where $(\s^\mu)^\alpha_\beta$ are the standard Pauli matrices, with $\mu=1,2,3$. All $SU(2)$ indices are raised and lowered by the corresponding $\epsilon$-tensor, 
with the convention $\epsilon^{12} = \epsilon_{21}=1$. The second term denotes a central extension of the algebra which are realized as the two sets of $\CN=4$ supersymmetry preserving deformations in the Lagrangian -- the triplet of masses and FI parameters, i.e.
\be
m^{(a'b')} =\begin{pmatrix} 2m_\BC & -m_\BR \\ - {m}_\BR & -2\overline{m}_\BC \end{pmatrix}, \qquad t^{(ab)} = \begin{pmatrix}  2t_\BC & -t_\BR \\ - {t}_\BR & -2\overline{t}_\BC  \end{pmatrix}.
\ee

We will split the Euclidean space-time $\BR^3 \cong \BC_z \times \BR_t$, and the SUSY algebra can then be rewritten as
\begin{align}
& \{ Q_{1}^{ aa'}, Q_{1}^{bb'}\}=-2 \epsilon^{ab} \epsilon^{a'b'} P_{\overline{z}}, \qquad \{ Q_{2}^{ aa'}, Q_{2}^{b b'}\}=2 \epsilon^{ab} \epsilon^{a' b'} P_{\overline{z}} ,\nn \\
& \{ Q_{1}^{ a a'}, Q_{2}^{b b'}\}= \{ Q_{2}^{ b b'}, Q_{1}^{a a'}\}=  \epsilon^{ab} \epsilon^{a'b'} P_t - i(\epsilon^{ab} m^{(a' b')} + \epsilon^{a'b'} t^{(ab)}).  
\end{align}
In this set-up, we will study half-BPS line operators inserted at $z=\bz=0$, preserving a 1d $\CN=4$ subalgebra (i.e. supercharges of the 
subalgebra anti-commute to translation along $\BR_t$) of the full 3d $\CN=4$ algebra. There are two inequivalent choices of these 
subalgebras, which will be referred to as ${\rm SQM}_A$ and ${\rm SQM}_B$. The associated line operators will be 
referred to as Type-$A$ and Type-$B$ line operators respectively.\\

The subalgebra ${\rm SQM}_A$ is defined as a 1d subalgebra which breaks the 
R-symmetry to $SU(2)_C \times U(1)_H$. There exists a $\BC \BP^1$ worth of such subalgebras corresponding to the choices of an unbroken 
$U(1)_H$ inside $SU(2)_H$, which in turn is related to choices of a complex structure on the Higgs branch. 
Following \cite{Dimofte:2019zzj}, we adopt the following choice for the ${\rm SQM}_A$ subalgebra:
\be
Q^{a'}_A= \delta^\alpha_a Q^{aa'}_\alpha, \qquad \wt{Q}^{a'}_A= (\s^3)^\alpha_a Q^{aa'}_\alpha,
\ee
where the non-vanishing anti-commutation relations are given as 
\be
\{ Q^{a'}_A, \wt{Q}^{b'}_A\}= 2\epsilon^{a'b'} (P_t - it_\BR), \qquad \{ Q^{a'}_A, {Q}^{b'}_A\}=\{\wt{Q}^{a'}_A, \wt{Q}^{b'}_A\}=2im^{(a'b')}.
\ee

The subalgebra ${\rm SQM}_B$ is defined as a 1d subalgebra which breaks the 
R-symmetry to $SU(2)_H \times U(1)_C$. Again, there exists a $\BC \BP^1$ worth of such subalgebras corresponding to the choices of an unbroken 
$U(1)_C$ inside $SU(2)_C$, which in turn is related to choices of a complex structure on the Coulomb branch. 
We adopt the following choice for the ${\rm SQM}_B$ subalgebra:
\be
Q^{a}_B= \delta^\alpha_{a'} Q^{aa'}_\alpha, \qquad \wt{Q}^{a}_B= (\s^3)^\alpha_{a'} Q^{aa'}_\alpha, 
\ee
with the non-vanishing anti-commutation relations
\be
\{ Q^{a}_B, \wt{Q}^{b}_B\}= 2\epsilon^{ab} (P_t - it_\BR), \qquad \{ Q^{a}_A, {Q}^{b}_A\}=\{\wt{Q}^{a}_A, \wt{Q}^{b}_A\}=2im^{(ab)}.
\ee
Finally, each subalgebra contains a topological supercharge:
\be
Q_A = Q^{1'}_A, \qquad Q_B = Q^{1}_B,
\ee
where $Q_A$ can be identified as the 3d reduction of the supercharge associated with Donaldson-Witten twisted 4d $\CN=2$ theories, 
and $Q_B$ is associated with the Rozansky-Witten twist of 3d $\CN=4$ sigma models. 

\subsubsection{Wilson defect on $\BR^3$}
A Wilson defect is a half-BPS Type-$B$ operator which can be defined on $\BR^3$ as follows.
The bosonic part of a 3d $\CN=4$ vector multiplet contains a gauge field $A_\mu$ and an $SU(2)_C$ triplet of real scalars in the adjoint representation of 
the gauge group $G$. Given a choice of complex structure on the Coulomb branch, the latter can be written as :
\be
\phi^{(a'b')} =\begin{pmatrix} 2\phi & \s \\ \s & -2\overline{\phi} \end{pmatrix},
\ee
where the real scalar $\s \in \frg$ and the complex scalar $\phi \in \frg_{\BC}$. The following combination of the gauge field and the real scalar $\s$:
\be
\CA_t := A_t - i\s
\ee
is preserved by the 1d subalgebra ${\rm SQM}_B$. A Wilson line in a finite-dimensional unitary representation $R$ of the gauge group $G$ can then be given as
\be \label{susywilson}
W_R={\rm Hol}\Big(\rho_R(\CA_t)\Big)= P\Big(\exp{\int_{\BR_t} \rho_R(\CA_t )\, dt}\Big)\,,
\ee
where $\rho_R$ is the map $\rho_R: \frg_\BC \to \frgl({\rm dim} R)$. As pointed out in \cite{Dimofte:2019zzj}, one can take the trace of the 
holonomy operator to give a gauge-invariant operator, if the non-compact line were replaced by a loop. For a non-compact line, gauge invariance 
requires appropriate boundary conditions at $t \to \pm \infty$. This, however, does not affect the local structure of these line operators studied in 
\cite{Dimofte:2019zzj} or in this work.

\subsubsection{Vortex  defect on $\BR^3$}
A vortex defect is a half-BPS operator Type-$A$ operator which can be defined on $\BR^3$ using any one of the 
two following equivalent constructions, i.e. 
\begin{enumerate}
\item Disorder operator
\item Coupled 3d-1d quiver.
\end{enumerate}
We review the two approaches briefly below, focusing mainly on the second approach, which will 
concern us for the rest of this work.\\

\noindent \textbf{Disorder operator:} Consider a 3d $\CN=4$ theory with a gauge group $G$ and $N$ hypermultiplets, 
where the complex hypermultiplet scalars are labelled as $(X_i, Y_i)$ with $i=1, \ldots, N$. The $N$ complex  
scalars $(X_i, Y_i)$ arrange themselves in a representation $\CR \oplus \CR^*$ of $G \times G_{H}$. 
Insertion of a vortex defect in the theory amounts to performing the path integral in a background of certain singular 
solutions of the BPS equations\footnote{The generalized vortex equations in the $\BC_z$-plane, i.e.
\begin{align}
F_{z \bz} =\mu_{\BR} + t_\BR,\qquad D_{\bz} X=D_{\bz} Y =0, \qquad \mu_{\BC}=-t_{\BC},
\end{align}
are obtained as a subset of the BPS equations for the subalgebra ${\rm SQM}_A$. We refer the reader to Section 4 and Appendix A of the paper \cite{Dimofte:2019zzj} (and the references therein) for the complete set of BPS equations.}. 
The solutions, in question, are characterized by a singular profile of the hypermultiplet scalars 
$(X,Y)$ at $z=0$ on the $\BC_z$-plane, as well as a compatible gauge symmetry breaking at $z=0$. A given  
disorder operator can be labelled by certain holomorphic/algebraic data -- we refer the reader to \cite{Dimofte:2019zzj} for details.\\

\noindent \textbf{Coupled 3d-1d system: } Alternatively, a vortex defect can be inserted by coupling the 3d theory to an appropriate 
1d $\CN=(2,2)$ quiver gauge theory \cite{Assel:2015oxa}. Such theories can be obtained by the dimensional reduction of a 2d 
$\CN=(2,2)$ SQM. An $\CN=(2,2)$ SQM associated with vortex defects in 3d $\CN=4$ theories
has the generic form:
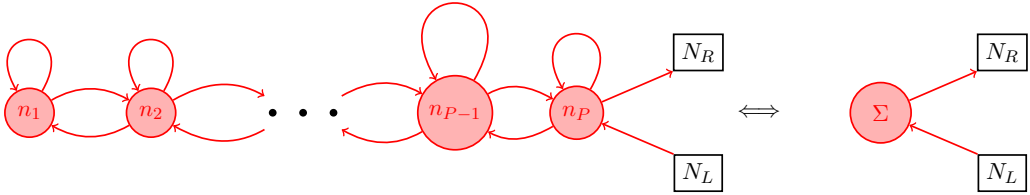
\begin{figure}[htbp]
\begin{center}
\scalebox{0.8}{\begin{tikzpicture}[
nnode/.style={circle,draw,thick, fill=blue,minimum size= 6mm},cnode/.style={circle,draw,thick,minimum size=4mm},snode/.style={rectangle,draw,thick,minimum size=6mm},rnode/.style={red, circle,draw,thick,fill=red!30 ,minimum size=4mm},rrnode/.style={red, circle,draw,thick,fill=red!30 ,minimum size=1.0cm}]
\node[rnode] (1) at (-4,0) {$n_1$};
\node[rnode] (2) at (-2,0) {$n_2$};
\node[] (3) at (0,0.2){};
\node[] (4) at (0,-0.2){};
\node[circle,draw,thick, fill, inner sep=1 pt] (5) at (0,0){} ;
\node[circle,draw,thick, fill, inner sep=1 pt] (6) at (0.5,0){} ;
\node[circle,draw,thick, fill, inner sep=1 pt] (7) at (1,0){} ;
\node[] (8) at (1,0.2){};
\node[] (9) at (1,-0.2){};
\node[rnode] (10) at (3,0) {$n_{P-1}$};
\node[rnode] (11) at (5,0) {$n_{P}$};
\node[snode] (12) at (7,1) {$N_R$};
\node[snode] (13) at (7,-1) {$N_L$};
\draw[red, thick, ->] (1) to [out=30,in=150] (2);
\draw[red, thick, ->] (2) to [out=210,in=330] (1);
\draw[red, thick, ->] (2) to [out=30,in=150] (3);
\draw[red, thick, ->] (4) to [out=210,in=330] (2);
\draw[red, thick, ->] (8) to [out=30,in=150] (10);
\draw[red, thick, ->] (10) to [out=210,in=330] (9);
\draw[red, thick, ->] (10) to [out=30,in=150] (11);
\draw[red, thick, ->] (11) to [out=210,in=330] (10);
\draw[red, thick, ->] (11) to (12.south west);
\draw[red, thick, ->] (13.north west) to (11);
\draw[red, thick, ->] (1) to [out=60,in=120,looseness=8] (1);
\draw[red, thick, ->] (2) to [out=60,in=120,looseness=8] (2);
\draw[red, thick, ->] (10) to [out=60,in=120,looseness=8] (10);
\draw[red, thick, ->] (11) to [out=60,in=120,looseness=8] (11);
\node[] (17) at (8,0) {$\Longleftrightarrow$};
\node[rrnode] (14) at (10,0) {$\Sigma$};
\node[snode] (15) at (12,1) {$N_R$};
\node[snode] (16) at (12,-1) {$N_L$};
\draw[red, thick, ->] (14) to (15.south west);
\draw[red, thick, ->] (16.north west) to (14);
\end{tikzpicture}}
\caption{\footnotesize{The general form of (2,2) SQM which realizes a vortex defect in a 3d $\CN=4$ theory.}}
\label{1dquiv-gen}
\end{center}
\end{figure}

The circular nodes in the 1d quiver (on the left) represent $(2,2)$ vector multiplets for unitary gauge groups, so that 
the full gauge group is $\prod^P_{i=1} U(n_i)$. The directed arrows connecting two circular nodes 
denote bifundamental chiral multiplets, with the incoming arrow denoting a fundamental chiral for the corresponding 
gauge node. The directed arrows beginning and ending on the same gauge node represent adjoint chiral multiplets for 
the corresponding gauge node. 
Finally, the rectangular nodes denote fundamental chiral/anti-chiral multiplets, with the incoming arrow 
(w.r.t. the circular node) denoting fundamental chirals, and the outgoing arrow denoting anti-chirals. 
On the right, we have a shorthand notation for presenting a generic quiver of this class.\\

\begin{figure}[htbp]
\begin{center}
\begin{tabular}{ccc}
\scalebox{0.8}{\begin{tikzpicture}[node distance=2cm, nnode/.style={circle,draw,thick, fill, inner sep=1 pt},cnode/.style={circle,draw,thick,minimum size=1.0 cm},snode/.style={rectangle,draw,thick,minimum size=1.0 cm},rnode/.style={red, circle,draw,thick,fill=red!30 ,minimum size=1.0cm}]
\node[nnode] (1) at (0,0){} ;
\node[nnode] (2) at (0.5,0){} ;
\node[nnode] (3) at (1,0){} ;
\node[](4) at (1.5,0){};
\node[cf-group] (5) at (3,0) {\rotatebox{-90}{$N_L$}};
\node[cf-group] (6) at (5,0) {\rotatebox{-90}{$N_R$}};
\node[rnode] (7) at (4,2) {$\Sigma$};
\draw[-] (4) -- (5);
\draw[thick, -] (5) to (6);
\draw[red, thick,->] (5) -- (7);
\draw[red, thick,->] (7) -- (6);
\node[] (9) at (6.5,0){} ;
\node[nnode] (10) at (7,0){} ;
\node[nnode] (11) at (7.5,0){} ;
\node[nnode](12) at (8,0){};
\draw[-] (6) -- (9);
\node[text width=0.2cm] (13) at (3,1){$q^a_i$};
\node[text width=0.2cm] (14) at (5,1){$\tq^j_a$};
\node[text width=0.2cm] (16) at (4,-0.75){};
\end{tikzpicture}}
&{\begin{tikzpicture}\node[] (17) at (0,0) {$\Longleftrightarrow$}; 
\node[] (18) at (0,-0.6) {};
\end{tikzpicture}}
&\scalebox{0.8}{\begin{tikzpicture}[node distance=2cm, nnode/.style={circle,draw,thick, fill, inner sep=1 pt},cnode/.style={circle,draw,thick,minimum size=1.0 cm},snode/.style={rectangle,draw,thick,minimum size=1.0 cm},rnode/.style={red, circle,draw,thick,fill=red!30 ,minimum size=1.0cm}]
\node[nnode] (1) at (0,0){} ;
\node[nnode] (2) at (0.5,0){} ;
\node[nnode] (3) at (1,0){} ;
\node[](4) at (1.5,0){};
\node[cf-group] (5) at (3,0) {\rotatebox{-90}{$N_L$}};
\node[cf-group] (6) at (5,0) {\rotatebox{-90}{$N_R$}};
\node[rnode] (7) at (4,2) {$\Sigma$};
\draw[-] (4) -- (5);
\draw[thick, ->] (5) to (6);
\draw[thick, ->] (4.5,0.25) to (3.5, 0.25);
\draw[red, thick,->] (5) -- (7);
\draw[red, thick,->] (7) -- (6);
\node[] (9) at (6.5,0){} ;
\node[nnode] (10) at (7,0){} ;
\node[nnode] (11) at (7.5,0){} ;
\node[nnode](12) at (8,0){};
\draw[-] (6) -- (9);
\node[text width=0.2cm] (13) at (3,1){$q^a_i$};
\node[text width=0.2cm] (14) at (5,1){$\tq^j_a$};
\node[text width=0.2cm] (15) at (4,0.5){$X^i_j$};
\node[text width=0.2cm] (16) at (4,-0.5){$Y^j_i$};
\end{tikzpicture}}
\end{tabular}
\caption{\footnotesize{The general form of a 3d-1d quiver that realizes a vortex defect in a 3d $\CN=4$ quiver gauge theory.}}
\label{3d1dquiv-gen}
\end{center}
\end{figure}
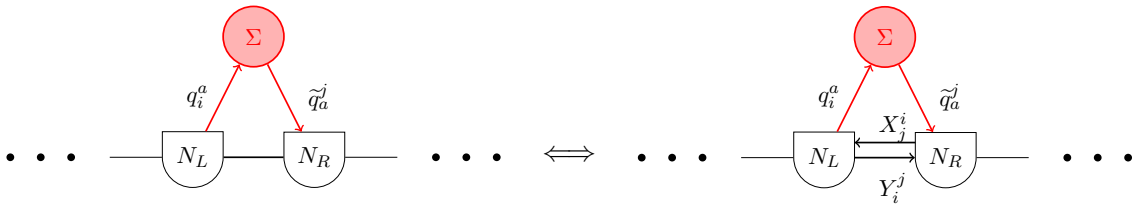

A generic 3d-1d quiver which realizes a vortex defect in 3d has the form given in \figref{3d1dquiv-gen}.
The 1d flavor symmetries (or their subgroups) which are gauged by 3d vector multiplets are represented by the nodes 
\scalebox{0.5}{\begin{tikzpicture} \node[cf-group] (5) at (3,0) {}; \end{tikzpicture}}. In the above quiver, the $(2,2)$ SQM 
is coupled to the 3d theory by gauging the $(U(N_L) \times U(N_R))/U(1)$ flavor symmetry with 3d vector multiplets. In general, 
one can also gauge a subgroup of the flavor symmetry with background 3d vector multiplets, such that the a 1d flavor 
symmetry subgroup is identified with a 3d flavor symmetry group. In the quiver on the RHS, we have decomposed the 
3d $\CN=4$ bifundamental hypermultiplet into a pair of $\CN=2$ chiral multiplets -- $X$ and $Y$, which transform in the 
bifundamental representation of $U(N_L) \times U(N_R)$ and $U(N_R) \times U(N_L)$ respectively\footnote{In general, one can 
have a hypermultiplet in a representation $\CR$ of $U(N_L) \times U(N_R)$, and the complex scalars $X$ and $Y$ will 
transform accordingly.}.
The 3d-1d coupling introduces the following cubic superpotential in the theory:
\be \label{3d-1dSup}
\wt{W}_0 = q^a_i\,X^i_j\, \tq^j_a + \ldots,
\ee
where the $\ldots$ in the superpotential denote additional terms which contain higher derivatives of the complex scalar $X$.
The indices run over $a=1,\ldots, n_P, \, i= 1,\ldots,N_L, \, j=1,\ldots,N_R$.\\

A vortex defect in a 3d $\CN=4$ quiver gauge theory is specified by the following two pieces of data:
\begin{enumerate}
\item A 3d-1d quiver of the form discussed above, including the cubic superpotential.
\item Signs of the FI parameters $\vec \xi$ for the gauge nodes of the SQM.
\end{enumerate}

A given vortex defect can be realized by multiple coupled 3d-1d systems. This was called the ``hopping duality" in the context of 3d linear quivers 
in \cite{Assel:2015oxa}, where a vortex defect generically can be realized by at least a pair of coupled 3d-1d systems.

\subsubsection{Mirror symmetry} 

Mirror symmetry is a special case of an IR duality in three dimensions for theories with eight supercharges, with the following properties:
\begin{itemize}

\item Given a pair of dual theories $X$ and $Y$, mirror symmetry exchanges the Coulomb and the Higgs branches in the deep IR, 
i.e. as $g^2_{YM} \to \infty$:
\be
\mathcal{M}^{(X)}_{\rm C}= \mathcal{M}^{(Y)}_{\rm H}, \qquad \mathcal{M}^{(X)}_{\rm H}= \mathcal{M}^{(Y)}_{\rm C}.
\ee

\item The duality exchanges $SU(2)_C$ and $SU(2)_H$, and therefore maps hypermultiplet masses on one side of the duality 
to FI parameters on the other. 

\end{itemize}
Mirror symmetry relates observables in theory $X$ with observables in theory $Y$, and the precise map is referred to as the 
``mirror map". In particular, it exchanges the subalgebra SQM$_A$ with the subalgebra SQM$_B$, as well as the associated 
topological charges $Q_A$ and $Q_B$. This implies that the half-BPS vortex defects, which are observables preserved by SQM$_A$ on 
one side of the duality, are mapped to the half-BPS Wilson defects, preserved by SQM$_B$, on the other. The vortex defect can be 
additionally embellished by Wilson defects for the 3d topological symmetries, which are also preserved by SQM$_A$. In the next section,
we will review a Type IIB construction for realizing these mirror maps in the class of linear quivers with unitary gauge groups. 
This will be the starting point for constructing mirror maps of line defects in more general quiver gauge theories, as we discuss in 
\Secref{SOps-def}.

\subsection{Brane construction for line defects in linear quivers}\label{LQ2B-main}
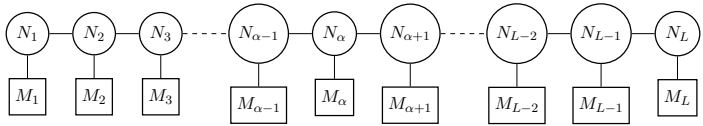
\begin{figure}[htbp]
\begin{center}
\scalebox{.6}{\begin{tikzpicture}[
cnode/.style={circle,draw,thick,minimum size=4mm},snode/.style={rectangle,draw,thick,minimum size=8mm},pnode/.style={rectangle,red,draw,thick,minimum size=1.0cm}]
\node[cnode] (1) {$N_1$};
\node[cnode] (2) [right=.5cm  of 1]{$N_2$};
\node[cnode] (3) [right=.5cm of 2]{$N_3$};
\node[cnode] (4) [right=1cm of 3]{$N_{\alpha-1}$};
\node[cnode] (5) [right=0.5cm of 4]{$N_{\alpha}$};
\node[cnode] (6) [right=0.5cm of 5]{$N_{\alpha + 1}$};
\node[cnode] (7) [right=1cm of 6]{{$N_{L-2}$}};
\node[cnode] (8) [right=0.5cm of 7]{$N_{L-1}$};
\node[cnode] (9) [right=0.5cm of 8]{$N_L$};
\node[snode] (10) [below=0.5cm of 1]{$M_1$};
\node[snode] (11) [below=0.5cm of 2]{$M_2$};
\node[snode] (12) [below=0.5cm of 3]{$M_3$};
\node[snode] (13) [below=0.5cm of 4]{$M_{\alpha-1}$};
\node[snode] (14) [below=0.5cm of 5]{$M_{\alpha}$};
\node[snode] (15) [below=0.5cm of 6]{$M_{\alpha+1}$};
\node[snode] (16) [below=0.5cm of 7]{$M_{L-2}$};
\node[snode] (17) [below=0.5cm of 8]{$M_{L-1}$};
\node[snode] (18) [below=0.5cm of 9]{$M_{L}$};
\draw[-] (1) -- (2);
\draw[-] (2)-- (3);
\draw[dashed] (3) -- (4);
\draw[-] (4) --(5);
\draw[-] (5) --(6);
\draw[dashed] (6) -- (7);
\draw[-] (7) -- (8);
\draw[-] (8) --(9);
\draw[-] (1) -- (10);
\draw[-] (2) -- (11);
\draw[-] (3) -- (12);
\draw[-] (4) -- (13);
\draw[-] (5) -- (14);
\draw[-] (6) -- (15);
\draw[-] (7) -- (16);
\draw[-] (8) -- (17);
\draw[-] (9) -- (18);
\end{tikzpicture}}
\caption{\footnotesize{A generic linear quiver with $L$ gauge nodes.}}
\label{fig: LQGen}
\end{center}
\end{figure}

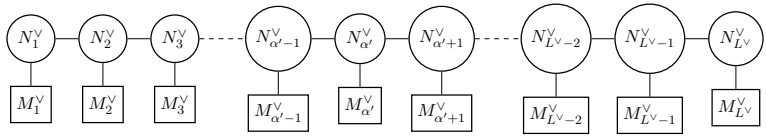
\begin{figure}[htbp]
\begin{center}
\scalebox{.6}{\begin{tikzpicture}[
cnode/.style={circle,draw,thick,minimum size=4mm},snode/.style={rectangle,draw,thick,minimum size=8mm},pnode/.style={rectangle,red,draw,thick,minimum size=1.0cm}]
\node[cnode] (1) {$N^{\vee}_1$};
\node[cnode] (2) [right=.5cm  of 1]{$N^{\vee}_2$};
\node[cnode] (3) [right=.5cm of 2]{$N^{\vee}_3$};
\node[cnode] (4) [right=1cm of 3]{$N^{\vee}_{\alpha'-1}$};
\node[cnode] (5) [right=0.5cm of 4]{$N^{\vee}_{\alpha'}$};
\node[cnode] (6) [right=0.5cm of 5]{$N^{\vee}_{\alpha' + 1}$};
\node[cnode] (7) [right=1cm of 6]{{$N^{\vee}_{L^{\vee}-2}$}};
\node[cnode] (8) [right=0.5cm of 7]{$N^{\vee}_{L^{\vee}-1}$};
\node[cnode] (9) [right=0.5cm of 8]{$N^{\vee}_{L^{\vee}}$};
\node[snode] (10) [below=0.5cm of 1]{$M^{\vee}_1$};
\node[snode] (11) [below=0.5cm of 2]{$M^{\vee}_2$};
\node[snode] (12) [below=0.5cm of 3]{$M^{\vee}_3$};
\node[snode] (13) [below=0.5cm of 4]{$M^{\vee}_{\alpha'-1}$};
\node[snode] (14) [below=0.5cm of 5]{$M^{\vee}_{\alpha'}$};
\node[snode] (15) [below=0.5cm of 6]{$M^{\vee}_{\alpha'+1}$};
\node[snode] (16) [below=0.5cm of 7]{$M^{\vee}_{L^{\vee}-2}$};
\node[snode] (17) [below=0.5cm of 8]{$M^{\vee}_{L^{\vee}-1}$};
\node[snode] (18) [below=0.5cm of 9]{$M^{\vee}_{L^{\vee}}$};
\draw[-] (1) -- (2);
\draw[-] (2)-- (3);
\draw[dashed] (3) -- (4);
\draw[-] (4) --(5);
\draw[-] (5) --(6);
\draw[dashed] (6) -- (7);
\draw[-] (7) -- (8);
\draw[-] (8) --(9);
\draw[-] (1) -- (10);
\draw[-] (2) -- (11);
\draw[-] (3) -- (12);
\draw[-] (4) -- (13);
\draw[-] (5) -- (14);
\draw[-] (6) -- (15);
\draw[-] (7) -- (16);
\draw[-] (8) -- (17);
\draw[-] (9) -- (18);
\end{tikzpicture}}
\caption{\footnotesize{The linear quiver which is mirror dual to the generic linear quiver in \figref{fig: LQGen}. The total number of gauge 
nodes is $L^{\vee}$.}}
\label{fig: LQGenDual}
\end{center}
\end{figure}

In this section, we present a brief review of the Type IIB construction of the vortex and the Wilson line defects in 
linear quivers with unitary gauge groups, and the map of such defects under mirror symmetry. 
A generic linear quiver gauge theory with $L$ gauge nodes is shown in \figref{fig: LQGen}, 
while its mirror dual (which is also a linear quiver gauge theory) is shown in \figref{fig: LQGenDual}. We first discuss the Type IIB construction of 
linear quivers without defects in \Secref{LQ2B}, followed by incorporation of line defects in \Secref{LQ2BD}, and discussion of the mirror map of 
line defects in \Secref{MM2BD}.

\subsubsection{Brane construction for linear quivers without defects}\label{LQ2B}

The linear quivers have a very simple realization in terms of a Type IIB brane construction of the Hanany-Witten type \cite{Hanany:1996ie}.
For a modern review and generalization of the construction, we refer the reader to \cite{Gaiotto:2008ak,Gaiotto:2008sa}. 
A large class of 3d $\CN=4$ Lagrangian theories can be obtained by considering D3 branes extending along a line segment $L$, with 1/2-BPS 
boundary conditions at the two ends \cite{Gaiotto:2008ak}. The simplest of these boundary conditions correspond to D3-branes ending on NS5- branes 
and/or D5-branes, and these are the only ingredients one needs for constructing a linear quiver.\\

\begin{table}[htbp]
\begin{center}
\begin{tabular}{|c|c|c|c|c|c|c|c|c|c|c|}
\hline
         & 0 & 1 &  2 & 3 & 4 & 5 & 6 & 7 & 8 & 9 \\
\hline \hline
NS5 & x &  x &  x & $\cdot$  & $\cdot$  & $\cdot$ & $\cdot$   &  x     &  x & x  \\
\hline
D5    & x &  x & x  &  $\cdot$ & x  & x  & x &  $\cdot$  &  $\cdot$   &  $\cdot$  \\
\hline
D3    & x &  x & x & x &  $\cdot$  &  $\cdot$   & $ \cdot$  & $\cdot$  &   $\cdot$   &  $\cdot$  \\
\hline
\end{tabular}
\caption{\footnotesize{Basic Type IIB brane construction.}} 
\label{tab:HWbranes}
\end{center}
\end{table}

The respective world-volumes of D3, D5 and NS5-branes are specified in Table \ref{tab:HWbranes}. A more precise way of 
writing this data is as follows:
\begin{align}\label{IIB-1}
& \text{D3:}\quad \BR^{2,1} \times L \times \{\vec X\}_{4,5,6} \times \{\vec Y\}_{7,8,9} \nn\\
& \text{D5:}\quad \BR^{2,1} \times \{X_3\}  \times \BR^{3}_{4,5,6} \times \{ \vec Y'\}_{7,8,9} \nn \\
& \text{NS5:}\quad \BR^{2,1} \times \{X'_3\}  \times  \{\vec X'\}_{4,5,6} \times \BR^3_{7,8,9},
\end{align}
where $\{X_3\},\{\vec X\},  \{\vec Y\}$ (and $\{X'_3\},\{\vec X'\},  \{\vec Y'\}$) are points in $L, \BR^{3}_{4,5,6}, \BR^3_{7,8,9}$ respectively. 
To a given configuration of D3-NS5-D5-branes, one can associate a set of linking numbers of the 5-branes 
$(l^{\rm NS5}_{\gamma}, l^{\rm D5}_\beta)$ which are defined as follows:
\begin{align}
& l^{\rm NS5}_{\gamma} = n_{\rm left}({\rm D5}) - \wt{n}_{\rm left}({\rm D3}) + \wt{n}_{\rm right}({\rm D3}),\qquad \gamma =1,\ldots, L+1,  \\
& l^{\rm D5}_\beta = n_{\rm left}({\rm NS5}) - \wt{n}_{\rm left}({\rm D3}) + \wt{n}_{\rm right}({\rm D3}), \qquad \beta=1, \ldots, L^\vee +1,
\end{align}
where $L^\vee=\sum^L_{\gamma=1} M_\gamma -1$, $n_{\rm left,right}({\rm D5/NS5})$ denotes the number of D5/NS5-branes to the left or right of the 5-brane in question, while $\wt{n}_{\rm left,right}({\rm D3})$ denotes the number of D3 branes ending on the 5-brane from the left and the right 
respectively. One can move the D5 and the NS5-branes past each other such that the linking numbers remain the same in the initial and the final configuration. Such moves, known as Hanany-Witten moves, create/annihilate D3-branes stretched between pairs of NS5-D5 branes. We will also refer to a given configuration of D3-NS5-D5-branes, among all the possible configurations for a given set of linking numbers, as a \textit{Hanany-Witten frame}.\\

From the perspective of the 3d world volume, the gauge theory data can be read off from the Hanany-Witten frame where all 
D3-branes end on NS5-branes, using the following rules:
\begin{itemize}
\item The $\gamma$-th NS5 chamber containing $N_\gamma$ D3 branes gives a $U(N_\gamma)$ vector multiplet. 
This arises from the massless modes of the D3-D3 open strings which survive the NS5 boundary conditions.

\item The $\gamma$-th NS5 chamber, containing $M_\gamma$ D5 branes, gives $M_\gamma$ hypermultiplets in the 
fundamental representation of $U(N_\gamma)$. These arise from the massless modes of the D3-D5 open strings.

\item  There are hypermultiplets in the bifundamental of $U(N_\gamma) \times U(N_{\gamma+1})$, which arise from 
D3-D3 open strings running between the $\gamma$-th and the $\gamma+1$-th NS5 chambers.
\end{itemize}

The $\CN=4$ supersymmetry preserving deformations are encoded in the brane construction as follows.
The triplet of FI parameters are $\eta^Y_{\gamma}= t^Y_{\gamma} - t^Y_{\gamma +1}$, with $\gamma=1,\ldots,L$, 
where the parameter ${t}^Y_\gamma$ corresponds to the relative position of the NS5 brane in $\BR^{3}_{4,5,6}$ 
w.r.t. D3-branes in the chamber $\gamma$. The triplet of fundamental mass parameters $m^Z_\beta$, with 
$\beta=1,\ldots, \sum^L_{\gamma=1} M_\gamma$, correspond to the relative positions 
of the D5 branes in $\BR^{3}_{7,8,9}$ w.r.t the D3-brane in a given chamber. Given the translational 
symmetry on $\BR^3$, both sets of moduli should be counted up to an overall shift. The bifundamental 
masses correspond to the relative position in $\BR^{3}_{7,8,9}$ of the D3-branes in the $\gamma$-chamber 
w.r.t. the D3-branes in the $\gamma+1$-chamber. In the above Hanany-Witten frame, one can choose to 
put all D3-branes at the same point in $\BR^{3}_{7,8,9}$, which sets all the bifundamental masses to zero and 
leaves the fundamental masses unconstrained up to an overall shift.

\begin{figure}[h]
\begin{center}
\scalebox{0.8}{\begin{tikzpicture}[
nnode/.style={circle,draw,thick, fill=red},cnode/.style={circle,draw,thick,minimum size=4mm},snode/.style={rectangle,draw,thick,minimum size=6mm}]
\node[nnode] (1) at (1,0) {} ;
\node[nnode] (2) at (3,0) {};
\node[nnode] (3) at (5,0) {};
\node[nnode] (4) at (7,0) {};
\draw[blue, thick,-] (0, 2) -- (0, -2);
\draw[blue,thick, -] (8, 2) -- (8, -2);
\node[cnode] (5) at (11,0) {2};
\node[snode] (6) at (11, -1.5) {4};
\draw[-] (5)--(6);
\draw[->] (0,-4) -- (1, -4);
\draw[->] (0,-4) -- (0,-3);
\draw[-] (0,0.5) -- (8,0.5);
\draw[-] (0,-0.5) -- (8,-0.5);
\node[text width=.2cm](9) at (1.2, -4){$x_3$};
\node[text width=1cm](10) at (0, -2.5){$\BR^3_{7,8,9}$};
\end{tikzpicture}}
\caption{\footnotesize{The figure on the left shows the Type IIB brane construction for the linear quiver on the right. 
The red nodes represent D5 branes, the horizontal black lines are D3 branes, and the vertical blue lines 
represent NS5 branes.}}
\label{Fig:HWEx1X}
\end{center}
\end{figure}
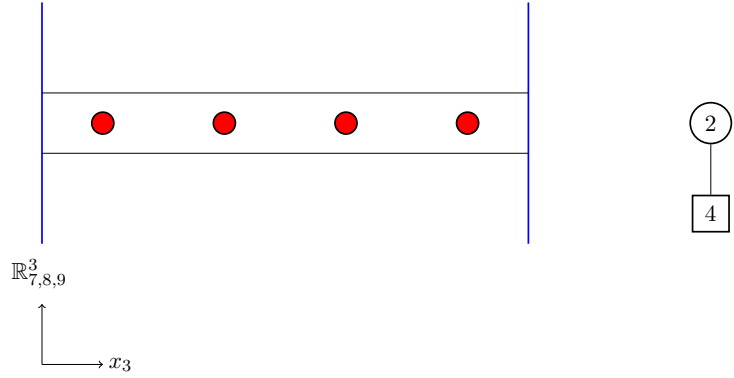

\begin{figure}[h]
\begin{center}
\begin{tikzpicture}[
nnode/.style={circle,draw,thick, fill=blue},cnode/.style={circle,draw,thick,minimum size=4mm},snode/.style={rectangle,draw,thick,minimum size=6mm}]
\node[nnode] (1) at (0,0) {} ;
\node[nnode] (2) at (2,0) {};
\node[nnode] (3) at (4,0) {};
\node[nnode] (4) at (6,0) {};
\draw[thick, -] (1)--(2);
\draw[thick,-] (2.north east)--(3.north west);
\draw[thick,-] (2.south east)--(3.south west);
\draw[thick, -] (3)--(4);
\draw[red, thick, -] (2.5, 2) -- (2.5, -2);
\draw[red, thick, -] (3.5, 2) -- (3.5, -2);
\node[cnode] (5) at (10,0) {1};
\node[cnode] (6) at (12,0) {2};
\node[cnode] (7) at (14,0) {1};
\node[snode] (8) at (12, -1.5) {2};
\draw[-] (5)--(6);
\draw[-] (6)--(7);
\draw[-] (6)--(8);
\draw[->] (0,-2) -- (1, -2);
\draw[->] (0,-2) -- (0,-1);
\node[text width=.2cm](9) at (1.2, -2){$x_3$};
\node[text width=1cm](10) at (-0.3, -0.7){$\BR^3_{4,5,6}$};
\end{tikzpicture}
\caption{\footnotesize{The figure on the left shows the Type IIB brane construction for the linear quiver on the right. 
The blue nodes represent NS5 branes, the horizontal blck lines are D3 branes, and the vertical red lines 
represent D5 branes.}}
\label{Fig:HWEx1Y}
\end{center}
\end{figure}

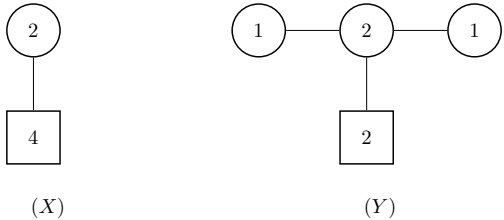
\begin{figure}[htbp]
\begin{center}
\scalebox{0.7}{\begin{tikzpicture}[cnode/.style={circle,draw,thick, minimum size=1.0cm},snode/.style={rectangle,draw,thick,minimum size=1cm}]
\node[cnode] (1) {2};
\node[snode] (2) [below=1cm of 1]{4};
\draw[-] (1) -- (2);
\node[text width=0.1cm](20)[below=0.5 cm of 2]{$(X)$};
\end{tikzpicture}
\qquad \qquad \qquad \qquad 
\begin{tikzpicture}[
cnode/.style={circle,draw,thick, minimum size=1.0cm},snode/.style={rectangle,draw,thick,minimum size=1cm}]
\node[cnode] (1) {1};
\node[cnode] (2) [right=1cm  of 1]{2};
\node[cnode] (3) [right=1cm  of 2]{1};
\node[snode] (4) [below=1cm of 2]{2};
\node[text width=0.1cm](20)[below=0.5 cm of 4]{$(Y)$};
\draw[-] (1) -- (2);
\draw[-] (2)-- (3);
\draw[-] (2)-- (4);
\end{tikzpicture}}
\caption{\footnotesize{An example of a pair of linear quivers with unitary gauge groups which are 3d mirrors.}}
\label{fig: LQEx1}
\end{center}
\end{figure}

\figref{Fig:HWEx1X} gives an illustrative example of how one can read off the gauge theory content from the brane set up, 
using the rules listed above. 
Mirror symmetry in three dimensions can be understood as an S-duality of the above brane configuration, followed by a 
rotation $R: \vec x^{7,8,9} \to - \vec {x}^{4,5,6}, \, \vec x^{4,5,6} \to \vec x^{7,8,9}$. NS5 and D5 branes are exchanged 
under S-duality, while D3 branes are self-dual. To read off the dual gauge theory from the rotated S-dual brane system,
one has to move to a Hanany-Witten frame where the all D3 branes end on NS5 branes. This can be done by performing a series 
of Hanany-Witten moves, where NS5 and D5 branes are moved past each other along the compact direction $x^3$ involving 
creation/annihilation of D3 branes. In the simple example of \figref{Fig:HWEx1X}, the rotated S-duality transformation followed by 
appropriate Hanany-Witten moves lead to the brane 
configuration in \figref{Fig:HWEx1Y}, from which the dual quiver gauge theory can again be read off, using the same rules as before. The mirror pair is 
shown in \figref{fig: LQEx1}.

\subsubsection{Line defects: Wilson and Vortex}\label{LQ2BD}
In this section, we review the Type IIB construction of line defects in linear quivers -- the Wilson and the vortex defects, which
are defects of the Type-B and the Type-A respectively, as reviewed in \Secref{Review}. Our discussion largely follows 
\cite{Assel:2015oxa}, which is closely related to the discussion of Wilson defects in the 4d $\CN=4$ theory
\cite{Gomis:2006im, Gomis:2006sb}.\\

\subsubsection*{Wilson defects}
In addition to the standard D3-D5-NS5 system for a linear quiver, the insertion of Wilson defects in linear quivers 
requires additional D5-branes and F1-strings connecting these D5-branes with the D3-branes. These additional 
D5-branes can either have the same world-volume as the original D5-branes or be rotated in a specific fashion 
as discussed below. In the latter case, we will refer to them as D5'-branes, using the same notation as \cite{Assel:2015oxa}. 

The respective world-volumes of the D3, D5, D5', NS5-branes, and the F1-strings are specified in 
Table \ref{tab:HWbranesWL}. A more precise way of writing this data is as follows:
\begin{align}
& \text{D3:}\quad \BR^{2,1} \times L \times \{\vec X\}_{4,5,6} \times \{\vec Y\}_{7,8,9} \nn\\
& \text{D5:}\quad \BR^{2,1} \times \{X_3\}  \times \BR^{3}_{4,5,6} \times \{ \vec Y'\}_{7,8,9} \nn \\
& \text{D5':}\quad \BR^{0,1} \times \{X_1\} \times \{X_2\} \times \{X'_3\}  \times \BR^{3}_{4,5,6} \times \BR^2_{7,8} \times \{X_9\} \nn \\
& \text{NS5:}\quad \BR^{2,1} \times \{X'_3\}  \times  \{\vec X'\}_{4,5,6} \times \BR^3_{7,8,9} \nn\\
& \text{F1:}\quad \BR^{0,1} \times \{X'_1\} \times \{X'_2\} \times \{X''_3\}  \times \{\vec X'' \}_{4,5,6} \times \{X_7\} \times \{X_8\} \times \BR_9, \nn \\
\end{align}
where we use the notation that $\{\cdot \}$ indicates a point in respective direction(s). 

\begin{table}[htbp]
\begin{center}
\begin{tabular}{|c|c|c|c|c|c|c|c|c|c|c|}
\hline
         & 0 & 1 &  2 & 3 & 4 & 5 & 6 & 7 & 8 & 9 \\
\hline \hline
NS5 & x &  x &  x & $\cdot$  & $\cdot$  & $\cdot$ & $\cdot$   &  x     &  x & x  \\
\hline
D5    & x &  x & x  &  $\cdot$ & x  & x  & x &  $\cdot$  &  $\cdot$   &  $\cdot$  \\
\hline
D3    & x &  x & x & x &  $\cdot$  &  $\cdot$   & $ \cdot$  & $\cdot$  &   $\cdot$   &  $\cdot$  \\
\hline
D5'    & x &  $\cdot$ & $\cdot$  &  $\cdot$ & x  & x  & x &  x  &  x   &  $\cdot$  \\
\hline
F1   & x &  $\cdot$ & $\cdot$  &  $\cdot$ & $\cdot$  & $\cdot$  & $\cdot$ &  $\cdot$  &  $\cdot$   &  x  \\
\hline
\end{tabular}
\caption{\footnotesize{Type IIB brane construction for a Wilson defect.}}
\label{tab:HWbranesWL}
\end{center}
\end{table}

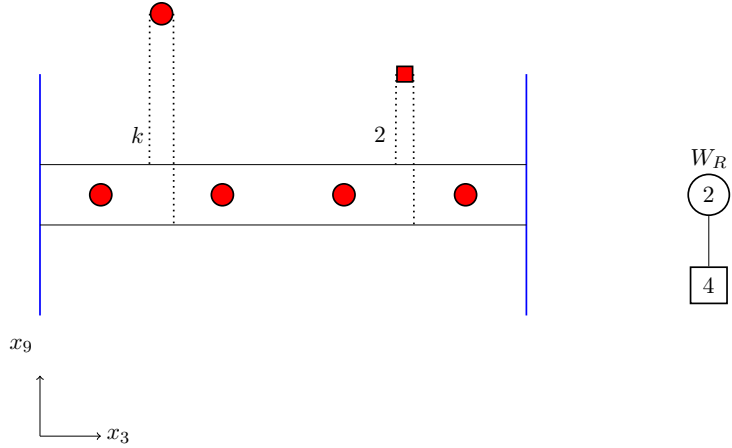
\begin{figure}[h]
\begin{center}
\scalebox{0.8}{\begin{tikzpicture}[
nnode/.style={circle,draw,thick, fill=red},cnode/.style={circle,draw,thick,minimum size=4mm},snode/.style={rectangle,draw,thick,minimum size=6mm}]
\node[nnode] (1) at (1,0) {} ;
\node[nnode] (2) at (3,0) {};
\node[nnode] (3) at (5,0) {};
\node[nnode] (4) at (7,0) {};
\node[nnode] (7) at (2,3) {};
\node[rectangle,draw,thick, fill=red] (8) at (6,2) {};
\node[text width=0.1] (9) at (1.5, 1) {$k$};
\node[text width=0.1] (10) at (5.5, 1) {2};
\draw[blue, thick,-] (0, 2) -- (0, -2);
\draw[blue,thick, -] (8, 2) -- (8, -2);
\node[cnode] (5) at (11,0) {2};
\node[snode] (6) at (11, -1.5) {4};
\node[text width=0.1] (11) at (10.7,0.6){$W_R$};
\draw[-] (5)--(6);
\draw[->] (0,-4) -- (1, -4);
\draw[->] (0,-4) -- (0,-3);
\draw[-] (0,0.5) -- (8,0.5);
\draw[-] (0,-0.5) -- (8,-0.5);
\draw[black, thick, dotted] (7.west) -- (1.8,0.5);
\draw[black, thick, dotted] (7.east) -- (2.2,-0.5);
\draw[black, thick, dotted] (8.west) -- (5.85,0.5);
\draw[black, thick, dotted] (8.east) -- (6.15,-0.5);
\node[text width=.2cm](9) at (1.2, -4){$x_3$};
\node[text width=1cm](10) at (0, -2.5){$x_9$};
\end{tikzpicture}}
\caption{\footnotesize{Type IIB brane construction for a gauge Wilson defect for $U(2)$ in a 
representation $R=\CS_k \otimes \CA_2$. The red rectangular nodes denote D5'-branes, while the dotted vertical lines 
denote F1-strings.}}
\label{Fig:HWEx1XWLA}
\end{center}
\end{figure}

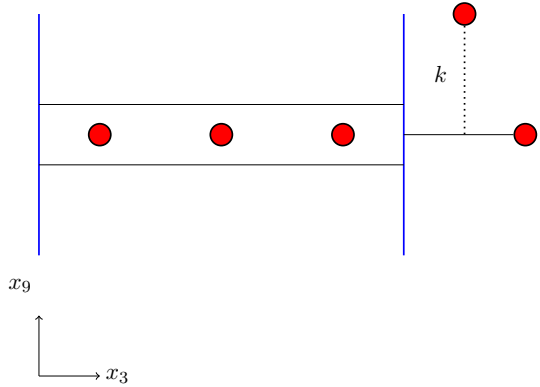
\begin{figure}[h]
\begin{center}
\scalebox{0.8}{\begin{tikzpicture}[
nnode/.style={circle,draw,thick, fill=red},cnode/.style={circle,draw,thick,minimum size=4mm},snode/.style={rectangle,draw,thick,minimum size=6mm}]
\node[nnode] (1) at (1,0) {} ;
\node[nnode] (2) at (3,0) {};
\node[nnode] (3) at (5,0) {};
\node[nnode] (4) at (8,0) {};
\node[nnode] (7) at (7,2) {};
\draw[blue, thick,-] (0, 2) -- (0, -2);
\draw[blue,thick, -] (6, 2) -- (6, -2);
\draw[->] (0,-4) -- (1, -4);
\draw[->] (0,-4) -- (0,-3);
\draw[-] (0,0.5) -- (6,0.5);
\draw[-] (0,-0.5) -- (6,-0.5);
\draw[-] (6,0) -- (4);
\draw[black, thick, dotted] (7) -- (7,0);
\node[text width=0.1] (9) at (6.5, 1) {$k$};
\node[text width=.2cm](9) at (1.2, -4){$x_3$};
\node[text width=1cm](10) at (0, -2.5){$x_9$};
\end{tikzpicture}}
\caption{\footnotesize{Type IIB brane construction for a flavor Wilson defect of charge $k$.}}
\label{Fig:HWEx1XWLF}
\end{center}
\end{figure}

Let us briefly list the important features of the Type IIB construction that will be relevant for our work. Unless otherwise stated, 
we will always work with the canonical configuration where all D3-branes end on NS5-branes.
\begin{itemize}
\item In the $\gamma$-th NS5 chamber, the presence of a D5 or D5'-brane, away from the main stack of D5-branes in the $x^9$-direction, 
introduces a gauge Wilson defect for the $U(N_\gamma)$ factor. The defect arises from F1-strings stretching between the D5/D5'-brane and the 
D3-branes in the given chamber. 

\item A stack of $k$ F1-strings connecting a D5-brane and $N_\gamma$ D3-branes introduces a Wilson defect 
in the $k$-th symmetric representation $\CS_k$ of the gauge group $U(N_\gamma)$. The set of non-negative integers 
$\{ k_j\}_{j=1,\ldots,N_\gamma}$, where $k_j$ is the number of F1-strings ending on the $j$-th D3-brane, generate the weights of the 
representation $\CS_k$ in the orthogonal basis, i.e. given a weight $w=(w_1, w_2,\ldots, w_{N_\gamma})$ of $\CS_k$, we have $w_j=k_j$, 
with $\sum^{N_\gamma}_{j=1} k_j = k$. Similarly, a stack of $k'$ F1-strings connecting a D5'-brane and $N_\gamma$ D3-branes introduces a Wilson defect in the $k$-th antisymmetric representation $\CA_{k'}$ of the gauge group $U(N_\gamma)$. Similar to the case of the 
symmetric representation, the set of integers $\{ k_j\}_{j=1,\ldots,N_\gamma}$, where $k_j$ is the number of F1-strings 
ending on the $j$-th D3-brane, generate the weights of the representation $\CA_{k'}$ in the orthogonal basis. In this case, 
the integers are constrained to be $k_j=0,1$ $\forall j$, by the s-rule \cite{Hanany:1996ie}.

\item In general, there can be multiple stacks F1-strings in a given NS5 chamber, connecting $l$ D5-branes and $l'$ D5'-branes with 
the $N_\gamma$ D3-branes, such that the D5 and D5'-branes are all separated in the $x^3$ and the $x^9$ directions.
This generates a Wilson defect in the representation 
$R= \otimes^l_{a=1} \CS_{k^{(a)}}\otimes^{l'}_{b=1} \CA_{k'^{(b)}} $ of $U(N_\gamma)$, with the weights $w$ being given as 
$w_j= \sum^l_{a=1} k^{(a)}_j + \sum^{l'}_{b=1} k'^{(b)}_j$.
An illustrative example of Wilson defects in the quiver gauge theory $X$ of \figref{fig: LQEx1} is given in \figref{Fig:HWEx1XWLA}. 
The defect involves a stack of F1-strings connecting a D5 and a D5' respectively with D3-branes, introducing a gauge Wilson defect 
for $U(2)$ in the representation $R= \CS_k \otimes \CA_2$.

\item The Hanany-Witten moves for the Type IIB construction with defects are substantially restricted. Moving the F1-strings past an 
NS5-brane will change the Wilson defect and is therefore not allowed. However, one can move them past the D5-branes without 
changing the IR physics, provided that the equal number of D3-branes end on the D5 from the left and the right. This is always the
case in the canonical configuration, where no D3-branes end on the D5-branes.

\item Introducing flavor Wilson defects requires moving to a Hanany-Witten frame, where a stack of D3-branes 
is stretched between an NS5 and a D5-brane. One can then introduce an additional D5-brane and F1-strings connecting the former 
and the D3-branes, leading to an Abelian defect in some representation of the maximal torus of the flavor symmetry group associated 
with these D3-branes. An example of a flavor Wilson loop in the theory $X$ of \figref{fig: LQEx1} is shown in \figref{Fig:HWEx1XWLF}.

\end{itemize}

\subsubsection*{Vortex defects}
In addition to the standard D3-D5-NS5 system for a linear quiver, the insertion of vortex defects in linear quivers 
requires additional NS5-branes and D1-branes connecting these NS5-branes with the NS5-branes in the original 
D3-D5-NS5 system. These additional NS5-branes can either have the same world-volume as the original NS5-branes 
or be rotated in a specific fashion as discussed below. 
In the latter case, we will refer to them as NS5'-branes, using the same notation as the D5-branes that appeared for the 
Wilson defects described above. 

The respective world-volumes of the D3, D5, NS5, NS5', and the D1-branes are specified in 
Table \ref{tab:HWbranesVL}. A more precise way of writing this data is as follows:
\begin{align}
& \text{D3:}\quad \BR^{2,1} \times L \times \{\vec X\}_{4,5,6} \times \{\vec Y\}_{7,8,9} \nn\\
& \text{D5:}\quad \BR^{2,1} \times \{X_3\}  \times \BR^{3}_{4,5,6} \times \{ \vec Y'\}_{7,8,9} \nn \\
& \text{NS5:}\quad \BR^{2,1} \times \{X'_3\}  \times  \{\vec X'\}_{4,5,6} \times \BR^3_{7,8,9} \nn\\
& \text{NS5':}\quad \BR^{0,1} \times \{X_1\} \times \{X_2\} \times \{X'_3\}  \times \BR^{3}_{4,5} \times \{X_6\} \times \BR^3_{7,8,9} \nn \\
& \text{D1:}\quad \BR^{0,1} \times \{X'_1\} \times \{X'_2\} \times \{X''_3\}  \times \{\vec X'' \}_{4,5,6} \times \BR_6 \times \{\vec Y'' \}_{7,8,9}, \nn \\
\end{align}
where we use the notation that $\{\cdot \}$ indicates a point in respective direction(s).

\begin{table}[htbp]
\begin{center}
\begin{tabular}{|c|c|c|c|c|c|c|c|c|c|c|}
\hline
         & 0 & 1 &  2 & 3 & 4 & 5 & 6 & 7 & 8 & 9 \\
\hline \hline
NS5 & x &  x &  x & $\cdot$  & $\cdot$  & $\cdot$ & $\cdot$   &  x     &  x & x  \\
\hline
D5    & x &  x & x  &  $\cdot$ & x  & x  & x &  $\cdot$  &  $\cdot$   &  $\cdot$  \\
\hline
D3    & x &  x & x & x &  $\cdot$  &  $\cdot$   & $ \cdot$  & $\cdot$  &   $\cdot$   &  $\cdot$  \\
\hline
NS5'    & x &  $\cdot$ & $\cdot$  &  $\cdot$ & x  & x  & $\cdot$ &  x  &  x   &  x  \\
\hline
D1   & x &  $\cdot$ & $\cdot$  &  $\cdot$ & $\cdot$  & $\cdot$  & x &  $\cdot$  &  $\cdot$   &  $\cdot$  \\
\hline
\end{tabular}
\caption{\footnotesize{Type IIB brane construction for a vortex line defect.} }
\label{tab:HWbranesVL}
\end{center}
\end{table}

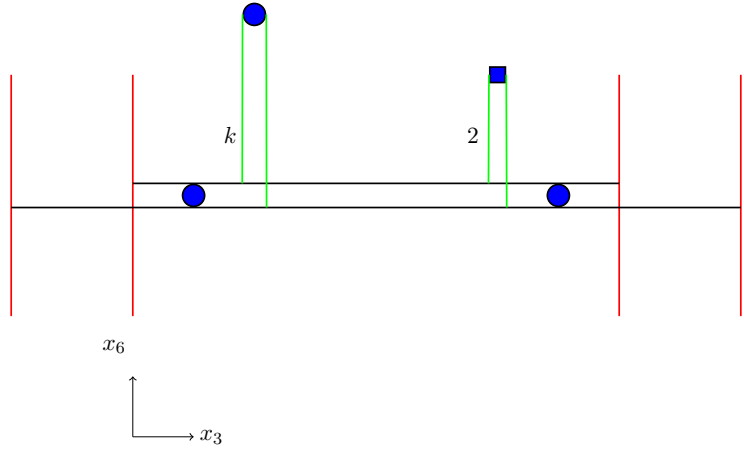
\begin{figure}[h]
\begin{center}
\scalebox{0.8}{\begin{tikzpicture}[
nnode/.style={circle,draw,thick, fill=blue},cnode/.style={circle,draw,thick,minimum size=4mm},snode/.style={rectangle,draw,thick,minimum size=6mm}]
\node[nnode] (1) at (1,0) {} ;
\node[nnode] (4) at (7,0) {};
\node[nnode] (7) at (2,3) {};
\node[rectangle,draw,thick, fill=blue] (8) at (6,2) {};
\node[text width=0.1] (9) at (1.5, 1) {$k$};
\node[text width=0.1] (10) at (5.5, 1) {2};
\draw[red, thick,-] (-2, 2) -- (-2, -2);
\draw[red, thick,-] (0, 2) -- (0, -2);
\draw[red,thick, -] (8, 2) -- (8, -2);
\draw[red, thick,-] (10, 2) -- (10, -2);
\draw[->] (0,-4) -- (1, -4);
\draw[->] (0,-4) -- (0,-3);
\draw[thick, -] (0,0.2) -- (8,0.2);
\draw[thick, -] (-2,-0.2) -- (10,-0.2);
\draw[green, thick] (7.west) -- (1.8,0.2);
\draw[green, thick] (7.east) -- (2.2,-0.2);
\draw[green, thick] (8.west) -- (5.85,0.2);
\draw[green, thick] (8.east) -- (6.15,-0.2);
\node[text width=.2cm](9) at (1.2, -4){$x_3$};
\node[text width=1cm](10) at (0, -2.5){$x_6$};
\end{tikzpicture}}
\caption{\footnotesize{Type IIB brane construction for a gauge vortex defect for $U(2)$ in a 
representation $R=\CS_k \otimes \CA_2$. The blue rectangular nodes denote NS5'-branes, while the green vertical lines 
denote D1-strings.}}
\label{Fig:HWEx1XVLA}
\end{center}
\end{figure}

We briefly list the important features of the Type IIB construction that will be relevant for the rest of the paper. 
Unless otherwise stated, we will always work with the canonical configuration where all D3-branes end on NS5-branes.
\begin{itemize}
\item In the $\gamma$-th NS5 chamber, the presence of an N5 or N5'-brane, away from the main stack of NS5-branes in the $x^6$-direction, 
introduces a gauge vortex defect for the $U(N_\gamma)$ factor. The defect arises from D1-branes stretching between the NS5/NS5'-brane 
and the D3-branes in the chamber $\gamma$. 

\item A stack of $k$ D1-branes connecting an NS5-brane and $N_\gamma$ D3-branes introduces a gauge vortex defect 
in the $k$-th symmetric representation $\CS_k$ of the gauge group $U(N_\gamma)$. The set of non-negative integers 
$\{ k_j\}_{j=1,\ldots,N_\gamma}$, where $k_j$ is the number of D1-branes ending on the $j$-th D3-brane, generate the weights of the 
representation $\CS_k$ in the orthogonal basis, i.e. given a weight $w=(w_1, w_2,\ldots, w_{N_\gamma})$ of $\CS_k$, we have $w_j=k_j$, 
with $\sum^{N_\gamma}_{j=1} k_j = k$. Similarly, a stack of $k'$ D1-branes connecting an NS5'-brane and $N_\gamma$ D3-branes introduces 
a vortex defect in the $k$-th antisymmetric representation $\CA_{k'}$ of the gauge group $U(N_\gamma)$. 
In this case, the set of integers $\{ k_j\}_{j=1,\ldots,N_\gamma}$, where $k_j=0,1$ is the number of D1-branes 
ending on the $j$-th D3-brane, generate the weights of the representation $\CA_{k'}$ in the orthogonal basis.

\item In general, there can be multiple stacks D1-branes in a given NS5 chamber, connecting $l$ NS5-branes and $l'$ NS5'-branes with 
the $N_\gamma$ D3-branes, such that the NS5 and NS5'-branes are all separated in the $x^3$ and the $x^9$ directions.
This generates a vortex defect in the representation $R$ of $U(N_\gamma)$:
\be \label{gen-R}
R= \otimes^l_{a=1} \CS_{k^{(a)}}\otimes^{l'}_{b=1} \CA_{k'^{(b)}},
\ee
with the weights $w$ being given as 
$w_j= \sum^l_{a=1} k^{(a)}_j + \sum^{l'}_{b=1} k'^{(b)}_j$. Different relative orderings of the NS5/NS5'-branes in the $x^6$ and 
$x^3$ directions give the same vortex defect.
An illustrative example of a vortex defect in the quiver gauge theory $X$ of \figref{fig: LQEx1} is given in \figref{Fig:HWEx1XVLA}. 
The defect involves a stack of D1-branes connecting an NS5 and an NS5' respectively with D3-branes, introducing a gauge vortex defect 
for $U(2)$ in the representation $R= \CS_k \otimes \CA_2$. 

\item The Hanany-Witten moves for the Type IIB construction with defects are again restricted. Moving the D1-branes past an 
D5-brane are not allowed. However, one can move them past the NS5-branes without 
changing the IR physics, provided that the equal number of D3-branes end on the NS5 from the left and the right. Therefore, 
the position of the D1-stack in the $x^3$-direction w.r.t the D5-branes in a given NS5-chamber is important. Generically, distinct 
insertion points (i.e. different number of D5-branes on the left/right of the D1-stack in the chamber) will lead to different vortex 
defect. Without loss of generality, let us assume that there are $Q$ D5-branes to the right of the D1-stack. We will denote the 
vortex defect as $V_{Q,\,R}$, where $R$ is defined above. 

\item A flavor Wilson defect for an Abelian subgroup of the 3d Coulomb branch symmetry can be introduced when 
a stack of D1-branes ends on an NS5-brane.

\end{itemize}

\begin{figure}[h]
\begin{center}
\scalebox{0.8}{\begin{tikzpicture}[
nnode/.style={circle,draw,thick, fill=blue},cnode/.style={circle,draw,thick,minimum size=1.0cm},snode/.style={rectangle,draw,thick,minimum size=1.0cm},
rnode/.style={red, circle,draw,thick,fill=red!30 ,minimum size=1.0cm}]
\node[nnode] (1) at (1,0) {} ;
\node[nnode] (4) at (7,0) {};
\node[nnode] (7) at (7,3) {};
\node[rectangle,draw,thick, fill=blue] (8) at (7,2) {};
\node[text width=0.1] (9) at (6.5, 2.5) {$k$};
\node[text width=0.1] (10) at (6.5, 1) {2};
\draw[red, thick,-] (-2, 2) -- (-2, -2);
\draw[red, thick,-] (0, 2) -- (0, -2);
\draw[red,thick, -] (8, 2) -- (8, -2);
\draw[red, thick,-] (10, 2) -- (10, -2);
\draw[->] (0,-4) -- (1, -4);
\draw[->] (0,-4) -- (0,-3);
\draw[thick, -] (0,0.2) -- (8,0.2);
\draw[thick, -] (-2,-0.2) -- (10,-0.2);
\draw[green, thick] (7) -- (8);
\draw[green, thick] (7) -- (8);
\draw[green, thick] (8) -- (4);
\draw[green, thick] (8) -- (4);
\node[text width=.2cm](9) at (1.2, -4){$x_3$};
\node[text width=1cm](10) at (0, -2.5){$x_6$};
\node[snode] (11) at (12,0) {2} ;
\node[cf-group] (12) at (14,0) {\rotatebox{-90}{2}};
\node[snode] (13) at (16,0) {2};
\node[rnode] (14) at (15,2) {$k+2$};
\node[rnode] (15) at (15,4) {$k$};
\draw[red, thick, ->] (12)--(14);
\draw[red, thick, ->] (14)--(13);
\draw[-] (11) -- (12);
\draw[-] (12) -- (13);
\draw[red, thick, ->] (15) to [out=210,in=150] (14);
\draw[red, thick, ->] (14) to [out=30,in=330] (15);
\end{tikzpicture}}
\caption{\footnotesize{Type IIB brane construction for the right 3d-1d quiver that realizes a vortex defect $V_{2,R}$ for the 
$U(2)$ gauge group in the representation $R=\CS_k \otimes \CA_2$.}}
\label{Fig:HWEx1XVLB}
\end{center}
\end{figure}
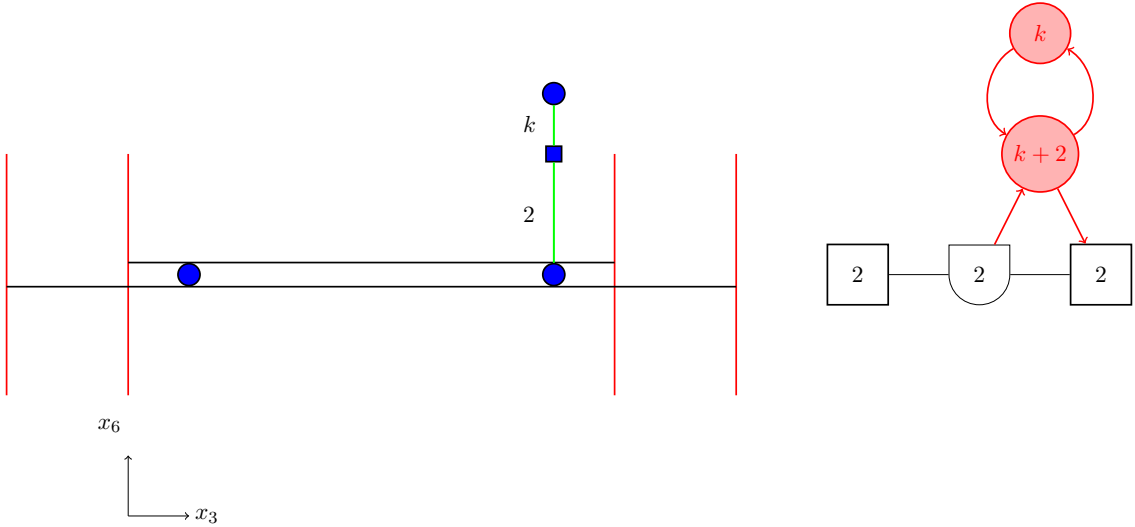

The above realization of a vortex defect can be regarded as a deformation of at least two different coupled 3d-1d system. 
This corresponds to making all the D1-branes end on the NS5-brane to the left, or making all of them end on the 
NS5-brane to the right. In the former case, we will refer to the associated 1d quiver as the ``left SQM", and in the latter 
case as the ``right SQM". The rule for obtaining the left/right SQM from the set of D1-brane stacks, described above,
is as follows. Suppose we have a set of $P$ stacks where the $x^9$ positions of the NS5/NS5'-branes are in a 
decreasing order from left to right, and let $k_i$ denote the number of D1-branes in the $i$-th stack. 
To begin with, one brings the NS5/NS5'-branes at the same point in $x^3$ direction (coincident with the left/right
NS5-brane in main stack) but different points in $x^9$, with D1-branes stretching between them. Before performing 
this operation, one needs to move the D5-branes in the chamber to the left/right of the NS5-brane so that the D-branes 
do not cross each other. In this configuration, the number of 
D1-branes in the $i$-th chamber, computed by demanding that the linking numbers associated with 
the NS5/NS5'-branes remain unchanged, is given by $n_i = \sum^{i}_{j=1} k_j$, with $i=1, \ldots, P$. 
For the defect in \figref{Fig:HWEx1XVLA}, this brane configuration is shown in 
\figref{Fig:HWEx1XVLB}, where the two D1-brane stacks are moved to the right NS5-brane. Given such a 
brane configuration, the 1d quiver can be obtained by the following rules:
\begin{enumerate}
\item $k$ D1-branes stretched between an NS5 and an NS5'-brane give a $U(k)$ vector multiplet, 
while $k$ D1-branes between a pair of NS5 or NS5'-branes give a $U(k)$ vector multiplet plus a 
single chiral multiplet in the adjoint representation. The gauge group is therefore $\prod^P_{i=1} U(n_i)$.

\item A given pair of consecutive 5-brane chambers, labelled $i$ and $i+1$, gives a bifundamental 
chiral and an anti-bifundamental chiral for $U(n_i) \times U(n_{i+1})$.

\item If $N_L$ and $N_R$ respectively denote the number of D3-branes ending on the NS5-brane 
in the main stack, then one has one bifundamental chiral for $U(n_P) \times U(N_L)$ and another 
bifundamental chiral for $U(N_R) \times U(n_P)$. 

\end{enumerate}
The 1d FI parameters can be related to the relative 
position of the NS5/NS5'-branes in the $x^3$ direction, i.e. $\eta_i= x^3_i - x^3_{i-1}$, where $\eta_i$ is the FI parameter 
associated with the 1d gauge group $U(n_i)$ with $i=1,\ldots,P$, and $x^3_{0}$ denotes the position of the NS5-brane in the main 
stack. This implies that the original Type IIB configuration can be thought of as a deformation of the left SQM with a positive 
FI parameter $\eta_i$ for all the $U(n_i)$ factors, or equivalently as a deformation of the right SQM with negative FI parameters 
for the $U(n_i)$ factors.

The coupling of the SQM, constructed above, with the 3d quiver gauge theory can now be read off from the brane picture. 
The SQM has a $U(N_L) \times U(N_R)$ flavor symmetry, a subgroup of which will be gauged by the 3d bulk fields living on 
the D3-branes. Let us focus on the right SQM which implies that $Q$ of the $M_\gamma$ D5-branes in the $\gamma$-th NS5 chamber 
have to be moved to the right of the $\gamma+1$-th NS5-brane, and let $N_R= Q + n_R$. The SQM is then coupled 
to the 3d theory by gauging the $U(N_L) \times U(n_R)$ subgroup of the full flavor symmetry by dynamical 3d $\CN=4$ vector multiplets. 
In addition, the $U(Q)$ subgroup of $U(N_R)$ is weakly gauged by background 3d $\CN=4$ vector multiplets, 
completing the construction of the 3d-1d coupled system. We will denote the right SQM as $\Sigma_r^{Q,R}$ and the 
left SQM as $\Sigma_l^{Q,R}$.

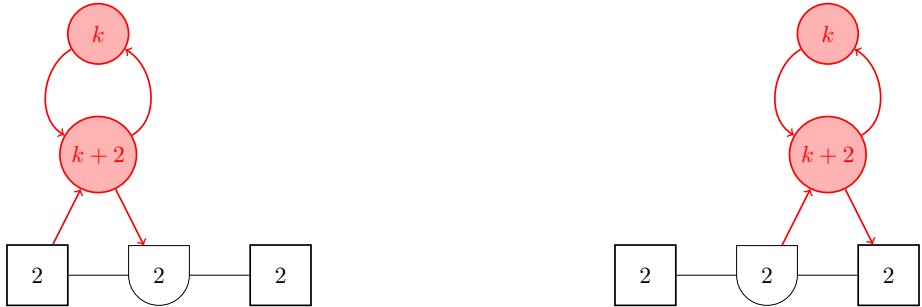
\begin{figure}[h]
\begin{center}
\scalebox{0.8}{\begin{tikzpicture}[
nnode/.style={circle,draw,thick, fill=blue},cnode/.style={circle,draw,thick,minimum size=1.0cm},snode/.style={rectangle,draw,thick,minimum size=1.0cm},
rnode/.style={red, circle,draw,thick,fill=red!30 ,minimum size=1.0cm}]
\node[snode] (1) at (2,0) {2} ;
\node[cf-group] (2) at (4,0) {\rotatebox{-90}{2}};
\node[snode] (3) at (6,0) {2};
\node[rnode] (4) at (3,2) {$k+2$};
\node[rnode] (5) at (3,4) {$k$};
\draw[red, thick, ->] (1)--(4);
\draw[red, thick, ->] (4)--(2);
\draw[-] (1) -- (2);
\draw[-] (2) -- (3);
\draw[red, thick, ->] (5) to [out=210,in=150] (4);
\draw[red, thick, ->] (4) to [out=30,in=330] (5);
\node[snode] (11) at (12,0) {2} ;
\node[cf-group] (12) at (14,0) {\rotatebox{-90}{2}};
\node[snode] (13) at (16,0) {2};
\node[rnode] (14) at (15,2) {$k+2$};
\node[rnode] (15) at (15,4) {$k$};
\draw[red, thick, ->] (12)--(14);
\draw[red, thick, ->] (14)--(13);
\draw[-] (11) -- (12);
\draw[-] (12) -- (13);
\draw[red, thick, ->] (15) to [out=210,in=150] (14);
\draw[red, thick, ->] (14) to [out=30,in=330] (15);
\end{tikzpicture}}
\caption{\footnotesize{The right and the left 3d-1d quivers that realize a vortex defect $V_{2,R}$ for the 
$U(2)$ gauge group in a representation $R=\CS_k \otimes \CA_2$.}}
\label{Fig:Ex1XVLHop}
\end{center}
\end{figure}

To summarize, a gauge vortex defect in a 3d quiver gauge theory is completely (although not uniquely) specified by 
a 3d-1d coupled quiver along with a choice of the signs of the 1d FI parameters. In the construction presented above, 
they are either all positive (left SQM) or all negative (right SQM). The two 3d-1d systems associated with the vortex defect 
of \figref{Fig:HWEx1XVLA} are shown in \figref{Fig:Ex1XVLHop}.\\

\subsubsection{S-duality and mirror map}\label{MM2BD}
\begin{figure}[htbp]
\begin{center}
\begin{tabular}{ccc}
\scalebox{0.6}{\begin{tikzpicture}[
nnode/.style={circle,draw,thick, fill=blue},cnode/.style={circle,draw,thick,minimum size=4mm},snode/.style={rectangle,draw,thick,minimum size=6mm}]
\node[nnode] (1) at (1,0) {} ;
\node[nnode] (4) at (7,0) {};
\node[nnode] (7) at (2,3) {};
\node[rectangle,draw,thick, fill=blue] (8) at (6,2) {};
\node[text width=0.1] (9) at (1.5, 1) {$k$};
\node[text width=0.1] (10) at (5.5, 1) {2};
\draw[red, thick,-] (-2, 2) -- (-2, -2);
\draw[red, thick,-] (0, 2) -- (0, -2);
\draw[red,thick, -] (8, 2) -- (8, -2);
\draw[red, thick,-] (10, 2) -- (10, -2);
\draw[thick, -] (0,0.2) -- (8,0.2);
\draw[thick, -] (-2,-0.2) -- (10,-0.2);
\draw[green, thick] (7.west) -- (1.8,0.2);
\draw[green, thick] (7.east) -- (2.2,-0.2);
\draw[green, thick] (8.west) -- (5.85,0.2);
\draw[green, thick] (8.east) -- (6.15,-0.2);
\end{tikzpicture}}
&\begin{tikzpicture}
\node[text width=1cm] (1) at (0,0) {$\xrightarrow{S}$}; 
\node[] (2) at (0,-1){};
\end{tikzpicture}
& \scalebox{0.6}{\begin{tikzpicture}[
nnode/.style={circle,draw,thick, fill=red},cnode/.style={circle,draw,thick,minimum size=4mm},snode/.style={rectangle,draw,thick,minimum size=6mm}]
\node[nnode] (1) at (1,0) {} ;
\node[nnode] (4) at (7,0) {};
\node[nnode] (7) at (2,3) {};
\node[rectangle,draw,thick, fill=red] (8) at (6,2) {};
\node[text width=0.1] (9) at (1.5, 1) {$k$};
\node[text width=0.1] (10) at (5.5, 1) {2};
\draw[blue, thick,-] (-2, 2) -- (-2, -2);
\draw[blue, thick,-] (0, 2) -- (0, -2);
\draw[blue,thick, -] (8, 2) -- (8, -2);
\draw[blue, thick,-] (10, 2) -- (10, -2);
\draw[thick, -] (0,0.2) -- (8,0.2);
\draw[thick, -] (-2,-0.2) -- (10,-0.2);
\draw[black, thick, dotted] (7.west) -- (1.8,0.2);
\draw[black, thick, dotted] (7.east) -- (2.2,-0.2);
\draw[black, thick, dotted] (8.west) -- (5.85,0.2);
\draw[black, thick, dotted] (8.east) -- (6.15,-0.2);
\end{tikzpicture}}\\
\qquad & \qquad 
& \scalebox{0.5}{\begin{tikzpicture}
\draw[->] (0,0) -- (0,-2);
\node[text width=0.1cm] (1) at (0.2, -1) {HW};
\end{tikzpicture}}  \\
\scalebox{0.6}{\begin{tikzpicture}[
cnode/.style={circle,draw,thick, minimum size=1.0cm},snode/.style={rectangle,draw,thick,minimum size=1cm}]
\node[cnode] (1) {1};
\node[cnode] (2) [right=1cm  of 1]{2};
\node[cnode] (3) [right=1cm  of 2]{1};
\node[snode] (4) [below=1cm of 2]{2};
\draw[-] (1) -- (2);
\draw[-] (2)-- (3);
\draw[-] (2)-- (4);
\node[text width=.2cm](5)[above= 0.2 cm of 2] {$W_R$};
\end{tikzpicture}} 
& \begin{tikzpicture}
\node[text width=1cm] (1) at (0,0) {$\longleftarrow$}; 
\node[] (2) at (0,-1){};
\end{tikzpicture} 
& \scalebox{0.6}{\begin{tikzpicture}[
nnode/.style={circle,draw,thick, fill=red},cnode/.style={circle,draw,thick,minimum size=4mm},snode/.style={rectangle,draw,thick,minimum size=6mm}]
\node[nnode] (1) at (1,0) {} ;
\node[nnode] (4) at (7,0) {};
\node[nnode] (7) at (2,3) {};
\node[rectangle,draw,thick, fill=red] (8) at (6,2) {};
\node[text width=0.1] (9) at (1.5, 1) {$k$};
\node[text width=0.1] (10) at (5.5, 1) {2};
\draw[blue, thick,-] (-2, 2) -- (-2, -2);
\draw[blue, thick,-] (0, 2) -- (0, -2);
\draw[blue,thick, -] (8, 2) -- (8, -2);
\draw[blue, thick,-] (10, 2) -- (10, -2);
\draw[thick, -] (0,0.2) -- (8,0.2);
\draw[thick, -] (0,-0.2) -- (8,-0.2);
\draw[thick, -] (8,-0.5) -- (10,-0.5);
\draw[thick, -] (0,-0.5) -- (-2,-0.5);
\draw[black, thick, dotted] (7.west) -- (1.8,0.2);
\draw[black, thick, dotted] (7.east) -- (2.2,-0.2);
\draw[black, thick, dotted] (8.west) -- (5.85,0.2);
\draw[black, thick, dotted] (8.east) -- (6.15,-0.2);
\end{tikzpicture}}
\end{tabular}
\caption{\footnotesize{Type IIB configuration for a vortex defect mapping to the corresponding configuration for a Wilson defect.}}
\label{Fig:HWEx1MirrWV}
\end{center}
\end{figure}
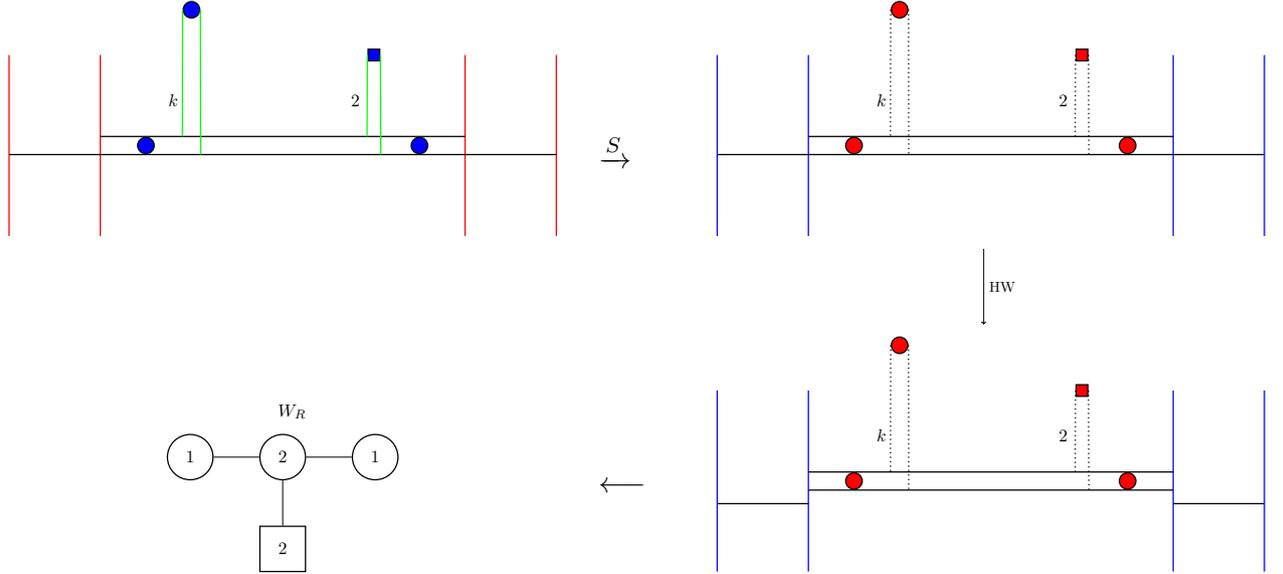

Given the Type IIB construction of the Wilson and the vortex defects in linear quivers, the mirror maps for such defects
can be understood in the standard fashion.
One implements an S-duality operation on a given brane configuration with defects, followed by a 
rotation $\vec x^{7,8,9} \to - \vec {x}^{4,5,6}, \, \vec x^{4,5,6} \to \vec x^{7,8,9}$, and finally a set of 
Hanany-Witten moves to read off the mirror theory with the dual defects.
NS5 and NS5'-branes are exchanged with D5 and D5'-branes under S-duality, while D3-branes are self-dual, as before. 
In addition, F1-strings and D1-branes are exchanged under the operation. As mentioned earlier, the Hanany-Witten 
moves are restricted, since F1-strings cannot cross NS5-branes while D1-branes cannot cross the D5-branes. 
In \figref{Fig:HWEx1MirrWV}, we show a concrete example of a vortex defect in quiver $X$ mapping to a Wilson defect in 
quiver $Y$, where the quivers $X,Y$ are given in \figref{Fig:HWEx1XVLA}. We refer the reader to \cite{Assel:2015oxa} 
for the more general cases.

\subsection{Wilson/vortex defects on $S^3$ and the Localization formulae}\label{LocForm-defects}
The localization formulae for the expectation values of Wilson and vortex defects for 3d $\CN=4$ theories on $S^3$ 
will be our basic tool for performing computations on the QFT side. In this section, we give a brief review of these 
formulae, emphasizing aspects that will be important in the rest of the paper. 
For a detailed study of the $\CN=4$ supersymmetry algebra on $S^3$, and the inequivalent 1d $\CN=4$ subalgebras, 
we refer the reader to \cite{Dedushenko:2016jxl, Dedushenko:2017avn}. The study of Wilson defects for $\CN \geq 2$ theories 
was initiated in \cite{Kapustin:2009kz}, while vortex defects for Abelian $\CN=4$ theories were addressed in 
\cite{Drukker:2012sr, Kapustin:2012iw}, and extended to non-Abelian linear quivers in \cite{Assel:2015oxa}.\\

Let us summarize a few important and relevant features of $\CN=4$ supersymmetry on $S^3$ :
\begin{itemize}

\item The 3d $\CN=4$ supersymmetry algebra on $S^3$ is $\frsu(2|1)_l \oplus \frsu(2|1)_r$. 
The even generators consist of the generators of $\frsu(2)_l \oplus \frsu(2)_r$, which is the Lie algebra of the
isometry group of $S^3$, and the $\fru(1)_l \oplus \fru(1)_r$ generators. 
The $\fru(1)_{l,r}$ generators are given by the linear combinations $R_C \pm R_H$, where $R_{C}, R_{H}$ 
are the Cartan generators of the $\frsu(2)_C, \frsu(2)_H$ Lie algebras respectively. The odd generators have the following anticommutators:
\begin{align}
& \{Q_{l \, \alpha}, \bQ_{l \, \beta} \}=\frac{2}{L}\, \Big[ \s^\mu_{\alpha \beta} J^l_\mu + \frac{\epsilon_{\alpha \beta}}{2}(R_C + R_H) \Big] ,\\
& \{ Q_{r \, \alpha}, \bQ_{r \, \beta}\}= \frac{2}{L}\, \Big[ \s^\mu_{\alpha \beta} J^r_\mu + \frac{\epsilon_{\alpha \beta}}{2}(R_C - R_H) \Big],
\end{align}
where $J^l_\mu, J^r_\mu$ are the $\frsu(2)_l, \frsu(2)_r$ generators respectively. The odd generators reduce to the 
$\CN=4$ Poincare supercharges in the flat space limit $L \to \infty$. 

\item 3d $\CN=4$ theories admit two types of supersymmetry deformation -- hypermultiplet masses and FI parameters. 
Unlike flat space, supersymmetry permits turning on a single scalar in a given background (twisted) vector multiplet on $S^3$, as opposed 
to a triplet. Observables like the partition function with/without defects are therefore written as functions of a set of real variables, one 
for every Cartan generator in the Lie algebra of $G_H \times G_C$.

\item There are two inequivalent subalgebras inside the 3d $\CN=4$ supersymmetry algebra, which correspond to two different 
embeddings of $\frsu(1|1)_l \oplus \frsu(1|1)_r$ inside $\frsu(2|1)_l \oplus \frsu(2|1)_r$. Analogous to the flat space case, these 
subalgebras are exchanged under 3d mirror symmetry. Observables preserving one of the subalgebras is mapped to observables 
preserving the other, under mirror symmetry.

\end{itemize}

\subsubsection*{Bare partition function}
Consider a 3d $\CN = 4$ quiver gauge theory $\CT$ with gauge group $G$ and a Higgs branch global symmetry group $G_{H}$, 
such that the hypermultiplets transform in a representation $\CR$ of $G \times G_{H}$.
The partition function without defects \cite{Kapustin:2010xq} can be written as a matrix integral in terms of a single real scalar $s$  
which lives in the Cartan subalgebra of the gauge group. The real scalar $s$ is the zero-mode associated with 
the real adjoint scalar field that sits inside the 3d $\mathcal{N}=2$ vector multiplet, which in turn sits inside the 3d $\CN=4$ vector multiplet. 
As mentioned earlier, the partition function is a function of the real masses $\vec m$ and real FI parameters $\vec \eta$, 
which are real scalars inside background 3d $\CN=4$ vector and twisted vector multiplets respectively. 
The matrix integral assumes the following form: 
\begin{align}\label{PF-main}
Z^{(\CT)}(\vec m; \vec \eta)=  \int  \Big[d\vec s\Big] \, Z_{\rm int}(\vec s, \vec m, \vec \eta) = : \int  \Big[d\vec s\Big] \, Z_{\rm FI}(\vec s, \vec \eta)\, Z^{\rm vector}_{\rm{1-loop}}(\vec s)\,Z^{\rm hyper}_{\rm{1-loop}}(\vec s, \vec m),
\end{align}
where the measure $\Big[d\vec s\Big]=\frac{d^k \vec{s}}{|{W}(G)|}$, and $|{W}(G)|$ is the order of the Weyl group of $G$. 
The individual terms in the integrand on the RHS consist of classical and one-loop contributions, and are given as follows. 
\begin{itemize}

\item The classical contribution arises from the $l$ $U(1)$ factors in the gauge group:
\begin{equation} \label{PF-main-FI}
Z_{\rm FI}(\vec s, \vec \eta) = \prod^l_{\gamma=1} e^{2\pi i \eta_\gamma \, \text{Tr}(\vec{s}^\gamma)}\,,
\end{equation}
where $\gamma$ runs over the $l$ gauge nodes with a $U(1)$ factor and $\eta_\gamma$ is the associated FI parameter.

\item The  $\mathcal{N}=4$ vector multiplet contributes a one-loop term:
\begin{equation}\label{PF-main-vec}
Z^{\rm vector}_{\text{1-loop}}(\vec s)=\prod_{\alpha} \sinh{\pi \alpha(\vec s)}\,,
\end{equation}
where the product extends over the roots of the Lie algebra of $G$. In fact, this is precisely the contribution of an $\mathcal{N}=2$ vector multiplet since contribution of the adjoint chiral which is part of the $\mathcal{N}=4$ vector multiplet is trivial \cite{Kapustin:2010xq}.

\item The one-loop contribution from $\mathcal{N}=4$ hypermultiplets transforming in a representation  $\CR$ of $G \times G_{H}$ :
\begin{equation} \label{PF-main-hyper}
Z^{\rm hyper}_{\text{1-loop}}(\vec s, \vec m)=\prod_{\rho(\CR)} \frac{1}{\cosh{\pi \rho(\vec s, \vec m)}}\,,
\end{equation}
where the product extends over the weights of the representation $\CR$.
\end{itemize}

Given a mirror pair of theories $X$ and $Y$, the partition functions of the theories are related as:
\begin{align}
&Z^{(X)}(\vec m ; \vec \eta) = C_{XY}(\vec m, \vec \eta) \, Z^{(Y)}(\vec m'(\vec \eta) ; \vec \eta' (\vec m)), \\
& C_{XY}(\vec m, \vec \eta)= e^{2\pi i \sum_{k,l}\, a^{kl} m_k \eta_l}, \label{def-CXY}
\end{align}
where $(\vec m, \vec \eta)$, $(\vec m', \vec \eta' )$ are the masses and FI parameters for $X$ and $Y$ respectively. 
The mass parameters $\vec m'$ and the FI parameters $\vec \eta'$ of $Y$ are linear functions of the FI parameters  
and the mass parameters of $X$ respectively. The partition functions agree up to an overall phase factor, whose 
exponent is linear in $\vec m$ and $\vec \eta$ with integer coefficients $a^{kl}$. These phase factors arise from the 
three-dimensional contact terms as discussed in \cite{Closset:2012vg}. For the special case where $(X,Y)$ are linear 
quivers with unitary gauge groups, the mirror map simply exchanges the mass and FI parameters up to a sign, and 
the partition functions are related as
\be \label{pf-LQmirror}
Z^{(X)}(\vec m; \vec t)= e^{2\pi i a^{kl} m_k t_l} \, Z^{(Y)}(\vec t; -\vec m),
\ee
where the parameters $t$ and $m$ are as defined in \Secref{LQ2B}.

\subsubsection*{Partition function with Wilson defects}
A Wilson defect in a representation $R$ of the gauge group $G$ can be defined as:
\be
W_R = \tr_R \Big(P\, \exp{\oint_\gamma} \, (iA_\mu \dot{x}^\mu - |\dot{x}| \s)\,d\tau \Big),
\ee
where $\gamma$ is a curve on $S^3$ and $\tau$ is a coordinate along $\gamma$. The curve $\gamma$, 
if chosen to wrap a Hopf fiber at the north pole of the $S^2$ base of $S^3$, gives a half-BPS Wilson defect 
\cite{Assel:2015oxa}. The preserved subalgebra corresponds to a certain embedding of $\frsu(1|1)_l \oplus \frsu(1|1)_r$ 
inside $\frsu(2|1)_l \oplus \frsu(2|1)_r$. The partition function of the theory $\CT$ with a half-BPS Wilson defect $W_R$ 
inserted along the curve $\gamma \in S^3$, is given as
\begin{align} \label{PF-main-defect}
& Z^{(\CT[W_R])}(\vec m; \vec \eta)
=\int  \Big[d\vec s\Big] \, Z_{\rm Wilson}(\vec s, R)\, Z_{\rm FI}(\vec s, \vec \eta)\, 
Z^{\rm vector}_{\rm{1-loop}}(\vec s)\,Z^{\rm hyper}_{\rm{1-loop}}(\vec s, \vec m), \\
& Z_{\rm Wilson}(\vec s, R)= \text{Tr}_{R} \Big(e^{2\pi \vec s}\Big) = \sum_{w \in R} e^{2\pi w\cdot \vec s},
\end{align}
where $w$ are the weights of the representation $R$, and the functions $Z_{\rm FI}$, $Z^{\rm vector}_{\rm{1-loop}}$, 
$Z^{\rm hyper}_{\rm{1-loop}}$ are defined above. The expectation value of the defect is then given as
\be
\langle W_R \rangle_{\CT}(\vec m; \vec \eta)
=\frac{1}{Z^{(\CT)}(\vec m; \vec \eta)}\, \int  \Big[d\vec s\Big] \, Z_{\rm Wilson}(\vec s, R)\, Z_{\rm FI}(\vec s, \vec \eta)\, 
Z^{\rm vector}_{\rm{1-loop}}(\vec s)\,Z^{\rm hyper}_{\rm{1-loop}}(\vec s, \vec m).
\ee

The matrix integral for a generic Wilson defect is divergent, 
with the divergence arising from the region(s) $s_i \to \pm \infty$. The integral can be regularized by deforming the contour 
of integration for each eigenvalue $s_i$ as follows. Note that the matrix integral poles, which exclusively come from the 
function $Z^{\rm hyper}_{\rm{1-loop}}$, are located in the upper and lower half-planes of the complex variable $s_i$, 
along lines parallel to the imaginary axis. For $\eta >0$ ($\eta <0$), the contour is deformed such that it encloses all 
the poles of the integrand in the upper (lower) half-plane, and moves off the real axis along an imaginary line at a 
finite point (see figure 37 of \cite{Assel:2015oxa} for a schematic version of such a contour). 

\subsubsection*{Partition function with vortex defects}

Half-BPS vortex defects can be defined as disorder operators in a 3d $\CN=4$ theory on $S^3$ \cite{Drukker:2012sr}, 
where the defect is supported on a Hopf fiber at the north/south pole of the $S^2$ base of $S^3$. The preserved supersymmetry corresponds 
to an inequivalent embedding of $\frsu(1|1)_l \oplus \frsu(1|1)_r$ inside $\frsu(2|1)_l \oplus \frsu(2|1)_r$, compared to the 
Wilson defect.\\

In order to realize a vortex defect as a coupled 3d-1d quiver with the SQM living on the Hopf fiber, 
one has to first specify which class of 1d theories preserves the same supersymmetry as that of a vortex defect. 
It was shown in \cite{Assel:2015oxa} that :
\begin{itemize}
\item The 3d half-BPS vortex subalgebra can be written as a 1d (2,2) 
supersymmetry algebra deformed by background gauge fields for $J_-$ and a $U(1)$ flavor symmetry $G_F$, 
where $J_\pm$ are the Cartan generators of the Lie algebra of the 1d R-symmetry. 
The 1d generators can be related to the 3d generators as 
\begin{align}
& J_- =: J^l_3 + J^r_3 -R_C,\\
& G_F=: J^l_3 + J^r_3 -R_H.
\end{align}

\item The identification works only for special values of the associated background vector multiplet fields:
\begin{align}
& J_- :  a_-=-i/L, \quad m_-=0,\label{bg-J}\\
& G_F: a_F=0, \quad m_F= 1/L. \label{bg-F}
\end{align}

\end{itemize}

Given this identification, one can construct a generic coupled 3d-1d quiver which preserves the appropriate supersymmetry algebra. 
The general procedure involves identifying 1d flavor symmetries (other than $G_F$) with 3d flavors symmetries and gauging 
them with dynamical or background 3d vector multiplets. At the level of the UV Lagrangian, this identification is imposed by turning on 
cubic superpotentials of the form \eref{3d-1dSup}, involving 1d chiral multiplets and 3d half-hypermultiplets. The SQM will have 
additional superpotential couplings consistent with (2,2) supersymmetry. Given this 3d-1d Lagrangian, one can now proceed to 
compute its partition function on $S^3$, by performing a localization analysis \cite{Assel:2015oxa}. 
The answer passes various consistency checks, and agrees with the known dualities from String Theory. 
Below, we summarize the main steps of the recipe, and refer the reader to \cite{Assel:2015oxa} for further details.
\begin{itemize}
\item \textbf{1d Witten index:} The first step is to compute the twisted partition function (Witten index) of the SQM $\Sigma$
with a gauge group $G_{1d}$ and a flavor symmetry $G_F \times G'_F$. The twisted Witten index is formally defined as,
\be
\CI^\Sigma (\vec \frm, z, \mu| \vec \xi) = \tr_{\CH} (-1)^F e^{2\pi i z J_-}\, e^{2\pi i \mu G_F}\, e^{2\pi i \frm_i J_i},
\ee
where we have isolated the $J_-$ and $G_F$ generators with complex chemical potentials $z=L(i a_- + m_-)$ and $\mu = L(i a_F + m_F)$ respectively. The remaining flavor symmetry generators are collectively denoted as $\{J_i\}$, with complex chemical potentials $\{\frm_i\}$.
The parameters $\vec \xi$ are the FI parameters associated with unitary factors in $G_{1d}$. On a circle, the path integral can be worked 
out using localization and can be written in terms of JK-residue formula \cite{Hori:2014tda}, i.e.
\begin{align}
& \CI^\Sigma(\vec \frm, z, \mu| \vec \xi) = \oint_{JK-\vec \xi} \, \prod^{{\rm{rank}}(G_{1d})}_{I=1}\, dx_I \, g_{\rm vector} (\vec x, z) 
\cdot g_{\rm chiral} (\vec x, \vec \frm, \mu, z),\\
& g_{\rm vector} (\vec x, z)= \Big(\frac{\pi}{\sh{z}}\Big)^{{\rm{rank}}(G_{1d})}\, \prod_{\alpha} \frac{\sh{(-\alpha(x))}}{\sh{(\alpha(x) -z)}} ,\\
& g_{\rm chiral} (\vec x, \vec \frm, \mu, z)= \prod_{\rho \in \CR} \frac{\sh{[-(\rho(\vec x, \vec{\frm}) + q\mu + (\frac{r}{2} -1)z)]}}{\sh{(\rho(\vec x, \vec{m})+ q\mu+\frac{r}{2} z)}}.
\end{align}
In the above formula, $\alpha$ is a root of the Lie algebra of $G_{1d}$, $\rho$ is a weight of the representation 
$\CR$ of $G_{1d} \times G'_{F}$ in which the chiral multiplet transforms, $q$ is the $G_F$ charge and $r$ is the charge under 
$J_{-}$. The $G_F$ and $R$-charges of the chiral multiplets are subject to various superpotential constraints. 
The convention for the JK residue for a meromorphic $k$-form $g(\vec{x}, \vec \frm, \mu, z)$ is given as:
\begin{align}
{\rm JK-Res}_{\xi}[x^*]g(\vec{x}, \vec \frm, \mu, z) = \begin{cases}
|{\rm Res}[x^*]g(\vec{x}, \vec \frm, \mu, z)| ,& \text{if } - \xi \in C(w^I),\\
0, & \text{otherwise},
\end{cases}
\end{align}
where a given set of poles $x^*= \{ x^*_I\}$ is given by the intersection of $k$ hyperplanes in $\BC^k$, 
parametrized by the equations $w^{(I)}\cdot x^* + \vec \tq^{(I)} \cdot \vec\frm + q^{(I)}\cdot \mu + \frac{r^{(I)}}{2} z=0$, with $I=1,\ldots,k$.
$C(w^I)$ is a positive cone spanned by the $k$ weight vectors: $C(w^I) = \{ \sum^k_{I} c_I w^{(I)} | c_I >0 \}$, 
and $\xi = \eta(1,1,\ldots,1)$ should be treated as a vector in $\BC^k$.

\item \textbf{3d-1d matrix model:} The next step is to write the combined partition function of the SQM $\Sigma$ coupled 
with a 3d quiver gauge theory $\CT$, as a matrix model. At the level of the matrix model, identification of 1d flavor symmetries 
with 3d flavor symmetries can be performed in two steps. Firstly, one restricts the complex chemical potentials $\vec \frm$ 
to take real values, and secondly, one identifies those real scalars with appropriate real scalars from the background 
vector multiplet associated with the 3d flavor symmetry. Finally, one can gauge a subgroup of the flavor symmetry by 
integrating over some of the scalars with the right measure. This reasoning leads to the following expression for the 
3d-1d partition function on $S^3$:
\begin{align}\label{Z-3d1d-gen1}
Z^{\CT[\Sigma]}(\vec m; \vec \eta) = \lim_{\substack{z\to 1 \\ \mu \to 1}} \int  \Big[d\vec s\Big] \, Z^{(\CT)}_{\rm int}(\vec s, \vec m, \vec \eta) \cdot 
\CI^{\Sigma}(\vec s, \vec m, \mu, z| \vec \xi),
\end{align}
where the limits on $z$ and $\mu$ arise from the fact that one needs special values of the $J_-$ and $G_F$ background gauge 
fields to preserve supersymmetry, as given in \eref{bg-J}-\eref{bg-F}. 

\item \textbf{Analytic continuation and $z \to 1$ limit :} The final step of the recipe is to give a prescription about how the limits on the RHS 
of \eref{Z-3d1d-gen1} should be taken. It turns out that the $\mu \to 1$ can be taken in a straightforward fashion. However, for $z$, the 
correct prescription for taking the limit, is to first analytically continue $z \in i \BR$ in the integrand, perform the integration, and then 
finally set $z=1$. 
One can then write down the final form of the 3d-1d partition function:
\begin{align}\label{Z-3d1d-gen2}
Z^{\CT[\Sigma]}(\vec m; \vec \eta) = \lim_{z\to 1} \int  \Big[d\vec s\Big] \, Z^{(\CT)}_{\rm int}(\vec s, \vec m, \vec \eta) \cdot 
\CI^{\Sigma}(\vec s, \vec m, z| \vec \xi),
\end{align}
where the Witten index $\CI^{\Sigma}(\vec s, \vec m, z| \vec \xi) = \lim_{\mu \to 1} \CI^{\Sigma}(\vec s, \vec m, \mu, z| \vec \xi)$. 
While doing an actual computation, we will only retain the $z$-dependence of the matrix model integrand in places where it changes the location 
of a pole. The $z \to 1$ limit can be taken trivially everywhere else.

\end{itemize}

Given the above recipe, one can compute the partition function (and therefore the expectation value) of a vortex defect 
inserted in a 3d theory on $S^3$. The only input needed to perform this exercise is the precise 3d-1d quiver that corresponds 
to a given vortex defect, including the chamber in which the Witten index should be computed. 
Assuming that the 3d-1d quiver which realizes a vortex defect $V_{\CD,R}$ (with 
$R$ being a representation of the 3d gauge group $G$ and $\CD$ being some additional data) is known, the expectation value of the defect is given by the general formula:
\begin{align}\label{vev-V-gen}
\boxed{\langle{V_{\CD, R}}\rangle_{\CT}= W_{\rm b.g.}(\vec \eta) \times \frac{1}{Z^{(\CT)}(\vec m; \vec \eta)} \times \lim_{z\to 1} \int  \Big[d\vec s\Big] \, Z^{(\CT)}_{\rm int}(\vec s, \vec m, \vec \eta) \, \CI^{\Sigma^{\CD,R}}(\vec s, \vec m, z| \vec \xi),}
\end{align}
where $W_{\rm b.g.}(\vec \eta)$ are Wilson defects for 3d topological symmetries that can also be turned on.\\

\begin{figure}[htbp]
\begin{center}
\begin{tabular}{ccc}
\scalebox{0.8}{\begin{tikzpicture}[node distance=2cm, nnode/.style={circle,draw,thick, fill, inner sep=1 pt},cnode/.style={circle,draw,thick,minimum size=1.0 cm},snode/.style={rectangle,draw,thick,minimum size=1.0 cm},rnode/.style={red, circle,draw,thick,fill=red!30 ,minimum size=1.0cm}]
\node[nnode] (1) at (-1.5,0){} ;
\node[nnode] (2) at (-1,0){} ;
\node[](3) at (-0.5,0){};
\node[cf-group] (4) at (1,0) {\rotatebox{-90}{$P$}};
\node[cf-group] (5) at (3,0) {\rotatebox{-90}{$N$}};
\node[cf-group] (6) at (5,0) {\rotatebox{-90}{$M$}};
\node[snode] (12) at (2,-1.5) {$K-Q$};
\node[snode] (13) at (4,-1.5) {$Q$};
\node[rnode] (7) at (4,2) {$\Sigma^{Q,R}$};
\draw[thick,-] (3) -- (4);
\draw[thick,-] (4) -- (5);
\draw[thick, -] (5) to (6);
\draw[thick, -] (5) to (12);
\draw[thick, -] (5) to (13);
\draw[red, thick,->] (5) -- (7);
\draw[red, thick,->] (7) -- (6);
\draw[red, thick,->] (7) -- (13);
\node[] (9) at (6.5,0){} ;
\node[nnode] (10) at (7,0){} ;
\node[nnode] (11) at (7.5,0){} ;
\draw[-] (6) -- (9);
\node[text width=0.1cm](20)[below=2.2 cm of 5]{$(\CT[\Sigma^{Q,R}_r])$};
\end{tikzpicture}}
& \qquad \qquad
& \scalebox{0.8}{\begin{tikzpicture}[node distance=2cm, nnode/.style={circle,draw,thick, fill, inner sep=1 pt},cnode/.style={circle,draw,thick,minimum size=1.0 cm},snode/.style={rectangle,draw,thick,minimum size=1.0 cm},rnode/.style={red, circle,draw,thick,fill=red!30 ,minimum size=1.0cm}]
\node[nnode] (2) at (0.5,0){} ;
\node[nnode] (3) at (1,0){} ;
\node[](4) at (1.5,0){};
\node[cf-group] (5) at (3,0) {\rotatebox{-90}{$P$}};
\node[cf-group] (6) at (5,0) {\rotatebox{-90}{$N$}};
\node[cf-group] (8) at (7,0) {\rotatebox{-90}{$M$}};
\node[snode] (12) at (4,-1.5) {$K-Q$};
\node[snode] (13) at (6,-1.5) {$Q$};
\node[rnode] (7) at (4,2) {$\Sigma^{Q,R}$};
\draw[thick,-] (4) -- (5);
\draw[thick, -] (5) to (6);
\draw[thick, -] (8) to (6);
\draw[thick, -] (6) to (12);
\draw[thick, -] (6) to (13);
\draw[red, thick,->] (5) -- (7);
\draw[red, thick,->] (7) -- (6);
\draw[red, thick,->] (12) -- (7);
\node[] (9) at (8.5,0){} ;
\node[nnode] (10) at (9,0){} ;
\node[nnode] (11) at (9.5,0){} ;
\draw[-] (8) -- (9);
\node[text width=0.1cm](20)[below=2.2 cm of 6]{$(\CT[\Sigma^{Q,R}_l])$};
\end{tikzpicture}}
\end{tabular}
\caption{\footnotesize{The right SQM $\Sigma^{Q,R}_r$ and the left SQM $\Sigma^{Q,R}_l$ coupled to a 3d linear quiver, 
which realize a vortex defect $V_{Q, R}$ of the gauge group $U(N)$. Note that $\Sigma^{Q,R}_r$ and $\Sigma^{Q,R}_l$ 
only differ in the numbers of fundamental and anti-fundamental chiral multiplets for the terminal gauge node.}}
\label{LR-1d-gen}
\end{center}
\end{figure}
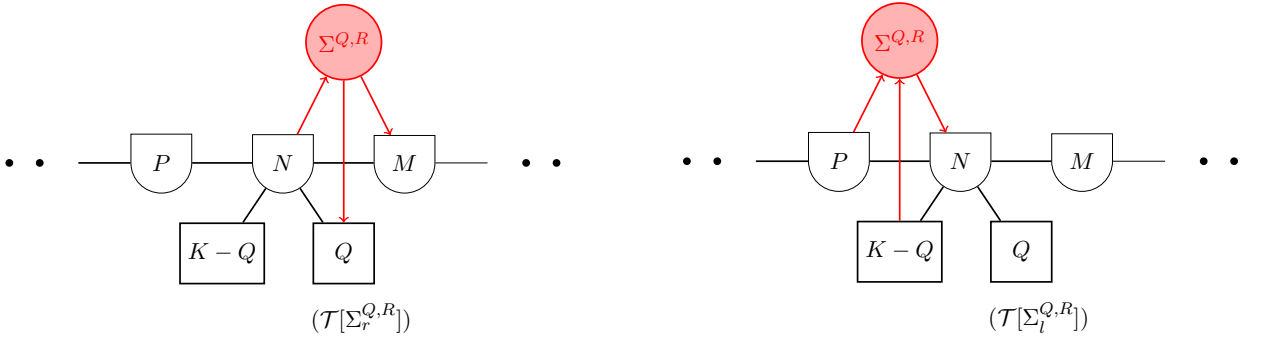

Using the Type IIB construction, reviewed in \Secref{LQ2B-main}, it is possible to 
obtain the coupled 3d-1d quivers which realize the vortex defects for linear quivers with unitary gauge groups. The general form of the 
1d quivers that arise from this construction is shown in \figref{1dquiv-gen} and the associated 3d-1d quivers are of the generic form 
in \figref{3d1dquiv-gen}. As we noted in our discussion in \Secref{LQ2BD}, there are at least two different 3d-1d quivers that can describe 
the same vortex defect in a linear quiver, involving the ``right" and the ``left" SQM. A generic example of this hopping duality 
is shown in \figref{LR-1d-gen}. 

Therefore, for a linear quiver $\CT$, one can explicitly write down the expectation values of the vortex defect in a representation $R$ of the 
gauge group $U(N) \subset G$:
\begin{align}\label{vev-V-LQ}
\langle{V^{l,r}_{Q,R}}\rangle_{\CT}= W^{l,r}_{\rm b.g.}(\vec t) \times \frac{1}{Z^{(\CT)}(\vec m; \vec t)} \times \lim_{z\to 1} \int  \Big[d\vec s\Big] \, Z^{(\CT)}_{\rm int}(\vec s, \vec m, \vec t) \cdot \CI^{\Sigma^{Q,R}_{l,r}}(\vec s, \vec m, z| \vec \xi \gtrless 0),
\end{align}
where the superscripts $l,r$ denote the choice of the specific 3d-1d system (left or right) that realizes the vortex defect. Note that 
the left/right SQM realization comes with a prescription for the signs of the FI parameters, which determines the specific chamber 
in which the Witten index should be evaluated. As discussed in \Secref{LQ2BD}, this piece of information can also be read off from 
the Type IIB construction. The Witten indices for the right and the left SQM associated with the vortex defect in \figref{LR-1d-gen} 
have the general expression
\begin{align}
& \CI^{\Sigma^{Q,R}_{r}}=\sum_{w \in R}\, \CF^r(\vec s, z)\, \prod^N_{j=1} \prod^M_{i=1} \frac{\ch{(s^{(N)}_j - s^{(M)}_i)}}{\ch{(s^{(N)}_j  + iw_jz -s^{(M)}_i)}}\,\prod^Q_{a=1}\frac{\ch{(s^{(N)}_j - m_a)}}{\ch{(s^{(N)}_j  + iw_jz -m_a)}} , \label{Z1d-r}\\
& \CI^{\Sigma^{Q,R}_{l}}= \sum_{w \in R}\, \CF^l(\vec s, z)\, \prod^N_{j=1} \prod^P_{l=1} \frac{\ch{(s^{(N)}_j - s^{(P)}_l)}}{\ch{(s^{(N)}_j  - iw_jz -s^{(P)}_l)}}\,\prod^{K-Q}_{b=1}\frac{\ch{(s^{(N)}_j - m_b)}}{\ch{(s^{(N)}_j  - iw_jz -m_b)}} ,\label{Z1d-l}
\end{align}
where $w= (w_1, \ldots, w_N)$ are the weights of the representation $R$. Subgroups of the $U(Q)$ and/or $U(K-Q)$ flavor symmetry groups 
can be promoted to gauge groups, which would amount to integrating over a subset of the $\{m_a\}$ and/or the $\{m_b\}$ parameters with 
appropriate measure in the matrix model. The functions $\CF^r(\vec s, z), \CF^l(\vec s, z)$ have poles which give zero residues in the limit 
$z \to 1$, and therefore one can take the limit for this term before performing the integral. Therefore, one can drop 
these functions in the formulae for Witten indices in the rest of the paper\footnote{For the precise forms of these functions, see Appendix B of 
\cite{Assel:2015oxa}.}.

Finally, the background Wilson defects for the left and the right SQM are given by
\be
W^{r}_{\rm b.g.}(\vec t)= e^{2\pi |R| t_r}, \qquad W^{l}_{\rm b.g.}(\vec t)= e^{2\pi |R| t_l},
\ee
where the FI parameter of the $U(N)$ gauge group is given by $\eta^{(N)}= t_{l} -t_r$.  
The integer $|R|$ is the number of boxes in the Young diagram associated with the representation $R$.\\

At the level of the matrix model, the hopping duals are related by a simple change of variables. Starting 
from the partition function of the right 3d-1d quiver, given by \eref{vev-V-LQ}-\eref{Z1d-r}, one can obtain the partition 
function of the left 3d-1d quiver, given by \eref{vev-V-LQ}-\eref{Z1d-l}, by the following transformation:
\be
s^{(N)}_j \to s^{(N)}_j - iw_j\,z ,\qquad \forall j =1,\ldots, N,
\ee
keeping all the other integration variables fixed. Since we have analytically continued $z \in i\BR$, this operation will 
not introduce or omit any poles. 

\subsubsection*{Mirror Map}
Given that 3d mirror symmetry exchanges the 1d supersymmetry subalgebras SQM$_A$ and SQM$_B$, one 
expects the half-BPS vortex defects to map to the half-BPS Wilson defects. The precise map, however, is 
difficult to determine for a generic pair of mirror duals. For linear quivers, the Type IIB construction allows one 
to explicitly construct these mirror maps, which could be independently checked by the sphere partition functions 
discussed above. Given a pair of linear quiver theories $X$ and $Y$, the vortex-to-Wilson mirror map has the 
schematic form
\be
\langle V^{r,l}_{Q,R} \rangle_X (\vec m; \vec t) = \langle \wt{W}_{R'} \rangle_Y (\vec t; -\vec m),
\ee
where $\wt{W}_{R'}$ is Wilson defect in representation $R'$ of the gauge and flavor symmetry groups of the 
theory $Y$. An explicit mirror map for defects in the dual linear quiver pair of \figref{fig: LQEx1} is discussed in 
\Appref{U24-V2W}. Similarly, a Wilson-to-vortex mirror map has the schematic form
\be
\langle W_{R} \rangle_{X}(\vec m; \vec t) = \sum_i \langle \wt{W}^{(i)}_{\rm b.g.} \cdot \wt{V}^{r,l}_{Q_i,R_i} \rangle_{Y}(-\vec t; \vec m),
\ee
where the RHS is a sum over vortex defects in the theory $Y$, combined with Wilson defects for 3d topological symmetries.
An explicit mirror map of this type is discussed in \Appref{U24-W2V} for the dual pair in \figref{fig: LQEx1}.

\section{Elementary $S$-type operations and mirror map of line defects}\label{SOps-def}
In this section, we discuss the construction of half-BPS defects and their mirror maps in theories beyond linear quivers, 
using the elementary $S$-type operations introduced in \cite{Dey:2020hfe}. In addition, we present a systematic procedure 
to compute the associated mirror maps of line defects. 

\subsection{Elementary $S$-type operations on theories without defects: Review}\label{Rev-SOps}
Consider a pair of mirror dual quiver gauge theories $(X,Y)$ where $X$ belongs to class $\CU$, 
i.e. a class of 3d CFTs with a UV Lagrangian that has a manifest global symmetry 
subgroup $G^{\rm sub}_{\rm global}=\prod_\gamma U(M_\gamma)$. 
We will additionally assume that $X$ and $Y$ are good quivers in the Gaiotto-Witten sense \cite{Gaiotto:2008ak}. 
Given such a quiver $X$, one can define a set of four basic operations \cite{Dey:2020hfe}:
\begin{enumerate}
\item {\bf{Gauging ($G^\alpha_\CP$):}} A \textit{gauging operation} $G^\alpha_\CP$ at a flavor node $\alpha$ of the theory $X$ (shown in the schematic figure below)
involves the following two steps :
\begin{itemize}
\item The flavor node $U(M_\alpha)$ is split into two flavor nodes, corresponding to a $U(r_\alpha) \times U(M_\alpha - r_\alpha)$ global symmetry. 
The $U(1)^{M_\alpha}$ background vector multiplets are identified as the $U(1)^{r_\alpha} \times U(1)^{M_\alpha-r_\alpha}$ background vector multiplets 
by the following map:
\be \label{uvdef0}
\overrightarrow{m^\alpha}_{i_\alpha} = \CP_{i_\alpha i} \, \overrightarrow{u}^\alpha_i + \CP_{i_\alpha \, r_\alpha + j} \, \overrightarrow{v}^\alpha_j, \quad( i_\alpha=1,\ldots, M_\alpha,\quad i=1,\ldots, r_\alpha, \quad j=1,\ldots, M_\alpha - r_\alpha),
\ee
where any such map is parametrized by a choice of $\CP$ -- a permutation matrix of order $M_\alpha$. The resultant theory deformed by the $U(r_\alpha) \times U(M_\alpha - r_\alpha)$ mass parameters will be denoted as $(X, \CP)$.

\item Given the theory $(X, \CP)$, the flavor symmetry node $U(r_\alpha) $ is promoted to a gauge node, i.e. the vector multiplets for $U(r_\alpha)$ 
are now dynamical, and a background twisted vector multiplet for the $U(1)_J$ topological symmetry is turned on.
\end{itemize}

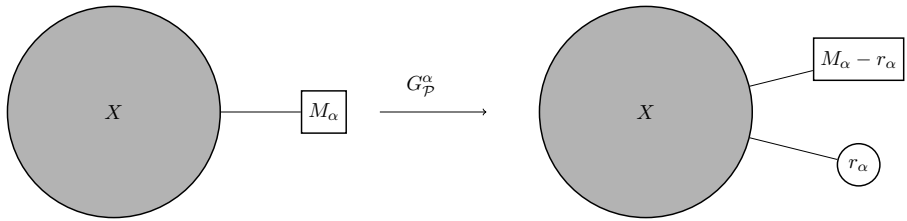
\begin{center}
\scalebox{0.7}{\begin{tikzpicture}[
cnode/.style={circle,draw,thick,minimum size=4mm},snode/.style={rectangle,draw,thick,minimum size=8mm},pnode/.style={rectangle,red,draw,thick,minimum size=1.0cm}, bnode/.style={circle,draw, thick, fill=black!30,minimum size=4cm}]
\node[bnode] (1) at (0,0){$X$} ;
\node[snode] (2) [right=1.5cm  of 1]{$M_\alpha$} ;
\draw[-] (1)--(2);
\draw[->] (5,0) -- (7,0); 
\node[bnode] (3) at (10,0){$X$} ;
\node[snode] (4) at (14, 1){$M_\alpha- r_\alpha$} ;
\node[cnode] (5) at (14,-1){$r_\alpha$} ;
\draw[-] (3)--(4);
\draw[-] (3)--(5);
\node[text width=1cm](6) at (6,0.5){$G^\alpha_\CP$};
\end{tikzpicture}}
\end{center}

\item {\bf{Flavoring ($F^\alpha_\CP$):}} A \textit{flavoring operation} $F^\alpha_\CP$ at a flavor node $\alpha$ of the quiver gauge theory $X$ (shown schematically in the figure below) involves the following two steps:
\begin{itemize}
\item Firstly, the theory $(X, \CP)$ is defined as above.

\item Given the theory $(X, \CP)$, a flavor node labelled by ${G}^\alpha_{ F}$ is attached to the flavor node $U(r_\alpha)$, as shown on the RHS of the figure below. 
This introduces some free hypermultiplets in the theory which transform under some representation of the global symmetry group $U(r_\alpha) \times {G}^\alpha_{ F}$.
\end{itemize}

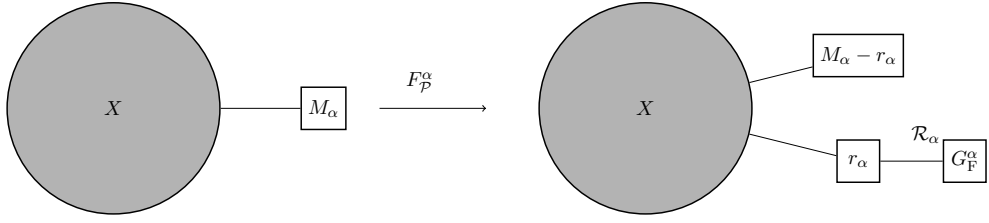
\begin{center}
\scalebox{0.7}{\begin{tikzpicture}[
cnode/.style={circle,draw,thick,minimum size=4mm},snode/.style={rectangle,draw,thick,minimum size=8mm},pnode/.style={rectangle,red,draw,thick,minimum size=1.0cm}, bnode/.style={circle,draw, thick, fill=black!30,minimum size=4cm}]
\node[bnode] (1) at (0,0){$X$} ;
\node[snode] (2) [right=1.5cm  of 1]{$M_\alpha$} ;
\draw[-] (1)--(2);
\draw[->] (5,0) -- (7,0); 
\node[bnode] (3) at (10,0){$X$} ;
\node[snode] (4) at (14, 1){$M_\alpha- r_\alpha$} ;
\node[snode] (5) at (14,-1){$r_\alpha$} ;
\node[snode] (6) at (16,-1){$G^\alpha_{\rm F}$} ;
\draw[-] (3)--(4);
\draw[-] (3)--(5);
\draw[-] (5)--(6);
\node[text width=1cm](10) at (6,0.5){$F^\alpha_\CP$};
\node[text width=1cm](11) at (15.5, - 0.5){$\CR_\alpha$};
\end{tikzpicture}}
\end{center}

\item \textbf{Identification($I^\alpha_{\vec \CP}$):} Given the quiver gauge theory $X$ with a flavor symmetry subgroup $\prod^L_{\gamma=1}U(M_\gamma)$,
let  $N^{\beta}_{p,r_\alpha}$ denote a set of (not necessarily consecutive) $p \leq L$ flavor nodes labelled by $\beta$, where $\beta=\gamma_1, \ldots, \gamma_p$.
The parameter $r_\alpha$ is a positive integer such that $r_\alpha \leq {\rm{Min}}(\{M_\beta\}| \beta \in N^{\beta}_{p,r_\alpha})$. Let $\alpha$ be a chosen node in 
$N^{\beta}_{p,r_\alpha}$. The identification operation $I^\alpha_{\vec \CP}$, shown in a figure below for $\beta=2$, can be performed at a chosen 
node $\alpha$ in two steps:
\begin{itemize}

\item For all $\beta \in N^{\beta}_{p,r_\alpha}$, we split the corresponding flavor node $U(M_\beta) \to U(r_\alpha)_\beta \times U(M_\beta -r_\alpha)$ symmetry.  
For a given $\beta$, the choice of the $U(1)^{r_\alpha} \times U(1)^{M_\beta-r_\alpha}$ background vector multiplets in terms of the original $U(1)^{M_\beta}$ background vector multiplets can be encoded in the map: 
\be \label{uvdef1}
\overrightarrow{m^\beta}_{i_\beta} = \CP_{i_\beta i} \, \overrightarrow{u}^\beta_i + \CP_{i_\beta \, r_\alpha + j} \, \overrightarrow{v}^\beta_j, \quad ( i_\beta=1,\ldots, M_\beta,\quad i=1,\ldots, r_\alpha, \quad j=1,\ldots, M_\beta - r_\alpha),
\ee
where each such choice is parametrized by a permutation matrix $\CP_\beta$ of order $M_\beta$. We denote the resultant theory as $(X, \{\CP_\beta \})$.

\item Given the theory $(X, \{\CP_\beta \})$, we identify the flavor nodes $U(r_\alpha)_\beta$ for all $\beta \neq \alpha$ to the flavor node $U(r_\alpha)_\alpha$, as shown on the RHS of the figure below.
\end{itemize}

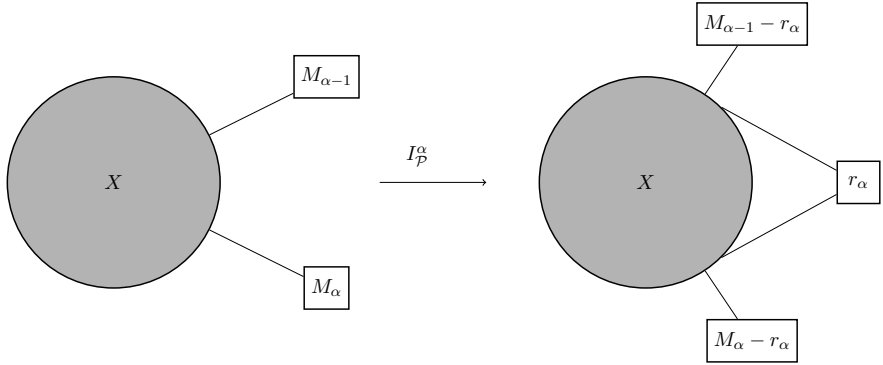
\begin{center}
\scalebox{0.7}{\begin{tikzpicture}[
cnode/.style={circle,draw,thick,minimum size=4mm},snode/.style={rectangle,draw,thick,minimum size=8mm},pnode/.style={rectangle,red,draw,thick,minimum size=1.0cm}, bnode/.style={circle,draw, thick, fill=black!30,minimum size=4cm}]
\node[bnode] (1) at (0,0){$X$} ;
\node[snode] (2) at (4,-2) {$M_\alpha$} ;
\node[snode] (3) at (4, 2) {$M_{\alpha -1}$} ;
\draw[-] (1)--(2);
\draw[-] (1)--(3);
\draw[->] (5,0) -- (7,0); 
\node[bnode] (3) at (10,0){$X$} ;
\node[snode] (4) at (12, 3){$M_{\alpha-1}- r_\alpha$} ;
\node[snode] (5) at (14,0){$r_\alpha$} ;
\node[snode] (6) at (12,-3){$M_\alpha- r_\alpha$} ;
\draw[-] (3)--(4);
\draw[-] (3.north east) to  (5);
\draw[-] (3.south east) to (5);
\draw[-] (3)--(6);
\node[text width=1cm](6) at (6,0.5){$I^\alpha_\CP$};
\end{tikzpicture}}
\end{center}

\item {\bf{Defect ($D^\alpha_\CP$):}} Given a $U(M_\alpha)$ flavor node of a quiver gauge theory $X$, the operation $D^\alpha_\CP$ can be implemented 
at the node $\alpha$ in two steps:
\begin{itemize}

\item The theory $(X, \CP)$ is defined as above.

\item A Type-A or Type-B defect is turned on for the flavor node $U(r_\alpha)$.

\end{itemize}

\end{enumerate}

\textbf{Definition.} An elementary $S$-type operation $\CO^\alpha_{\vec \CP}$ on $X$ at a flavor node $\alpha$, is defined by the action of any possible combination of the identification ($I^\alpha_{\vec \CP}$), the  flavoring ($F^\alpha_{\CP}$), and the defect ($D^\alpha_{\CP}$) operations followed by a \textit{single} gauging operation $G^\alpha_{\CP}$. 
\be \label{Sbasic-def}
\boxed{\CO^\alpha_{\vec \CP}(X) := (G^\alpha_{\vec \CP}) \circ (D^\alpha_{\vec \CP})^{n_3} \circ (F^\alpha_{\vec \CP})^{n_2} \circ (I^\alpha_{\vec \CP})^{n_1}(X), 
\quad (n_i=0,1, \,\, \forall i).}
\ee
In the special case where $G^\alpha_{\vec \CP}$ involves gauging a $U(1)$, we will refer to the corresponding $S$-type operation $\CO^\alpha_{\vec \CP}$ as an \textit{elementary Abelian $S$-type operation}.
The order of the different constituent operations in the above definition is important to emphasize. The composition of the identification and the flavoring operations is commutative, while none of the operations commute with the gauging operation. For a Wilson defect, the defect operation also commutes with the identification and the flavoring operations. However, for a vortex defect, the defect operation may or may not commute with $F^\alpha_{\vec \CP}$ and $I^\alpha_{\vec \CP}$ -- it depends on the details of the defect in question. In general, therefore, the order of the operations constituting 
$\CO^\alpha_{\vec \CP}$ is important.\\

An elementary $S$-type operation $\CO^\alpha_{\vec \CP}$, therefore, will map a quiver gauge theory $X$ to a new quiver gauge theory $X'$, 
decorated by a defect $D$, i.e.
\be 
\CO^\alpha_{\vec \CP} : X \mapsto X'[D].
\ee
Given this operation, one can define a dual operation $\wt{\CO}^\alpha_{\vec \CP}$ which acts on the quiver gauge theory $Y$.
The dual operation maps $Y$ to a new theory $Y'$ (not necessarily Lagrangian) decorated by a line defect $D^\vee$, i.e.
\be
\wt{\CO}^\alpha_{\vec \CP} : Y \mapsto Y'[D^\vee], 
\ee
such that the pair $(X'[D], Y'[D^\vee])$ is IR dual.\\

Let us now describe the realization of the operation $\CO^\alpha_{\vec \CP}$ and its dual $\wt{\CO}^\alpha_{\vec \CP}$ 
in terms of a three-sphere partition function. 
For the dual pair $(X,Y)$, we denote as $(Y, \{\CP_\beta \})$ the theory $Y$ where the $U(1)^{r_\alpha} \times U(1)^{M_\beta-r_\alpha}$ (for all $\beta$) background twisted vector multiplets have been relabelled as given in \eref{uvdef1}. The partition function of the theory $(Y, \{\CP_\beta \})$ 
has the form:
\be \label{Z-int-Y}
Z^{(Y, \{\CP_\beta\})}(\vec{m}^Y(\vec{\eta}) ; \vec{\eta}^Y(\{\vec{u}^\beta\}, \{\vec{v}^\beta\},\ldots))= 
\int \prod_{\gamma'} \Big[d\vec{\s}^{\gamma'} \Big]\, Z^{(Y,\{\CP_\beta\})}_{\rm int}(\{\vec \s^{\gamma'} \}, \vec{m}^Y(\vec{\eta}), \vec{\eta}^Y(\{\vec{u}^\beta\},\ldots)),
\ee
where $\gamma'$ labels the gauge nodes of the theory $Y$, and $\vec{m}^Y, \vec{\eta}^Y$ denote the masses and FI parameters. 
Mirror symmetry implies that the partition functions of $X$ and $Y$ are related as:
\begin{align} \label{MS-XY}
Z^{(X,\{ P_\beta\})} (\{\vec{u}^\beta\}, \{\vec{v}^\beta\},\ldots;\vec{\eta}) 
 = e^{2\pi i \sum_{i,l,\beta}\,b^{il}_\beta u^\beta_i \eta_l} \cdot  C_{XY}(\{\vec u^\beta=0\},\ldots, \vec \eta) \cdot Z^{(Y, \{\CP_\beta\})}(\vec{m}^Y; \vec{\eta}^Y),
\end{align}
where $b^{il}_\beta$ has integer entries, and $C_{XY}$ is defined in \eref{def-CXY}.
We will now address the two cases, where $\CO^\alpha_{\vec \CP}$ 
doesn't and does involve defect operations, separately. \\

\subsubsection*{$S$-operation not involving a defect operation} 
In this case, the partition function of the theory $X'=\CO^\alpha_{\vec \CP}(X)$ is given as 
\begin{align}\label{PF-OP-woD}
Z^{\CO^\alpha_{\vec \CP}(X)} (\vec{m}^{\CO^\alpha_{\vec \CP}}, \ldots; \vec \eta, \eta_\alpha)= \int \Big[d\vec{u}^\alpha\Big] \, \CZ_{\CO^\alpha_{\vec \CP}(X)}(\vec u^{\alpha}, \{\vec{u}^\beta\}_{\beta \neq \alpha}, \eta_\alpha, \vec{m}^{\CO^\alpha_{\vec \CP}})
\cdot Z^{(X,\{ P_\beta\})} (\{\vec{u}^\beta\},\ldots; \vec \eta),
\end{align}
where $\eta_\alpha$ is an FI parameter associated with gauging, and $\vec{m}^{\CO^\alpha_{\vec \CP}}$ are hypermultiplet masses associated 
with flavoring and/or identification operations. The operator $\CZ_{\CO^\alpha_{\vec \CP}(X)}$ can be constituted from the partition function contributions of gauging ($G^\alpha_\CP$) , flavoring ($F^\alpha_\CP$), and identification ($I^\alpha_\CP$) operations as follows:
\be \label{CZ-OP}
\CZ_{\CO^\alpha_{\vec \CP}(X)}(\vec u^{\alpha}, \{\vec{u}^\beta\}_{\beta \neq \alpha}, \eta_\alpha, \vec{m}^{\CO^\alpha_{\vec \CP}}) =\CZ_{G^\alpha_\CP(X)} \cdot \Big(\CZ_{F^\alpha_\CP(X)}\Big)^{n_2} \cdot \Big(\CZ_{I^\alpha_\CP(X)}\Big)^{n_1}, \quad (n_1,n_2=0,1),
\ee
where the individual constituents are:
\begin{align}
& \CZ_{G^\alpha_\CP(X)}(\vec{u}^\alpha,\eta_\alpha) = Z_{\rm FI} (\vec{u}^\alpha,\eta_\alpha) \,Z_{\rm 1-loop} ^{\rm vector} (\vec{u}^\alpha)=e^{2\pi i \eta_\alpha \sum_i u^\alpha_i} \, \prod_{i < j} \sinh^2{\pi (u^\alpha_i - u^\alpha_j)}, \label{CZ-gauging}\\
& \CZ_{F^\alpha_\CP(X)}(\vec{u}^\alpha, \vec{m}^\alpha_F)= Z_{\rm 1-loop} ^{\rm hyper} (\vec{u}^\alpha, \vec{m}^\alpha_F)=\prod_{\rho(\CR_\alpha)}\frac{1}{ 
\ch{\rho(\vec{u}^\alpha, \vec{m}^\alpha_F)}}, \label{CZ-flavoring}\\
&\CZ_{I^\alpha_\CP(X)}(\vec u^{\alpha}, \{\vec{u}^\beta\}, \vec \mu) = \int \prod^{p}_{j=1} \prod^{r_\alpha}_{i=1} d {u^{\gamma_j}}_i \, \prod^{p}_{j=1} \delta^{(r_\alpha)}\Big(\vec{u}^{\alpha} - \vec{u}^{\gamma_{j}} + {\mu}^{\gamma_j} \Big). \label{CZ-identification}
\end{align}

The partition function of the dual theory $Y'=\wt{\CO}^\alpha_{\vec \CP}(Y)$ can be computed in the following fashion. 
First, one can isolate the $\vec{u}^\beta$-dependent part of the matrix model integrand of the theory $(Y,\{\CP_\beta\})$, i.e.
\be
\frac{Z^{(Y,\{\CP_\beta\})}_{\rm int}(\{\vec \s^{\gamma'} \}, \vec{m}^Y(\vec{\eta}), \vec{\eta}^Y(\{\vec{u}^\beta\},\ldots))}{Z^{(Y,\{\CP_\beta\})}_{\rm int}(\{\vec \s^{\gamma'} \}, \vec{m}^Y(\vec{\eta}), \vec{\eta}^Y(\{\vec{u}^\beta =0 \},\ldots))}
=e^{2\pi i \,\sum_{i,\beta}g^i_\beta (\{\vec \s^{\gamma'} \}, \CP_\beta)\,u^{\beta}_i}, 
\ee
where $g^i_\beta$ is a set of linear functions of its arguments and is entirely determined 
by the mirror map of mass and FI parameters for the dual pair $(X,Y)$.
The IR duality, along with the assumption that both $\CO^\alpha_{\vec \CP}(X)$ and $\wt{\CO}^\alpha_{\vec \CP}(Y)$ are 
good theories, will imply that the two partition functions are related as
\begin{align}\label{IR-OP}
Z^{\CO^\alpha_{\vec \CP}(X)} ( \vec{m}^{\CO^\alpha_{\vec \CP}}, \ldots; \vec \eta, \eta_\alpha) = Z^{\wt{\CO}^\alpha_{\vec \CP}(Y)}(\vec m'(\vec \eta, \eta_\alpha) ; \vec \eta'(\vec m^{\CO^\alpha_{\vec \CP}}, \ldots)),
\end{align}
up to some contact terms, where $(\vec m,\vec \eta)$ and $(\vec m',\vec \eta')$ collectively denote the 
$\CN=4$ preserving masses and FI parameters of $\CO^\alpha_{\vec \CP}(X)$ and 
$\wt{\CO}^\alpha_{\vec \CP}(Y)$ respectively. 
From \eref{IR-OP} and \eref{PF-OP-woD}, using \eref{MS-XY}-\eref{Z-int-Y} and changing the order of integration, 
the partition function of the dual theory can then be cast in the form:
\begin{empheq}[box=\widefbox]{align}\label{PF-wtOPgen}
Z^{\wt{\CO}^\alpha_{\vec \CP}(Y)}(\vec{m}'; \vec \eta')
= \int \prod_{\gamma'} \Big[d\vec{\s}^{\gamma'} \Big]\, & \CZ_{\wt{\CO}^\alpha_{\vec \CP}(Y)}(\{\s^{\gamma'}\},\vec{m}^{{\CO}^\alpha_{\vec\CP}}, \eta_{\alpha},\vec \eta)\,\cdot C_{XY}(\{\vec u^\beta=0\},\ldots, \vec \eta) \nn \\
& \times Z^{(Y,\{\CP_\beta\})}_{\rm int}(\{\vec \s^{\gamma'} \}, \vec{m}^Y(\vec{\eta}), \vec{\eta}^Y(\{\vec{u}^\beta =0 \},\ldots)), 
\end{empheq}
where $\vec m'=\vec{m}'(\vec{\eta},\eta_{\alpha}), \vec \eta'=\vec \eta'(\vec{m}^{{\CO}^\alpha_{\vec\CP}},\ldots)$, 
and the function $\CZ_{\wt{\CO}^\alpha_{\vec \CP}(Y)}$ can be formally written as a Fourier-transform of the operator 
$\CZ_{\CO^\alpha_{\vec \CP}(X)}$:
\begin{align} \label{CZ-wtOP}
\CZ_{\wt{\CO}^\alpha_{\vec \CP}(Y)}
= \int \Big[d\vec{u}^\alpha\Big] \, \CZ_{\CO^\alpha_{\vec \CP}(X)}(\vec u^{\alpha}, \{\vec{u}^\beta\}_{\beta \neq \alpha}, \eta_\alpha, \vec{m}^{\CO^\alpha_{\vec \CP}})\, \cdot \,e^{2\pi i \,\sum_{i,\beta}(g^i_\beta (\{\vec \s^{\gamma'} \}, \CP_\beta) + \sum_l b^{il}_\beta \eta_l)\,u^{\beta}_i},
\end{align}
where $g^i_\beta$ is the set of functions defined above, and $b^{il}_\beta$ are integers defined in \eref{MS-XY}. 
The expressions \eref{PF-wtOPgen}-\eref{CZ-wtOP} give a working definition of the dual of the $S$-type operation on the 
theory $Y$. Note that although the RHS of \eref{PF-wtOPgen} is written as an operation on a Lagrangian theory $Y$, the 
theory $Y'$ is not guaranteed to be Lagrangian. However, if $Y'$ \textit{is} Lagrangian, one can rewrite the RHS of 
\eref{PF-wtOPgen} in the standard form of \eref{PF-main}, from which the gauge group and matter content of the theory 
can be easily read off. The contact term involved can also be read off from \eref{PF-wtOPgen}.

\subsubsection*{$S$-operation involving a defect operation} 

In this case, we will need to distinguish between the vortex and the Wilson defects to write down concrete formulae. 
For the Wilson defects, the partition function of the theory $X'=\CO^\alpha_{\vec \CP}(X)$ is given by \eref{PF-OP-woD}, 
where the operator $\CZ_{\CO^\alpha_{\vec \CP}(X)}$ can be constructed in the following fashion:
\begin{align} 
&\CZ_{\CO^\alpha_{\vec \CP}(X)}(\vec u^{\alpha}, \{\vec{u}^\beta\}_{\beta \neq \alpha}, \eta_\alpha, \vec{m}^{\CO^\alpha_{\vec \CP}}| R') =\CZ_{G^\alpha_\CP(X)} \cdot \Big(\CZ_{D^\alpha_\CP(X)}\Big) \cdot \Big(\CZ_{F^\alpha_\CP(X)}\Big)^{n_2} \cdot \Big(\CZ_{I^\alpha_\CP(X)}\Big)^{n_1},  \label{CZ-OP-wDB}\\
& \CZ_{D^\alpha_\CP(X)}(\vec{u}^\alpha) = Z_{\rm Wilson}(\vec{u}^\alpha, R') = \tr_{R'} e^{2\pi \vec{u}^\alpha}, \label{CZ-defectB}
\end{align}
where $n_i=0,1,$ $\forall i$, and the contribution of the defect to the partition function depends on $\vec{u}^\alpha$ and the representation $R'$. 
The resultant theory is the quiver $X'$ with a Wilson defect insertion. The dual theory will involve the quiver $Y'$ with a vortex 
defect insertion and its partition function will be given by \eref{PF-wtOPgen}-\eref{CZ-wtOP}, with the operator $\CZ_{\CO^\alpha_{\vec \CP}}$ 
given by the expression in \eref{CZ-OP-wDB}.\\

As discussed earlier, vortex defects are realized by coupled 3d-1d systems. We will consider defects where the associated (2,2) SQM $\Sigma'$ 
generically couples to both the new gauge node and the new flavor node. The action of an $S$-type operation $\CO^\alpha_{\vec \CP}$, 
involving a vortex defect, on the quiver $X$ gives the theory $X'$ with a vortex defect insertion. Its partition function can be written as:
\begin{align}\label{PF-OP-wDA}
Z^{\CO^\alpha_{\vec \CP}(X)} (\vec{m}^{\CO^\alpha_{\vec \CP}}, \ldots; \vec \eta, \eta_\alpha)= \lim_{z\to 1} \int \Big[d\vec{u}^\alpha\Big] \, & \CZ_{\CO^\alpha_{\vec \CP}(X)}(\vec u^{\alpha}, \{\vec{u}^\beta\}_{\beta \neq \alpha}, \eta_\alpha, \vec{m}^{\CO^\alpha_{\vec \CP}}, z | \Sigma') \nn \\
& \times \, Z^{(X,\{ P_\beta\})} (\{\vec{u}^\beta\},\ldots; \vec \eta),
\end{align}
where the expression on the RHS is computed by the analytic continuation procedure as before.
In this case, the operator $\CZ_{\CO^\alpha_{\vec \CP}(X)}$ can be constructed in the following fashion:
\begin{align} 
&\CZ_{\CO^\alpha_{\vec \CP}(X)}(\vec u^{\alpha}, \{\vec{u}^\beta\}_{\beta \neq \alpha}, \eta_\alpha, \vec{m}^{\CO^\alpha_{\vec \CP}}, z | \Sigma') 
=\CZ_{G^\alpha_\CP(X)} \cdot \Big(\CZ_{D^\alpha_\CP(X)}\Big) \cdot \Big(\CZ_{F^\alpha_\CP(X)}\Big)^{n_2} \cdot \Big(\CZ_{I^\alpha_\CP(X)}\Big)^{n_1}, 
 \label{CZ-OP-wDA}\\
& \CZ_{D^\alpha_\CP(X)}(\vec{u}^\alpha, \vec{m}^\alpha_F, \eta_\alpha, z) = W_{\rm b.g.}(\eta_\alpha, \vec \eta) \cdot \CI^{\Sigma'} (\vec{u}^\alpha, \vec{m}^\alpha_F, z| \vec \xi'), 
\label{CZ-defectA}
\end{align}
where $n_i=0,1,$ $\forall i$. In the above expression, $\CI^{\Sigma'}$ is the Witten index of the SQM $\Sigma'$ for a 
certain choice of the signs of the 1d FI parameters $\vec \xi'$, which is needed to specify the data of the 3d defect completely.  
The dual theory will involve the quiver $Y'$ with a Wilson defect insertion. The dual defect partition function, following the same logic as before, 
will be given by an expression analogous to \eref{PF-wtOPgen}, i.e.
\begin{empheq}[box=\widefbox]{align}\label{PF-wtOPgenDA}
 Z^{\wt{\CO}^\alpha_{\vec \CP}(Y)}(\vec{m}'; \vec \eta')
=\lim_{z\to 1} \int \prod_{\gamma'} \Big[d\vec{\s}^{\gamma'} \Big]\, & \CZ_{\wt{\CO}^\alpha_{\vec \CP}(Y)}(\{\s^{\gamma'}\},\vec{m}^{{\CO}^\alpha_{\vec\CP}}, \eta_{\alpha},\vec \eta, z)\,\cdot C_{XY}(\{\vec u^\beta=0\},\ldots, \vec \eta) \nn \\
& \times Z^{(Y,\{\CP_\beta\})}_{\rm int}(\{\vec \s^{\gamma'} \}, \vec{m}^Y(\vec{\eta}), \vec{\eta}^Y(\{\vec{u}^\beta =0 \},\ldots)), 
\end{empheq}
where the function $\CZ_{\wt{\CO}^\alpha_{\vec \CP}(Y)}(\{\s^{\gamma'}\},\vec{m}^{{\CO}^\alpha_{\vec\CP}}, \eta_{\alpha},\vec \eta,z)$ is related to the 
operator $\CZ_{\CO^\alpha_{\vec \CP}(X)}$ in \eref{CZ-OP-wDA}-\eref{CZ-defectA} in the following fashion:
\begin{align} \label{CZ-wtOPDA}
\CZ_{\wt{\CO}^\alpha_{\vec \CP}(Y)}
= \int \Big[d\vec{u}^\alpha\Big] \, \CZ_{\CO^\alpha_{\vec \CP}(X)}(\vec u^{\alpha}, \{\vec{u}^\beta\}_{\beta \neq \alpha}, \eta_\alpha, \vec{m}^{\CO^\alpha_{\vec \CP}}, z| \Sigma')\, \cdot e^{2\pi i \,\sum_{i,\beta}(g^i_\beta (\{\vec \s^{\gamma'} \}, \CP_\beta) + \sum_l b^{il}_\beta \eta_l)\,u^{\beta}_i}.
\end{align}
For a Lagrangian theory $Y'$, the Wilson defect can be read off from \eref{PF-wtOPgenDA}, after some 
straightforward change of variables.\\

\subsubsection*{Generic $S$-type operation}

A generic $S$-type operation on the quiver gauge theory $X$ is defined as the action of successive elementary $S$-type operations at nodes 
$\alpha_1, \alpha_2, \ldots, \alpha_l$:
\be \label{Comp-OP}
\CO^{(\alpha_1, \ldots, \alpha_l)}_{({\vec \CP_1},\ldots,{\vec \CP_l})}(X) :={\CO^{\alpha_l}_{\vec \CP_l}} \circ {\CO^{\alpha_{l-1}}_{\vec \CP_{l-1}}} \circ \ldots \circ {\CO^{\alpha_2}_{\vec \CP_2}} \circ  {\CO^{\alpha_1}_{\vec \CP_1}}(X).
\ee
The partition function of the theory $\CO^{(\alpha_1, \ldots, \alpha_l)}_{({\vec \CP_1},\ldots,{\vec \CP_l})}(X)$ is given by using 
\eref{PF-OP-woD} (or \eref{PF-OP-wDA} for operations with vortex defects) iteratively, while the partition function of the dual theory 
$\wt{\CO}^{(\alpha_1, \ldots, \alpha_l)}_{({\vec \CP_1},\ldots,{\vec \CP_l})}(Y)$ is given by using the formula \eref{PF-wtOPgen}-\eref{CZ-wtOP} 
(or \eref{PF-wtOPgenDA}-\eref{CZ-wtOPDA} for operators with vortex defects) iteratively.\\

Starting from a pair of dual linear quiver gauge theories, where the mirror map between the mass and FI parameters are known, one can generate new 
dual pairs of theories (with or without defects) by a sequence of elementary $S$-type operations.

\subsection{Abelianization of the non-Abelian $S$-type operation}\label{NonAb}

One of the important results obtained in \cite{Dey:2020hfe} was that the action of an Abelian $S$-type operation (consisting of a sequence of elementary 
operations) on a pair of dual quiver gauge theories $(X,Y)$ produces another dual pair of demonstrably Lagrangian theories $(X',Y')$. 
This procedure is substantially harder to implement if the $S$-type operation in question is non-Abelian\footnote{In general, there is no guarantee that the 
dual of a non-Abelian operation produces a Lagrangian theory, in contrast to the Abelian operation.}. 
However, at the level of the three-sphere partition function, an elementary non-Abelian $S$-type operation (which does not include any defect operation) can be written in terms of a set of Abelian $S$-type operations, which involve certain Wilson defect operations. We will refer to this procedure as the \textit{abelianization} of 
a non-Abelian $S$-type operation. We would like to emphasize that this procedure should be thought of as a convenient way of writing the matrix model
integral for the sphere partition function, and not as a QFT operation. In particular, this procedure does not extend naturally to other supersymmetric observable, 
like the 3d superconformal index.

For concreteness, we will discuss the procedure in detail for the case of a gauging 
operation. The argument can be readily extended to the other elementary operations, i.e. flavoring-gauging, identification-gauging, and 
identification-flavoring-gauging operations with/without defects.\\

\begin{figure}[htbp]
\begin{center}
\scalebox{.7}{\begin{tikzpicture}[
cnode/.style={circle,draw,thick,minimum size=4mm},snode/.style={rectangle,draw,thick,minimum size=8mm}]
\node[cnode] (1) {$N_1$};
\node[cnode] (2) [right=.5cm  of 1]{$N_2$};
\node[cnode] (3) [right=.5cm of 2]{$N_3$};
\node[cnode] (4) [right=1cm of 3]{$N_{\alpha-1}$};
\node[cnode] (5) [right=0.5cm of 4]{$N_{\alpha}$};
\node[cnode] (6) [right=0.5cm of 5]{$N_{\alpha + 1}$};
\node[cnode] (7) [right=1cm of 6]{{$N_{l-2}$}};
\node[cnode] (8) [right=0.5cm of 7]{$N_{l-1}$};
\node[cnode] (9) [right=0.5cm of 8]{$N_l$};
\node[snode] (10) [below=0.5cm of 1]{$M_1$};
\node[snode] (11) [below=0.5cm of 2]{$M_2$};
\node[snode] (12) [below=0.5cm of 3]{$M_3$};
\node[snode] (13) [above=0.5cm of 4]{$M_{\alpha-1}$};
\node[snode] (19) [below=0.5cm of 5]{$M_\alpha$};
\node[snode] (15) [above=0.5cm of 6]{$M_{\alpha+1}$};
\node[snode] (16) [below=0.5cm of 7]{$M_{l-2}$};
\node[snode] (17) [below=0.5cm of 8]{$M_{l-1}$};
\node[snode] (18) [below=0.5cm of 9]{$M_{l}$};
\draw[-] (1) -- (2);
\draw[-] (2)-- (3);
\draw[dashed] (3) -- (4);
\draw[-] (4) --(5);
\draw[-] (5) --(6);
\draw[dashed] (6) -- (7);
\draw[-] (7) -- (8);
\draw[-] (8) --(9);
\draw[-] (1) -- (10);
\draw[-] (2) -- (11);
\draw[-] (3) -- (12);
\draw[-] (4) -- (13);
\draw[-] (5) -- (19);
\draw[-] (6) -- (15);
\draw[-] (7) -- (16);
\draw[-] (8) -- (17);
\draw[-] (9) -- (18);
\draw[->] (6.5,-3) -- (6.5,-5);
\node[text width=0.2cm](22) at (7, -4){$\overline{\CO}^{N, \alpha}_{\CP}$};
\end{tikzpicture}}
\scalebox{.7}{\begin{tikzpicture}[
cnode/.style={circle,draw,thick,minimum size=4mm},snode/.style={rectangle,draw,thick,minimum size=8mm}]
\node[cnode] (1) {$N_1$};
\node[cnode] (2) [right=.5cm  of 1]{$N_2$};
\node[cnode] (3) [right=.5cm of 2]{$N_3$};
\node[cnode] (4) [right=1cm of 3]{$N_{\alpha-1}$};
\node[cnode] (5) [right=0.5cm of 4]{$N_{\alpha}$};
\node[cnode] (6) [right=0.5cm of 5]{$N_{\alpha + 1}$};
\node[cnode] (7) [right=1cm of 6]{{$N_{l-2}$}};
\node[cnode] (8) [right=0.5cm of 7]{$N_{l-1}$};
\node[cnode] (9) [right=0.5cm of 8]{$N_l$};
\node[snode] (10) [below=0.5cm of 1]{$M_1$};
\node[snode] (11) [below=0.5cm of 2]{$M_2$};
\node[snode] (12) [below=0.5cm of 3]{$M_3$};
\node[snode] (13) [above=0.5cm of 4]{$M_{\alpha-1}$};
\node[cnode] (14) at (5,-1.5){$1$};
\node[cnode] (20) at (6,-2){$1$};
\node[cnode] (21) at (8,-2){$1$};
\node[snode] (19)  [above=0.5cm of 5] {$M_\alpha -N$};
\node[snode] (15) [above=0.5cm of 6]{$M_{\alpha+1}$};
\node[snode] (16) [below=0.5cm of 7]{$M_{l-2}$};
\node[snode] (17) [below=0.5cm of 8]{$M_{l-1}$};
\node[snode] (18) [below=0.5cm of 9]{$M_{l}$};
\draw[-] (1) -- (2);
\draw[-] (2)-- (3);
\draw[dashed] (3) -- (4);
\draw[-] (4) --(5);
\draw[-] (5) --(6);
\draw[dashed] (6) -- (7);
\draw[-] (7) -- (8);
\draw[-] (8) --(9);
\draw[-] (1) -- (10);
\draw[-] (2) -- (11);
\draw[-] (3) -- (12);
\draw[-] (4) -- (13);
\draw[-] (5) -- (14);
\draw[-] (5) -- (19);
\draw[-] (6) -- (15);
\draw[-] (7) -- (16);
\draw[-] (8) -- (17);
\draw[-] (9) -- (18);
\draw[thick,dotted] (20)  to [bend right=40] (21);
\draw[-] (5) -- (20);
\draw[-] (5) -- (21);
\node[text width=0.2cm](22) at (4.7, -2.1){$q_1$};
\node[text width=0.2cm](22) at (5.7, -2.6){$q_2$};
\node[text width=0.2cm](22) at (7.7, -2.6){$q_N$};
\end{tikzpicture}}
\caption{\footnotesize{Construction of the theory $\overline{\CO}^{N, \alpha}_{\CP}(X)$ for a linear quiver gauge theory $X$.}}
\label{fig: barO(X)}
\end{center}
\end{figure}

Consider the action of an elementary gauging operation on a quiver gauge theory $X$ at a flavor node $\alpha$.
From equations \eref{CZ-OP} -\eref{CZ-gauging}, the partition function of the theory $G^\alpha_\CP(X)$ is given as
 \begin{align}
 Z^{G^\alpha_\CP(X)} (\vec{v}^\alpha,\ldots;\vec{\eta},{\eta}_\alpha)=&  \frac{1}{|\mathcal{W}_\alpha|} \int \prod^{r_\alpha}_{i=1} \de {u^\alpha_i} \, Z_{\rm FI} (\vec{u}^\alpha,\eta_\alpha) \, Z_{\rm 1-loop} ^{\rm vector} (\vec{u}^\alpha) \,Z^{(X,\CP)} (\vec{u}^\alpha, \vec{v}^\alpha\ldots;\vec{\eta}), \nn \\
=& \frac{1}{r_\alpha !} \int \prod^{r_\alpha}_{i=1} \de {u^\alpha_i} \, e^{2\pi i \eta_\alpha \sum_i u^\alpha_i} \, \prod_{i < j} \sinh^2{\pi (u^\alpha_i - u^\alpha_j)} \,Z^{(X,\CP)} (\vec{u}^\alpha, \vec{v}^\alpha\ldots;\vec{\eta}).
\end{align}
Recall the following identity, 
\be \label{WeylvectorId}
\prod_{i < j} 2 \sinh{\pi (u^\alpha_i - u^\alpha_j)}= \sum_{\rho \in \CS_{r_\alpha}} (-1)^\rho e^{2 \pi \omega^{(\rho)}_{r_\alpha} \cdot \vec{u}^\alpha}, 
\ee
where $\CS_N$ is the permutation group of $N$ distinct objects, $\omega_N =: \Big(\frac{N-1}{2}, \frac{N-3}{2}, \ldots, - \frac{N-1}{2}\Big)$, and 
$\omega^{(\rho)}_N$ denotes a vector whose entries are obtained from those of $\omega_N$ by an action of an element $\rho \in \CS_N$. 
Using this identity, the partition function can be written as
\begin{align}\label{NAG-AG}
Z^{G^\alpha_\CP(X)} (\vec{v}^\alpha,\ldots;\vec{\eta},{\eta}_\alpha) = & \frac{1}{r_\alpha!}  \sum_{\rho,\rho'} (-1)^{\rho + \rho'} \, \int \prod^{r_\alpha}_{i=1} \de {u^\alpha_i} \, e^{2\pi i \eta_\alpha \sum_i u^\alpha_i} \,e^{2 \pi (\omega^{(\rho)}_{r_\alpha} + \omega^{(\rho')}_{r_\alpha})\cdot \vec{u}^\alpha}\,Z^{(X,\CP)} (\vec{u}^\alpha, \vec{v}^\alpha\ldots;\vec{\eta}) \nn \\ 
 = & \frac{1}{r_\alpha!}  \sum_{\rho,\rho'} (-1)^{\rho + \rho'} \, \int \prod^{r_\alpha}_{i=1} \de {u^\alpha_i} \, \Big(\prod ^{r_\alpha}_{i=1} \, e^{2\pi i \eta_\alpha u^\alpha_i} \,e^{2 \pi \Omega^i_{(\rho, \rho')} {u}_i^\alpha}\Big)\,Z^{(X,\CP)} (\vec{u}^\alpha, \vec{v}^\alpha\ldots;\vec{\eta}),
\end{align}
where the vector $\Omega_{(\rho, \rho')} =\omega^{(\rho)}_{r_\alpha} + \omega^{(\rho')}_{r_\alpha}$, with $\rho,\rho' \in \CS_{\rho_\alpha}$. The last equation can be rewritten as
\begin{align}\label{NAG-AG}
Z^{G^\alpha_\CP(X)} =\frac{1}{r_\alpha!}  \sum_{\rho,\rho'} (-1)^{\rho + \rho'} \, \int \prod^{r_\alpha}_{i=1} \de {u^\alpha_i} \, \Big(\prod ^{r_\alpha}_{i=1} \ Z_{\rm FI} (u^\alpha_i, \eta_\alpha) \, Z_{\rm Wilson}(u^\alpha_i, \Omega^i_{(\rho, \rho')})\Big) \,Z^{(X,\CP)} (\vec{u}^\alpha, \vec{v}^\alpha\ldots;\vec{\eta}),
\end{align}
where $Z_{\rm Wilson}(u^\alpha_i, \Omega^i_{(\rho, \rho')})$ is the partition function contribution of an Abelian Wilson defect of charge 
$\Omega^i_{(\rho, \rho')}$ for a $U(1)$ gauge group parametrized by $u^\alpha_i$. \\

To write it more clearly, let $\overline{\CO}^\beta_\CP$ denote a defect-gauging Abelian elementary $S$-type operation at a given node $\beta$ of $X$:
\be
\overline{\CO}^\beta_\CP = G^\beta_\CP \circ \CD^\beta_\CP(X),
\ee
where the defect operation involves turning on an Abelian Wilson line of charge $q$ for the $U(1)$ flavor node in $U(M_\beta) \to U(1) \times U(M_\beta -1)$, 
and is followed by gauging the $U(1)$ flavor symmetry. 
Let us define the $S$-type operation $\overline{\CO}^{N, \beta}_{\CP}(X)$:
\be \label{Ab-defg}
\overline{\CO}^{N, \beta}_{\CP}(X) =: \overline{\CO}^{\beta_N}_{\CP_N} \circ \overline{\CO}^{\beta_{N-1}}_{\CP_{N-1}} \circ \cdots \circ \overline{\CO}^{\beta_1}_{\CP_1} (X),
\ee
such that $\beta_1=\beta$, $\beta_2$ corresponds to the flavor node $U(M_\beta -1)$, $\beta_3$ corresponds to the flavor node $U(M_\beta -2)$, 
and so on, and the Abelian defect charge introduced in the $i$-th step is $q^i$.
Also, $\CP$ denotes the effective permutation matrix which determines the choice of $U(1)^{N}$ mass parameters from the original
$U(1)^{M_\beta}$ mass parameters. For a linear quiver $X$, the quiver operation is shown in \figref{fig: barO(X)}.
The partition function of $\overline{\CO}^{N, \beta}_{\CP}(X)$ is therefore given in terms of the partition 
function of $X$ as :
\be  \label{Ab-defg}
Z^{\overline{\CO}^{N, \beta}_{\CP}(X)}(\vec{v}^\beta,\ldots;\vec{\eta}, \vec {\eta}_\beta| \vec{q}) = \int  \, \frac{\prod^{N}_{i=1} \de {u^\beta_i}}{N!}\,\prod^{N}_{i=1} \, Z_{\rm FI} (u^\beta_i, \eta^i_\beta) \, Z_{\rm Wilson}(u^\beta_i, q^i) \,Z^{(X,\CP)} (\vec{u}^\beta, \vec{v}^\beta \ldots;\vec{\eta}),
\ee
where $\{\eta^i_\beta\}_{i=1,\ldots,N}$ denote the FI parameters associated with $N$ Abelian gauging operations. 
The partition function of the theory $G^\alpha_\CP(X)$, as given in \eref{NAG-AG}, can then be written as:
\begin{align}\label{NAG-AG1}
\boxed{Z^{G^\alpha_\CP(X)} =  \sum_{\rho,\rho'} (-1)^{\rho + \rho'} \, Z^{\overline{\CO}^{r_\alpha, \alpha}_{\CP}(X)}(\vec{v}^\alpha,\ldots;\vec{\eta},{\eta}_\alpha |\vec{\Omega_{(\rho, \rho')}}),}
\end{align}
where the FI parameters associated with the $r_\alpha$ Abelian gauging operations are identical, and set equal to $\eta_\alpha$.
We would like to emphasize that all terms in the sum correspond to the partition function of the same Lagrangian theory (i.e. with the same gauge group and matter content) and differ only in the precise Wilson defect for the $r_\alpha$ $U(1)$ gauge nodes.

\subsection{Elementary $S$-type operations on theories with defects}\label{SOps-defects}
In this section, we discuss the action of $S$-type operations on quiver gauge theories with half-BPS defects.  
Let $X[\CD]$ denote the quiver gauge theory $X$ in class $\CU$, decorated by a line defect $\CD$ of Type-A (Type-B). 
Also, let $Y[\CD^\vee]$ denote the mirror dual quiver gauge theory $Y$, decorated by the dual line defect $\CD^\vee$ 
of Type-B (Type-A). 
Relabelling the mass parameters of $X[\CD]$ as in \eref{uvdef1}, we denote the resultant theory as 
$(X[\CD], \{P_\beta \})$. The mirror, with the relabelled FI parameters, will be denoted as 
$(Y[\CD^\vee], \{P_\beta \})$. We will treat the two cases -- $\CD$ being a vortex and a Wilson defect, separately.

\subsubsection*{Vortex defect $\CD$}
Let $\CD = V^{\Sigma}_{R(G)}$ denote the vortex defect in a representation $R$ 
of the gauge group $G$ for the theory $X$, which is realized by a specific 3d-1d system $\Sigma$. 
The partition function of the coupled 3d-1d quiver is given by:
\begin{align}
Z^{(X[V^{\Sigma}_{R(G)}], \{P_\beta \})}(\{ \vec{u}^\beta \}, \ldots; \vec \eta)=& \lim_{z\to 1} \int \,\Big[d\vec s\Big] \, Z^{(X[V^{\Sigma}_{R(G)}], \{P_\beta \})}_{\rm int}(\vec s, \{ \vec{u}^\beta \}, \ldots, \vec \eta,z) \label{Z-XV-1}\\
=: & \lim_{z\to 1} Z^{(X[V^{\Sigma}_{R(G)}], \{P_\beta \})}(\{ \vec{u}^\beta \}, \ldots; \vec \eta | z), \label{Z-XV-2}
\end{align}
where the precise form of the function $Z^{(X[V^{\Sigma}_{R(G)}], \{P_\beta \})}_{\rm int}$ can be read off 
from the general formula \eref{Z-3d1d-gen2}:
\be \label{Z-int-XV}
Z^{(X[V^{\Sigma}_{R(G)}], \{P_\beta \})}_{\rm int} = W_{\rm b.g.}(\vec \eta)\,Z^{(X, \{P_\beta \})}_{\rm int}(\vec s, \{ \vec{u}^\beta \}, \ldots, \vec \eta) \, \CI^{\Sigma}(\vec s, \{ \vec{u}^\beta \}, \ldots, z| \vec \xi).
\ee
The action of an $S$-type operation on the 3d-1d quiver $X[\CD]$ at a flavor node $\alpha$ gives a new 3d-1d quiver $X'[\CD']$:
\be \label{S-map-0}
\CO^\alpha_{\vec \CP} \,:\, X[\CD] \mapsto X'[\CD'].
\ee
Let us first consider the case where $\CO^\alpha_{\vec \CP}$ 
does not involve any defect operation. In this case, the $S$-operation is realized at the level of the partition function as follows:
\begin{empheq}[box=\widefbox]{align}\label{S-Op-A}
&Z^{\CO^\alpha_{\vec \CP}(X[\CD])} (\vec{m}^{\CO^\alpha_{\vec \CP}}, \ldots; \vec \eta, \eta_\alpha)
=: Z^{(X'[\CD'])} (\vec{m}^{\CO^\alpha_{\vec \CP}}, \ldots; \vec \eta, \eta_\alpha) \nn \\
&= \lim_{z\to 1} \int \Big[d\vec{u}^\alpha\Big] \, \CZ_{\CO^\alpha_{\vec \CP}(X)}(\vec u^{\alpha}, \{\vec{u}^\beta\}_{\beta \neq \alpha}, \eta_\alpha, \vec{m}^{\CO^\alpha_{\vec \CP}}) \cdot Z^{(X[\CD],\{ P_\beta\})} (\{\vec{u}^\beta\},\ldots; \vec \eta | z),
\end{empheq}
where the operator $\CZ_{\CO^\alpha_{\vec \CP}(X)}$ is constructed from \eref{CZ-OP}, and $Z^{(X[\CD],\{ P_\beta\})} (\ldots | z)$ is 
defined in \eref{Z-XV-2}. The new 3d-1d system $X'[\CD']$ can be read off from the matrix model formula in the second line. 
Note that the RHS of the above formula should be computed using the analytic continuation prescription 
for $z \in i\BR$, and taking the $z \to 1$ limit in the final step. \\

Now, consider the case where the operation $\CO^\alpha_{\vec \CP}$ involves a defect operation $D$, which needs to be of  
Type-A, so that the combined 3d-1d system still preserves the subalgebra SQM$_A$. Let $D = V^{\Sigma'}_{R'(U(r_\alpha))}$ 
denote a vortex defect in a representation $R'$ of $U(r_\alpha)$ realized by a 3d-1d system $\Sigma'$. 
Then, the map \eref{S-map-0} can be realized at the level of the parition function as follows:
\begin{empheq}[box=\widefbox]{align}\label{S-Op-Agen}
&Z^{\CO^\alpha_{\vec \CP}(X[\CD])} (\vec{m}^{\CO^\alpha_{\vec \CP}}, \ldots; \vec \eta, \eta_\alpha)
=: Z^{(X'[\CD'])} (\vec{m}^{\CO^\alpha_{\vec \CP}}, \ldots; \vec \eta, \eta_\alpha) \nn \\
=& \lim_{z\to 1} \int \Big[d\vec{u}^\alpha\Big] \, \CZ_{\CO^\alpha_{\vec \CP}(X)}(\vec u^{\alpha}, \{\vec{u}^\beta\}_{\beta \neq \alpha}, \eta_\alpha, \vec{m}^{\CO^\alpha_{\vec \CP}},z| \Sigma') \, \cdot \, Z^{(X[\CD], \{P_\beta \})}( \{ \vec{u}^\beta \}, \ldots, \vec \eta | z),
\end{empheq}
where the $\CZ_{\CO^\alpha_{\vec \CP}(X)}$ operator is given in \eref{CZ-OP-wDA}, and the function 
$Z^{(X[V^{\Sigma}_{R(G)}], \{P_\beta \})}_{\rm int}$ is given in \eref{Z-XV-2}. As before, the new 3d-1d system $X'[\CD']$ 
can be read off from the matrix model formula in the second line.

\subsubsection*{Wilson defect $\CD$}
Let $\CD = W_{R(G)}$ denote the vortex defect in a representation $R$ of the gauge group $G$ for the 
theory $X$. The defect partition function has the following form:
\begin{align}
Z^{(X[W_{R(G)}], \{P_\beta \})}(\{ \vec{u}^\beta \}, \ldots; \vec \eta)= \int \,\Big[d\vec s\Big] \, Z^{(X[W_{R(G)}], \{P_\beta \})}_{\rm int}(\vec s, \{ \vec{u}^\beta \}, \ldots, \vec \eta) \label{Z-XW-1},
\end{align}
where the function $Z^{(X[W_{R(G)}], \{P_\beta \})}_{\rm int}$ can be read off from \eref{PF-main-defect}:
\be \label{Z-int-XW}
Z^{(X[W_{R(G)}], \{P_\beta \})}_{\rm int}= Z^{(X, \{P_\beta \})}_{\rm int}(\vec s, \{ \vec{u}^\beta \}, \ldots, \vec \eta) \, Z_{\rm Wilson}(\vec s, R).
\ee
The action of an $S$-type defect on a 3d quiver with a Wilson defect $X[\CD] $ produces another 3d quiver with a Wilson defect $X'[\CD']$, i.e.
\be \label{S-map-1}
\CO^\alpha_{\vec \CP} \,:\, X[\CD] \mapsto X'[\CD'].
\ee
For an $S$-type operation $\CO^\alpha_{\vec \CP}$ which does not involve a defect operation, the operation is realized 
at the level of the partition function as follows:
\begin{empheq}[box=\widefbox]{align}\label{S-Op-B}
&Z^{\CO^\alpha_{\vec \CP}(X[\CD])} (\vec{m}^{\CO^\alpha_{\vec \CP}}, \ldots; \vec \eta, \eta_\alpha)
=: Z^{(X'[\CD'])} (\vec{m}^{\CO^\alpha_{\vec \CP}}, \ldots; \vec \eta, \eta_\alpha) \nn \\
&= \int \Big[d\vec{u}^\alpha\Big] \, \CZ_{\CO^\alpha_{\vec \CP}(X)}(\vec u^{\alpha}, \{\vec{u}^\beta\}_{\beta \neq \alpha}, \eta_\alpha, \vec{m}^{\CO^\alpha_{\vec \CP}}) \cdot Z^{(X[\CD],\{ P_\beta\})} (\{\vec{u}^\beta\},\ldots; \vec \eta),
\end{empheq}
where the operator $\CZ_{\CO^\alpha_{\vec \CP}(X)}$ is constructed from \eref{CZ-OP}, and the defect partition function 
$Z^{(X[\CD],\{ P_\beta\})}$ is defined in \eref{Z-XW-1}. The new 3d quiver with a Wilson defect $X'[\CD']$ can be read off 
from the matrix model formula in the second line. For an $S$-type operation $\CO^\alpha_{\vec \CP}$, which involves a 
Wilson defect $W_{R'(U(r_\alpha))}$ for the new gauge node $U(r_\alpha)$ in a representation $R'$, 
the above formula can be generalized to the following form:
\begin{empheq}[box=\widefbox]{align}\label{S-Op-Bgen}
&Z^{\CO^\alpha_{\vec \CP}(X[\CD])} (\vec{m}^{\CO^\alpha_{\vec \CP}}, \ldots; \vec \eta, \eta_\alpha)
= Z^{(X'[\CD'])} (\vec{m}^{\CO^\alpha_{\vec \CP}}, \ldots; \vec \eta, \eta_\alpha) \nn \\
=&\int \Big[d\vec{u}^\alpha\Big] \, \CZ_{\CO^\alpha_{\vec \CP}(X)}(\vec u^{\alpha}, \{\vec{u}^\beta\}_{\beta \neq \alpha}, \eta_\alpha, \vec{m}^{\CO^\alpha_{\vec \CP}}| R') \, \cdot \, Z^{(X[\CD], \{P_\beta \})}( \{ \vec{u}^\beta \}, \ldots, \vec \eta),
\end{empheq}
where the $\CZ_{\CO^\alpha_{\vec \CP}(X)}$ operator is given in \eref{CZ-OP-wDB}, and the function 
$Z^{(X[\CD], \{P_\beta \})}$ is given in \eref{Z-XW-1}. 

To summarize: the action of an $S$-operation $\CO^\alpha_{\vec \CP}$ involving a vortex defect operation $D$, 
on a 3d theory with a vortex defect $X[\CD]$ is another 3d theory with a vortex defect $X'[\CD']$, where 
$\CD'$ is a line defect built out of the pair of defects $(D, \CD)$. Schematically, we can represent the 
action of the $S$-operation as:
\be \label{S-map-2}
\CO^\alpha_{\vec \CP} \,: \, X[\CD] \mapsto X'[\CD'], \qquad \CD'= L(D,\CD).
\ee
For the system to preserve the SQM$_A$ or SQM$_B$ supersymmetry, the line defects $(D,\CD)$ must be 
of the same type. The action can be realized at the level of the 3d partition function :
\be
Z^{\CO^\alpha_{\vec \CP}(X[\CD])} (\vec{m}^{\CO^\alpha_{\vec \CP}}, \ldots; \vec \eta, \eta_\alpha)=: Z^{(X'[L(D,\CD)])} (\vec{m}^{\CO^\alpha_{\vec \CP}}, \ldots; \vec \eta, \eta_\alpha),
\ee
as discussed above in the individual cases, and the line defect $L(D,\CD)$ can be read off from the partition function expressions.

\subsection{Dual operations and the new mirror map of defects}\label{SOps-defects-d} 
Let the theory dual to $X[\CD]$ is denoted by $Y[{\CD}^\vee]$, which is the quiver gauge theory $Y$ decorated by a line defect 
${\CD}^\vee$ of Type-A (Type-B), given that $\CD$ is of Type-B (Type-A). An $S$-type operation acts on a 3d 
quiver with a defect to give another 3d quiver with a defect, as given in \eref{S-map-2}.
Given an operation $\CO^\alpha_{\vec \CP}$ on $X[\CD]$, one can define a dual operation on $Y[{\CD}^\vee]$:
\be
\wt{\CO}^\alpha_{\vec \CP} :  Y[{\CD}^\vee] \mapsto Y'[\CD'^\vee], \qquad \CD'^\vee= L^\vee(D^\vee,\CD^\vee),
\ee
such that the pair $(X'[\CD'], Y'[D'^\vee])$ is IR dual. In this section, we discuss a systematic procedure to read off the 
dual line defect $L^\vee(D^\vee,\CD^\vee)$, and thereby write down the mirror map for the new defects $L$ and $L^\vee$ 
constructed above. As before, we will treat the case of a vortex defect and that of a Wilson defect, separately.\\

To begin with, the expectation values of the line defects in theory $X$ and theory $Y$ are generically related as follows:
\begin{align}\label{Basic-MM}
 \langle \CD \rangle_{X}(\{\vec{u}^\beta\},\ldots; \vec \eta) =  
 \langle {\CD}^\vee\rangle_{Y}(\vec{m}^Y(\vec{\eta}) ; \vec{\eta}^Y(\{\vec{u}^\beta\}, \ldots)),
\end{align} 
which implies that the defect partition functions are related as follows:
\begin{align}\label{MS-XYD}
Z^{(X[\CD],\{ P_\beta\})} (\{\vec{u}^\beta\},\ldots; \vec \eta)
 = C_{XY}(\{\vec{u}^\beta\},\ldots, \vec \eta)\, Z^{(Y[{\CD}^\vee],\{ P_\beta\})} (\vec{m}^Y(\vec{\eta}) ; \vec{\eta}^Y(\{\vec{u}^\beta\}, \ldots)).
\end{align}
To work out the dual partition functions, we will need a slightly refined version of the above equality. 
Implementing the $S$-type operations involve performing integrals over $\{\vec u^\beta\}$ with an appropriate measure. 
Therefore, one needs to make sure that the limit $z \to 1$ is not taken trivially for any $z$-dependent term that 
involves a pole for the variables $\{\vec u^\beta\}$. This consideration is obviously not important for writing identities 
of the form \eref{MS-XYD}, where the $\{\vec u^\beta\}$ are merely background gauge fields.

Let $\CD$ be a vortex defect and $\CD^\vee$ be a Wilson defect. In this case, the mirror symmetry implies a $z$-dependent equality:
\begin{empheq}[box=\widefbox]{align}\label{MS-XYD-1}
Z^{(X[\CD],\{ P_\beta\})} (\{\vec{u}^\beta\},\ldots; \vec \eta |z) 
& = C_{XY}(\{\vec{u}^\beta\},\ldots, \vec \eta)\,Z^{(Y[{\CD}^\vee],\{ P_\beta\})} (\vec{m}^Y(\vec{\eta}) ; \vec{\eta}^Y(\{\vec{u}^\beta\}, \ldots)|z) \nn \\
&= C_{XY}\cdot \,\int \prod_{\gamma'}  \Big[d\vec{\s}^{\gamma'} \Big]\,Z^{(Y[{\CD}^\vee],\{ P_\beta\})}_{\rm int} (\{\vec{\s}^{\gamma'} \}, \vec{m}^Y(\vec{\eta}), \vec{\eta}^Y(\{\vec{u}^\beta\}, \ldots), z),
\end{empheq}
where the function $Z^{(X[\CD],\{ P_\beta\})} (\ldots |z) $ is defined in \eref{Z-XV-2} in terms of the vortex defect 
partition function. 
The function $Z^{(Y[{\CD}^\vee],\{ P_\beta\})} (\ldots |z)$ has the property that it reduces to the Wilson defect partition function in the limit $z \to 1$, i.e.
\be
\lim_{z\to 1}\,Z^{(Y[{\CD}^\vee],\{ P_\beta\})} (\vec{m}^Y(\vec{\eta}) ; \vec{\eta}^Y(\{\vec{u}^\beta\}, \ldots)|z)
= Z^{(Y[{\CD}^\vee],\{ P_\beta\})} (\vec{m}^Y(\vec{\eta}) ; \vec{\eta}^Y(\{\vec{u}^\beta\}, \ldots)).
\ee
\\
Similarly, for the case where $\CD$ is a Wilson defect and $\CD^\vee$ is a vortex defect, one can write 
a $z$-dependent equality of the same form as \eref{MS-XYD-1}, where $Z^{(Y[{\CD}^\vee],\{ P_\beta\})} (\ldots |z)$ 
is now defined by \eref{Z-XV-2}, and the function $Z^{(X[\CD],\{ P_\beta\})} (\ldots |z)$ has the property that it reduces 
to the Wilson defect partition function in the limit $z \to 1$, i.e.
\be 
\lim_{z\to 1}\, Z^{(X[\CD],\{ P_\beta\})} (\{\vec{u}^\beta\},\ldots; \vec \eta |z)= Z^{(X[\CD],\{ P_\beta\})} (\{\vec{u}^\beta\},\ldots; \vec \eta).
\ee
Using the result \eref{MS-XYD-1}, one can now write explicit formulae for the dual partition functions in the two cases.\\

\subsubsection*{Dual of a vortex defect} 
Let $\CD = V^{\Sigma}_{R(G)}$ denote the vortex defect in a representation $R$ 
of the gauge group $G$ for the theory $X$, which is realized by a specific 3d-1d system $\Sigma$. 
The dual Wilson defect ${\CD}^\vee$ is of the generic form:
\be
{\CD}^\vee = \sum_\kappa c_\kappa \, \wt{W}^{\rm flavor}_\kappa \cdot \wt{W}_{\wt{R}_\kappa(\wt{G})}, 
\ee 
where $\wt{W}_{\wt{R}_\kappa(\wt{G})}$ is a Wilson defect in a representation $\wt{R}_\kappa$ of the gauge 
group $\wt{G}$ of the theory $Y$, and $\wt{W}^{\rm flavor}_\kappa$ is a flavor Wilson defect associated 
with the hypermultiplets. The requirement of IR duality for the pair $(X'[\CD'], Y'[\CD'^\vee])$ implies 
that the defect partition functions are related, up to certain contact terms, as follows:
\begin{align}\label{IR-OP-1}
Z^{\CO^\alpha_{\vec \CP}(X'[\CD'])} (\vec{m}^{\CO^\alpha_{\vec \CP}}, \ldots; \vec \eta, \eta_\alpha) = Z^{\wt{\CO}^\alpha_{\vec \CP}(Y'[\CD'^\vee])}(\vec m'(\vec \eta, \eta_\alpha) ; \vec \eta'(\vec m^{\CO^\alpha_{\vec \CP}}, \ldots)).
\end{align}

Consider the case where the $S$-type operation does not involve any defect operation. The operation is 
realized at the level of the partition function by \eref{S-Op-A}. Using \eref{S-Op-A} and \eref{IR-OP-1}, along 
with the mirror symmetry relation \eref{MS-XYD-1} and changing the order of integration, the dual partition function is given by 
the following formula (up to certain contact terms): 
\begin{empheq}[box=\widefbox]{align}\label{PF-wtOPgenD-A2B}
Z^{\wt{\CO}^\alpha_{\vec \CP}(Y[{\CD}^\vee])}(\vec{m}'; \vec \eta')
=& \sum_\kappa c_\kappa \lim_{z\to 1} \int \prod_{\gamma'}  \Big[d\vec{\s}^{\gamma'} \Big]\,  \CZ^\kappa_{\wt{\CO}^\alpha_{\vec \CP}(Y)}(\{\s^{\gamma'}\},\vec{m}^{{\CO}^\alpha_{\vec\CP}}, \eta_{\alpha},\vec \eta) \cdot C_{XY}(\{\vec{u}^\beta=0 \},\ldots, \vec \eta)  \nn \\
\times & Z_{\wt{W}^{\rm flavor}_\kappa}(\vec{m}^Y(\vec{\eta})|z) \cdot Z^{(Y[\wt{W}_{\wt{R}_\kappa(\wt{G})}],\{\CP_\beta\})}_{\rm int}(\{\vec \s^{\gamma'} \}, \vec{m}^Y(\vec{\eta}), \vec{\eta}^Y(\{\vec{u}^\beta =0 \},\ldots), z),
\end{empheq}
where the integrand $Z^{(Y[\wt{W}_{\wt{R}_\kappa(\wt{G})}],\{\CP_\beta\})}_{\rm int}$ and the function $Z_{\wt{W}^{\rm flavor}_\kappa}$ 
can be read off from the second line of \eref{MS-XYD-1}. The functions $\CZ^\kappa_{\wt{\CO}^\alpha_{\vec \CP}(Y)}$ are given by a 
formal Fourier transformation of the operator $\CZ_{\CO^\alpha_{\vec \CP}(X)}$ defined in \eref{CZ-OP} :
\begin{align} \label{CZ-wtOPD-A2B}
\boxed{\CZ^\kappa_{\wt{\CO}^\alpha_{\vec \CP}(Y)}
= \int \Big[d\vec{u}^\alpha\Big] \, \CZ_{\CO^\alpha_{\vec \CP}(X)}(\vec u^{\alpha}, \{\vec{u}^\beta\}_{\beta \neq \alpha}, \eta_\alpha, \vec{m}^{\CO^\alpha_{\vec \CP}}) \cdot e^{2\pi i \sum_{i,\beta}(g^i_\beta (\{\vec \s^{\gamma'} \}, \CP_\beta) +\sum_l b'^{il}_\beta \eta_l)\,u^{\beta}_i}.}
\end{align}
If the dual theory $Y'$ is Lagrangian, then the expression on the RHS of \eref{PF-wtOPgenD-A2B} can again be identified as the partition function 
of a quiver gauge theory $Y'$, decorated by a Wilson defect, which we can formally denote as ${L}^\vee( \BU, {\CD}^\vee)$. The 
formula \eref{PF-wtOPgenD-A2B} can be readily generalized to the case where the $S$-type operation involves a defect $D$, by 
simply replacing in the integrand:
\be
 \CZ^\kappa_{\wt{\CO}^\alpha_{\vec \CP}(Y)}(\{\s^{\gamma'}\},\vec{m}^{{\CO}^\alpha_{\vec\CP}}, \eta_{\alpha},\vec \eta)
 \to \CZ^\kappa_{\wt{\CO}^\alpha_{\vec \CP}(Y)}(\{\s^{\gamma'}\},\vec{m}^{{\CO}^\alpha_{\vec\CP}}, \eta_{\alpha},\vec \eta, z).
\ee
The latter functions are defined by the formal Fourier transform: 
\begin{align} \label{CZ-wtOPDD-A2B}
\CZ^\kappa_{\wt{\CO}^\alpha_{\vec \CP}(Y)}
= \int \Big[d\vec{u}^\alpha\Big] \, \CZ_{\CO^\alpha_{\vec \CP}(X)}(\vec u^{\alpha}, \{\vec{u}^\beta\}_{\beta \neq \alpha}, \eta_\alpha, \vec{m}^{\CO^\alpha_{\vec \CP}}, z| \Sigma')\, \cdot e^{2\pi i \sum_{i,\beta}(g^i_\beta (\{\vec \s^{\gamma'} \}, \CP_\beta) +\sum_l b'^{il}_\beta \eta_l)\,u^{\beta}_i},
\end{align}
where the operator $\CZ_{\CO^\alpha_{\vec \CP}(X)}$ is defined in \eref{CZ-OP-wDA}. The expression \eref{PF-wtOPgenD-A2B} will 
serve as the working definition for the dual $S$-type operation acting on the quiver gauge theory $Y$, decorated by a Wilson 
defect $\CD^\vee$. For a Lagrangian $Y'$, the RHS can be identified as a Wilson defect partition function, and we denote the 
resultant Wilson defect as ${L}^\vee(D^\vee, {\CD}^\vee)$.\\

\subsubsection*{Dual of a Wilson defect}
Let $\CD = W_{R(G)}$ denote the vortex defect in a representation $R$ 
of the gauge group $G$ for the theory $X$. The dual vortex defect ${\CD}^\vee$ will have the generic form:
\be
{\CD}^\vee = \sum_\kappa c_\kappa \, \wt{W}^{\rm b.g.}_\kappa \, \wt{V}^{\Sigma_\kappa}_{\wt{R}_\kappa(\wt{G})}, 
\ee
where $\wt{V}^{\Sigma_\kappa}_{\wt{R}_\kappa(\wt{G})}$ is a vortex defect in a representation $\wt{R}_\kappa$
for the gauge group $\wt{G}$ of the theory $Y$, and $\wt{W}^{\rm b.g.}_\kappa$ is a Wilson defect for a combination 
of 3d topological symmetries of $Y$.

Let us first consider the case where $S$-operation does not involve any defect operation.
Following the same reasoning as before, the dual partition function can be written in the form (up to contact terms):
\begin{empheq}[box=\widefbox]{align}\label{PF-wtOPgenD-B2A}
Z^{\wt{\CO}^\alpha_{\vec \CP}(Y[{\CD}^\vee])} &(\vec{m}'; \vec \eta')
= \sum_\kappa c_\kappa  \Big( \lim_{z\to 1} \, \int \prod_{\gamma'}  \Big[d\vec{\s}^{\gamma'} \Big]\,  \CZ^\kappa_{\wt{\CO}^\alpha_{\vec \CP}(Y)}(\{\s^{\gamma'}\},\vec{m}^{{\CO}^\alpha_{\vec\CP}}, \eta_{\alpha},\vec \eta) \cdot  Z_{\wt{W}^{\rm b.g.}_\kappa}(\{\vec{u}^\beta =0 \}, \ldots) \nn \\
& \times C_{XY}(\{\vec{u}^\beta=0 \},\ldots, \vec \eta)\,\, Z^{(Y[\wt{V}^{\Sigma_\kappa}_{\wt{R}_\kappa(\wt{G})}],\{\CP_\beta\})}_{\rm int}(\{\vec \s^{\gamma'} \}, \vec{m}^Y(\vec{\eta}), \vec{\eta}^Y(\{\vec{u}^\beta =0 \},\ldots),z) \Big),
\end{empheq}
where the integrand $Z^{(Y[\wt{V}^{\Sigma_\kappa}_{\wt{R}_\kappa(\wt{G})}],\{\CP_\beta\})}_{\rm int}$ is given by the formula \eref{Z-int-XV},
and the functions $\CZ^\kappa_{\wt{\CO}^\alpha_{\vec \CP}(Y)}$ are given by the following formula:
\begin{align} \label{CZ-wtOPD-B2A}
\boxed{\CZ^\kappa_{\wt{\CO}^\alpha_{\vec \CP}(Y)}
= \int \Big[d\vec{u}^\alpha\Big] \, \CZ_{\CO^\alpha_{\vec \CP}(X)}(\vec u^{\alpha}, \{\vec{u}^\beta\}_{\beta \neq \alpha}, \eta_\alpha, \vec{m}^{\CO^\alpha_{\vec \CP}})\, \cdot e^{2\pi i \sum_{i,\beta}(g^i_\beta (\{\vec \s^{\gamma'} \}, \CP_\beta) +\sum_l b'^{il}_\beta \eta_l)\,u^{\beta}_i}\, Z^\kappa_{\rm b.g.}(\{\vec{u}^\beta\},\ldots),}
\end{align}
where $Z^\kappa_{\rm b.g.}$ denotes the combined $\{\vec{u}^\beta\}$-dependent part of $Z_{\wt{W}^{\rm b.g.}_\kappa}$ and any
$\{\vec{u}^\beta\}$-dependent background Wilson defect term in $Z^{(Y[\wt{V}^{\Sigma_\kappa}_{\wt{R}_\kappa(\wt{G})}],\{\CP_\beta\})}_{\rm int}$. 
If the dual theory $Y'$ is Lagrangian, then the expression on the RHS of \eref{PF-wtOPgenD-B2A} can be identified as the partition function of a quiver gauge theory $Y'$ decorated by a vortex defect, which we denote as ${L}^\vee(\BU, {\CD}^\vee)$. The formula \eref{PF-wtOPgenD-B2A} can be readily generalized to the case where the $S$-type operation involves a defect $D$, by simply replacing in the integrand:
\be
 \CZ^\kappa_{\wt{\CO}^\alpha_{\vec \CP}(Y)}(\{\s^{\gamma'}\},\vec{m}^{{\CO}^\alpha_{\vec\CP}}, \eta_{\alpha},\vec \eta)
 \to \CZ^\kappa_{\wt{\CO}^\alpha_{\vec \CP}(Y)}(\{\s^{\gamma'}\},\vec{m}^{{\CO}^\alpha_{\vec\CP}}, \eta_{\alpha},\vec \eta|R').
\ee
The latter functions are defined by :
\begin{align} \label{CZ-wtOPD-B2A-gen}
\CZ^\kappa_{\wt{\CO}^\alpha_{\vec \CP}(Y)}
= \int \Big[d\vec{u}^\alpha\Big] \, & \CZ_{\CO^\alpha_{\vec \CP}(X)}(\vec u^{\alpha}, \{\vec{u}^\beta\}_{\beta \neq \alpha}, \eta_\alpha, \vec{m}^{\CO^\alpha_{\vec \CP}}|R')\, \cdot Z^\kappa_{\rm b.g.}(\{\vec{u}^\beta\},\ldots)\nn\\
& \times e^{2\pi i \sum_{i,\beta}(g^i_\beta (\{\vec \s^{\gamma'} \}, \CP_\beta) +\sum_l b'^{il}_\beta \eta_l)\,u^{\beta}_i},
\end{align}
where the operator $\CZ_{\CO^\alpha_{\vec \CP}(X)}$ is given in \eref{CZ-OP-wDB}. The expression \eref{PF-wtOPgenD-B2A} will 
serve as the working definition for the dual $S$-type operation acting on the quiver gauge theory $Y$, decorated by a vortex 
defect $\CD^\vee$. For a Lagrangian $Y'$, the RHS can be identified as a vortex defect partition function, and we denote the 
resultant vortex defect as ${L}^\vee(D^\vee, {\CD}^\vee)$.\\

To summarize the above results, the dual of the $S$-type operation generates a theory $Y'$ decorated by a 
line defect ${L}^\vee$ built out of the pair $({D}^\vee,{\CD}^\vee)$, where ${L}^\vee, {D}^\vee,{\CD}^\vee$ are all defects of the 
same type, i.e. they all preserve either the subalgebra ${\rm SQM}_A$ or ${\rm SQM}_B$.
The line defect ${L}^\vee$ can be read off (in some cases, after some non-trivial manipulations of 
the partition function) if the dual theory $Y'$ is Lagrangian. Schematically, we have:
\be
Z^{\wt{\CO}^\alpha_{\vec \CP}(Y[{\CD}^\vee])}(\vec{m}'(\vec{\eta},\eta_{\alpha}); \vec \eta'(\vec{m}^{{\CO}^\alpha_{\vec\CP}},\ldots)) := Z^{(Y'[{L}^\vee({D}^\vee,{\CD}^\vee)])}(\vec{m}'(\vec{\eta},\eta_{\alpha}); \vec \eta'(\vec{m}^{{\CO}^\alpha_{\vec\CP}},\ldots)),
\ee
where the precise forms of the defect partition functions are given by \eref{PF-wtOPgenD-A2B} and \eref{PF-wtOPgenD-B2A} 
for $\CD$ being a vortex defect and a Wilson defect respectively. 

The action of the $S$-type operation on the dual pair $(X[\CD],Y[\CD^\vee])$ therefore leads to the new 
duality statement :
\be \label{MM-new-main}
\boxed{\langle L(D,\CD) \rangle_{X'}(\vec{m}^{\CO^\alpha_{\vec \CP}}, \ldots; \vec \eta, \eta_\alpha) = \langle {L}^\vee({D}^\vee,{\CD}^\vee) \rangle_{Y'}(\vec{m}'; \vec \eta'),}
\ee
where $\vec{m}'=\vec{m}'(\vec{\eta},\eta_{\alpha}), \vec \eta'=\vec \eta'(\vec{m}^{{\CO}^\alpha_{\vec\CP}},\ldots)$. 
Equation \eref{MM-new-main} allows one to directly read off the mirror map between the line defect $L(D,\CD)$ in the new quiver gauge theory $X'$ and the line defect ${L}^\vee({D}^\vee,{\CD}^\vee)$ in the dual theory $Y'$. We would like to emphasize that this procedure should work whenever the pair $(X',Y')$ are Lagrangian, regardless of whether the theories have a Hanany-Witten type description.\\

\subsection{Simple illustrative example: Abelian $S$-type operations on $T(U(2))$ } \label{S-Ex-def}

In this section, we work out an explicit example which illustrates the basic points of the construction described in 
\Secref{SOps-defects}-\Secref{SOps-defects-d}.
We will study an elementary Abelian flavoring-identification operation acting on a $T(U(2))$ theory\footnote{Recall that the theory $T(U(2))$ is obtained 
from $T(SU(2))$ by adding a decoupled theory which consists of a mixed CS term involving a background $U(1)$ vector multiplet and the $U(1)_J$ 
twisted vector multiplet.} with defects, which leads to another linear quiver with defects, and work out the mirror dual of the latter.  Obviously, this mirror map 
can be worked out using the Type IIB description of \cite{Assel:2015oxa}, but we present it here to illustrate how the general procedure of 
\Secref{SOps-defects}-\Secref{SOps-defects-d} works in simple examples. \\

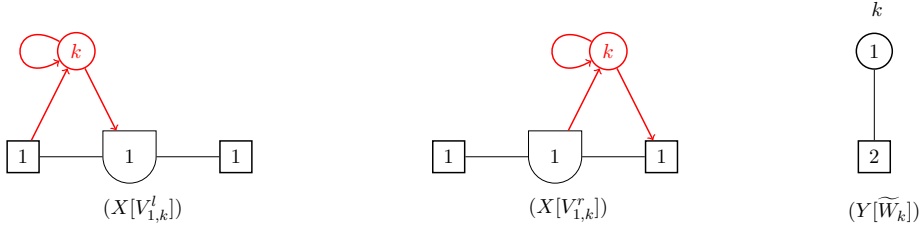
\begin{figure}[htbp]
\begin{center}
\scalebox{0.7}{\begin{tikzpicture}[
nnode/.style={circle,draw,thick, red},cnode/.style={circle,draw,thick,minimum size=4mm},snode/.style={rectangle,draw,thick,minimum size=6mm}]
\node[snode] (1) at (0,0) {1} ;
\node[cf-group] (2) at (2,0) {\rotatebox{-90}{1}};
\node[snode] (3) at (4,0) {1};
\node[nnode] (4) at (1,2) {$k$};
\draw[red, thick, ->] (1)--(4);
\draw[red, thick, ->] (4)--(2);
\draw[-] (1) -- (2);
\draw[-] (2) -- (3);
\draw[red, thick, ->] (4) to [out=150,in=210,looseness=8] (4);
\node[text width=1cm](20) at (2,-1) {$(X[V^{l}_{1,k}])$};
\node[snode] (5) at (8,0) {1} ;
\node[cf-group] (6) at (10,0) {\rotatebox{-90}{1}};
\node[snode] (7) at (12,0) {1};
\node[nnode] (8) at (11,2) {$k$};
\draw[red, thick, ->] (6)--(8);
\draw[red, thick, ->] (8)--(7);
\draw[-] (5) -- (6);
\draw[-] (6) -- (7);
\draw[red, thick, ->] (8) to [out=150,in=210,looseness=8] (8);
\node[text width=1cm](10) at (10,-1) {$(X[V^{r}_{1,k}])$};
\node[cnode] (11) at (16,2) {1} ;
\node[snode] (12) at (16,0) {2};
\draw[-] (11) -- (12);
\node[text width=0.1cm](3) at (16,2.8) {$k$};
\node[text width=1cm](4) at (16, -1) {$(Y[\wt{W}_k])$};
\end{tikzpicture}}
\caption{\footnotesize{The two 3d-1d systems that realize the 3d vortex operator $V_{1,k}$ in a $U(1)$ gauge theory with two hypermultiplets. The mirror dual is a gauge Wilson defect of charge $k$.}}
\label{fig:Ex1ab}
\end{center}
\end{figure}

$T(U(2))$ is a self-mirror, and its partition function obeys the following identity:
\be\label{MS-TU2}
Z^{T(U(2))}(\vec m; \vec t) =e^{2\pi i (m_1 t_1 - m_2 t_2)}\, Z^{T(U(2))}(\vec t; -\vec m).
\ee
Consider a vortex defect of charge $k$ for the $U(1)$ gauge group. The defect can realized as two different 3d-1d quivers, shown 
as $X[V^{l}_{1,k}]$ and $X[V^{r}_{1,k}]$ in \figref{fig:Ex1ab}. The defect partition functions for the two quivers are given as:
\begin{align}
& Z^{(X[V^{l}_{1,k}])}(\vec m; \vec t) = e^{2\pi k t_1 } \cdot \lim_{z \to 1}\, Z^{T(U(2))}(m_1 +ikz, m_2; \vec t),\\
& Z^{(X[V^{r}_{1,k}])}(\vec m; \vec t) = e^{2\pi k t_2 } \cdot \lim_{z \to 1}\, Z^{T(U(2))}(m_1, m_2 - ikz; \vec t).
\end{align}
Using the identity \eref{MS-TU2}, one can rewrite the $z$-dependent functions on the RHS of the first equation as:
\begin{align}
e^{2\pi k t_1 } \cdot Z^{T(U(2))}(m_1 +ikz, m_2; \vec t) =& e^{2\pi i (m_1 t_1 - m_2 t_2)}\,Z^{(Y[\wt{W}_k])}(\vec t; -\vec m | z)  \nn \\
= & e^{2\pi i (m_1 t_1 - m_2 t_2)}\, \int d\s \, \frac{e^{-2\pi i \s (m_1 - m_2)}\, e^{2\pi k\,\s \, z}}{\ch{(\s-t_1)}\,\ch{(\s -t_2)}}. \label{MS-XYD-TU2}
\end{align}
The above equality is a special case of the equation \eref{MS-XYD-1}, for the $T(U(2))$ theory and the 
vortex defect under consideration. Note that on taking the $z \to 1$ limit, one gets the result:
\begin{align}
Z^{(X[V^{l}_{1,k}])}(\vec m; \vec t) =& e^{2\pi i (m_1 t_1 - m_2 t_2)}\,\lim_{z \to 1}\,Z^{(Y[\wt{W}_k])}(\vec t; -\vec m | z) \nn\\
= & e^{2\pi i (m_1 t_1 - m_2 t_2)}\,Z^{(Y[\wt{W}_k])}(\vec t; -\vec m),
\end{align}
leading to the mirror map:
\be
\langle V^l_{1,k}\rangle_{X} (\vec m; \vec t)= \langle \wt{W}_{k}\rangle_{Y}(\vec t, -\vec m),
\ee
which can also be read off from the Type IIB construction. Proceeding in a similar fashion, one can similarly show that:
\be
\langle V^r_{1,k}\rangle_{X} (\vec m; \vec t)= \langle \wt{W}_{k}\rangle_{Y}(\vec t, -\vec m).
\ee

\begin{figure}[htbp]
\begin{center}
\begin{tabular}{ccc}
\scalebox{.7}{\begin{tikzpicture}[node distance=2cm,
cnode/.style={circle,draw,thick, minimum size=1.0cm},snode/.style={rectangle,draw,thick,minimum size=1cm}, nnode/.style={red, circle,draw,thick, minimum size=1.0cm}]
\node[snode] (1) at (0,0) {1} ;
\node[cf-group] (2) at (2,0) {\rotatebox{-90}{1}};
\node[snode] (3) at (4,0) {1};
\node[nnode] (4) at (1,2) {$k$};
\node[text width=1cm](5) at (2, -1) {$(X[V^{l}_{1,k}])$};
\draw[red, thick, ->] (1)--(4);
\draw[red, thick, ->] (4)--(2);
\draw[-] (1) -- (2);
\draw[-] (2) -- (3);
\draw[red, thick, ->] (4) to [out=150,in=210,looseness=8] (4);
\end{tikzpicture}}
& \qquad \qquad \qquad
& \scalebox{.7}{\begin{tikzpicture}[node distance=2cm,
cnode/.style={circle,draw,thick, minimum size=1.0cm},snode/.style={rectangle,draw,thick,minimum size=1cm}, pnode/.style={red,rectangle,draw,thick, minimum size=1.0cm}]
\node[cnode] (1) at (0,0) {1} ;
\node[snode] (2) at (0,-2) {2};
\draw[-] (1) -- (2);
\node[text width=0.1cm](3) at (0,0.8) {$k$};
\node[text width=1cm](4) at (0, -3) {$(Y[\wt{W}_k])$};
\end{tikzpicture}}\\
 \scalebox{.7}{\begin{tikzpicture}
\draw[thick, ->] (15,-3) -- (15,-5);
\node[text width=0.1cm](20) at (14.5, -4) {$\CO_\CP$};
\end{tikzpicture}}
&\qquad \qquad \qquad
& \scalebox{.7}{\begin{tikzpicture}
\draw[thick,->] (15,-3) -- (15,-5);
\node[text width=0.1cm](29) at (15.5, -4) {$\wt{\CO}_\CP$};
\end{tikzpicture}}\\
\scalebox{.7}{\begin{tikzpicture}[node distance=2cm, nnode/.style={circle,draw,thick, red, minimum size=1.0 cm},cnode/.style={circle,draw,thick,minimum size=1.0 cm},snode/.style={rectangle,draw,thick,minimum size=1.0 cm}]
\node[snode] (1) at (0,0) {1} ;
\node[cf-group] (2) at (2,0) {\rotatebox{-90}{1}};
\node[cnode] (3) at (4,0) {1};
\node[nnode] (4) at (1,2) {$k$};
\node[snode] (5) at (6,0) {1};
\draw[red, thick, ->] (1)--(4);
\draw[red, thick, ->] (4)--(2);
\draw[-] (1) -- (2);
\draw[-] (2) -- (3);
\draw[-] (3) -- (5);
\draw[red, thick, ->] (4) to [out=150,in=210,looseness=8] (4);
\node[text width=1cm](9) at (2, -1) {$(X'[V'^{(I)}_{1,k}])$};
\end{tikzpicture}}
&\qquad \qquad
& \scalebox{.7}{\begin{tikzpicture}[node distance=2cm,
cnode/.style={circle,draw,thick, minimum size=1.0cm},snode/.style={rectangle,draw,thick,minimum size=1cm}, pnode/.style={red,rectangle,draw,thick, minimum size=1.0cm}]
\node[cnode] (1) at (0,0) {1} ;
\node[snode] (2) at (0,-2) {3};
\draw[-] (1) -- (2);
\node[text width=0.1cm](3) at (0,0.8) {$k$};
\node[text width=1cm](4) at (0, -3) {$(Y'[\wt{W}'_k])$};
\end{tikzpicture}}
\end{tabular}
\caption{\footnotesize{The construction of a vortex defect and its dual Wilson defect, using a flavoring-gauging operation, with a pair of defects 
in $T(U(2))$ as a starting point.}}
\label{SimpAbEx1GFI}
\end{center}
\end{figure}
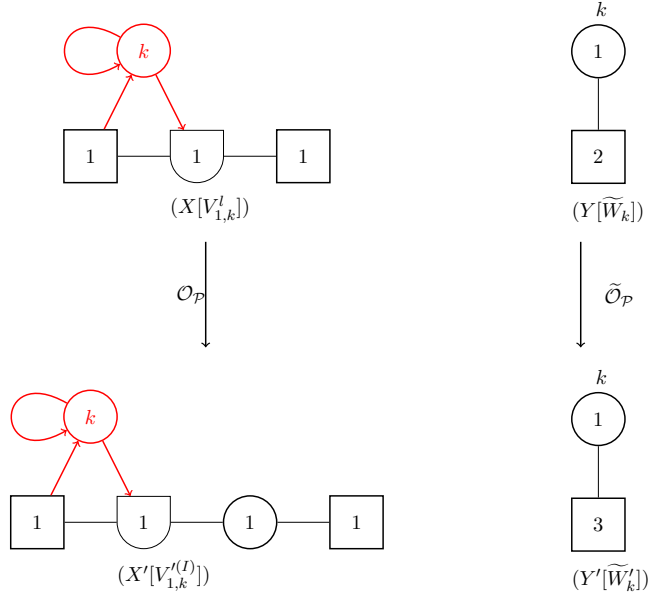

Now, one can take the 3d-1d system $(X[V^{l}_{1,k}])$ on the LHS of \figref{fig:Ex1ab}, and implement a flavoring-gauging operation on a $U(1)$ 
flavor node, as shown in \figref{SimpAbEx1GFI}, following the general procedure outlined in \Secref{SOps-defects}. This corresponds to the following 
choice of the parameters $(u,v)$ and the permutation matrix $\CP$:
\be
u=m_2, v=m_1, \qquad \CP = \begin{pmatrix} 0 & 1 \\ 1 & 0 \end{pmatrix}.
\ee
The resultant theory is another 3d-1d quiver, which we denote as $\CO_\CP(X[V^{l}_{1,k}])$. The defect 
partition function can be written down from the general expression \eref{S-Op-A} as follows:
\begin{align}
Z^{\CO_\CP(X[V^{l}_{1,k}])} = & e^{2\pi k t_1 } \times  \lim_{z \to 1}\, \Big(\int du \frac{e^{2\pi i \eta u}}{\ch{(u-m_f)}} \,Z^{T(U(2))}(v +ikz, u; \vec t)\Big)\nn\\
=& e^{2\pi k t_1 } \cdot \lim_{z \to 1}\,\int du\,ds\,\frac{e^{2\pi i \eta u}}{\ch{(u-m_f)}} \,Z^{T(U(2))}_{\rm int}(s, v, u, \vec t) \, \CI^{\Sigma^l}(s,v,z| \xi >0). 
\end{align}
The matrix model integral in the last step can be identified as the partition function of the 3d-1d quiver 
$X'[V'^{(I)}_{1,k}]$ in \figref{SimpAbEx1GFI}. 

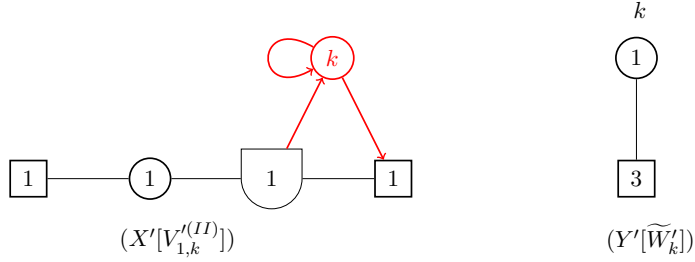
\begin{figure}[h]
\begin{center}
\scalebox{.8}{\begin{tikzpicture}[
nnode/.style={circle,draw,thick, red},cnode/.style={circle,draw,thick,minimum size=4mm},snode/.style={rectangle,draw,thick,minimum size=6mm}]
\node[snode] (1) at (0,0) {1} ;
\node[cnode] (2) at (2,0) {1};
\node[cf-group] (3) at (4,0) {\rotatebox{-90}{1}};
\node[nnode] (4) at (5,2) {$k$};
\node[snode] (5) at (6,0) {1};
\draw[red, thick, ->] (3)--(4);
\draw[red, thick, ->] (4)--(5);
\draw[-] (1) -- (2);
\draw[-] (2) -- (3);
\draw[-] (3) -- (5);
\draw[red, thick, ->] (4) to [out=150,in=210,looseness=8] (4);
\node[cnode] (6) at (10,2) {1} ;
\node[snode] (7) at (10,0) {3};
\draw[-] (6) -- (7);
\node[text width=0.1cm](3) at (10, 2.8) {$k$};
\node[text width=1cm](9) at (2, -1) {$(X'[V'^{(II)}_{1,k}])$};
\node[text width=1cm](10) at (10, -1) {$(Y'[\wt{W}'_k])$};
\end{tikzpicture}}
\caption{\footnotesize{The second realization of the vortex defect in the theory $X'$ via a 3d-1d system.}}
\label{fig:Ex1dual2}
\end{center}
\end{figure}

The dual defect partition function can be obtained from the 
general expressions \eref{PF-wtOPgenD-A2B}-\eref{CZ-wtOPD-A2B}. Firstly, one can compute the function
$\CZ_{\wt{\CO}_{\CP}(Y)}$ as follows:
\be
\CZ_{\wt{\CO}_{\CP}(Y)}(\s,v,m_f,\eta,\vec t)= \int du\,\frac{e^{2\pi i \eta u}}{\ch{(u-m_f)}}\,e^{2\pi i (\s -t_2) u}= \frac{e^{2\pi i m_f (\s +\eta -t_2)}}{\ch{(\s +\eta -t_2)}}.
\ee
Using the above expression and \eref{MS-XYD-TU2} in the general expression \eref{PF-wtOPgenD-A2B}, the 
defect partition function can be written as:
\begin{align}
Z^{\wt{\CO}_{\CP}(Y[\wt{W}_k])} = & \lim_{z \to 1}\, \int d\s \,\frac{e^{2\pi i m_f (\s +\eta -t_2)}}{\ch{(\s +\eta -t_2)}}\,  e^{2\pi i v t_1} \,
\frac{e^{-2\pi i \s v} \, e^{2\pi k\,\s \, z}}{\ch{(\s-t_1)}\,\ch{(\s -t_2)}} \nn \\
= & e^{2\pi i (v t_1 + m_f(\eta -t_2))} \, \int d\s \, \frac{e^{2\pi i (m_f -v) \s} \, e^{2\pi k \s}}{\ch{(\s-t_1)}\,\ch{(\s -t_2)}\, \ch{(\s +\eta - t_2)}}\\
=: & e^{2\pi i (v t_1 + m_f(\eta -t_2))} \, Z^{(Y'[\wt{W}'_k])}, 
\end{align}
where we can set $\eta = t_2 -t_3$, and $v=m_1$, $m_f =m_2$, to write the above partition function in a more standard form. 
The matrix model in the second step can be identified as the partition function of the theory $Y'$ -- 
a $U(1)$ gauge theory with three hypermultiplets of charge 1 -- with a gauge $U(1)$ Wilson line of charge $k$ inserted. 
The theory is denoted as $Z^{(Y'[\wt{W}'_k])}$ in \figref{SimpAbEx1GFI}. The above computation then implies the mirror map:
\be \label{MM-Ex1dual1}
\langle V'^{(I)}_{1,k}\rangle_{X'} = \langle \wt{W}'_{k}\rangle_{Y'}.
\ee
\\
One can similarly start from the 3d-1d system $X[V^{r}_{1,k}]$ on the RHS of \figref{fig:Ex1ab}, one can implement 
a flavoring-gauging operation with
\be
u=m_1, v=m_2, \qquad \CP = \begin{pmatrix} 1 & 0 \\ 0 & 1 \end{pmatrix}.
\ee
The resultant dual system is shown in \figref{fig:Ex1dual2}, which leads to the mirror map of the defect operators:
\be\label{MM-Ex1dual2}
\langle V'^{(II)}_{1,k}\rangle_{X'} = \langle \wt{W}'_{k}\rangle_{Y'}.
\ee

Finally, we can also take the 3d-1d quiver $(X[V^{l}_{1,k}])$ in \figref{fig:Ex1ab}, and implement a flavoring-gauging operation 
on the other  $U(1)$ flavor node, which corresponds to the following choice of the parameters $(u,v)$ and the permutation matrix $\CP$:
\be
u=m_1, v=m_2, \qquad \CP = \begin{pmatrix} 0 & 1 \\ 1 & 0 \end{pmatrix}.
\ee
The resultant dual system is shown in \figref{fig:Ex1dual2}, which leads to the mirror map of the defect operators:
\be\label{MM-Ex1dual3}
\langle V'^{(III)}_{1,k}\rangle_{X'} = \langle \wt{W}'_{k}\rangle_{Y'}.
\ee
From \eref{MM-Ex1dual1}, \eref{MM-Ex1dual2}, \eref{MM-Ex1dual3}, we observe the fact that the vertex operator in $X'$ dual to the 
gauge Wilson defect $\wt{W}'_{k}$ in $Y'$ has three different realizations as a 3d-1d quivers. This is also what one expects from the Type IIB description of the vortex operator in $X'$, where the stack of $k$ D1-branes can end on any one of the three NS5-branes, thereby leading to the three 
3d-1d systems.

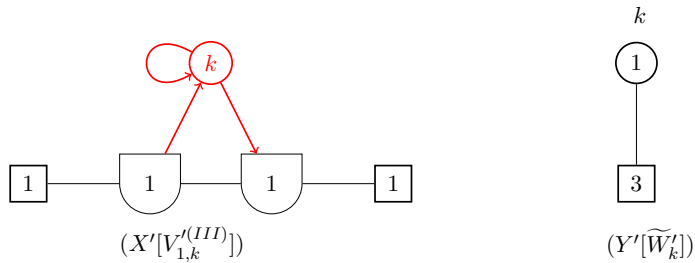
\begin{figure}[h]
\begin{center}
\scalebox{.8}{\begin{tikzpicture}[
nnode/.style={circle,draw,thick, red},cnode/.style={circle,draw,thick,minimum size=4mm},snode/.style={rectangle,draw,thick,minimum size=6mm}]
\node[snode] (1) at (0,0) {1} ;
\node[cf-group] (2) at (2,0) {\rotatebox{-90}{1}};
\node[cf-group] (3) at (4,0) {\rotatebox{-90}{1}};
\node[nnode] (4) at (3,2) {$k$};
\node[snode] (5) at (6,0) {1};
\draw[red, thick, ->] (2)--(4);
\draw[red, thick, ->] (4)--(3);
\draw[-] (1) -- (2);
\draw[-] (2) -- (3);
\draw[-] (3) -- (5);
\draw[red, thick, ->] (4) to [out=150,in=210,looseness=8] (4);
\node[cnode] (6) at (10,2) {1} ;
\node[snode] (7) at (10,0) {3};
\draw[-] (6) -- (7);
\node[text width=0.1cm](3) at (10, 2.8) {$k$};
\node[text width=1cm](9) at (2, -1) {$(X'[V'^{(III)}_{1,k}])$};
\node[text width=1cm](10) at (10, -1) {$(Y'[\wt{W}'_k])$};
\end{tikzpicture}}
\caption{\footnotesize{The third realization of the vortex defect in the theory $X'$ via a 3d-1d system.}}
\label{fig:Ex1dual3}
\end{center}
\end{figure}

\section{Beyond linear quivers : $D_n$ and $\wh{D}_n$ quiver gauge theories}\label{1-D}

In this section, we will focus on deriving the mirror map of defects in quiver gauge theories of the 
$D_n$ and $\wh{D}_n$ type, as a concrete application of the general construction developed 
in \Secref{SOps-defects}-\Secref{SOps-defects-d}. The $D_n$ and $\wh{D}_n$ quiver gauge theories 
discussed in this section can be realized by a Type IIB construction involving configurations of D3-D5-NS5 branes 
with one and two orbifold 5-planes respectively.

\subsection{Defects in a $D_4$ quiver}\label{SwD-D4}

In this section, we will discuss an example of mirror symmetry involving a $D_4$ quiver gauge theory 
with a gauge group $G=U(2) \times U(1)^3$ and a single fundamental hypermultiplet charged under 
the $U(2)$ factor, which is dual to an $SU(2)$ gauge theory with $N_f=4$ flavors. 

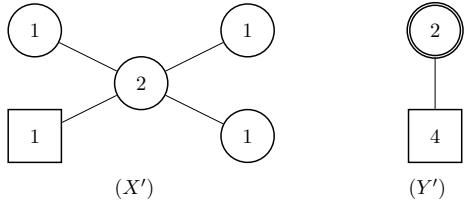
\begin{center}
 \scalebox{.7}{\begin{tikzpicture}[node distance=2cm, nnode/.style={circle,draw,thick, red, fill=red!30, minimum size=2.0 cm},cnode/.style={circle,draw,thick,minimum size=1.0 cm},snode/.style={rectangle,draw,thick,minimum size=1.0 cm}]
\node[cnode] (1) at (0,1) {1} ;
\node[cnode] (2) at (2,0) {2};
\node[snode] (3) at (0,-1) {1};
\node[cnode] (5) at (4, 1) {1};
\node[cnode] (6) at (4, -1) {1};
\draw[-] (1) -- (2);
\draw[-] (2) -- (3);
\draw[-] (2) -- (5);
\draw[-] (2) -- (6);
\node[text width=1cm](9) at (2, -2) {$(X')$};
\end{tikzpicture}}
\qquad \qquad
 \scalebox{.7}{\begin{tikzpicture}[node distance=2cm,
cnode/.style={circle,draw,thick, minimum size=1.0cm},snode/.style={rectangle,draw,thick,minimum size=1cm}, pnode/.style={circle,draw,double,thick, minimum size=1.0cm}, lnode/.style = {shape = rounded rectangle, minimum size=1.0cm, rotate=90, rounded rectangle right arc = none, draw, double}]
\node[pnode] (1) at (0,0) {2} ;
\node[snode] (2) at (0,-2) {4};
\draw[-] (1) -- (2);
\node[text width=1cm](4) at (0, -3) {$(Y')$};
\end{tikzpicture}}
\end{center}

Given the dual pair, we will explicitly construct the mirror map for a class of Wilson/vortex defects in 
the dual theories, starting from defects in a dual pair of linear quivers, using Abelian $S$-type operations.\\

\begin{figure}[htbp]
\begin{center}
\scalebox{.7}{\begin{tikzpicture}[cnode/.style={circle,draw,thick, minimum size=1.0cm},snode/.style={rectangle,draw,thick,minimum size=1cm}]
\node[cnode] (1) {2};
\node[snode] (2) [below=1cm of 1]{4};
\draw[-] (1) -- (2);
\node[text width=0.1cm](20)[below=0.5 cm of 2]{$(X)$};
\end{tikzpicture}
\qquad \qquad \qquad \qquad 
\begin{tikzpicture}[
cnode/.style={circle,draw,thick, minimum size=1.0cm},snode/.style={rectangle,draw,thick,minimum size=1cm}]
\node[cnode] (1) {1};
\node[cnode] (2) [right=1cm  of 1]{2};
\node[cnode] (3) [right=1cm  of 2]{1};
\node[snode] (4) [below=1cm of 2]{2};
\node[text width=0.1cm](20)[below=0.5 cm of 4]{$(Y)$};
\draw[-] (1) -- (2);
\draw[-] (2)-- (3);
\draw[-] (2)-- (4);
\end{tikzpicture}}
\end{center}
\caption{\footnotesize{A pair of dual linear quivers with unitary gauge groups.}}
\label{LQ-basic}
\end{figure}
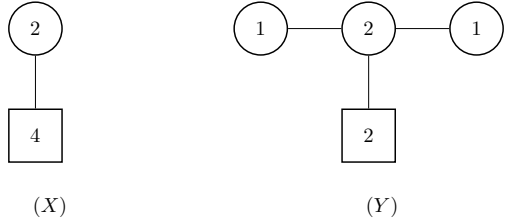

\begin{figure}[htbp]
\begin{center}
\scalebox{.7}{\begin{tikzpicture}[node distance=2cm,
cnode/.style={circle,draw,thick, minimum size=1.0cm},snode/.style={rectangle,draw,thick,minimum size=1.0cm}, 
nnode/.style={red, circle,draw,thick,fill=red!30 ,minimum size=2.0cm}]
\node[snode] (1) at (0,0) {$4-M$} ;
\node[cf-group] (2) at (2,0) {\rotatebox{-90}{2}};
\node[snode] (3) at (4,0) {$M$};
\node[nnode] (4) at (2,2) {$\Sigma^{M,R}$};
\node[text width=1cm](5) at (2, -1) {$(X[V^r_{M,R}])$};
\draw[red, thick, ->] (2)--(4);
\draw[red, thick, ->] (4)--(3);
\draw[-] (1) -- (2);
\draw[-] (2) -- (3);
\end{tikzpicture}}
\qquad \qquad
\scalebox{.7}{\begin{tikzpicture}[node distance=2cm,
cnode/.style={circle,draw,thick, minimum size=1.0cm},snode/.style={rectangle,draw,thick,minimum size=1.0cm}, 
nnode/.style={red, circle,draw,thick,fill=red!30 ,minimum size=2.0cm}]
\node[snode] (1) at (0,0) {$4-M$} ;
\node[cf-group] (2) at (2,0) {\rotatebox{-90}{2}};
\node[snode] (3) at (4,0) {$M$};
\node[nnode] (4) at (2,2) {$\Sigma^{M,R}$};
\node[text width=1cm](5) at (2, -1) {$(X[V^l_{M,R}])$};
\draw[red, thick, ->] (1)--(4);
\draw[red, thick, ->] (4)--(2);
\draw[-] (1) -- (2);
\draw[-] (2) -- (3);
\end{tikzpicture}}
\end{center}
\caption{\footnotesize{A family of vortex defects $V_{M,R}$ in the theory $X$ and the associated 3d-1d coupled quivers.}}
\label{LQ-VX}
\end{figure}
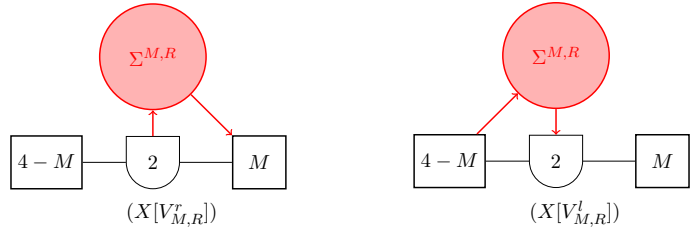

The starting point is the dual pair of linear quivers in \figref{LQ-basic}.
The mirror map of defect operators for the dual pair above was discussed in detail 
in \cite{Assel:2015oxa}, using the $S^3$ partition function as well as the Type IIB brane construction. A set of vortex defects in quiver gauge theory 
$X$, labelled as $V_{M, R}$, can be realized as the coupled 3d-1d systems in \figref{LQ-VX}, with $R$ being a representation of the gauge group 
$U(2)$ of the form \eref{gen-R}, and $0 \leq M \leq 4$.

The coupled 3d-1d systems above are two equivalent ways of describing the same defect $V_{M, R}$, and
represent a hopping duality associated with the vortex defect. We will refer to the two SQMs as $\Sigma_r^{M,R}$ 
and $\Sigma_l^{M,R}$ respectively.\\

We will denote the dual Wilson defect in the quiver gauge theory $Y$ generically as $\wt{W}_{\wt{R}}$, where $\wt{R}$ is related to the 
representation $R$, and the precise relation is part of the mirror map between the defect operators. 
The dual Wilson defects were worked out in \cite{Assel:2015oxa} and have been reviewed in \Appref{U24-V2W}.  
The mirror maps for various choices of the integer $M$ can be enumerated as follows:
\begin{itemize} 
\item {\bf {$M=0,4$:}} These vortex defects are mapped to flavor Wilson defects in the theory $Y$. Let $U(2)_f$ denote the flavor symmetry 
group of $Y$ associated with the fundamental hypermultiplets, and let  
$U(1)_1 \times U(1)_2 \subset U(2)_f$ be the maximal torus of $U(2)_f$. Also, let $\wt{W}^{\rm flavor}_{i, q}$ denote Wilson defects for $U(1)_{i=1,2}$ with charge $q$. The mirror maps
are then given as:
\begin{align}
&\langle V_{0,R} \rangle_X (\vec m;\vec t)= \langle \wt{W}^{\rm flavor}_{2, |R|} \rangle_Y (\vec t; -\vec m) ,\\
&\langle V_{4,R} \rangle_X (\vec m;\vec t)= \langle \wt{W}^{\rm flavor}_{1, |R|} \rangle_Y (\vec t; -\vec m).
\end{align}
\item {\bf{$M=1,3$:}} These vortex defects are mapped to a combination of flavor Wilson defects and gauge Wilson 
defect in a $U(1)$ subgroup of the gauge group $\wt{G}=U(1) \times U(2) \times U(1)$ of the theory $Y$. 
The mirror maps are given as:
\begin{align}
& \langle V_{1,R} \rangle_X (\vec m; \vec t) = \langle \sum_{\kappa \in \Delta} \wt{W}^{\rm flavor} _{2, q^\kappa_2}\, \wt{W}^{(3)}_{q^\kappa_1} \rangle_Y (\vec t; -\vec m), \nn \\
&  \langle V_{3,R} \rangle_X (\vec m; \vec t) = \langle  \sum_{\kappa \in \Delta} \wt{W}^{\rm flavor} _{1, q^\kappa_2}\, \wt{W}^{(1)}_{q^\kappa_1} \rangle_Y (\vec t; -\vec m),
\end{align}
where $\wt{W}^{(\gamma'=1,3)}_{q^\kappa_1}$ are gauge Wilson defects of charge $q^\kappa_1$ for the $U(1)$ gauge nodes, 
labelled $\gamma'=1,3$ in the quiver $Y$. The charges $(q^\kappa_1, q^\kappa_2)$ are obtained from the decomposition of 
the representation $R$ under the maximal torus $U(1) \times U(1) \subset U(2)$, and $\Delta$ is the set of such charge doublets 
counted with degeneracies.\\

\item {\bf{$M=2$:}} This vortex defect maps to a gauge Wilson defect $\wt{W}_R$ in the same representation $R$ of the central $U(2)$ gauge node of 
the theory $Y$, i.e.
\be
\langle V_{2,R} \rangle_X (\vec m; \vec t) = \langle \wt{W}_{R} \rangle_Y (\vec t; -\vec m).
\ee
\end{itemize}

One can similarly consider a Wilson defect $W_R$, labelled by a representation $R$ of the U(2) gauge group in the theory $X$. The dual object  
is a vortex defect $\wt{V}_{1,R}$, labelled by a representation $R$ of the central $U(2)$ gauge group in the quiver gauge theory $Y$. 
The coupled 3d-1d systems, worked out in \cite{Assel:2015oxa}, that realize the vortex defect are given in \figref{LQ-VY}.
\begin{figure}[htbp]
\begin{center}
\scalebox{.7}{\begin{tikzpicture}[node distance=2cm,
cnode/.style={circle,draw,thick, minimum size=1.0cm},snode/.style={rectangle,draw,thick,minimum size=1cm}, nnode/.style={red, circle,draw,thick,fill=red!30 ,minimum size=2.0cm}]
\node[cnode] (1) at (0,0) {1} ;
\node[cf-group] (2) at (2,0) {\rotatebox{-90}{2}};
\node[cf-group] (3) at (4,0) {\rotatebox{-90}{1}};
\node[nnode] (4) at (3,2) {$\wt{\Sigma}^{1,R}$};
\node[snode] (5) at (3,-2) {1};
\node[snode] (6) at (1,-2) {1};
\node[text width=1cm](7) at (2, -3) {$(Y[\wt{V}^r_{1,R}])$};
\draw[red, thick, ->] (2)--(4);
\draw[red, thick, ->] (4)--(3);
\draw[red, thick, ->] (4)--(5);
\draw[-] (1) -- (2);
\draw[-] (2) -- (3);
\draw[-] (2) -- (5);
\draw[-] (2) -- (6);
\end{tikzpicture}}
\qquad \qquad \qquad
\scalebox{.7}{\begin{tikzpicture}[node distance=2cm,
cnode/.style={circle,draw,thick, minimum size=1.0cm},snode/.style={rectangle,draw,thick,minimum size=1cm}, nnode/.style={red, circle,draw,thick,fill=red!30 ,minimum size=2.0cm}]
\node[cf-group] (1) at (0,0) {\rotatebox{-90}{1}} ;
\node[cf-group] (2) at (2,0) {\rotatebox{-90}{2}};
\node[cnode] (3) at (4,0) {1};
\node[nnode] (4) at (1,2) {$\wt{\Sigma}^{1,R}$};
\node[snode] (5) at (3,-2) {1};
\node[snode] (6) at (1,-2) {1};
\node[text width=1cm](7) at (2, -3) {$(Y[\wt{V}^l_{1,R}])$};
\draw[red, thick, ->] (1)--(4);
\draw[red, thick, ->] (6)--(4);
\draw[red, thick, ->] (4)--(2);
\draw[-] (1) -- (2);
\draw[-] (2) -- (3);
\draw[-] (2) -- (5);
\draw[-] (2) -- (6);
\end{tikzpicture}}
\end{center}
\caption{\footnotesize{A family of vortex defects $\wt{V}_{1,R}$ in the theory $Y$ and the associated 3d-1d coupled quivers.}}
\label{LQ-VY}
\end{figure}

The construction of the mirror map in terms of the partition function is reviewed in \Secref{U24-W2V}, and is given as:
\be
\langle W_{R} \rangle_{X}(\vec m; \vec t) = \langle  \wt{V}_{1,R} \rangle_{Y}(-\vec t; \vec m).
\ee

\subsubsection{A vortex defect for the central node}\label{V2W-D}

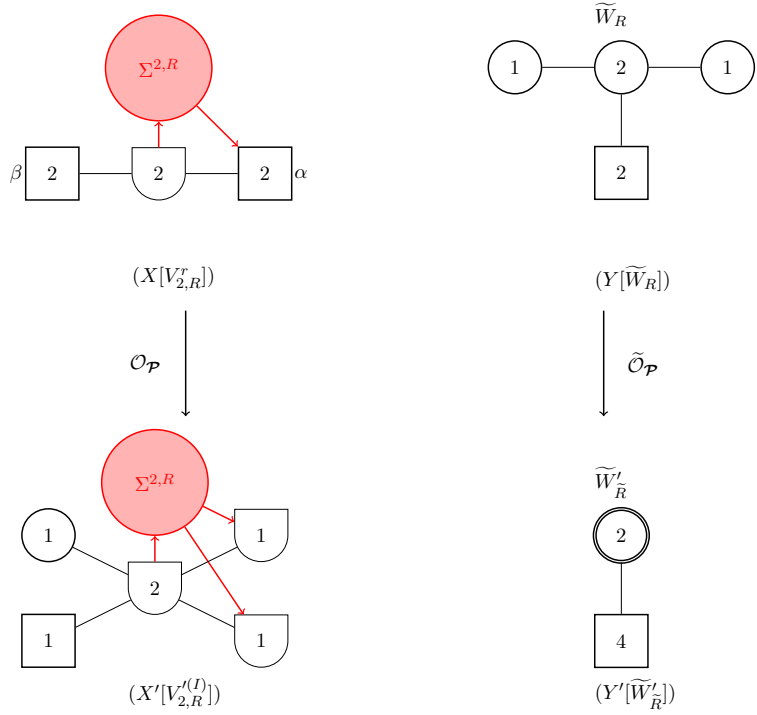
\begin{figure}[htbp]
\begin{center}
\begin{tabular}{ccc}
\scalebox{.7}{\begin{tikzpicture}[node distance=2cm,
cnode/.style={circle,draw,thick, minimum size=1.0cm},snode/.style={rectangle,draw,thick,minimum size=1.0cm}, nnode/.style={red, circle,draw,thick,fill=red!30 ,minimum size=2.0cm}]
\node[snode] (1) at (0,0) {2} ;
\node[cf-group] (2) at (2,0) {\rotatebox{-90}{2}};
\node[snode] (3) at (4,0) {2};
\node[nnode] (4) at (2,2) {$\Sigma^{2,R}$};
\node[text width=1cm](7) at (2, -2) {$(X[V^r_{2,R}])$};
\node[text width=0.1cm](8) at (4.6, 0) {$\alpha$};
\node[text width=0.1cm](9) at (-.75, 0) {$\beta$};
\draw[red, thick, ->] (2)--(4);
\draw[red, thick, ->] (4)--(3);
\draw[-] (1) -- (2);
\draw[-] (2) -- (3);
\end{tikzpicture}}
& \qquad \qquad \qquad
& \scalebox{.7}{\begin{tikzpicture}[node distance=2cm,
cnode/.style={circle,draw,thick, minimum size=1.0cm},snode/.style={rectangle,draw,thick,minimum size=1cm}, pnode/.style={red,rectangle,draw,thick, minimum size=1.0cm}]
\node[cnode] (1) at (0,0) {1} ;
\node[cnode] (2) at (2,0) {2} ;
\node[cnode] (3) at (4,0) {1} ;
\node[snode] (4) at (2,-2) {2};
\draw[-] (1) -- (2);
\draw[-] (2) -- (3);
\draw[-] (2) -- (4);
\node[text width=1cm](5) at (2, 1) {$\wt{W}_{R}$};
\node[text width=1cm](6) at (2, -4) {$(Y[\wt{W}_{R}])$};
\end{tikzpicture}}\\
 \scalebox{.7}{\begin{tikzpicture}
\draw[thick, ->] (15,-3) -- (15,-5);
\node[text width=0.1cm](20) at (14.0, -4) {$\CO_{\vec \CP}$};
\end{tikzpicture}}
&\qquad \qquad \qquad
& \scalebox{.7}{\begin{tikzpicture}
\draw[thick,->] (15,-3) -- (15,-5);
\node[text width=0.1cm](29) at (15.5, -4) {$\wt{\CO}_{\vec \CP}$};
\end{tikzpicture}}\\
\scalebox{.7}{\begin{tikzpicture}[node distance=2cm, nnode/.style={circle,draw,thick, red, fill=red!30, minimum size=2.0 cm},cnode/.style={circle,draw,thick,minimum size=1.0 cm},snode/.style={rectangle,draw,thick,minimum size=1.0 cm}]
\node[cnode] (1) at (0,1) {1} ;
\node[cf-group] (2) at (2,0) {\rotatebox{-90}{2}};
\node[snode] (3) at (0,-1) {1};
\node[cf-group] (5) at (4, 1) {\rotatebox{-90}{1}};
\node[cf-group] (6) at (4, -1) {\rotatebox{-90}{1}};
\node[nnode] (7) at (2,2) {$\Sigma^{2,R}$};
\draw[red, thick, ->] (2)--(7);
\draw[red, thick, ->] (7)--(5);
\draw[red, thick, ->] (7)--(6);
\draw[-] (1) -- (2);
\draw[-] (2) -- (3);
\draw[-] (2) -- (5);
\draw[-] (2) -- (6);
\node[text width=1cm](9) at (2, -2) {$(X'[V'^{(I)}_{2,R}])$};
\end{tikzpicture}}
&\qquad \qquad
& \scalebox{.7}{\begin{tikzpicture}[node distance=2cm,
cnode/.style={circle,draw,thick, minimum size=1.0cm},snode/.style={rectangle,draw,thick,minimum size=1cm}, pnode/.style={circle,draw,double,thick, minimum size=1.0cm}, lnode/.style = {shape = rounded rectangle, minimum size=1.0cm, rotate=90, rounded rectangle right arc = none, draw, double}]
\node[pnode] (1) at (0,0) {2} ;
\node[snode] (2) at (0,-2) {4};
\draw[-] (1) -- (2);
\node[text width=1cm](3) at (0,1) {$\wt{W}'_{\wt R}$};
\node[text width=1cm](4) at (0, -3) {$(Y'[\wt{W}'_{\wt R}])$};
\end{tikzpicture}}
\end{tabular}
\caption{\footnotesize{The construction of a vortex defect in a flavored ${D}_4$ quiver gauge theory and its dual Wilson defect, using an Abelian gauging operation.}}
\label{AbSU2}
\end{center}
\end{figure}

We begin with an example where a gauge vortex defect in the $D_4$ quiver gauge theory is mapped to a gauge Wilson defect in the 
$SU(2)$ theory. 
Consider the dual pair of defects $(V^r_{2,R}, \wt{W}_R)$ -- the coupled 3d-1d system (corresponding to the right SQM) 
and its dual are shown in the first line of \figref{AbSU2}. Let the fundamental masses in theory $X$ be labelled as 
$\{m_i| i=1,\ldots, 4\}$, such that the $U(2)$ flavor node identified with the 1d flavor symmetry in $X[V^r_{M,R}]$ is associated with the 
masses $m_3,m_4$. Following the general prescription of \Secref{SOps-defects}, we implement 
an Abelian $S$-type operation on the system $X[V^r_{2, R}]$, consisting of three elementary Abelian gauging operations, as follows:
\be \label{AbS-D}
\CO_{\vec \CP} (X[V^r_{2, R}]) = G^\beta_{\CP_3} \circ G^{\alpha_2}_{\CP_2} \circ G^{\alpha_1}_{\CP_1} (X[V^r_{2, R}]),
\ee
where $\alpha_1=\alpha$, and $\alpha_2$ is the residual $U(1)$ flavor node from $U(2)_\alpha$ in the theory $G^{\alpha_1}_{\CP_1} (X)$.
The mass parameters corresponding to the $U(1)^3$ global symmetry are chosen as:
\be \label{u-choice1}
u_1=m_3, \quad u_2=m_4, \quad u_3=m_1.
\ee 
Note that the $S$-type operation is splitting the $U(2)_\alpha$ flavor node into two 
$U(1)$ flavor nodes and gauging them. 
From \eref{S-Op-A}, the partition function of the resultant 3d-1d system can then be written down 
as follows:
\begin{align}
&Z^{\CO_{\vec \CP}(X[V^r_{2,R}])} =\lim_{z\to 1} \int \prod^3_{i=1} du_i \Big[d\vec s\Big]  \prod^3_{i=1} Z_{\rm FI} (u_i, \eta_i)
Z^{(X)}_{\rm int}(\vec s, \vec u, v, \vec t)\,W_{\rm b.g.} (\vec t, R)\, \CI^{\Sigma^{2,R}_r}(\vec s, \vec u, z | \vec \xi < 0), \label{ZAbS-D4}\\
& Z^{(X)}_{\rm int}(\vec s, \vec u, v, \vec t)= \Big\{Z^{(X)}_{\rm int}(\vec s, \vec m, \vec t)| m_3=u_1, m_4=u_2, m_1=u_3, m_2=v \Big\},\label{ZAbS-D4a}\\
&W_{\rm b.g.} (\vec t, R)= e^{2\pi t_2 |R|}, \label{ZAbS-D4b}\\
& \CI^{\Sigma^{2,R}_r}(\vec s, \vec u, z| \vec \xi < 0)= \sum_{w \in R} \prod^2_{j=1} \prod^2_{i=1} \frac{\ch{(s_j -u_i)}}{\ch{(s_j + i w_j z - u_i)}},
\label{ZAbS-D4c}
\end{align}
where $w=(w_1, w_2)$ is a weight of representation $R$ of $U(2)$.
The expression on the RHS of \eref{ZAbS-D4} can be readily identified as the partition function of a coupled 3d-1d system, where the 3d theory is a
${D}_4$ quiver gauge theory $X'$ with a gauge group $G'= U(2) \times U(1)^3$, and the SQM is $\Sigma^{2,R}_r$. 
This 3d-1d quiver is denoted as $X'[V'^{(I)}_{2,R}]$ in \figref{AbSU2} on the LHS of the second line. We claim that this 3d-1d quiver realizes a vortex defect in the representation $R$ of the central $U(2)$ gauge node in the theory $X'$. In fact, we will find it convenient to redefine the defect partition function by a global Wilson defect factor as follows:
\be
Z^{(X'[V'^{(I)}_{2,R}])}:= e^{-2\pi (t_2 -\frac{\eta_1+\eta_2}{2})|R|} \, Z^{\CO_{\vec \CP}(X[V^r_{2,R}])}(v; \vec t, \vec \eta),
\ee
where the superscript $(I)$ denotes the specific 3d-1d system that arises in the construction described above. \\

The dual system can be determined following the discussion of \Secref{SOps-defects-d}. The dual partition function
can be read off from the general expressions \eref{PF-wtOPgenD-A2B}-\eref{CZ-wtOPD-A2B} -- the details of the computation 
are summarized in \Appref{app:SU(2)-V2W}. The dual partition function can be written in the following form:
\begin{align}
& Z^{(Y'[(V'^{(I)}_{2,R})^\vee])}= e^{-2\pi (t_2 -\frac{\eta_1+\eta_2}{2})|R|} \, Z^{\wt{\CO}_{\vec \CP}(Y[\wt{W}_{R}])},\\
& Z^{(Y'[(V'^{(I)}_{2,R})^\vee])}=C(v, \vec \eta, \vec t)\,\int \,\Big[d\vec \s\Big] \, \delta(\tr \vec{\s}) \, Z^{U(2), N_f=4}_{\rm int} (\vec{\s}, \vec m'(\vec t, \vec\eta), \eta'=0)\, \sum_{w \in R} e^{2\pi \sum_j w_j \s_j}, \label{ZAbS-D4-d}\\
&C(v, \vec \eta, \vec t)= e^{2\pi i v(\eta_1 +\eta_2+\eta_3 + 2(t_1-t_2))},
\end{align}
where $C(v, \vec \eta, \vec t)$ is a contact term, $Z^{U(2), N_f=4}_{\rm int} (\vec{\s}, \vec m'(\vec t, \vec\eta), \eta')$ is the integrand for the partition function of a $U(2)$ gauge theory with $N_f=4$ hypermultiplets, $\vec m'(\vec t, \vec\eta)$ are the masses of the hypermultiplets, and $\eta'$ is the FI parameter for the $U(2)$ gauge group. \\

The RHS of \eref{ZAbS-D4-d} implies that the theory $Y'[(V'^{(I)}_{2,R})^\vee]$  
can be identified as an $SU(2)$ gauge theory with $N_f=4$ flavors, along with a Wilson defect in a representation $\wt{R}$ of the 
$SU(2)$ gauge group, 
where the representation $\wt{R}$ is the restriction of the $U(2)$ representation $R$ to an $SU(2)$ subgroup. More explicitly, we have
\begin{align}\label{def-wtR}
U(2) & \rightarrow  SU(2), \nn \\
R & \rightarrow \wt R.
\end{align}
We denote the theory $Y'$ with the Wilson defect insertion as $Y'[\wt{W}'_{\wt R}]$, as 
shown on the RHS of the second line in \figref{AbSU2}. Normalizing the integrals with appropriate partition functions, 
we obtain the following mirror map of line defects in the pair of dual theories $(X',Y')$:
\be \label{MM-D4-V2W1}
\langle V'^{(I)}_{2,R} \rangle_{X'}(\vec t, \vec \eta) = \langle  \wt{W}'_{\wt R} \rangle_{Y'}(\vec m'(\vec t, \vec \eta)),
\ee
where the functions $\vec m'(\vec t, \vec \eta)$ can be read off from \eref{mm-1}-\eref{mm-4}.\\

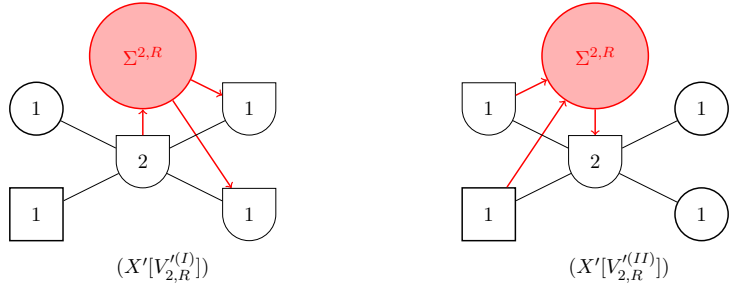
\begin{figure}[htbp]
\begin{center}
\scalebox{.7}{\begin{tikzpicture}[node distance=2cm, nnode/.style={circle,draw,thick, red, fill=red!30, minimum size=2.0 cm},cnode/.style={circle,draw,thick,minimum size=1.0 cm},snode/.style={rectangle,draw,thick,minimum size=1.0 cm}]
\node[cnode] (1) at (0,1) {1} ;
\node[cf-group] (2) at (2,0) {\rotatebox{-90}{2}};
\node[snode] (3) at (0,-1) {1};
\node[cf-group] (5) at (4, 1) {\rotatebox{-90}{1}};
\node[cf-group] (6) at (4, -1) {\rotatebox{-90}{1}};
\node[nnode] (7) at (2,2) {$\Sigma^{2,R}$};
\draw[red, thick, ->] (2)--(7);
\draw[red, thick, ->] (7)--(5);
\draw[red, thick, ->] (7)--(6);
\draw[-] (1) -- (2);
\draw[-] (2) -- (3);
\draw[-] (2) -- (5);
\draw[-] (2) -- (6);
\node[text width=1cm](9) at (2, -2) {$(X'[V'^{(I)}_{2,R}])$};
\end{tikzpicture}}
\qquad \qquad \qquad
\scalebox{.7}{\begin{tikzpicture}[node distance=2cm, nnode/.style={circle,draw,thick, red, fill=red!30, minimum size=2.0 cm},cnode/.style={circle,draw,thick,minimum size=1.0 cm},snode/.style={rectangle,draw,thick,minimum size=1.0 cm}]
\node[cf-group] (1) at (0,1) {\rotatebox{-90}{1}} ;
\node[cf-group] (2) at (2,0) {\rotatebox{-90}{2}};
\node[snode] (3) at (0,-1) {1};
\node[cnode] (5) at (4, 1) {1};
\node[cnode] (6) at (4, -1) {1};
\node[nnode] (7) at (2,2) {$\Sigma^{2,R}$};
\draw[red, thick, ->] (7)--(2);
\draw[red, thick, ->] (1)--(7);
\draw[red, thick, ->] (3)--(7);
\draw[-] (1) -- (2);
\draw[-] (2) -- (3);
\draw[-] (2) -- (5);
\draw[-] (2) -- (6);
\node[text width=1cm](9) at (2, -2) {$(X'[V'^{(II)}_{2,R}])$};
\end{tikzpicture}}
\caption{\footnotesize{Different 3d-1d systems which realize the same vortex defect $V'_{2,R}$ in the ${D}_4$ quiver gauge theory $X'$. 
The Witten indices in the two cases are computed in the chambers $\vec \xi < 0$ and $\vec \xi > 0$ respectively. 
The system $(I)$ is realized by starting with the right SQM for the defect $V_{2,R}$ in a $U(2)$ gauge theory with $N_f=4$, while $(II)$ 
is realized when one starts with the left SQM.}}
\label{AbEx1D4Hopping}
\end{center}
\end{figure}

A second coupled 3d-1d quiver $(X'[V'^{(II)}_{2,R}])$ which realizes the same vortex defect is shown in \figref{AbEx1D4Hopping}.
The Witten index for the coupled SQM in this case should be computed in the chamber $\vec \xi > 0$.
One can read off this 3d-1d quiver directly by a change of variables in the matrix integral of \eref{ZAbS-D4}: $s_j \to s_j - iw_j\,z$.
It can also be obtained by implementing the $S$-type operation \eref{AbS-D}, characterized by the choice of the 
$U(1)^3$ parameters \eref{u-choice1}, on the theory $(X[V^{l}_{2,R}])$. 
Note that the $U(2)$ flavor node identified with 1d flavor symmetry is, in this case, being split into 
two $U(1)$ flavor nodes, of which only one is being gauged. The defect partition function is again redefined by a global 
Wilson defect factor, i.e.
\be
Z^{(X'[V'^{(II)}_{2,R}])}:= e^{-2\pi (t_2 -\frac{\eta_1+\eta_2}{2})|R|} \, Z^{\CO_{\vec \CP}(X[V^l_{2,R}])}(v; \vec t, \vec \eta).
\ee

Proceeding as before, the dual partition function can be worked out from \eref{PF-wtOPgenD-A2B}-\eref{CZ-wtOPD-A2B} 
leading to the relation:
\be 
Z^{(Y'[(V'^{(II)}_{2,R})^\vee])} = Z^{(Y'[(V'^{(I)}_{2,R})^\vee])},
\ee
where $Z^{(Y'[(V'^{(I)}_{2,R})^\vee])}$ is given in \eref{ZAbS-D4-d}. 
Therefore, one concludes that the two 3d-1d coupled systems $X'[V'^{(I)}_{2,R}]$ and $X'[V'^{(II)}_{2,R}]$ are both realizations of the 
same vortex defect in the theory $X'$. This leads to the following mirror map:
\begin{empheq}[box=\widefbox]{align}
\langle V'^{(II)}_{2,R} \rangle_{X'}(\vec t, \vec \eta)= \langle V'^{(I)}_{2,R} \rangle_{X'}(\vec t, \vec \eta) = \langle  \wt{W}'_{\wt R} \rangle_{Y'}(\vec m'(\vec t, \vec \eta)).
\end{empheq}
We will discuss how this hopping duality arises from the Type IIB description of defects in 
$D$-type quivers in \Secref{D-TypeIIB-def}.

One can ask what is the interpretation of the additional 3d-1d quivers which can be obtained from $(X'[V^{r}_{2,R}])$ (or $(X'[V^{l}_{2,R}])$)
by an Abelian $S$-type operation of the same form as $\CO_{\vec \CP}$ in \eref{AbS-D} but involving a different choice of the $U(1)^3$ global symmetry to be gauged. These 3d-1d quivers can be shown to be identical to the quivers $X'[V'^{(I)}_{2,R}]$ or $X'[V'^{(II)}_{2,R}]$ up to a
relabelling of the FI parameters. 

\subsubsection{Other vortex defects in the $D_4$ theory}\label{VW-D1}
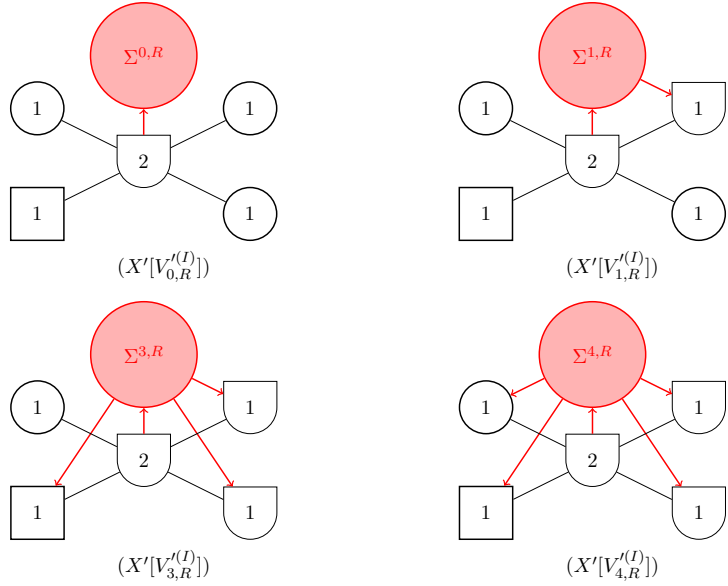
\begin{figure}[htbp]
\begin{center}
\begin{tabular}{ccc}
\scalebox{.7}{\begin{tikzpicture}[node distance=2cm, nnode/.style={circle,draw,thick, red, fill=red!30, minimum size=2.0 cm},cnode/.style={circle,draw,thick,minimum size=1.0 cm},snode/.style={rectangle,draw,thick,minimum size=1.0 cm}]
\node[cnode] (1) at (0,1) {1} ;
\node[cf-group] (2) at (2,0) {\rotatebox{-90}{2}};
\node[snode] (3) at (0,-1) {1};
\node[cnode] (5) at (4, 1) {{1}};
\node[cnode] (6) at (4, -1) {{1}};
\node[nnode] (7) at (2,2) {$\Sigma^{0,R}$};
\draw[red, thick, ->] (2)--(7);
\draw[-] (1) -- (2);
\draw[-] (2) -- (3);
\draw[-] (2) -- (5);
\draw[-] (2) -- (6);
\node[text width=1cm](9) at (2, -2) {$(X'[V'^{(I)}_{0,R}])$};
\end{tikzpicture}}
& \qquad \qquad \qquad
& \scalebox{.7}{\begin{tikzpicture}[node distance=2cm, nnode/.style={circle,draw,thick, red, fill=red!30, minimum size=2.0 cm},cnode/.style={circle,draw,thick,minimum size=1.0 cm},snode/.style={rectangle,draw,thick,minimum size=1.0 cm}]
\node[cnode] (1) at (0,1) {1} ;
\node[cf-group] (2) at (2,0) {\rotatebox{-90}{2}};
\node[snode] (3) at (0,-1) {1};
\node[cf-group] (5) at (4, 1) {\rotatebox{-90}{1}};
\node[cnode] (6) at (4, -1) {{1}};
\node[nnode] (7) at (2,2) {$\Sigma^{1,R}$};
\draw[red, thick, ->] (2)--(7);
\draw[red, thick, ->] (7)--(5);
\draw[-] (1) -- (2);
\draw[-] (2) -- (3);
\draw[-] (2) -- (5);
\draw[-] (2) -- (6);
\node[text width=1cm](9) at (2, -2) {$(X'[V'^{(I)}_{1,R}])$};
\end{tikzpicture}}\\
\scalebox{.7}{\begin{tikzpicture}[node distance=2cm, nnode/.style={circle,draw,thick, red, fill=red!30, minimum size=2.0 cm},cnode/.style={circle,draw,thick,minimum size=1.0 cm},snode/.style={rectangle,draw,thick,minimum size=1.0 cm}]
\node[cnode] (1) at (0,1) {1} ;
\node[cf-group] (2) at (2,0) {\rotatebox{-90}{2}};
\node[snode] (3) at (0,-1) {1};
\node[cf-group] (5) at (4, 1) {\rotatebox{-90}{1}};
\node[cf-group] (6) at (4, -1) {\rotatebox{-90}{1}};
\node[nnode] (7) at (2,2) {$\Sigma^{3,R}$};
\draw[red, thick, ->] (2)--(7);
\draw[red, thick, ->] (7)--(5);
\draw[red, thick, ->] (7)--(6);
\draw[red, thick, ->] (7)--(3);
\draw[-] (1) -- (2);
\draw[-] (2) -- (3);
\draw[-] (2) -- (5);
\draw[-] (2) -- (6);
\node[text width=1cm](9) at (2, -2) {$(X'[V'^{(I)}_{3,R}])$};
\end{tikzpicture}}
&\qquad \qquad
& \scalebox{.7}{\begin{tikzpicture}[node distance=2cm, nnode/.style={circle,draw,thick, red, fill=red!30, minimum size=2.0 cm},cnode/.style={circle,draw,thick,minimum size=1.0 cm},snode/.style={rectangle,draw,thick,minimum size=1.0 cm}]
\node[cnode] (1) at (0,1) {1} ;
\node[cf-group] (2) at (2,0) {\rotatebox{-90}{2}};
\node[snode] (3) at (0,-1) {1};
\node[cf-group] (5) at (4, 1) {\rotatebox{-90}{1}};
\node[cf-group] (6) at (4, -1) {\rotatebox{-90}{1}};
\node[nnode] (7) at (2,2) {$\Sigma^{4,R}$};
\draw[red, thick, ->] (2)--(7);
\draw[red, thick, ->] (7)--(5);
\draw[red, thick, ->] (7)--(6);
\draw[red, thick, ->] (7)--(1);
\draw[red, thick, ->] (7)--(3);
\draw[-] (1) -- (2);
\draw[-] (2) -- (3);
\draw[-] (2) -- (5);
\draw[-] (2) -- (6);
\node[text width=1cm](9) at (2, -2) {$(X'[V'^{(I)}_{4,R}])$};
\end{tikzpicture}}
\end{tabular}
\caption{\footnotesize{The construction of vortex defects in the flavored ${D}_4$ quiver gauge theory $X'$, using an Abelian $S$-type operation 
from $X[V^r_{M,R}]$ with $M=0,1,3,4$.}}
\label{VMneq2-D4}
\end{center}
\end{figure}

Let us now present a set of examples where a gauge vortex defect in the $D_4$ theory is mapped to a 
flavor Wilson defect in the $SU(2)$ theory. These can be constructed as follows.
One can implement the $S$-type operation \eref{AbS-D}-\eref{u-choice1} on the 3d-1d coupled quivers 
corresponding to the vortex defects $V_{M,R}$ in the theory $X$, for $M \neq 2$. The resultant 
3d-1d coupled quivers are shown in \figref{VMneq2-D4}, where we have again used the right 3d-1d  
quiver $X[V^r_{M,R}]$ as our starting point. 

For the case of generic $M$, the resultant vortex defect is defined as
\be
Z^{(X'[V'^{(I)}_{M,R}])}(v; \vec t, \vec \eta) := e^{-2\pi (t_2 -\frac{\eta_1+\eta_2}{2})|R|} \, Z^{\CO_{\vec \CP}(X[V^r_{M,R}])}(v; \vec t, \vec \eta),
\ee
where the defect partition function $Z^{\CO_{\vec \CP}(X[V^r_{M,R}])}$ being given as 
\begin{align}
Z^{\CO_{\vec \CP}(X[V^r_{M,R}])} = \lim_{z\to 1} \int \prod^3_{i=1} du_i \, \Big[d\vec s\Big] \, \prod^3_{i=1} Z_{\rm FI} (u_i, \eta_i)\,
Z^{(X)}_{\rm int}(\vec s, \vec u, v, \vec t)\, W_{\rm b.g.} (\vec t, R) \, \CI^{\Sigma^{M,R}_r}(\vec s, \vec u, z | \vec \xi < 0),
\end{align}
and the various constituent functions on the RHS are given in \eref{U24-V2W-main1}-\eref{U24-V2W-main3}. The dual 
Wilson defects can be worked out in a fashion similar to the $M=2$ case. We enumerate the results below:
\begin{itemize}
\item {\bf{$M=1$:}} The dual defect partition function is given as:
\begin{align}
& Z^{(Y'[(V'^{(I)}_{1,R})^\vee])}= e^{-2\pi (t_2 -\frac{\eta_1+\eta_2}{2})|R|} \, Z^{\wt{\CO}_{\vec \CP}(Y[\wt{W}_{R}])} \nn \\
& = C(v, \vec t, \vec \eta)\, \Big(\sum_w\, e^{2\pi w_1 m'_4} \, e^{2\pi w_2 m'_3}\Big)\, Z^{(Y')}(\vec m'(\vec t, \vec \eta)),
\end{align}
where $C(v, \vec t, \vec \eta)$ is the contact term, and $Z^{(Y')}$ is the partition function of the quiver $Y'$. The form 
of the partition function shows that the defect dual to $V'_{1,R}$ is a flavor Wilson defect. The mirror map can therefore
be written as:
\be
\langle V'_{1,R} \rangle_{X'}(\vec t, \vec \eta) = \langle \sum_{\kappa \in \Delta} \, \wt{W}'^{\rm flavor}_{4,q^{\kappa}_1}\, \wt{W}'^{\rm flavor}_{3,q^{\kappa}_2} \rangle_{Y'}(\vec m'(\vec t, \vec \eta)),
\ee
where $\wt{W}'^{\rm flavor}_{i,q^{\kappa}}$ denotes a flavor Wilson defect with charge $q^{\kappa}$ for the $U(1)_i$ subgroup of the full flavor group, embedded as $U(1)_1 \times U(1)_2 \times U(1)_3\times U(1)_4 \subset SO(8)$. The charges $(q^{\kappa}_1, q^{\kappa}_2)$
are given by decomposing the representation $R$ into $U(1)_1\times U(1)_2 \subset U(2)$, where $U(1)_1\times U(1)_2$ is the 
maximal torus of $U(2)$ and  $\Delta$ denotes the set of all 
the charge doublets counted with degeneracies.

\item {\bf{$M=3$:}} In this case, the dual defect partition function is given as:
\begin{align}
& Z^{(Y'[(V'^{(I)}_{3,R})^\vee])}= e^{-2\pi (t_2 -\frac{\eta_1+\eta_2}{2})|R|} \, Z^{\wt{\CO}_{\vec \CP}(Y[\wt{W}_{R}])} \nn \\
& = C(v, \vec t, \vec \eta)\, \Big(\sum_w\, e^{2\pi w_1 m'_1} \, e^{2\pi w_2 m'_2}\Big)\, Z^{(Y')}(\vec m'(\vec t, \vec \eta)),
\end{align}
which leads to the following mirror map:
\be
\langle V'_{3,R} \rangle_{X'}(\vec t, \vec \eta) = \langle \sum_{\kappa \in \Delta} \, \wt{W}'^{\rm flavor}_{1,q^{\kappa}_1}\, \wt{W}'^{\rm flavor}_{2,q^{\kappa}_2} \rangle_{Y'}(\vec m'(\vec t, \vec \eta)),
\ee
where the notation of operators on the RHS is the same as defined above for the $M=1$ case.

\item {\bf{$M=0,4$:}} The dual partition functions in these cases are respectively given as:
\begin{align}
& Z^{(Y'[(V'^{(I)}_{0,R})^\vee])}= C(v, \vec t, \vec \eta)\, \Big(\sum_w\, e^{2\pi (w_1 +w_2) m'_3}\Big)\, Z^{(Y')}(\vec m'(\vec t, \vec \eta)),\\
& Z^{(Y'[(V'^{(I)}_{4,R})^\vee])}= C(v, \vec t, \vec \eta)\, \Big(\sum_w\, e^{2\pi (w_1 + w_2) m'_2}\Big)\, Z^{(Y')}(\vec m'(\vec t, \vec \eta)),
\end{align}
leading to the following mirror maps:
\begin{align}
& \langle V'_{0,R} \rangle_{X'}(\vec t, \vec \eta) = \langle \, \wt{W}'^{\rm flavor}_{3,|R|}\, \rangle_{Y'}(\vec m'(\vec t, \vec \eta)),\\
& \langle V'_{4,R} \rangle_{X'}(\vec t, \vec \eta) =\langle \, \wt{W}'^{\rm flavor}_{2,|R|}\, \rangle_{Y'}(\vec m'(\vec t, \vec \eta)),
\end{align}
where the notation of the various flavor Wilson defects are the same as defined for the $M=1$ case.
\end{itemize}

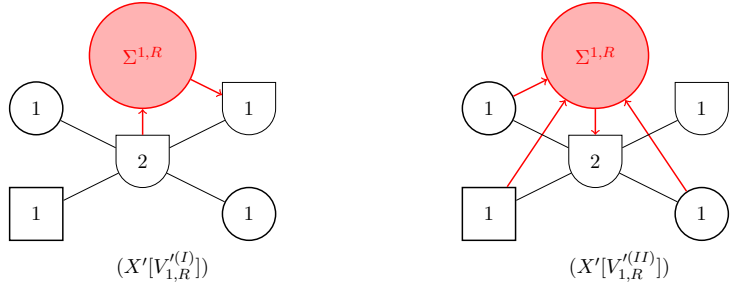
\begin{figure}[htbp]
\begin{center}
\scalebox{.7}{\begin{tikzpicture}[node distance=2cm, nnode/.style={circle,draw,thick, red, fill=red!30, minimum size=2.0 cm},cnode/.style={circle,draw,thick,minimum size=1.0 cm},snode/.style={rectangle,draw,thick,minimum size=1.0 cm}]
\node[cnode] (1) at (0,1) {1} ;
\node[cf-group] (2) at (2,0) {\rotatebox{-90}{2}};
\node[snode] (3) at (0,-1) {1};
\node[cf-group] (5) at (4, 1) {\rotatebox{-90}{1}};
\node[cnode] (6) at (4, -1) {{1}};
\node[nnode] (7) at (2,2) {$\Sigma^{1,R}$};
\draw[red, thick, ->] (2)--(7);
\draw[red, thick, ->] (7)--(5);
\draw[-] (1) -- (2);
\draw[-] (2) -- (3);
\draw[-] (2) -- (5);
\draw[-] (2) -- (6);
\node[text width=1cm](9) at (2, -2) {$(X'[V'^{(I)}_{1,R}])$};
\end{tikzpicture}}
\qquad \qquad \qquad
\scalebox{.7}{\begin{tikzpicture}[node distance=2cm, nnode/.style={circle,draw,thick, red, fill=red!30, minimum size=2.0 cm},cnode/.style={circle,draw,thick,minimum size=1.0 cm},snode/.style={rectangle,draw,thick,minimum size=1.0 cm}]
\node[cnode] (1) at (0,1) {1} ;
\node[cf-group] (2) at (2,0) {\rotatebox{-90}{2}};
\node[snode] (3) at (0,-1) {1};
\node[cf-group] (5) at (4, 1) {\rotatebox{-90}{1}};
\node[cnode] (6) at (4, -1) {{1}};
\node[nnode] (7) at (2,2) {$\Sigma^{1,R}$};
\draw[red, thick, ->] (7)--(2);
\draw[red, thick, ->] (1)--(7);
\draw[red, thick, ->] (3)--(7);
\draw[red, thick, ->] (6)--(7);
\draw[-] (1) -- (2);
\draw[-] (2) -- (3);
\draw[-] (2) -- (5);
\draw[-] (2) -- (6);
\node[text width=1cm](9) at (2, -2) {$(X'[V'^{(II)}_{1,R}])$};
\end{tikzpicture}}
\caption{\footnotesize{The hopping duals associated with the vortex defect $V'_{1,R}$ for the quiver gauge theory $X'$. The 
Witten indices in the two cases are computed in the $\vec \xi <0$ and $\vec \xi > 0$ chambers respectively.}}
\label{VMneq2-D4-hop}
\end{center}
\end{figure}

Each of the vortex defects listed above is realized by two inequivalent coupled 3d-1d quivers. In exact analogy to the case treated in \Secref{V2W-D}, 
the second quiver can be read off by implementing the $S$-type operation \eref{AbS-D}-\eref{u-choice1} on the left 3d-1d quiver 
$X[V^l_{2,R}]$. Alternatively, one can read it off from the partition function $Z^{(X'[V'^{(I)}_{M,R}])}$ by a simple change of variables 
$s_j \to s_j - iw_j z$. The 3d-1d quivers associated with the $M=1$ case are given explicitly in \figref{VMneq2-D4-hop}. The Witten index 
of the SQM for $V'^{(I)}_{1,R}$ is computed in the chamber $\vec \xi <0$, and that for $V'^{(II)}_{1,R}$ is computed in the chamber $\vec \xi >0$. 

\subsubsection{A Wilson defect for the central node}\label{W2V-D}
\begin{figure}[htbp]
\begin{center}
\begin{tabular}{ccc}
\scalebox{.7}{\begin{tikzpicture}[node distance=2cm,
cnode/.style={circle,draw,thick, minimum size=1.0cm},snode/.style={rectangle,draw,thick,minimum size=1cm}, pnode/.style={red,rectangle,draw,thick, minimum size=1.0cm}]
\node[cnode] (1) at (0,0) {2} ;
\node[snode] (2) at (0, -2) {4};
\draw[-] (1) -- (2);
\node[text width=1cm](5) at (0, 1) {$W_{R}$};
\node[text width=1cm](6) at (0, -4) {$(X[W_{R}])$};
\end{tikzpicture}}
& \qquad \qquad \qquad
& \scalebox{.7}{\begin{tikzpicture}[node distance=2cm,
cnode/.style={circle,draw,thick, minimum size=1.0cm},snode/.style={rectangle,draw,thick,minimum size=1cm}, nnode/.style={red, circle,draw,thick,fill=red!30 ,minimum size=2.0cm}]
\node[cnode] (1) at (0,0) {1} ;
\node[cf-group] (2) at (2,0) {\rotatebox{-90}{2}};
\node[cf-group] (3) at (4,0) {\rotatebox{-90}{1}};
\node[nnode] (4) at (3,2) {$\wt{\Sigma}^{1,R}$};
\node[snode] (5) at (3,-2) {1};
\node[snode] (6) at (1,-2) {1};
\node[text width=1cm](7) at (2, -3) {$(Y[\wt{V}^r_{1,R}])$};
\draw[red, thick, ->] (2)--(4);
\draw[red, thick, ->] (4)--(3);
\draw[red, thick, ->] (4)--(5);
\draw[-] (1) -- (2);
\draw[-] (2) -- (3);
\draw[-] (2) -- (5);
\draw[-] (2) -- (6);
\end{tikzpicture}}\\
 \scalebox{.5}{\begin{tikzpicture}
\draw[thick, ->] (15,-3) -- (15,-5);
\node[text width=0.1cm](20) at (14.0, -4) {$\CO_{\vec \CP}$};
\end{tikzpicture}}
&\qquad \qquad \qquad
& \scalebox{.5}{\begin{tikzpicture}
\draw[thick,->] (15,-3) -- (15,-5);
\node[text width=0.1cm](29) at (15.5, -4) {$\wt{\CO}_{\vec \CP}$};
\end{tikzpicture}}\\
\scalebox{.7}{\begin{tikzpicture}[node distance=2cm, nnode/.style={circle,draw,thick, red, fill=red!30, minimum size=2.0 cm},cnode/.style={circle,draw,thick,minimum size=1.0 cm},snode/.style={rectangle,draw,thick,minimum size=1.0 cm}]
\node[cnode] (1) at (0,1) {1} ;
\node[cnode] (2) at (2,0) {2};
\node[snode] (3) at (0,-1) {1};
\node[cnode] (5) at (4, 1) {1};
\node[cnode] (6) at (4, -1) {1};
\node[text width=1cm](15) at (2, 1) {$W'_{R}$};
\draw[-] (1) -- (2);
\draw[-] (2) -- (3);
\draw[-] (2) -- (5);
\draw[-] (2) -- (6);
\node[text width=1cm](9) at (2, -2) {$(X'[W'_{R}])$};
\node[text width=0.1cm] (10) at (4.8,1){$1$};
\node[text width=0.1cm] (11) at (4.8,-1){$2$};
\node[text width=0.1cm] (12) at (-0.8,1){$3$};
\end{tikzpicture}}
&\qquad \qquad
& \scalebox{.7}{\begin{tikzpicture}[node distance=2cm,
cnode/.style={circle,draw,thick, minimum size=1.0cm},snode/.style={rectangle,draw,thick,minimum size=1cm}, pnode/.style={circle,draw,double,thick, minimum size=1.0cm}, nnode/.style={circle,draw,thick, red, fill=red!30, minimum size=2.0 cm},lnode/.style = {shape = rounded rectangle, minimum size=1.0cm, rotate=90, rounded rectangle right arc = none, draw, double}]
\node[cf-group] (1) at (0,0) {\rotatebox{-90}{$SU(2)$}} ;
\node[snode] (3) at (2,0) {2};
\node[nnode] (4) at (1,2) {$\wt{\Sigma}^{1,R}$};
\node[snode] (5) at (-2,0) {2};
\draw[-] (1) -- (2);
\draw[-] (5) -- (1);
\draw[red, thick, ->] (1)--(4);
\draw[red, thick, ->] (4)--(3);
\node[text width=1cm](4) at (0, -1) {$(Y'[\wt{V}'_{2, \wt{R}}])$};
\end{tikzpicture}}\\
\end{tabular}
\caption{\footnotesize{The construction of a Wilson defect in a flavored ${D}_4$ quiver gauge theory and its dual vortex defect, using an Abelian gauging operation.}}
\label{AbEx2aD4}
\end{center}
\end{figure}
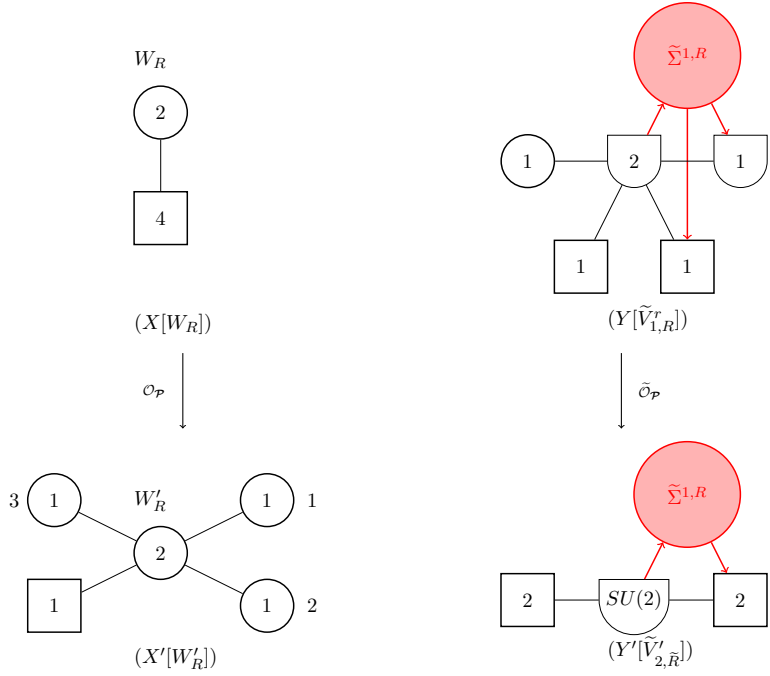

Let us now consider an example where a gauge Wilson defect in the $D_4$ theory is mapped to a gauge 
vortex defect in the $SU(2)$ theory. These defects and the associated mirror map can be constructed as 
follows.
Consider the dual pair of defects $({W}_R, \wt{V}^r_{1,R})$ -- the Wilson defect and the coupled 3d-1d system 
which realizes the dual vortex defect are shown in the first line of \figref{AbEx2aD4}. 
We implement the Abelian $S$-type operation $\CO_{\vec \CP}$ on the system $X[{W}_R]$, where $\CO_{\vec \CP}$ is 
given as:
\begin{align}\label{AbS-Da}
\CO_{\vec \CP} (X[{W}_R]) = G^{\alpha_3}_{\CP_3} \circ G^{\alpha_2}_{\CP_2} \circ G^{\alpha_1}_{\CP_1} (X[{W}_R]),
\end{align}
which corresponds to the following choice of the $U(1)^3$ global symmetry to be gauged:
\be \label{u-choice1a}
u_1=m_1, \quad u_2=m_2, \quad u_3=m_4.
\ee

The resultant defect partition function, following \eref{S-Op-Bgen}, is given as
\begin{align}
& Z^{\CO_{\vec \CP}(X[W_{R}])} = \int \prod^3_{i=1} du_i \, \Big[d\vec s\Big] \, \prod^3_{i=1} Z_{\rm FI} (u_i, \eta_i)\,
Z^{(X)}_{\rm int}(\vec s, \vec u, v, \vec t)\, Z_{\rm Wilson}(\vec s, R), \\
& Z^{(X)}_{\rm int}(\vec s, \vec u, v, \vec t)= \Big\{Z^{(X)}_{\rm int}(\vec s, \vec m, \vec t)| m_1=u_1, m_2=u_2, m_4=u_3, m_3=v \Big\},\\
& Z_{\rm Wilson}(\vec s, R)= \sum_{w \in R} e^{\sum_j w_j\, s_j}.
\end{align}
The defect partition function can be readily identified as the partition function of a $D$-type quiver gauge theory $X'$ with 
a gauge group $G'= U(2) \times U(1)^3$, with a Wilson defect labelled by the representation $R$ of $U(2)$. The 3d theory 
with the defect inserted is denoted by $X'[W'_R]$ and shown in \figref{AbEx2aD4}.\\

Following the general expressions \eref{PF-wtOPgenD-B2A}-\eref{CZ-wtOPD-B2A}, the dual partition function 
can be computed to read off the mirror map of the defects. The final answer takes the following form (the details of the 
computation can be found in \appref{app:SU(2)-W2V}):
\begin{align}
& Z^{(Y'[(W'_{R})^\vee])}= Z^{\wt{\CO}_{\vec \CP}(Y[\wt{V}^r_{1,R}])}, \\
&  Z^{(Y'[(W'_{R})^\vee])} =C\cdot W_{\rm b.g.} \cdot \lim_{z\to 1}\int\Big[d\vec \s\Big] \, \delta(\tr \vec\s)
Z^{U(2), N_f=4}_{\rm int} (\vec\s, \vec m'(\vec t, \vec\eta), \eta')\, \CI^{\wt{\Sigma}^{1,R}_r}(\vec \s, \vec m', z| \vec \xi <0), \label{ZAbS2-D4}\\
&C := C(v, \vec \eta, \vec t)= e^{2\pi i v(\eta_1 +\eta_2+\eta_3 + 2(t_1-t_2))}, \qquad W_{\rm b.g.}:= W_{\rm b.g.} (v, R)= e^{2\pi v |R|}, \\
& \CI^{\wt{\Sigma}^{1,R}_r}(\vec \s, \vec m', z| \vec \xi <0)= \sum_{w \in R}  \prod^2_{j=1} \prod^2_{i=1} \frac{\ch{(\s_j - m'_{i})}}{\ch{(\s_j + i w_j z - m'_{i})}},
\end{align}
where the masses $\vec m'(\vec t, \vec \eta)$ are given in \eref{mm-1}-\eref{mm-4}.
The coupled 3d-1d system so obtained is denoted as the quiver $Y'[\wt{V}'_{2,\wt{R}}]$ in \figref{AbEx2aD4}, 
where $\wt{R}$ is the restriction of the representation $R$ to $SU(2)$. We interpret this coupled 3d-1d quiver as
a vortex defect for the gauge group $SU(2)$ in the representation $\wt{R}$ (up to a global vortex defect), 
leading to the following mirror map:
\begin{empheq}[box=\widefbox]{align}
\langle W'_{R}\rangle_{X'} (v;\vec t, \vec \eta) = \langle \wt{V}'_{2,\wt{R}} \rangle_{Y'} (\vec m'(\vec t, \vec \eta);v).
\end{empheq}

If one implements the dual $S$-operation on the quiver $Y[\wt{V}^l_{1, R}]$ instead of $Y[\wt{V}^r_{1, R}]$, 
it leads to the same final 3d-1d quiver $Y'[\wt{V}'_{2,\wt{R}}]$, as shown in \appref{app:SU(2)-W2V}).
The fact that the above vortex defect does not admit 
a hopping dual, analogous to the case with 3d unitary gauge group, can also be seen from the 
Type IIB realization of these defects, as we discuss later.

\subsection{Generalization I : Mirror of an $Sp(N_c)$ theory with $N_f$ flavors}\label{1-D-Sp(2)}
In this section, we will generalize the results of \Secref{V2W-D}-\Secref{W2V-D} to a mirror pair involving a $D_{N_f}$-type quiver gauge 
theory $X'$ and an $Sp(N_c)$ gauge theory with $N_f$ flavors, $Y'$. For concreteness, we will focus on the case 
$N_c=2$ and $N_f=6$, but the extension of our 
results for generic $N_c, N_f$, obeying $N_f \geq 2N_c +1$ (so that $Y'$ is a good quiver), is straightforward.
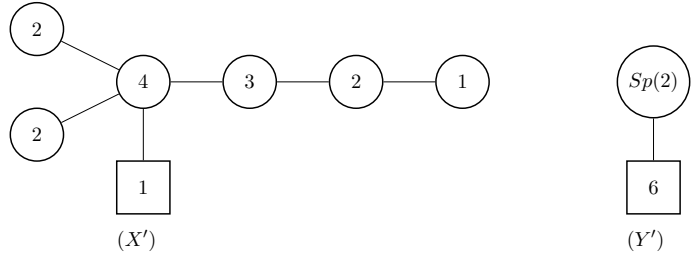
\begin{figure}[htbp]
\begin{center}
\scalebox{.7}{\begin{tikzpicture}[node distance=2cm,
cnode/.style={circle,draw,thick, minimum size=1.0cm},snode/.style={rectangle,draw,thick,minimum size=1cm}, pnode/.style={red,rectangle,draw,thick, minimum size=1.0cm},nnode/.style={red, circle,draw,thick,fill=red!30 ,minimum size=2.0cm}]
\node[cnode] (1) at (0,1) {2} ;
\node[cnode] (7) at (0,-1) {2} ;
\node[cnode] (2) at (2,0) {4};
\node[snode] (3) at (2,-2) {1};
\node[cnode] (4) at (4,0) {3};
\node[cnode] (5) at (6,0) {2};
\node[cnode] (6) at (8,0) {1};
\draw[-] (1) -- (2);
\draw[-] (2) -- (3);
\draw[-] (2) -- (4);
\draw[-] (4) -- (5);
\draw[-] (5) -- (6);
\draw[-] (2) -- (7);
\node[text width=1cm](8) at (2, -3) {$(X')$};
\end{tikzpicture}}
\qquad \qquad
\scalebox{.7}{\begin{tikzpicture}[node distance=2cm,
cnode/.style={circle,draw,thick, minimum size=1.0cm},snode/.style={rectangle,draw,thick,minimum size=1cm}, nnode/.style={red, circle,draw,thick,fill=red!30 ,minimum size=2.0cm}]
\node[cnode] (2) at (2,0) {$Sp(2)$};
\node[snode] (3) at (2,-2) {6};
\node[text width=1cm](7) at (2, -3) {$(Y')$};
\draw[-] (2)--(3);
\end{tikzpicture}}
\end{center}
\caption{\footnotesize{$Sp(2)$ gauge theory with $N_f=6$ flavors and its mirror dual.}}
\label{NAEx}
\end{figure}

Using the general discussion of \Secref{SOps-defects}-\Secref{SOps-defects-d} on constructing defects in generic quiver gauge theories, 
one can work out the mirror map for defects in the dual pair $(X',Y')$. As a starting point, we will choose the dual linear quiver pair $(X,Y)$ 
shown below, and engineer the pair $(X',Y')$ by implementing an elementary non-Abelian gauging operation $\CO_\CP$. 
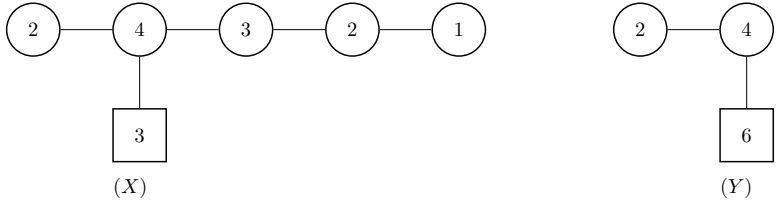
\begin{center}
\scalebox{.7}{\begin{tikzpicture}[node distance=2cm,
cnode/.style={circle,draw,thick, minimum size=1.0cm},snode/.style={rectangle,draw,thick,minimum size=1cm}, pnode/.style={red,rectangle,draw,thick, minimum size=1.0cm},nnode/.style={red, circle,draw,thick,fill=red!30 ,minimum size=2.0cm}]
\node[cnode] (1) at (0,0) {2} ;
\node[cnode] (2) at (2,0) {4};
\node[snode] (3) at (2,-2) {3};
\node[cnode] (4) at (4,0) {3};
\node[cnode] (5) at (6,0) {2};
\node[cnode] (6) at (8,0) {1};
\draw[-] (1) -- (2);
\draw[-] (2) -- (3);
\draw[-] (2) -- (4);
\draw[-] (4) -- (5);
\draw[-] (5) -- (6);
\node[text width=1cm](8) at (2, -3) {$(X)$};
\end{tikzpicture}}
\qquad \qquad
\scalebox{.7}{\begin{tikzpicture}[node distance=2cm,
cnode/.style={circle,draw,thick, minimum size=1.0cm},snode/.style={rectangle,draw,thick,minimum size=1cm}, nnode/.style={red, circle,draw,thick,fill=red!30 ,minimum size=2.0cm}]
\node[cnode] (1) at (0,0) {2} ;
\node[cnode] (2) at (2,0) {4};
\node[snode] (3) at (2,-2) {6};
\node[text width=1cm](7) at (2, -3) {$(Y)$};
\draw[-] (1)--(2);
\draw[-] (2)--(3);
\end{tikzpicture}}
\end{center}

One can also work out these mirror maps using a Type IIB construction presented, 
as we shall discuss later in \Secref{D-TypeIIB-def}.\\

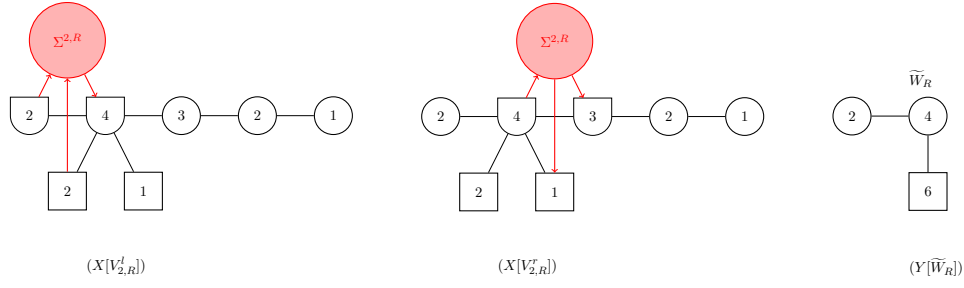
\begin{figure}[htbp]
\begin{center}
\scalebox{.5}{\begin{tikzpicture}[node distance=2cm,
cnode/.style={circle,draw,thick, minimum size=1.0cm},snode/.style={rectangle,draw,thick,minimum size=1cm}, pnode/.style={red,rectangle,draw,thick, minimum size=1.0cm},nnode/.style={red, circle,draw,thick,fill=red!30 ,minimum size=2.0cm}]
\node[cf-group] (1) at (0,0) {\rotatebox{-90}{2}} ;
\node[cf-group] (2) at (2,0) {\rotatebox{-90}{4}};
\node[snode] (3) at (3,-2) {1};
\node[snode] (11) at (1,-2) {2};
\node[cnode] (4) at (4,0) {3};
\node[cnode] (5) at (6,0) {2};
\node[cnode] (6) at (8,0) {1};
\node[nnode] (10) at (1,2) {$\Sigma^{2,R}$};
\draw[-] (1) -- (2);
\draw[-] (2) -- (3);
\draw[-] (2) -- (4);
\draw[-] (4) -- (5);
\draw[-] (5) -- (6);
\draw[-] (2) -- (11);
\draw[red, thick, ->] (1)--(10);
\draw[red, thick, ->] (11)--(10);
\draw[red, thick, ->] (10)--(2);
\node[text width=1cm](8) at (2, -4) {$(X[V^l_{2,R}])$};
\end{tikzpicture}}
\qquad 
\scalebox{.5}{\begin{tikzpicture}[node distance=2cm,
cnode/.style={circle,draw,thick, minimum size=1.0cm},snode/.style={rectangle,draw,thick,minimum size=1cm}, pnode/.style={red,rectangle,draw,thick, minimum size=1.0cm},nnode/.style={red, circle,draw,thick,fill=red!30 ,minimum size=2.0cm}]
\node[cnode] (1) at (0,0) {2} ;
\node[cf-group] (2) at (2,0) {\rotatebox{-90}{4}};
\node[snode] (3) at (3,-2) {1};
\node[snode] (11) at (1,-2) {2};
\node[cf-group] (4) at (4,0) {\rotatebox{-90}{3}};
\node[cnode] (5) at (6,0) {2};
\node[cnode] (6) at (8,0) {1};
\node[nnode] (10) at (3,2) {$\Sigma^{2,R}$};
\draw[-] (1) -- (2);
\draw[-] (2) -- (3);
\draw[-] (2) -- (4);
\draw[-] (4) -- (5);
\draw[-] (5) -- (6);
\draw[-] (2) -- (11);
\draw[red, thick, ->] (10)--(4);
\draw[red, thick, ->] (10)--(3);
\draw[red, thick, ->] (2)--(10);
\node[text width=1cm](8) at (2, -4) {$(X[V^r_{2,R}])$};
\end{tikzpicture}}
\qquad 
\scalebox{.5}{\begin{tikzpicture}[node distance=2cm,
cnode/.style={circle,draw,thick, minimum size=1.0cm},snode/.style={rectangle,draw,thick,minimum size=1cm}, nnode/.style={red, circle,draw,thick,fill=red!30 ,minimum size=2.0cm}]
\node[cnode] (1) at (0,0) {2} ;
\node[cnode] (2) at (2,0) {4};
\node[snode] (3) at (2,-2) {6};
\node[text width=1cm](15) at (2, 1) {$\wt{W}_{R}$};
\node[text width=1cm](7) at (2, -4) {$(Y[\wt{W}_{R}])$};
\draw[-] (1)--(2);
\draw[-] (2)--(3);
\end{tikzpicture}}
\end{center}
\caption{\footnotesize{The vortex defect $V_{2,R}$, realized by the deformation of two different 3d-1d quivers - 
$X[V^l_{2,R}]$ and $X[V^r_{2,R}]$. The dual quiver is $Y[\wt{W}_{R}]$.}}
\label{NAEx-LQdef}
\end{figure}

\noindent \textbf{Vortex defects:} Let us focus on a specific class of vortex defects in the $D_6$ quiver gauge theory $X'$ in \figref{NAEx}
-- vortex defects for the $U(4)$ gauge node at the bifurcated tail. 
Vortex defects for the gauge nodes away from the bifurcated tail can be handled in a fashion similar to defects in a linear quiver. 
Consider the vortex defect $V_{2,R}$ in $X$, realized by the 3d-1d quivers $X[V^l_{2,R}]$ and $X[V^r_{2,R}]$, and 
the dual Wilson defect $\wt{W}_{R}$ in $Y$, as shown in \figref{NAEx-LQdef}. 

Let the fundamental masses in theory $X$ be labelled as $\{m_i| i=1,\ldots, 3\}$, such that the $U(2)$ flavor node identified with the 1d flavor symmetry 
in $X[V^l_{M,R}]$ is associated with the masses $m_1,m_2$. The $U(1)$ flavor node identified with the 1d flavor symmetry in $X[V^r_{M,R}]$ is 
associated with the mass $m_3$\footnote{Here, we are considering the theory $X$ along with 
a background FI-type coupling to a $U(1) \times U(1)$ symmetry. The Higgs branch and the Coulomb branch symmetries of the combined theory 
are then $G_{\rm H} = U(3)$ and $G_{\rm C} = U(6)$ respectively. The $U(3)$ flavor node can then be split as $U(3) \to U(2) \times U(1)$, 
where the $U(2)$ factor talks to the SQM in $X[V^l_{M,R}]$ and the $U(1)$ factor does the same for $X[V^r_{M,R}]$.}. 

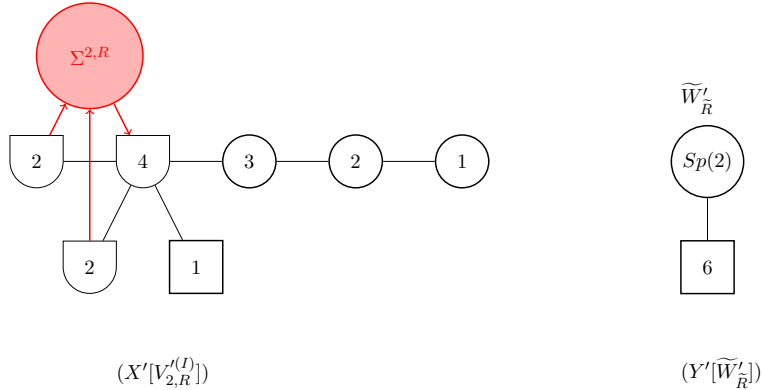
\begin{figure}[htbp]
\begin{center}
\begin{tabular}{ccc}
\scalebox{.7}{\begin{tikzpicture}[node distance=2cm,
cnode/.style={circle,draw,thick, minimum size=1.0cm},snode/.style={rectangle,draw,thick,minimum size=1cm}, pnode/.style={red,rectangle,draw,thick, minimum size=1.0cm},nnode/.style={red, circle,draw,thick,fill=red!30 ,minimum size=2.0cm}]
\node[cf-group] (1) at (0,0) {\rotatebox{-90}{2}} ;
\node[cf-group] (2) at (2,0) {\rotatebox{-90}{4}};
\node[snode] (3) at (3,-2) {1};
\node[cf-group] (11) at (1,-2) {\rotatebox{-90}{2}};
\node[cnode] (4) at (4,0) {3};
\node[cnode] (5) at (6,0) {2};
\node[cnode] (6) at (8,0) {1};
\node[nnode] (10) at (1,2) {$\Sigma^{2,R}$};
\draw[-] (1) -- (2);
\draw[-] (2) -- (3);
\draw[-] (2) -- (4);
\draw[-] (4) -- (5);
\draw[-] (5) -- (6);
\draw[-] (2) -- (11);
\draw[red, thick, ->] (1)--(10);
\draw[red, thick, ->] (11)--(10);
\draw[red, thick, ->] (10)--(2);
\node[text width=1cm](8) at (2, -4) {$(X'[V'^{(I)}_{2,R}])$};
\end{tikzpicture}}
&\qquad \qquad \qquad
& \scalebox{.7}{\begin{tikzpicture}[node distance=2cm,
cnode/.style={circle,draw,thick, minimum size=1.0cm},snode/.style={rectangle,draw,thick,minimum size=1cm}, pnode/.style={circle,draw,double,thick, minimum size=1.0cm}, nnode/.style={circle,draw,thick, red, fill=red!30, minimum size=2.0 cm}]
\node[cnode] (1) at (0,0) {$Sp(2)$} ;
\node[snode] (2) at (0,-2) {6};
\draw[-] (1) -- (2);
\node[text width=1cm](3) at (0, 1.2) {$\wt{W}'_{\wt{R}}$};
\node[text width=1cm](4) at (0, -4) {$(Y'[\wt{W}'_{\wt{R}}])$};
\end{tikzpicture}}
\end{tabular}
\caption{\footnotesize{The construction of a vortex defect in the bifurcated quiver gauge theory $X'$, labelled by a representation $R$ of the 
gauge node $U(4)$. It is dual to a Wilson defect in the theory $Y'$ (an $Sp(2)$ gauge theory with $N_f=6$), labelled by a 
representation $\wt{R}$ of $Sp(2)$, where $\wt{R}$ is the restriction of the representation $R$ for $Sp(2) \subset U(4)$. The 
$S$-type operation $\CO_\CP$ is a non-Abelian gauging operation.}}
\label{NAbEx1Sp}
\end{center}
\end{figure}

Following the general prescription of \Secref{SOps-defects}, let us implement 
an $S$-type operation $\CO_{\CP} $ on the quiver $X[V^l_{2, R}]$, where $\CO_{\CP} $ consists of a single gauging 
operation:
\be \label{NAbS-D}
\CO_{ \CP} (X[V^l_{M,R}]) = G^{\alpha}_{\CP} (X[V^l_{M,R}]),
\ee
with the operation $G^{\alpha}_{\CP}$ gauging the $U(2)$ flavor node labelled $\alpha$ in $X[V^l_{M,R}]$.
The mass parameters associated with $U(2)$ factor to be gauged are therefore chosen as:
\be \label{u-choice4}
u_1=m_1, \quad u_2=m_2.
\ee  
From the general formula \eref{S-Op-A}, the partition function $Z^{\CO_{\vec \CP}(X[V^l_{2,R}])}$ 
can be identified as the partition function of a coupled 3d-1d quiver $X'[V'^{(I)}_{2,R}]$ in \figref{NAbEx1Sp}. 
The 3d quiver is a $D_6$ quiver gauge theory and the SQM is $\Sigma^{2,R}_l$, with the Witten index being computed 
in the chamber $\vec \xi >0$. As before, we will redefine the partition function of the new 3d-1d quiver by 
a global Wilson defect factor, i.e.
\be \label{ZNAbS-d0}
Z^{(X'[V'^{(I)}_{2,R}])}:= e^{-2\pi ( \frac{\eta_\alpha}{2})|R|} \, Z^{\CO_{\vec \CP}(X[V^l_{2,R}])}(v; \vec t, \eta).
\ee

The dual partition function
\be
Z^{(Y'[(V'^{(I)}_{2,R})^\vee])}= e^{-2\pi (\frac{\eta_\alpha}{2})|R|} \, Z^{\wt{\CO}_{\vec \CP}(Y[\wt{W}_{R}])}
\ee
can be computed from the general expressions \eref{PF-wtOPgenD-A2B}-\eref{CZ-wtOPD-A2B}, and the abelianization formula
\eref{NAG-AG1}\footnote{It is also possible to perform the computation without resorting to the abelianization formula, as we show in 
\appref{app: NAG-Ex-2}.}. The computation, although somewhat tedious, is a straightforward exercise, and we will only state the result here 
(see \appref{app: NAG-Ex} for details): 
\begin{align}
Z^{(Y'[(V'^{(I)}_{2,R})^\vee])}=C(v, \vec \eta, \vec t)\,\int \,\Big[d\vec \s\Big] \, & \frac{\delta({\s}_1 +{\s}_2 )\, \delta({\s}_3 +{\s}_4)\,Z^{U(4), N_f=6}_{\rm int} (\vec{\s}, \vec m'(\vec t, \vec\eta), \eta'=0)}{\sh{\s_{13}}\, \sh{\s_{14}}\, \sh{\s_{23}}\, \sh{\s_{24}}}\,\nn\\
& \times \Big( \sum_{w \in R} e^{2\pi \sum_j w_j \s_j} \Big),  \nn \\
= C(v, \vec \eta, \vec t)\,\int \,\Big[d\vec \s\Big] \, & \delta({\s}_1 +{\s}_2 )\, \delta({\s}_3 +{\s}_4)\,Z^{Sp(2), N_f=6}_{\rm int} (\{{\s}_1,\s_3 \}, \vec m') \nn \\
& \times \Big( \sum_{w \in R} e^{2\pi \sum_j w_j \s_j} \Big),\label{ZNAbS-d}
\end{align}
where $C(v, \vec \eta, \vec t)$ is a contact term and $\s_{ij}=\s_i -\s_j$. The function $Z^{U(4), N_f=6}_{\rm int}$ is the matrix model integrand for a 
$U(4)$ gauge theory with $N_f=6$ flavors, and $Z^{Sp(2), N_f=6}_{\rm int}$ is the analogous function for an $Sp(2)$ gauge theory with $N_f=6$ flavors.
The dual of the 3d-1d coupled system can now be read off from the partition function above. It is given by the 
quiver $Y'[\wt{W}'_{\wt R}]$ in \figref{NAbEx1Sp}, where $Y'$ is an $Sp(2)$ gauge theory with $N_f=6$ flavors and $\wt{W}'_{\wt R}$ is 
a Wilson defect for the gauge group $Sp(2)$. The representation ${\wt R}$ is the restriction of the representation $R$ of 
$U(4)$ to the subgroup $Sp(2)$, i.e.
\begin{align}\label{def-wtR-Sp}
U(4) & \rightarrow Sp(2), \nn \\
R & \rightarrow \bigoplus_{\kappa \in \Delta}\, R_\kappa = {\wt R},
\end{align}
where $\Delta$ denotes the set of all representations $R_\kappa$ that appear in the decomposition of $R$ with degeneracies.
Normalizing the integrals with appropriate partition functions, we obtain the following mirror map of line defects in the pair of dual theories $(X',Y')$:
\be
\langle V'^{(I)}_{2,R} \rangle_{X'}(\vec t,  \eta) = \langle \sum_{\kappa \in \Delta} \wt{W}'_{R_\kappa} \rangle_{Y'}(\vec m'(\vec t,  \eta)).
\ee

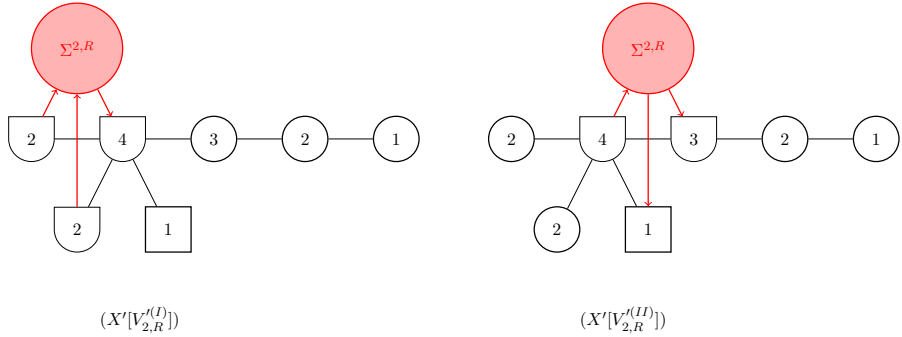
\begin{figure}[htbp]
\begin{center}
\scalebox{.6}{\begin{tikzpicture}[node distance=2cm,
cnode/.style={circle,draw,thick, minimum size=1.0cm},snode/.style={rectangle,draw,thick,minimum size=1cm}, pnode/.style={red,rectangle,draw,thick, minimum size=1.0cm},nnode/.style={red, circle,draw,thick,fill=red!30 ,minimum size=2.0cm}]
\node[cf-group] (1) at (0,0) {\rotatebox{-90}{2}} ;
\node[cf-group] (2) at (2,0) {\rotatebox{-90}{4}};
\node[snode] (3) at (3,-2) {1};
\node[cf-group] (11) at (1,-2) {\rotatebox{-90}{2}};
\node[cnode] (4) at (4,0) {3};
\node[cnode] (5) at (6,0) {2};
\node[cnode] (6) at (8,0) {1};
\node[nnode] (10) at (1,2) {$\Sigma^{2,R}$};
\draw[-] (1) -- (2);
\draw[-] (2) -- (3);
\draw[-] (2) -- (4);
\draw[-] (4) -- (5);
\draw[-] (5) -- (6);
\draw[-] (2) -- (11);
\draw[red, thick, ->] (1)--(10);
\draw[red, thick, ->] (11)--(10);
\draw[red, thick, ->] (10)--(2);
\node[text width=1cm](8) at (2, -4) {$(X'[V'^{(I)}_{2,R}])$};
\end{tikzpicture}}
\qquad 
\scalebox{.6}{\begin{tikzpicture}[node distance=2cm,
cnode/.style={circle,draw,thick, minimum size=1.0cm},snode/.style={rectangle,draw,thick,minimum size=1cm}, pnode/.style={red,rectangle,draw,thick, minimum size=1.0cm},nnode/.style={red, circle,draw,thick,fill=red!30 ,minimum size=2.0cm}]
\node[cnode] (1) at (0,0) {2} ;
\node[cf-group] (2) at (2,0) {\rotatebox{-90}{4}};
\node[snode] (3) at (3,-2) {1};
\node[cnode] (11) at (1,-2) {2};
\node[cf-group] (4) at (4,0) {\rotatebox{-90}{3}};
\node[cnode] (5) at (6,0) {2};
\node[cnode] (6) at (8,0) {1};
\node[nnode] (10) at (3,2) {$\Sigma^{2,R}$};
\draw[-] (1) -- (2);
\draw[-] (2) -- (3);
\draw[-] (2) -- (4);
\draw[-] (4) -- (5);
\draw[-] (5) -- (6);
\draw[-] (2) -- (11);
\draw[red, thick, ->] (10)--(4);
\draw[red, thick, ->] (10)--(3);
\draw[red, thick, ->] (2)--(10);
\node[text width=1cm](8) at (2, -4) {$(X'[V'^{(II)}_{2,R}])$};
\end{tikzpicture}}
\caption{\footnotesize{Different coupled 3d-1d quivers realizing the same vortex defect in the bifurcated quiver $X'$, which is dual to the 
Wilson defect $\wt{W}'_{\wt{R}}$ in the $Sp(2)$ gauge theory $Y'$. The Witten indices for the SQMs in the two cases are computed in the 
$\vec \xi > 0$ and $\vec \xi <0$ chambers respectively.}}
\label{NAbEx1SpHop}
\end{center}
\end{figure}

A second coupled 3d-1d quiver $(X'[V'^{(II)}_{2,R}])$ which realizes the same vortex defect 
can be obtained by implementing the $S$-type operation \eref{NAbS-D} on the quiver  
$(X[V^{r}_{2,R}])$ instead. The resultant quiver is shown in \figref{NAbEx1SpHop}, 
where the Witten index of the SQM should be computed in the chamber $\vec \xi <0$.
Following the same procedure as above, one can show that the dual defect is given 
by $Y'[\wt{W}'_{\wt{R}}]$. This leads to the mirror map:
\begin{empheq}[box=\widefbox]{align} \label{MM-D6-V2W1}
\langle V'^{(I)}_{2,R} \rangle_{X'}( \vec t,  \eta) = \langle V'^{(II)}_{2,R} \rangle_{X'}(\vec t,  \eta)= \langle \sum_{\kappa \in \Delta} \wt{W}'_{R_\kappa} \rangle_{Y'}(\vec m'(\vec t,  \eta)).
\end{empheq}
Alternatively, the 3d-1d quiver $(X'[V'^{(II)}_{2,R}])$ can be read off by a change of variables in the matrix integral on the RHS of \eref{ZNAbS-d0}: 
$s^{(2)}_j \to s^{(2)}_j + iw_j\,z$, where $\vec{s}^{(2)}$ denote the matrix model integration variables in the Cartan of the $U(4)$ gauge group 
of the theory $X'$.\\

\noindent \textbf{Wilson defects:} Analogous to the case of the vortex defect, we will focus on Wilson defects for the $U(4)$
gauge node. The starting point is the Wilson defect $W_R$ in the theory $X$ and the dual vortex defect $\wt{V}_{4,R}$ in 
$Y$ -- realized by the two 3d-1d quivers $Y[\wt{V}^r_{4,R}]$ and $Y[\wt{V}^l_{4,R}]$, as shown in \figref{NAEx-LQdef2}.

\begin{figure}[htbp]
\begin{center}
\scalebox{.6}{\begin{tikzpicture}[node distance=2cm,
cnode/.style={circle,draw,thick, minimum size=1.0cm},snode/.style={rectangle,draw,thick,minimum size=1cm}, pnode/.style={red,rectangle,draw,thick, minimum size=1.0cm}]
\node[cnode] (1) at (0,0) {2} ;
\node[cnode] (2) at (2,0) {4};
\node[snode] (3) at (2,-2) {3};
\node[cnode] (4) at (4,0) {3};
\node[cnode] (5) at (6,0) {2};
\node[cnode] (6) at (8,0) {1};
\draw[-] (1) -- (2);
\draw[-] (2) -- (3);
\draw[-] (2) -- (4);
\draw[-] (4) -- (5);
\draw[-] (5) -- (6);
\node[text width=1cm](7) at (2, 1) {$W_R$};
\node[text width=1cm](8) at (2, -4) {$(X[W_{R}])$};
\end{tikzpicture}}
\qquad 
\scalebox{.6}{\begin{tikzpicture}[node distance=2cm,
cnode/.style={circle,draw,thick, minimum size=1.0cm},snode/.style={rectangle,draw,thick,minimum size=1cm}, nnode/.style={red, circle,draw,thick,fill=red!30 ,minimum size=2.0cm}]
\node[cnode] (1) at (0,0) {2} ;
\node[cf-group] (2) at (2,0) {\rotatebox{-90}{4}};
\node[snode] (3) at (4,0) {4};
\node[nnode] (4) at (3,2) {$\wt{\Sigma}^{4, R}$};
\node[snode] (5) at (2,-2) {2};
\node[text width=1cm](7) at (2, -4) {$(Y[\wt{V}^r_{4, R}])$};
\draw[red, thick, ->] (2)--(4);
\draw[red, thick, ->] (4)--(3);
\draw[-] (1) -- (2);
\draw[-] (2) -- (3);
\draw[-] (2) -- (5);
\end{tikzpicture}}
\qquad 
\scalebox{.6}{\begin{tikzpicture}[node distance=2cm,
cnode/.style={circle,draw,thick, minimum size=1.0cm},snode/.style={rectangle,draw,thick,minimum size=1cm}, nnode/.style={red, circle,draw,thick,fill=red!30 ,minimum size=2.0cm}]
\node[cf-group] (1) at (0,0) {\rotatebox{-90}{2}} ;
\node[cf-group] (2) at (2,0) {\rotatebox{-90}{4}};
\node[snode] (3) at (4,0) {4};
\node[nnode] (4) at (1,2) {$\wt{\Sigma}^{4, R}$};
\node[snode] (5) at (1,-2) {2};
\node[text width=1cm](7) at (2, -4) {$(Y[\wt{V}^l_{4, R}])$};
\draw[red, thick, ->] (1)--(4);
\draw[red, thick, ->] (5)--(4);
\draw[red, thick, ->] (4)--(2);
\draw[-] (1) -- (2);
\draw[-] (2) -- (3);
\draw[-] (2) -- (5);
\end{tikzpicture}}
\caption{\footnotesize{A gauge Wilson defect in the theory $X$ and the dual vortex defect, realized by the deformation of two 
different 3d-1d quivers - $Y[\wt{V}^r_{4, R}]$ and $Y[\wt{V}^l_{4, R}]$.}}
\label{NAEx-LQdef2}
\end{center}
\end{figure}
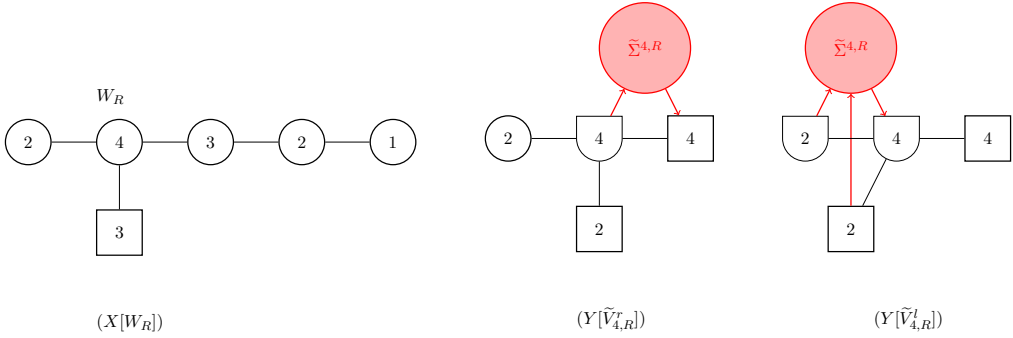

\begin{figure}[htbp]
\begin{center}
\scalebox{.7}{\begin{tikzpicture}[node distance=2cm,
cnode/.style={circle,draw,thick, minimum size=1.0cm},snode/.style={rectangle,draw,thick,minimum size=1cm}, pnode/.style={red,rectangle,draw,thick, minimum size=1.0cm}]
\node[cnode] (1) at (0,1) {2} ;
\node[cnode] (7) at (0,-1) {2} ;
\node[cnode] (2) at (2,0) {4};
\node[snode] (3) at (2,-2) {1};
\node[cnode] (4) at (4,0) {3};
\node[cnode] (5) at (6,0) {2};
\node[cnode] (6) at (8,0) {1};
\draw[-] (1) -- (2);
\draw[-] (7) -- (2);
\draw[-] (2) -- (3);
\draw[-] (2) -- (4);
\draw[-] (4) -- (5);
\draw[-] (5) -- (6);
\node[text width=1cm](7) at (2, 1) {$W'_R$};
\node[text width=1cm](8) at (2, -4) {$(X'[W'_{R}])$};
\end{tikzpicture}}
\qquad \qquad
\scalebox{.7}{\begin{tikzpicture}[node distance=2cm,
cnode/.style={circle,draw,thick, minimum size=1.0cm},snode/.style={rectangle,draw,thick,minimum size=1cm}, nnode/.style={red, circle,draw,thick,fill=red!30 ,minimum size=2.0cm}]
\node[] (1) at (2,4){};
\node[cf-group] (2) at (2,0) {\rotatebox{-90}{$Sp(2)$}};
\node[snode] (3) at (4,0) {4};
\node[nnode] (4) at (3,2) {$\wt{\Sigma}^{4, R}$};
\node[snode] (5) at (0,0) {2};
\node[text width=1cm](7) at (2, -4) {$(Y'[\wt{V}'_{4, R}])$};
\draw[red, thick, ->] (2)--(4);
\draw[red, thick, ->] (4)--(3);
\draw[-] (2) -- (3);
\draw[-] (2) -- (5);
\end{tikzpicture}}
\caption{\footnotesize{The construction of a Wilson defect in the bifurcated quiver gauge theory $X'$, labelled by a representation $R$ of the 
gauge node $U(4)$. It is dual to a vortex defect in the theory $Y'$ (an $Sp(2)$ gauge theory with $N_f=6$), realized by a 3d-1d system 
labelled by the representation $R$ for $U(4)$. The Witten index of the SQM in $(Y'[\wt{V}'_{4, R}])$ should be computed in the chamber $\vec \xi <0$.}}
\label{NAbEx2Sp}
\end{center}
\end{figure}
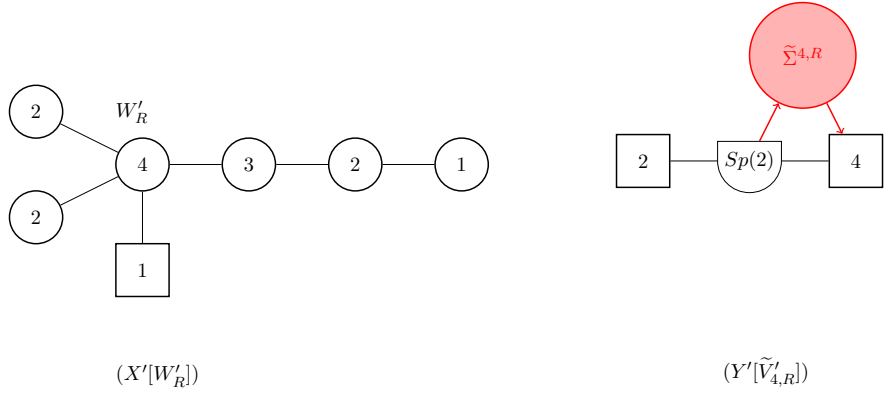

We now implement the non-Abelian $S$-type operation $\CO_\CP$ on the quiver $X[W_{R}]$:
\be
\CO_\CP(X[W_{R}]) = G^\beta_\CP(X[W_{R}]),
\ee
where $\beta$ labels the $U(3)$ flavor node of $X[W_{R}]$, and $G^\beta_\CP$ is a gauging operation which splits the $U(3)$ node
as a $U(2) \times U(1)$ and gauges the $U(2)$. The mass parameters associated with the gauged $U(2)$ are chosen (by an appropriate 
choice of $\CP$) as:
\be
u_1= m_1, \quad u_2= m_2.
\ee
This $S$-operation on the quiver $X[W_{R}]$ leads to the defect quiver $X'[W'_{R}]$ in \figref{NAbEx2Sp}. 
The dual operation on the 3d-1d quiver $Y[\wt{V}^r_{4,R}]$ leads to a 
vortex defect in the theory $Y'$. The defect can be read off from the dual partition function following the general expressions 
\eref{PF-wtOPgenD-B2A}-\eref{CZ-wtOPD-B2A}, and the abelianization formula \eref{NAG-AG1}. After a straightforward but tedious 
computation, we find 
\begin{align}
Z^{(Y'[(W'_{R})^\vee])} =C\cdot W_{\rm b.g.} \cdot \lim_{z\to 1}\int\Big[d\vec \s\Big] \, & \frac{\delta({\s}_1 +{\s}_2 )\, \delta({\s}_3 +{\s}_4)\,Z^{U(4), N_f=6}_{\rm int} (\vec{\s}, \vec m'(\vec t, \vec\eta), \eta'=0)}{\sh{\s_{13}}\, \sh{\s_{14}}\, \sh{\s_{23}}\, \sh{\s_{24}}} \nn \\
& \times \, \CI^{\wt{\Sigma}^{4,R}_r}(\vec \s, \vec m', z| \vec \xi <0), \nn \\
=C\cdot W_{\rm b.g.} \cdot \lim_{z\to 1}\int\Big[d\vec \s\Big] \, &  \delta({\s}_1 +{\s}_2 )\, \delta({\s}_3 +{\s}_4)\,Z^{Sp(2), N_f=6}_{\rm int} (\{{\s}_1,\s_3 \}, \vec m')\nn \\
& \times \, \CI^{\wt{\Sigma}^{4,R}_r}(\vec \s, \vec m', z| \vec \xi <0),
\label{ZNAbS2-d}
\end{align}
where $C$ and $W_{\rm b.g.}$ are a contact term and a background Wilson defect respectively, while the 1d index $\CI^{\wt{\Sigma}^{4,R}_r}$ is 
given by
\begin{align}\label{ZNAbS2-d1}
\CI^{\wt{\Sigma}^{4,R}_r}(\vec \s, \vec m', z| \vec \xi <0)= \sum_{w \in R}  \prod^4_{j=1} \prod^6_{i=3} \frac{\ch{(\s_j - m'_{i})}}{\ch{(\s_j + i w_j z - m'_{i})}}.
\end{align}

The coupled 3d-1d quiver which realizes the dual vortex defect can be read off from \eref{ZNAbS2-d}-\eref{ZNAbS2-d1}, and is given by  
$Y'[\wt{V}'_{4, R}]$ in \figref{NAbEx2Sp}. If one implements the dual $S$-operation on the quiver $Y[\wt{V}^l_{4, R}]$ instead, it leads to the same 
final 3d-1d quiver $Y'[\wt{V}'_{4, R}]$, as was the case in the $D_4$ example studied earlier. One therefore has a mirror map of the following 
form:
\begin{empheq}[box=\widefbox]{align} \label{MM-D6-W2V1}
\langle W'_R \rangle_{X'}(v;\vec t,  \eta) =  \langle  \wt{V}'_{4, R} \rangle_{Y'}(\vec m'(\vec t,  \eta);v).
\end{empheq}

\subsection{Adding flavors to the $D_4$ quiver}\label{SwD-F}
In this section, we incorporate the Abelian flavoring-gauging operation in our discussion. This $S$-operation will allow us 
to study defects in more general $D_4$ quiver gauge theories, by adding hypermultiplets to one or more of the gauged 
flavor nodes. As a simple example, we will study defects for the following dual pair:
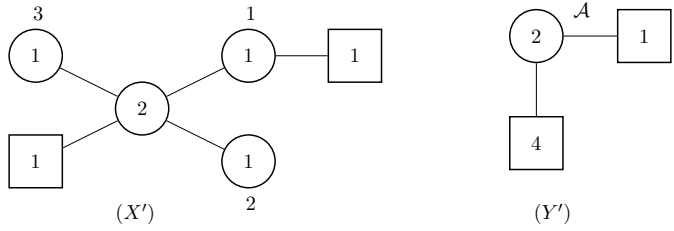
\begin{center}
\scalebox{0.7}{\begin{tikzpicture}[node distance=2cm, nnode/.style={circle,draw,thick, red, fill=red!30, minimum size=2.0 cm},cnode/.style={circle,draw,thick,minimum size=1.0 cm},snode/.style={rectangle,draw,thick,minimum size=1.0 cm}]
\node[cnode] (1) at (0,1) {1} ;
\node[cnode] (2) at (2,0) {2};
\node[snode] (3) at (0,-1) {1};
\node[snode] (4) at (6,1) {1};
\node[cnode] (5) at (4, 1) {1};
\node[cnode] (6) at (4, -1) {1};
\draw[-] (1) -- (2);
\draw[-] (2) -- (3);
\draw[-] (5) -- (4);
\draw[-] (2) -- (5);
\draw[-] (2) -- (6);
\node[text width=1cm](9) at (2, -2) {$(X')$};
\node[text width=0.1cm](10) at (4, 1.8) {1};
\node[text width=0.1cm](11) at (4, -1.8) {2};
\node[text width=0.1cm](12) at (0, 1.8) {3};
\end{tikzpicture}}
\qquad \qquad
\scalebox{0.7}{\begin{tikzpicture}[
cnode/.style={circle,draw,thick, minimum size=1.0cm},snode/.style={rectangle,draw,thick,minimum size=1cm}]
\node[cnode] (9) at (10,0){2};
\node[snode] (10) [below=1cm of 9]{4};
\node[snode] (11) [right=1cm of 9]{1};
\draw[-] (9) -- (10);
\draw[-] (9) -- (11);
\node[text width=0.1cm](12) at (10.75,0.5){$\CA$};
\node[text width=0.1cm](21)[below=0.5 cm of 10]{$(Y')$};
\end{tikzpicture}}
\end{center}
The label $\CA$ in the quiver $Y'$ denotes a hypermultiplet in the rank-2 antisymmetric representation of $U(2)$. 
Note that in contrast to the examples encountered previously in this section, the dual of the $D$-type quiver 
gauge theory is a $U(2)$ gauge theory (as opposed to $SU(2)$ or $Sp(N)$ more generally).\\

\begin{figure}[htbp]
\begin{center}
\begin{tabular}{ccc}
\scalebox{0.6}{\begin{tikzpicture}[node distance=2cm, nnode/.style={circle,draw,thick, red, fill=red!30, minimum size=2.0 cm},cnode/.style={circle,draw,thick,minimum size=1.0 cm},snode/.style={rectangle,draw,thick,minimum size=1.0 cm}]
\node[cnode] (1) at (0,1) {1} ;
\node[cf-group] (2) at (2,0) {\rotatebox{-90}{2}};
\node[snode] (3) at (0,-1) {1};
\node[snode] (4) at (6,1) {1};
\node[cf-group] (5) at (4, 1) {\rotatebox{-90}{1}};
\node[cf-group] (6) at (4, -1) {\rotatebox{-90}{1}};
\node[nnode] (7) at (2,2) {$\Sigma^{2,R}$};
\draw[red, thick, ->] (2)--(7);
\draw[red, thick, ->] (7)--(5);
\draw[red, thick, ->] (7)--(6);
\draw[-] (1) -- (2);
\draw[-] (2) -- (3);
\draw[-] (5) -- (4);
\draw[-] (2) -- (5);
\draw[-] (2) -- (6);
\node[text width=1cm](9) at (2, -2) {$(X'[V'^{(I)}_{2,R}])$};
\end{tikzpicture}}
& \qquad \qquad
& \scalebox{0.6}{\begin{tikzpicture}[
cnode/.style={circle,draw,thick, minimum size=1.0cm},snode/.style={rectangle,draw,thick,minimum size=1cm}]
\node[cnode] (9) at (10,0){2};
\node[snode] (10) [below=1cm of 9]{4};
\node[snode] (11) [right=1cm of 9]{1};
\draw[-] (9) -- (10);
\draw[-] (9) -- (11);
\node[text width=0.1cm](12) at (10.75,0.5){$\CA$};
\node[text width=0.1cm](13) at (9.8, .8){$\wt{W}'_{R}$};
\node[text width=0.1cm](21)[below=0.5 cm of 10]{$(Y'[\wt{W}'_{R}])$};
\end{tikzpicture}}
\end{tabular}
\caption{\footnotesize{A 3d-1d system which realizes a vortex defect in the flavored ${D}_4$ quiver, and the dual 
Wilson defect in the $U(2)$ gauge theory.}}
\label{AbEx1aD4}
\end{center}
\end{figure}
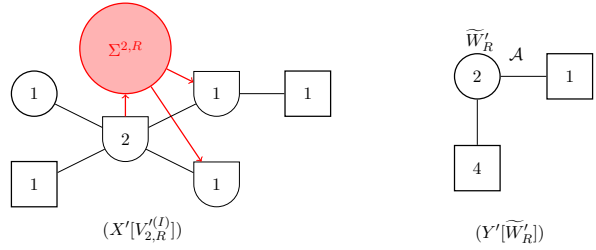

\noindent \textbf{Vortex defects:} As before, we will focus on vortex defects for the central $U(2)$ gauge node of the $D_4$ quiver. 
Consider again the dual pair of defect quivers -- $X[V^r_{2,R}]$ and $Y[\wt{W}_R]$, as shown in the first line of \figref{AbSU2}, 
and let us implement the following Abelian $S$-operation $\CO_{\vec \CP}$ on $X[V^r_{2,R}]$:
\be \label{AbS-D2}
\CO_{\vec \CP} (X[V^r_{2,R}]) = G^\beta_{\CP_3} \circ G^{\alpha_2}_{\CP_2} \circ (G \circ F)^{\alpha_1}_{\CP_1} (X[V^r_{2,R}]),
\ee
where $\alpha_1=\alpha$, and $\alpha_2$ is the residual $U(1)$ flavor node from $U(2)_\alpha$ in the theory $G^{\alpha_1}_{\CP_1} (X)$.
The mass parameters corresponding to the $U(1)^3$ global symmetry are chosen as:
\be \label{u-choice5}
u_1=m_3, \quad u_2=m_4, \quad u_3=m_1.
\ee 
From the general formula \eref{S-Op-A}, the partition function 
$Z^{\CO_{\vec \CP}(X[V^r_{2,R}])}$ can be identified as the partition function of a coupled 3d-1d quiver $X'[V'^{(I)}_{2,R}]$, 
where the 3d quiver is the $D_4$ quiver gauge theory $X'$ and the SQM is shown explicitly
in \figref{AbEx1aD4}. The Witten index for the SQM is computed in the chamber $\vec \xi < 0$.
Redefining the partition function of the new 3d-1d quiver by a global Wilson defect factor, i.e.
\be
Z^{(X'[V'^{(I)}_{2,R}])}:=  e^{-2\pi (t_2 -\frac{\eta_1+\eta_2}{2})|R|} \, Z^{\CO_{\vec \CP}(X[V^r_{2,R}])}(v, m^{(1)}_F; \vec t, \vec \eta),
\ee
where $v=m_2$ and $m^{(1)}_F$ are masses of the fundamental hypers in $X'$, the dual partition function
\be
Z^{(Y'[(V'^{(I)}_{2,R})^\vee])}= e^{-2\pi (t_2 -\frac{\eta_1+\eta_2}{2})|R|} \, Z^{\wt{\CO}_{\vec \CP}(Y[\wt{W}_{R}])}
\ee
can be computed from the general expressions \eref{PF-wtOPgenD-A2B}-\eref{CZ-wtOPD-A2B}, as we did in the previous examples.
The dual 3d-1d coupled system, which can be read off from the partition function, is given by the 
quiver $Y'[\wt{W}'_{R}]$ in \figref{AbEx1aD4}, where $Y'$ is a $U(2)$ gauge theory with $N_f=4$ fundamental hypers and a 
single rank-2 antisymmetric hyper, and $\wt{W}'_{R}$ is a Wilson defect in a representation $R$ for the gauge group $U(2)$.
Normalizing the integrals with appropriate partition functions, we obtain the following mirror map of line defects in the pair of dual theories $(X',Y')$:
\be \label{MM-D42-V2W1}
\langle V'^{(I)}_{2,R} \rangle_{X'}(v, m^{(1)}_F; \vec t, \vec \eta) = \langle \wt{W}'_{R} \rangle_{Y'}(\vec m'(\vec t, \vec \eta),\eta'(v, m^{(1)}_F)).
\ee

\begin{figure}[htbp]
\begin{center}
\begin{tabular}{ccc}
\scalebox{0.6}{\begin{tikzpicture}[node distance=2cm, nnode/.style={circle,draw,thick, red, fill=red!30, minimum size=2.0 cm},cnode/.style={circle,draw,thick,minimum size=1.0 cm},snode/.style={rectangle,draw,thick,minimum size=1.0 cm}]
\node[cnode] (1) at (0,1) {1} ;
\node[cf-group] (2) at (2,0) {\rotatebox{-90}{2}};
\node[snode] (3) at (0,-1) {1};
\node[snode] (4) at (6,1) {1};
\node[cf-group] (5) at (4, 1) {\rotatebox{-90}{1}};
\node[cf-group] (6) at (4, -1) {\rotatebox{-90}{1}};
\node[nnode] (7) at (2,2) {$\Sigma^{2,R}$};
\draw[red, thick, ->] (2)--(7);
\draw[red, thick, ->] (7)--(5);
\draw[red, thick, ->] (7)--(6);
\draw[-] (1) -- (2);
\draw[-] (2) -- (3);
\draw[-] (5) -- (4);
\draw[-] (2) -- (5);
\draw[-] (2) -- (6);
\node[text width=1cm](9) at (2, -2) {$(X'[V'^{(I)}_{2,R}])$};
\end{tikzpicture}}
& \qquad \qquad
& \scalebox{0.6}{\begin{tikzpicture}[node distance=2cm, nnode/.style={circle,draw,thick, red, fill=red!30, minimum size=2.0 cm},cnode/.style={circle,draw,thick,minimum size=1.0 cm},snode/.style={rectangle,draw,thick,minimum size=1.0 cm}]
\node[cf-group] (1) at (0,1) {\rotatebox{-90}{1}} ;
\node[cf-group] (2) at (2,0) {\rotatebox{-90}{2}};
\node[snode] (3) at (0,-1) {1};
\node[snode] (4) at (6,1) {1};
\node[cnode] (5) at (4, 1) {1};
\node[cnode] (6) at (4, -1) {1};
\node[nnode] (7) at (2,2) {$\Sigma^{2,R}$};
\draw[red, thick, ->] (7)--(2);
\draw[red, thick, ->] (1)--(7);
\draw[red, thick, ->] (3)--(7);
\draw[-] (1) -- (2);
\draw[-] (2) -- (3);
\draw[-] (5) -- (4);
\draw[-] (2) -- (5);
\draw[-] (2) -- (6);
\node[text width=1cm](9) at (2, -2) {$(X'[V'^{(II)}_{2,R}])$};
\end{tikzpicture}}
\end{tabular}
\caption{\footnotesize{Two different realizations of a vortex defect as 3d-1d systems in a 
flavored ${D}_4$ quiver. The Witten index for the SQM in the first case should be computed in the 
chamber $\vec \xi < 0$, while for the second case it should be computed in $\vec \xi >0$.}}
\label{AbEx1aD4hop}
\end{center}
\end{figure}
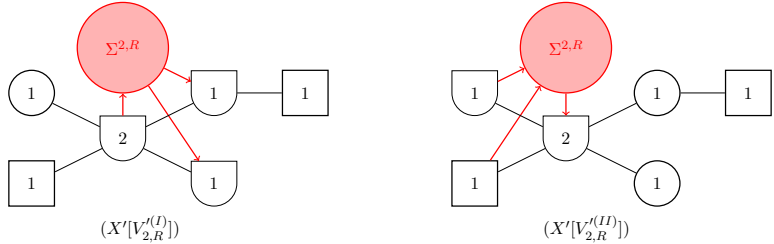

A second coupled 3d-1d quiver $(X'[V'^{(II)}_{2,R}])$ which realizes the same vortex defect 
can be obtained by implementing the $S$-type operation \eref{AbS-D2} on the quiver  
$(X[V^{l}_{2,R}])$ instead. The resultant quiver is shown in \figref{AbEx1aD4hop}, and the Witten index should 
be computed in the chamber $\vec \xi >0$. Following the same procedure as above, one can show that the dual defect is given 
by $Y'[\wt{W}'_{R}]$. The mirror map therefore assumes the final form:
\begin{empheq}[box=\widefbox]{align} \label{MM-D42-V2W2}
\langle V'^{(II)}_{2,R} \rangle_{X'}(v, m^{(1)}_F; \vec t, \vec \eta)=\langle V'^{(I)}_{2,R} \rangle_{X'}(v, m^{(1)}_F; \vec t, \vec \eta) = \langle \wt{W}'_{R} \rangle_{Y'}(\vec m'(\vec t, \vec \eta),\eta'(v, m^{(1)}_F)).
\end{empheq}

\noindent \textbf{Wilson defects:} We will focus on Wilson defects for central $U(2)$ gauge node of the $D_4$ 
quiver $X'$. The starting point is the dual pair of defects -- the Wilson defect $W_R$ in the theory $X$ and the 
vortex defect $\wt{V}_{1,R}$ in the theory $Y$, where the latter is realized by as a deformation of the right 3d-1d 
quiver. The pair $({W}_R, \wt{V}^r_{1,R})$ is shown in the first line of \figref{AbEx2aD4}. 
We implement the Abelian $S$-type operation $\CO_{\vec \CP}$ on the system $X[{W}_R]$, where $\CO_{\vec \CP}$ is 
given as:
\begin{align}\label{AbS-D2a}
\CO_{\vec \CP} (X[{W}_R]) = G^{\alpha_3}_{\CP_3} \circ G^{\alpha_2}_{\CP_2} \circ (G \circ F)^{\alpha_1}_{\CP_1} (X[{W}_R]),
\end{align}
which corresponds to the following choice of the $U(1)^3$ global symmetry to be gauged:
\be \label{u-choice5a}
u_1=m_1, \quad u_2=m_2, \quad u_3=m_4.
\ee

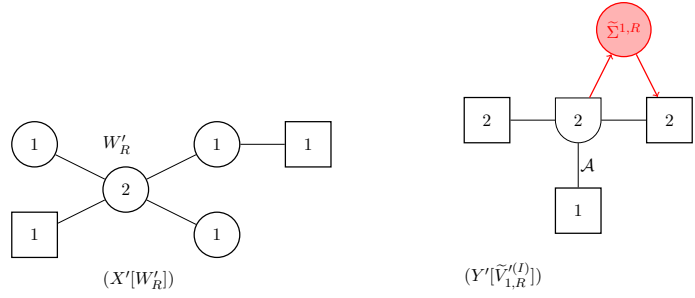
\begin{figure}[htbp]
\begin{center}
\scalebox{0.6}{\begin{tikzpicture}[node distance=2cm, nnode/.style={circle,draw,thick, red, fill=red!30, minimum size=2.0 cm},cnode/.style={circle,draw,thick,minimum size=1.0 cm},snode/.style={rectangle,draw,thick,minimum size=1.0 cm}]
\node[cnode] (1) at (0,1) {1} ;
\node[cnode] (2) at (2,0) {{2}};
\node[snode] (3) at (0,-1) {1};
\node[snode] (4) at (6,1) {1};
\node[cnode] (5) at (4, 1) {{1}};
\node[cnode] (6) at (4, -1) {{1}};
\node[text width=1cm](10) at (2, 1) {$W'_{R}$};
\draw[-] (1) -- (2);
\draw[-] (2) -- (3);
\draw[-] (5) -- (4);
\draw[-] (2) -- (5);
\draw[-] (2) -- (6);
\node[text width=1cm](9) at (2, -2) {$(X'[W'_{R}])$};
\end{tikzpicture}}
\qquad \qquad
\scalebox{0.6}{\begin{tikzpicture}[
cnode/.style={circle,draw,thick, minimum size=1.0cm},snode/.style={rectangle,draw,thick,minimum size=1cm}, nnode/.style={circle,draw,thick, red, fill=red!30, minimum size=5 mm}]
\node[snode] (9) at (8,0){2};
\node[cf-group] (10) at (10,0){\rotatebox{-90}{2}};
\node[snode] (11) at (12,0){2};
\node[nnode] (13) at (11,2) {$\wt{\Sigma}^{1, R}$};
\node[snode] (14) at (10,-2){1};
\draw[-] (9) -- (10);
\draw[-] (10) -- (11);
\draw[-] (10) -- (14);
\draw[red, thick, ->] (10)--(13);
\draw[red, thick, ->] (13)--(11);
\node[text width=0.1cm](15) at (10.1,-1){$\CA$};
\node[text width=1cm](21)[below=2.5 cm of 9]{$(Y'[\wt{V}'^{(I)}_{1, R}])$};
\end{tikzpicture}}
\caption{\footnotesize{Wilson defect in a flavored ${D}_4$ quiver, and its dual vortex defect in a $U(2)$ gauge theory.}}
\label{AbEx1bD4}
\end{center}
\end{figure}

\begin{figure}[htbp]
\begin{center}
\scalebox{0.6}{\begin{tikzpicture}[
cnode/.style={circle,draw,thick, minimum size=1.0cm},snode/.style={rectangle,draw,thick,minimum size=1cm}, nnode/.style={circle,draw,thick, red, fill=red!30, minimum size=5 mm}]
\node[snode] (9) at (8,0){2};
\node[cf-group] (10) at (10,0){\rotatebox{-90}{2}};
\node[snode] (11) at (12,0){2};
\node[nnode] (13) at (11,2) {$\wt{\Sigma}^{1, R}$};
\node[snode] (14) at (10,-2){1};
\draw[-] (9) -- (10);
\draw[-] (10) -- (11);
\draw[-] (10) -- (14);
\draw[red, thick, ->] (10)--(13);
\draw[red, thick, ->] (13)--(11);
\node[text width=0.1cm](15) at (10.1,-1){$\CA$};
\node[text width=1cm](21)[below=2.5 cm of 9]{$(Y'[\wt{V}'^{(I)}_{1, R}])$};
\end{tikzpicture}}
\qquad \qquad
\scalebox{0.6}{\begin{tikzpicture}[
cnode/.style={circle,draw,thick, minimum size=1.0cm},snode/.style={rectangle,draw,thick,minimum size=1cm}, nnode/.style={circle,draw,thick, red, fill=red!30, minimum size=5 mm}]
\node[snode] (9) at (8,0){2};
\node[cf-group] (10) at (10,0){\rotatebox{-90}{2}};
\node[snode] (11) at (12,0){2};
\node[nnode] (13) at (9,2) {$\wt{\Sigma}^{1, R}$};
\node[snode] (14) at (9,-2){1};
\draw[-] (9) -- (10);
\draw[-] (10) -- (11);
\draw[-] (10) -- (14);
\draw[red, thick, ->] (9)--(13);
\draw[red, thick, ->] (14)--(13);
\draw[red, thick, ->] (13)--(10);
\node[text width=0.1cm](15) at (9.1,-1){$\CA$};
\node[text width=1cm](21)[below=2.5 cm of 9]{$(Y'[\wt{V}'^{(II)}_{1, R}])$};
\end{tikzpicture}}
\caption{\footnotesize{Two different realizations of the vortex defect dual to the Wilson defect in the flavored ${D}_4$ quiver. The Witten index for the SQM in the first case should be computed in the chamber $\vec \xi < 0$, while for the second case it should be computed in $\vec \xi >0$.}}
\label{AbEx1bD4hop}
\end{center}
\end{figure}
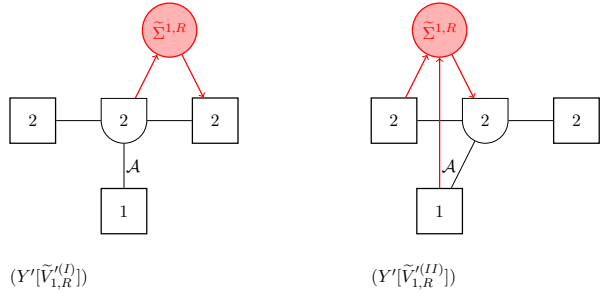

This $S$-operation on the quiver $X[W_{R}]$ leads to the defect quiver $X'[W'_{R}]$ in \figref{AbEx1bD4}. 
The dual $S$-operation acting on the 3d-1d quiver $Y[\wt{V}^r_{1,R}]$ leads to another 3d-1d quiver,
which can be read off from the dual partition function computed from \eref{PF-wtOPgenD-B2A}-\eref{CZ-wtOPD-B2A}.
\begin{align}
& Z^{(Y'[(W'_{R})^\vee])}= Z^{\wt{\CO}_{\vec \CP}(Y[\wt{V}^r_{1,R}])}, \\
&  Z^{(Y'[(W'_{R})^\vee])} =C\cdot W_{\rm b.g.} \cdot \lim_{z\to 1}\int\Big[d\vec \s\Big] \,
Z^{(Y')}_{\rm int} (\vec\s, \vec m'(\vec t, \vec\eta), \eta'(v, m^{(1)}_F))\, \CI^{\wt{\Sigma}'^{1,R}_r}(\vec \s, \vec m', z| \vec \xi <0), \label{ZAbS3-D4}\\
&C := C(v, \vec \eta, \vec t)= e^{2\pi i v(\eta_1 +\eta_2+\eta_3 + 2(t_1-t_2))}, \qquad W_{\rm b.g.}:= W_{\rm b.g.} (v, R)= e^{2\pi v |R|}, \\
& Z^{(Y')}_{\rm int} = \frac{\sinh^2{\pi (\s_1-\s_2)}\, e^{2\pi i (m^{(1)}_F -v)\tr\s}}{\ch{(\tr \vec\s)}\,\prod_{j}\prod^4_{a=1} \ch{(\s_j - m'_a(\vec t, \vec\eta))}},\\
& \CI^{\wt{\Sigma}'^{1,R}_r}(\vec \s, \vec m', z| \vec \xi <0)= \sum_{w \in R} \prod^2_{j=1} \prod^2_{a=1} \frac{\ch{(\s_j - m'_{a}(\vec t, \vec\eta))}}{\ch{(\s_j + i w_j z - m'_{a}(\vec t, \vec\eta))}}.
\end{align}
The resultant 3d-1d quiver can be read off from the RHS of \eref{ZAbS3-D4}, and is given by $Y'[\wt{V}'^{(I)}_{1, R}]$ in \figref{AbEx1bD4}.\\

Again, one could have used the dual pair $({W}_R, \wt{V}^l_{1,R})$ instead as a starting point. The dual partition function can be written directly 
from \eref{PF-wtOPgenD-B2A}-\eref{CZ-wtOPD-B2A}, or by a change of variables $\s_j \to \s_j - iw_j\,z$ in the matrix integral \eref{ZAbS3-D4}:
\begin{align}
& Z^{(Y'[(W'_{R})^\vee])}= Z^{\wt{\CO}_{\vec \CP}(Y[\wt{V}^l_{1,R}])}, \\
&  Z^{(Y'[(W'_{R})^\vee])} =C\cdot W'_{\rm b.g.} \cdot \lim_{z\to 1}\int\Big[d\vec \s\Big] \,
Z^{(Y')}_{\rm int} (\vec\s, \vec m'(\vec t, \vec\eta), \eta'(v, m^{(1)}_F))\, \CI^{\wt{\Sigma}'^{1,R}_l}(\vec \s, \vec m', z| \vec \xi > 0), \label{ZAbS4-D4}\\
&C := C(v, \vec \eta, \vec t)= e^{2\pi i v(\eta_1 +\eta_2+\eta_3 + 2(t_1-t_2))}, \qquad W'_{\rm b.g.}:= W'_{\rm b.g.} (m^{(1)}_F, R)= e^{2\pi m^{(1)}_F |R|}, \\
& Z^{(Y')}_{\rm int} = \frac{\sinh^2{\pi (\s_1-\s_2)}\, e^{2\pi i (m^{(1)}_F -v)\tr\s}}{\ch{(\tr \vec\s)}\,\prod_{j}\prod^4_{a=1} \ch{(\s_j - m'_a(\vec t, \vec\eta))}},\\
& \CI^{\wt{\Sigma}'^{1,R}_l}(\vec \s, \vec m', z| \vec \xi <0)= \sum_{w \in R} \frac{\ch{(\tr \vec\s)}}{\ch{(\tr \vec\s - i|R|\,z)}} \prod^2_{j=1}\prod^4_{a=3} \frac{\ch{(\s_j - m'_{a}(\vec t, \vec\eta))}}{\ch{(\s_j - i w_j z - m'_{a}(\vec t, \vec\eta))}}.
\end{align}
The dual defect is given by the 3d-1d quiver $Y'[\wt{V}'^{(II)}_{1, R}]$, as shown in \figref{AbEx1bD4hop}. This leads to the following mirror map:
\be
\boxed{\langle W'_R \rangle_{X'} (v, m^{(1)}_F; \vec t, \vec \eta) = \langle \wt{V}'^{(I)}_{1, R} \rangle_{Y'}(- \vec m'(\vec t, \vec \eta); \eta'(v, m^{(1)}_F))
= \langle \wt{V}'^{(II)}_{1, R} \rangle_{Y'}(- \vec m'(\vec t, \vec \eta); \eta'(v, m^{(1)}_F)).}
\ee

\subsection{Generalization II : Affine $D$-type quivers}\label{SwD-Dhat}
In this section, we extend our analysis to affine $D$-type quiver gauge theories with additional flavors. 
In our previous examples involving $D$-type quivers, we studied vortex and Wilson defects for gauge nodes 
located at the bifurcated tail. An affine $D$-type quiver consists of two such bifurcated tails, and therefore defects 
in these theories can be studied without introducing any new ingredient. For concreteness, we will study a 
$\wh{D}_4$ quiver $X'$ with gauge group $G=U(2) \times U(1)^4$ and two fundamental hypermultiplets 
and its mirror dual $Y'$, which is a $U(2)$ gauge theory with $N_f=4$ fundamental flavors and two 
additional hypermultiplets in the rank-2 antisymmetric representation of $U(2)$. The quiver gauge theories are 
shown below.

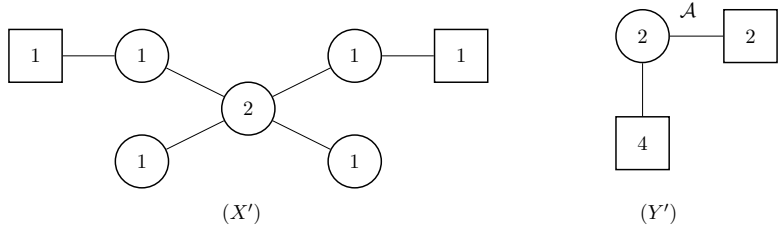
\begin{center}
\scalebox{0.7}{\begin{tikzpicture}[node distance=2cm, nnode/.style={circle,draw,thick, red, fill=red!30, minimum size=2.0 cm},cnode/.style={circle,draw,thick,minimum size=1.0 cm},snode/.style={rectangle,draw,thick,minimum size=1.0 cm}]
\node[cnode] (1) at (0,1) {1} ;
\node[cnode] (2) at (2,0) {2};
\node[cnode] (3) at (0,-1) {1};
\node[snode] (4) at (6,1) {1};
\node[cnode] (5) at (4, 1) {1};
\node[cnode] (6) at (4, -1) {1};
\node[snode] (8) at (-2, 1) {1};
\draw[-] (1) -- (2);
\draw[-] (2) -- (3);
\draw[-] (5) -- (4);
\draw[-] (2) -- (5);
\draw[-] (2) -- (6);
\draw[-] (8) -- (1);
\node[text width=1cm](9) at (2, -2) {$(X')$};
\end{tikzpicture}}
\qquad \qquad 
 \scalebox{0.7}{\begin{tikzpicture}[
cnode/.style={circle,draw,thick, minimum size=1.0cm},snode/.style={rectangle,draw,thick,minimum size=1cm}]
\node[cnode] (9) at (10,0){2};
\node[snode] (10) [below=1cm of 9]{4};
\node[snode] (11) [right=1cm of 9]{2};
\draw[-] (9) -- (10);
\draw[-] (9) -- (11);
\node[text width=0.1cm](12) at (10.75,0.5){$\CA$};
\node[text width=0.1cm](21)[below=0.5 cm of 10]{$(Y')$};
\end{tikzpicture}}
\end{center}

We will focus on vortex defects and Wilson defects for the central $U(2)$ gauge node of the 
$\wh{D}_4$ quiver and their mirror duals. Other defects can be addressed using our general construction 
in a similar fashion. The partition function computations are very similar to the analogous ones in \Secref{SwD-F},
and therefore we simply state the final results.\\

\noindent \textbf{Vortex defect:} The starting point is the dual pair of defect quivers -- $X[V^r_{2,R}]$ and $Y[\wt{W}_R]$, 
as shown in the first line of \figref{AbSU2}. The fundamental masses in theory 
$X$ be labelled as $\{m_i| i=1,\ldots, 4\}$, such that the $U(2)_\beta$  and the $U(2)_\alpha$ flavor nodes  
in $X[V^r_{2,R}]$ are associated with the masses $(m_1,m_2)$ and $(m_3,m_4)$. 
Let us implement the following Abelian $S$-operation $\CO_{\vec \CP}$ on 
$X[V^r_{2,R}]$:
\be \label{AbS-Dhat1}
\CO_{\vec \CP} (X[V^r_{2,R}]) = G^{\beta_2}_{\CP_4} \circ (G \circ F)^{\beta_1}_{\CP_3} \circ (G)^{\alpha_2}_{\CP_2} \circ (G\circ F)^{\alpha_1}_{\CP_1} (X[V^r_{2,R}]),
\ee
where $\alpha_1=\alpha$, and $\alpha_2$ is the residual $U(1)$ flavor node from $U(2)_\alpha$ in the theory $G^{\alpha_1}_{\CP_1} (X)$.
Similarly, $\beta_1=\beta$, and  $\beta_2$ is the residual $U(1)$ flavor node from $U(2)_\beta$. The mass parameters corresponding 
to the $U(1)^4$ global symmetry to be gauged are chosen as:
\be \label{u-choice6}
u_1=m_3, \quad u_2=m_4, \quad u_3=m_1, \quad u_4=m_2.
\ee 
The partition function $Z^{\CO_{\vec \CP}(X[V^r_{2,R}])}$ can be identified as the partition function of a 
coupled 3d-1d quiver $X'[V'^{(I)}_{2,R}]$, where the 3d quiver is the $\wh{D}_4$ quiver gauge theory $X'$, and 
the SQM is shown in \figref{AbEx1aD4}. The Witten index for the SQM is computed in the chamber $\vec \xi < 0$.
We redefine the partition function of the new 3d-1d quiver by a global Wilson defect factor, i.e.
\be
Z^{(X'[V'^{(I)}_{2,R}])}:= e^{-2\pi (t_1 + \frac{\eta_3+\eta_4}{2})|R|} \, Z^{\CO_{\vec \CP}(X[V^r_{2,R}])}(m^{(1)}_F, m^{(3)}_F; \vec t, \vec \eta),
\ee
where $m^{(1)}_F, m^{(3)}_F$ are masses of the fundamental hypers in $X'$ for the $U(1)_1$ and the $U(1)_3$ 
respectively. The dual defect is given by the quiver $Y'[\wt{W}'_{R}]$ in \figref{AbEx3aD4}, where 
$\wt{W}'_{R}$ is a Wilson defect in a representation $R$ for the gauge group $U(2)$. 
A second coupled 3d-1d quiver $(X'[V'^{(II)}_{2,R}])$ which realizes the same vortex defect 
can be obtained by implementing the $S$-type operation \eref{AbS-Dhat1} on the quiver  
$(X[V^{l}_{2,R}])$ instead. The 3d-1d quiver is shown in \figref{AbEx3bD4hop}, where the 
Witten index for the SQM should be computed in the chamber $\vec \xi > 0$. 
Computing the dual partition function, one can again show that the dual defect is given 
by $Y'[\wt{W}'_{R}]$. This leads to the following mirror map:
\be \label{MM-D4hat-V2W}
\boxed{\langle V'^{(II)}_{2,R} \rangle_{X'}(m^{(1)}_F, m^{(3)}_F; \vec t, \vec \eta)=\langle V'^{(I)}_{2,R} \rangle_{X'}(m^{(1)}_F, m^{(3)}_F; \vec t, \vec \eta) = \langle \wt{W}'_{R} \rangle_{Y'}(\vec m'(\vec t, \vec \eta),\eta'(m^{(1)}_F, m^{(3)}_F)).}
\ee

\begin{figure}[htbp]
\begin{center}
\scalebox{0.7}{\begin{tikzpicture}[node distance=2cm, nnode/.style={circle,draw,thick, red, fill=red!30, minimum size=2.0 cm},cnode/.style={circle,draw,thick,minimum size=1.0 cm},snode/.style={rectangle,draw,thick,minimum size=1.0 cm}]
\node[cnode] (1) at (0,1) {1} ;
\node[cf-group] (2) at (2,0) {\rotatebox{-90}{2}};
\node[cnode] (3) at (0,-1) {1};
\node[snode] (4) at (6,1) {1};
\node[cf-group] (5) at (4, 1) {\rotatebox{-90}{1}};
\node[cf-group] (6) at (4, -1) {\rotatebox{-90}{1}};
\node[snode] (8) at (-2, 1) {1};
\node[nnode] (7) at (2,2) {$\Sigma^{2,R}$};
\draw[red, thick, ->] (2)--(7);
\draw[red, thick, ->] (7)--(5);
\draw[red, thick, ->] (7)--(6);
\draw[-] (1) -- (2);
\draw[-] (2) -- (3);
\draw[-] (5) -- (4);
\draw[-] (2) -- (5);
\draw[-] (2) -- (6);
\draw[-] (8) -- (1);
\node[text width=1cm](9) at (2, -2) {$(X'[V'^{(I)}_{2,R}])$};
\end{tikzpicture}}
\qquad \qquad 
\scalebox{0.7}{\begin{tikzpicture}[
cnode/.style={circle,draw,thick, minimum size=1.0cm},snode/.style={rectangle,draw,thick,minimum size=1cm}]
\node[cnode] (9) at (10,0){2};
\node[snode] (10) [below=1cm of 9]{4};
\node[snode] (11) [right=1cm of 9]{2};
\draw[-] (9) -- (10);
\draw[-] (9) -- (11);
\node[text width=0.1cm](12) at (10.75,0.5){$\CA$};
\node[text width=0.1cm](13) at (9.8, .8){$\wt{W}'_{R}$};
\node[text width=0.1cm](21)[below=0.5 cm of 10]{$(Y'[\wt{W}'_{R}])$};
\end{tikzpicture}}
\caption{\footnotesize{A 3d-1d system which realizes a vortex defect in the flavored $\wh{D}_4$ quiver, and the dual 
Wilson defect in a $U(2)$ gauge theory.}}
\label{AbEx3aD4}
\end{center}
\end{figure}

\begin{figure}[htbp]
\begin{center}
\scalebox{0.7}{\begin{tikzpicture}[node distance=2cm, nnode/.style={circle,draw,thick, red, fill=red!30, minimum size=2.0 cm},cnode/.style={circle,draw,thick,minimum size=1.0 cm},snode/.style={rectangle,draw,thick,minimum size=1.0 cm}]
\node[cnode] (1) at (0,1) {1} ;
\node[cf-group] (2) at (2,0) {\rotatebox{-90}{2}};
\node[cnode] (3) at (0,-1) {1};
\node[snode] (4) at (6,1) {1};
\node[cf-group] (5) at (4, 1) {\rotatebox{-90}{1}};
\node[cf-group] (6) at (4, -1) {\rotatebox{-90}{1}};
\node[snode] (8) at (-2, 1) {1};
\node[nnode] (7) at (2,2) {$\Sigma^{2,R}$};
\draw[red, thick, ->] (2)--(7);
\draw[red, thick, ->] (7)--(5);
\draw[red, thick, ->] (7)--(6);
\draw[-] (1) -- (2);
\draw[-] (2) -- (3);
\draw[-] (5) -- (4);
\draw[-] (2) -- (5);
\draw[-] (2) -- (6);
\draw[-] (8) -- (1);
\node[text width=1cm](9) at (2, -2) {$(X'[V'^{(I)}_{2,R}])$};
\end{tikzpicture}}
\qquad \qquad
\scalebox{0.7}{\begin{tikzpicture}[node distance=2cm, nnode/.style={circle,draw,thick, red, fill=red!30, minimum size=2.0 cm},cnode/.style={circle,draw,thick,minimum size=1.0 cm},snode/.style={rectangle,draw,thick,minimum size=1.0 cm}]
\node[cf-group] (1) at (0,1) {\rotatebox{-90}{1}} ;
\node[cf-group] (2) at (2,0) {\rotatebox{-90}{2}};
\node[cf-group] (3) at (0,-1) {\rotatebox{-90}{1}};
\node[snode] (4) at (6,1) {1};
\node[cnode] (5) at (4, 1) {1};
\node[cnode] (6) at (4, -1) {1};
\node[snode] (8) at (-2, 1) {1};
\node[nnode] (7) at (2,2) {$\Sigma^{2,R}$};
\draw[red, thick, ->] (7)--(2);
\draw[red, thick, ->] (1)--(7);
\draw[red, thick, ->] (3)--(7);
\draw[-] (1) -- (2);
\draw[-] (2) -- (3);
\draw[-] (5) -- (4);
\draw[-] (2) -- (5);
\draw[-] (2) -- (6);
\draw[-] (8) -- (1);
\node[text width=1cm](9) at (2, -2) {$(X'[V'^{(II)}_{2,R}])$};
\end{tikzpicture}}
\caption{\footnotesize{Two different realizations of a vortex defect as 3d-1d systems in a flavored $\wh{D}_4$ quiver, dual to a 
Wilson defect in the $U(2)$ gauge theory of \figref{AbEx3aD4}. The Witten index for the SQM in the first case should be computed in the 
chamber $\vec \xi < 0$, while for the second case it should be computed in $\vec \xi >0$.}}
\label{AbEx3bD4hop}
\end{center}
\end{figure}
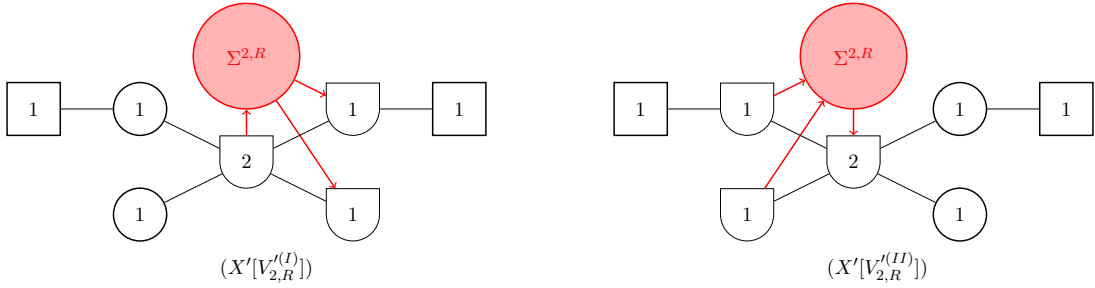

\noindent \textbf{Wilson defect:} Consider Wilson defects for central $U(2)$ gauge node of the $\wh{D}_4$ 
quiver $X'$. The starting point is the dual pair of defects -- the Wilson defect $W_R$ in the theory $X$ and the 
vortex defect $\wt{V}_{1,R}$ in the theory $Y$, where the latter is realized by as a deformation of two 3d-1d 
quivers $\wt{V}^r_{1,R}$ and $\wt{V}^l_{1,R}$. The pair $({W}_R, \wt{V}^r_{1,R})$ is shown in the first line 
of \figref{AbEx2aD4}. 
We implement the Abelian $S$-type operation $\CO_{\vec \CP}$ on the system $X[{W}_R]$, where $\CO_{\vec \CP}$ is 
given as:
\begin{align}\label{AbS-Dhata}
\CO_{\vec \CP} (X[{W}_R]) = G^{\alpha_4}_{\CP_4}\circ (G \circ F)^{\alpha_3}_{\CP_3} \circ G^{\alpha_2}_{\CP_2} \circ (G \circ F)^{\alpha_1}_{\CP_1} (X[{W}_R]),
\end{align}
which corresponds to the following choice of the $U(1)^4$ global symmetry to be gauged:
\be \label{u-choice6a}
u_1=m_1, \quad u_2=m_2, \quad u_3=m_3, \quad u_4=m_4.
\ee

This $S$-operation on the quiver $X[W_{R}]$ leads to the defect quiver $X'[W'_{R}]$ in \figref{AbEx4D4hat}. 
The dual vortex defect can be read off from the dual partition function as before, and is given by the 3d-1d 
quiver $Y'[\wt{V}'^{(I)}_{1, R}]$, where the Witten index of the 1d quiver is computed in the chamber 
$\vec \xi < 0$. 
Starting from the dual pair $({W}_R, \wt{V}^l_{1,R})$ instead, the dual defect is given by the 3d-1d 
quiver $Y'[\wt{V}'^{(II)}_{1, R}]$, as shown in \figref{AbEx4D4hatHop}. In this case, the Witten index of the 1d quiver 
is computed in the chamber $\vec \xi > 0$. This leads to the following mirror map:
\be
\boxed{\langle W'_R \rangle_{X'} (v, \vec m_F; \vec t, \vec \eta) = \langle \wt{V}'^{(I)}_{1, R} \rangle_{Y'}(- \vec m'(\vec t, \vec \eta); \eta(v, \vec m_F))
= \langle \wt{V}'^{(II)}_{1, R} \rangle_{Y'}(- \vec m'(\vec t, \vec \eta); \eta(v, \vec m_F)),}
\ee
where $\vec m_F= (m^{(1)}_F, m^{(3)}_F)$ collectively denotes the fundamental masses in the $\wh{D}_4$ quiver gauge theory.

\begin{figure}[htbp]
\begin{center}
\scalebox{0.7}{\begin{tikzpicture}[node distance=2cm, nnode/.style={circle,draw,thick, red, fill=red!30, minimum size=2.0 cm},cnode/.style={circle,draw,thick,minimum size=1.0 cm},snode/.style={rectangle,draw,thick,minimum size=1.0 cm}]
\node[cnode] (1) at (0,1) {1} ;
\node[cnode] (2) at (2,0) {{2}};
\node[cnode] (3) at (0,-1) {1};
\node[snode] (4) at (6,1) {1};
\node[cnode] (5) at (4, 1) {{1}};
\node[cnode] (6) at (4, -1) {{1}};
\node[snode] (8) at (-2, 1) {1};
\node[text width=1cm](10) at (2, 1) {$W'_{R}$};
\draw[-] (1) -- (2);
\draw[-] (2) -- (3);
\draw[-] (5) -- (4);
\draw[-] (2) -- (5);
\draw[-] (2) -- (6);
\draw[-] (8) -- (1);
\node[text width=1cm](9) at (2, -2) {$(X'[W'_{R}])$};
\end{tikzpicture}}
\qquad \qquad
\scalebox{0.7}{\begin{tikzpicture}[
cnode/.style={circle,draw,thick, minimum size=1.0cm},snode/.style={rectangle,draw,thick,minimum size=1cm}, nnode/.style={circle,draw,thick, red, fill=red!30, minimum size=5 mm}]
\node[cf-group] (9) at (10,0){\rotatebox{-90}{2}};
\node[snode] (10) at (11,-2){1};
\node[snode] (11) at (12,0){2};
\node[snode] (12) at (8,0){2};
\node[snode] (14) at (9,-2){1};
\node[nnode] (13) at (11,2) {$\wt{\Sigma}^{1, R}$};
\draw[-] (9) -- (10);
\draw[-] (9) -- (11);
\draw[-] (9) -- (12);
\draw[-] (9) -- (14);
\draw[red, thick, ->] (9)--(13);
\draw[red, thick, ->] (13)--(11);
\draw[red, thick, ->] (13)--(10);
\node[text width=0.1cm](15) at (9.1,-0.75){$\CA$};
\node[text width=0.1cm](16) at (10.7, -0.75){$\CA$};
\node[text width=1cm](21)[below=2.5 cm of 9]{$(Y'[\wt{V}'^{(I)}_{1, R}])$};
\end{tikzpicture}}
\caption{\footnotesize{Wilson defect in a flavored $\wh{D}_4$ quiver, and its dual vortex defect in a $U(2)$ gauge theory.}}
\label{AbEx4D4hat}
\end{center}
\end{figure}

\begin{figure}[htbp]
\begin{center}
\scalebox{0.7}{\begin{tikzpicture}[
cnode/.style={circle,draw,thick, minimum size=1.0cm},snode/.style={rectangle,draw,thick,minimum size=1cm}, nnode/.style={circle,draw,thick, red, fill=red!30, minimum size=5 mm}]
\node[cf-group] (9) at (10,0){\rotatebox{-90}{2}};
\node[snode] (10) at (11,-2){1};
\node[snode] (11) at (12,0){2};
\node[snode] (12) at (8,0){2};
\node[snode] (14) at (9,-2){1};
\node[nnode] (13) at (11,2) {$\wt{\Sigma}^{1, R}$};
\draw[-] (9) -- (10);
\draw[-] (9) -- (11);
\draw[-] (9) -- (12);
\draw[-] (9) -- (14);
\draw[red, thick, ->] (9)--(13);
\draw[red, thick, ->] (13)--(11);
\draw[red, thick, ->] (13)--(10);
\node[text width=0.1cm](15) at (9.1,-0.75){$\CA$};
\node[text width=0.1cm](16) at (10.7, -0.75){$\CA$};
\node[text width=1cm](21)[below=2.5 cm of 9]{$(Y'[\wt{V}'^{(I)}_{1, R}])$};
\end{tikzpicture}}
\qquad  \qquad 
\scalebox{0.7}{\begin{tikzpicture}[
cnode/.style={circle,draw,thick, minimum size=1.0cm},snode/.style={rectangle,draw,thick,minimum size=1cm}, nnode/.style={circle,draw,thick, red, fill=red!30, minimum size=5 mm}]
\node[cf-group] (9) at (10,0){\rotatebox{-90}{2}};
\node[snode] (10) at (11,-2){1};
\node[snode] (11) at (12,0){2};
\node[snode] (12) at (8,0){2};
\node[snode] (14) at (9,-2){1};
\node[nnode] (13) at (9,2) {$\wt{\Sigma}^{1, R}$};
\draw[-] (9) -- (10);
\draw[-] (9) -- (11);
\draw[-] (9) -- (12);
\draw[-] (9) -- (14);
\draw[red, thick, ->] (12)--(13);
\draw[red, thick, ->] (13)--(9);
\draw[red, thick, ->] (14)--(13);
\node[text width=0.1cm](15) at (9.1,-0.75){$\CA$};
\node[text width=0.1cm](16) at (10.7, -0.75){$\CA$};
\node[text width=1cm](21)[below=2.5 cm of 9]{$(Y'[\wt{V}'^{(II)}_{1, R}])$};
\end{tikzpicture}}
\caption{\footnotesize{Hopping duals for the vortex defect mirror dual to the Wilson defect $W'_R$ in the flavored $\wh{D}_4$ quiver $X'$. 
The Witten index for the SQM in the first case should be computed in the chamber $\vec \xi < 0$, while for the second case it should 
be computed in $\vec \xi >0$.}}
\label{AbEx4D4hatHop}
\end{center}
\end{figure}

\subsection{Adding defects to the $D_4$ quiver}\label{SwD-D}

In this section, we incorporate $S$-type operations involving defect operations in our discussion. 
These $S$-type operations can act non-trivially on the coupled SQM, in addition to the 3d quiver 
and its coupling to the SQM.
We will focus on an example where we 
introduce a vortex defect in the $D_4$ quiver gauge theory encountered in \Secref{SwD-F} 
and then use the general prescription of \Secref{SOps-defects-d} to find the dual defect. 
A Wilson defect can be addressed in a similar fashion, as discussed in \Secref{SOps-defects-d}.\\

\begin{figure}[htbp]
\begin{center}
\begin{tabular}{ccc}
\scalebox{.6}{\begin{tikzpicture}[node distance=2cm,
cnode/.style={circle,draw,thick, minimum size=1.0cm},snode/.style={rectangle,draw,thick,minimum size=1.0cm}, nnode/.style={red, circle,draw,thick,fill=red!30 ,minimum size=2.0cm}]
\node[snode] (1) at (0,0) {2} ;
\node[cf-group] (2) at (2,0) {\rotatebox{-90}{2}};
\node[snode] (3) at (4,0) {2};
\node[nnode] (4) at (2,2) {$\Sigma^{2,R}$};
\node[text width=1cm](7) at (2, -2) {$(X[V^r_{2,R}])$};
\node[text width=0.1cm](8) at (4.6, 0) {$\alpha$};
\node[text width=0.1cm](9) at (-.75, 0) {$\beta$};
\draw[red, thick, ->] (2)--(4);
\draw[red, thick, ->] (4)--(3);
\draw[-] (1) -- (2);
\draw[-] (2) -- (3);
\end{tikzpicture}}
& \qquad \qquad \qquad
& \scalebox{.6}{\begin{tikzpicture}[node distance=2cm,
cnode/.style={circle,draw,thick, minimum size=1.0cm},snode/.style={rectangle,draw,thick,minimum size=1cm}, pnode/.style={red,rectangle,draw,thick, minimum size=1.0cm}]
\node[cnode] (1) at (0,0) {1} ;
\node[cnode] (2) at (2,0) {2} ;
\node[cnode] (3) at (4,0) {1} ;
\node[snode] (4) at (2,-2) {2};
\draw[-] (1) -- (2);
\draw[-] (2) -- (3);
\draw[-] (2) -- (4);
\node[text width=1cm](5) at (2, 1) {$\wt{W}_{R}$};
\node[text width=1cm](6) at (2, -4) {$(Y[\wt{W}_{R}])$};
\end{tikzpicture}}\\
 \scalebox{.5}{\begin{tikzpicture}
\draw[thick, ->] (15,-3) -- (15,-5);
\node[text width=0.1cm](20) at (14.0, -4) {$\CO_{\vec \CP}$};
\end{tikzpicture}}
&\qquad \qquad \qquad
& \scalebox{.5}{\begin{tikzpicture}
\draw[thick,->] (15,-3) -- (15,-5);
\node[text width=0.1cm](29) at (15.5, -4) {$\wt{\CO}_{\vec \CP}$};
\end{tikzpicture}}\\
\scalebox{0.6}{\begin{tikzpicture}[node distance=2cm, nnode/.style={circle,draw,thick, red, fill=red!30, minimum size=2.0 cm},cnode/.style={circle,draw,thick,minimum size=1.0 cm},snode/.style={rectangle,draw,thick,minimum size=1.0 cm}]
\node[cnode] (1) at (0,1) {1} ;
\node[cf-group] (2) at (2,0) {\rotatebox{-90}{2}};
\node[snode] (3) at (0,-1) {1};
\node[snode] (4) at (6,1) {1};
\node[cf-group] (5) at (4, 1) {\rotatebox{-90}{1}};
\node[cf-group] (6) at (4, -1) {\rotatebox{-90}{1}};
\node[nnode] (7) at (2,2) {$\Sigma^{2,R}$};
\node[draw, circle,red, minimum size= 1.0 cm] (9) at (5, 3) {$k$};
\draw[red, thick, ->] (9) to [out=150,in=210,looseness=8] (9);
\draw[red, thick, ->] (2)--(7);
\draw[red, thick, ->] (7)--(5);
\draw[red, thick, ->] (7)--(6);
\draw[red, thick, ->] (5)--(9);
\draw[red, thick, ->] (9)--(4);
\draw[-] (1) -- (2);
\draw[-] (2) -- (3);
\draw[-] (5) -- (4);
\draw[-] (2) -- (5);
\draw[-] (2) -- (6);
\node[text width=1cm](9) at (2, -2) {$(X'[V'^{(I)}_{R,k}])$};
\end{tikzpicture}}
& \qquad \qquad \qquad \qquad
& \scalebox{0.6}{\begin{tikzpicture}[
cnode/.style={circle,draw,thick, minimum size=1.0cm},snode/.style={rectangle,draw,thick,minimum size=1cm}]
\node[cnode] (9) at (10,0){2};
\node[snode] (10) [below=1cm of 9]{4};
\node[snode] (11) [right=1cm of 9]{1};
\draw[-] (9) -- (10);
\draw[-] (9) -- (11);
\node[text width=0.1cm](12) at (10.75,0.25){$\CA$};
\node[text width=0.1 cm](13) at (10, .8){$\wt{W}'_{R,k}$};
\node[text width=0.1cm](21)[below=0.5 cm of 10]{$(Y'[\wt{W}'_{R,k}])$};
\end{tikzpicture}}
\end{tabular}
\caption{\footnotesize{Construction of a 3d-1d system which realizes a vortex defect in the flavored ${D}_4$ quiver and the dual 
Wilson defect in the mirror $U(2)$ gauge theory, using an Abelian $S$-type operation $\CO_\CP$. The $S$-type operation, 
in this case, involves a Type-A defect operation.}}
\label{GenDef1-D4}
\end{center}
\end{figure}
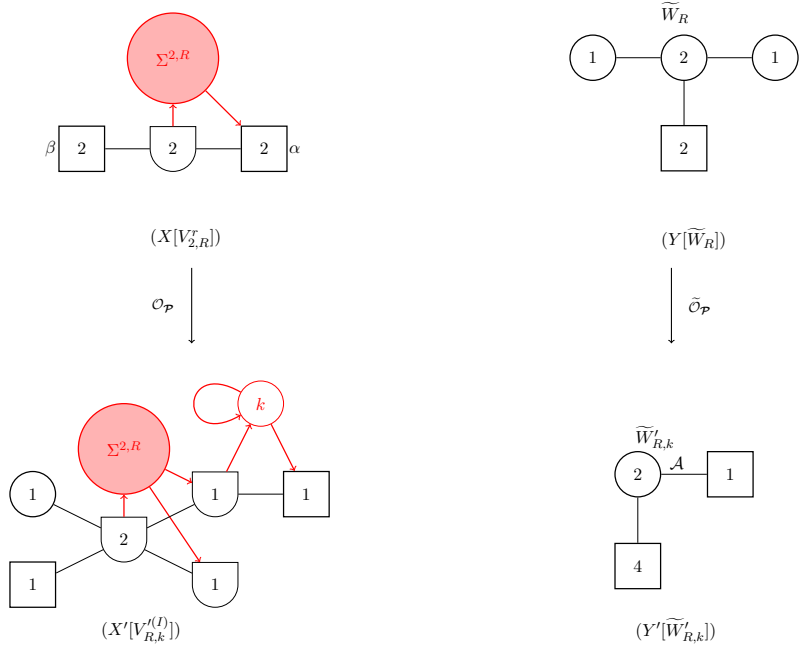

The starting point is the dual pair of defects $(V^r_{2,R}, \wt{W}_R)$ for the theories $X$ and $Y$ respectively,
as shown in the first line of \figref{GenDef1-D4}. Let the fundamental masses in theory $X$ be labelled as 
$\{m_i| i=1,\ldots, 4\}$, such that the $U(2)$ flavor node labelled $\alpha$ in $X[V^r_{M,R}]$ is associated with the 
masses $m_3,m_4$. Following the general prescription of \Secref{SOps-defects}, we implement 
an Abelian $S$-type operation $\CO_{\vec \CP}$ on the system $X[V^r_{2, R}]$, consisting of three elementary Abelian ones 
-- two gauging operations and a single flavoring-defect-gauging operation: 
\be \label{AbS-GenD}
\CO_{\vec \CP} (X[V^r_{2, R}]) = G^\beta_{\CP_3} \circ G^{\alpha_2}_{\CP_2} \circ (G \circ D \circ F)^{\alpha_1}_{\CP_1} (X[V^r_{2, R}]),
\ee
where $\alpha_1=\alpha$, and $\alpha_2$ is the residual $U(1)$ flavor node from $U(2)_\alpha$ in the theory 
$(G \circ D \circ F)^{\alpha_1}_{\CP_1} (X)$. The mass parameters corresponding to the $U(1)^3$ global symmetry are chosen as:
\be \label{u-choice7}
u_1=m_3, \quad u_2=m_4, \quad u_3=m_1.
\ee 

We will choose the defect operation $D$ such that it introduces a vortex defect of charge $k$ for 
the $U(1)_1$ gauge node of the $D_4$ quiver -- we denote it as $V'^{(1)}_{1,k}$. 
From the equation \eref{S-Op-Agen} (where both $D$ and $\CD$ are now non-trivial), the partition function of 
the 3d-1d quiver $\CO_{\vec \CP}(X[V^r_{2,R}])$ can then be written down as follows:
\begin{align}
& Z^{\CO_{\vec \CP}(X[V^r_{2,R}])} = \lim_{z\to 1} \int \prod^3_{i=1} du_i \Big[d\vec s\Big] 
\, \CZ_{\CO_{\vec \CP}}(\vec u, \vec \eta, \vec m_F, z | \Sigma'^k)\,
Z^{(X[V^r_{2, R}], \vec \CP)}_{\rm int}(\vec s, \vec u, v, \vec t,z), \label{ZAbS-GenD4}\\
& Z^{(X[V^r_{2, R}], \vec \CP)}_{\rm int}(\vec s, \vec u, v, \vec t,z)= Z^{(X)}_{\rm int}(\vec s, \vec u, v, \vec t)W_{\rm b.g.} (\vec t, R) \CI^{\Sigma^{2,R}_r}(\vec s, \vec u, z | \vec \xi < 0),
\end{align}
where the functions $Z^{(X)}_{\rm int}$, $W_{\rm b.g.} $, and $\CI^{\Sigma^{2,R}_r}$ are given in \eref{ZAbS-D4a}-\eref{ZAbS-D4c}.
The function $\CZ_{\CO_{\vec \CP}}$ can be assembled from \eref{CZ-gauging}, \eref{CZ-flavoring} and \eref{CZ-defectA}, as follows:
\be
\CZ_{\CO_{\vec \CP}}(\vec u, \vec \eta, \vec m_F, z | \Sigma'^k_r)= \prod^3_{i=1} Z_{\rm FI} (\vec u, \vec \eta)\, 
Z^{\rm hyper}_{\rm 1-loop} (u_1, m^{(1)}_F)\, W_{\rm b.g.}(\vec t, \vec \eta, k)\, \CI^{\Sigma'^k_r} (u_1, m^{(1)}_F, z | \xi' <0),
\ee
where $W_{\rm b.g.}$ is a global Wilson defect, and $\CI^{\Sigma'^k_r}$ is the Witten index of a SQM $\Sigma'^k_r$ -- a $U(k)$ 
gauge theory with a single adjoint chiral and a single fundamental and anti-fundamental chiral. Explicitly, we have
\begin{align}
& W_{\rm b.g.}(\vec t, \vec \eta, k)= e^{2\pi k(2t_2 - \eta_1 -\eta_2)},\\
& \CI^{\Sigma'^k_r} (u_1, m^{(1)}_F, z | \xi' <0)=\frac{\ch{(u_1 - m^{(1)}_F)}}{\ch{(u_1 - m^{(1)}_F + i k z)}}.
\end{align}
The expression on the RHS of \eref{ZAbS-GenD4} can be readily identified as the partition function of a 3d-1d system, 
where the 3d theory is the ${D}_4$ quiver gauge theory $X'$ with a gauge group $G'= U(2) \times U(1)^3$, with a pair 
of coupled SQM $\Sigma^{2,R}_r$ and $\Sigma'^k_r$. Let us denote the resultant vortex defect as 
$V'^{(I)}_{2,R} \cdot V'^{(1)}_{1,k} := V'^{(I)}_{R,k}$, and define the defect partition function as :
\be
Z^{(X'[V'^{(I)}_{R,k}])}:= e^{-2\pi (t_2 -\frac{\eta_1+\eta_2}{2})|R|} \, Z^{\CO_{\vec \CP}(X[V^r_{2,R}])}(v, m^{(1)}_F; \vec t, \vec \eta).
\ee

The dual partition function can be computed following the general expressions \eref{PF-wtOPgenD-A2B}-\eref{CZ-wtOPD-A2B},
and can be written as:
\begin{align}
& Z^{(Y'[(V'^{(I)}_{R,k})^\vee])}=C_{X'Y'}\cdot \int \,\Big[d\vec \s\Big] \, e^{2\pi i (m^{(1)}_f -v) \tr \vec \s} \,
Z^{(Y')}_{\rm 1-loop} (\vec{\s}, \vec m'(\vec t, \vec\eta))\,e^{2\pi k \tr \vec \s} \,\Big(\sum_{w \in R} e^{2\pi \sum_j w_j \s_j}\Big), \label{ZAbS-GenD4-d}
\end{align}
where $Y'$ is the quiver in \figref{GenDef1-D4}, with $C_{X'Y'}(v, m^{(1)}_f, \vec \eta, \vec t)$ and $Z^{(Y')}_{\rm 1-loop}$ being a contact term 
and the 1-loop contribution to the partition function of $Y'$. The dual defect can now be read off from the RHS of the above equation: it is a 
product of a Wilson defect $\wt{W}'_{R}$ for the gauge group $U(2)$ of $Y'$ in a representation $R$ and a Wilson defect $\wt{W}'_k$ of charge $k$ in the $U(1)$ subgroup of $U(2)$. We denote the combined Wilson defect as $\wt{W}'_{R,k}$.
This leads to the following mirror map:
\be
\boxed{\langle V'^{(I)}_{R,k} \rangle_{X'}(v, m^{(1)}_F; \vec t, \vec \eta) =\langle \wt{W}'_{R,k} \rangle_{Y'}(\vec m'(\vec t, \vec \eta),\eta'(v, m^{(1)}_F)).}
\ee

The above 3d-1d quiver can be recast in the standard form, where the 3d theory is coupled with a single SQM of the form of given in 
\figref{1dquiv-gen}. This can be seen by the following change of variables in 
the matrix integral \eref{ZAbS-GenD4}:
\be
u_1 \to u_1 - ik\,z, \qquad u_2 \to u_2 - ik\,z,
\ee
such that the partition function can be recast in the form: 
\begin{align}
Z^{\CO_{\vec \CP}(X[V^r_{2,R}])} = \lim_{z\to 1} \int \prod^3_{i=1} du_i \Big[d\vec s\Big]\,\prod^3_{i=1} Z_{\rm FI} (\vec u, \vec \eta)\, 
Z^{\rm hyper}_{\rm 1-loop} (u_1, m^{(1)}_F)\,& Z^{(X)}_{\rm int}(\vec s, \vec u, v, \vec t)\,W'_{\rm b.g.} (\vec t, k, R) \nn \\
& \times \CI^{\Sigma^{2,R,k}_r}(\vec s, \vec u, z | \vec \xi < 0).
\end{align}
The background Wilson defect and the Witten index contribution on the RHS are given as:
\begin{align}
& W'_{\rm b.g.} (\vec t, k, R)= e^{2\pi (2k+|R|)\,t_2}, \\
& \CI^{\Sigma^{2,R,k}_r}(\vec s, \vec u, z | \vec \xi < 0) = \sum_{w \in R} \, \prod^2_{j=1} \prod^2_{i=1} \frac{\ch{(s_j -u_i)}}{\ch{(s_j + i (w_j+k)\, z - u_i)}}.
\end{align}
There are multiple ways of writing down the resultant SQM $\Sigma^{2,R,k}_r$ from the original SQM $\Sigma^{2,R}_r$ as a quiver, 
where the different quivers are related by 1d Seiberg duality. A convenient choice for the quiver would be the following. Given 
$\Sigma^{2,R}_r$ of the form \figref{1dquiv-gen}, the SQM $\Sigma'^{2,R}_r$ can be obtained by substituting 
$n_P \to n_P + 2k$, keeping all the other $\{n_i\}$ fixed. Analogous to the previous examples, this 3d-1d quiver will have 
a standard hopping dual, which can be read off at the level of the matrix integral by the change of variables $s_j \to s_j -i(w_j+k)\,z$.

\section{Interlude: Type IIB construction of defects in quiver gauge theories}\label{LQ-TypeIIB}

In this section, we discuss the Type IIB construction of vortex and Wilson defects in $D$-type quivers 
with unitary gauge groups, and the S-dual configurations which allow one to read off the mirror map 
of line defects. In \Secref{D-TypeIIB}, we discuss configurations without any defects, before 
incorporating them in \Secref{D-TypeIIB-def}.

\subsection{Type IIB configurations involving orbifold/orientifold planes}\label{D-TypeIIB}

\subsubsection{D3-branes ending on orbifolds}\label{D1-TypeIIB}

Let us extend our discussion of \Secref{LQ2B} to 3d quivers for which the Type IIB configuration involves 
D3-branes ending on orbifold 5-planes, in addition to the NS5 and D5-branes. Generically, these 
boundary conditions lead to bifurcated/D-type 3d quiver gauge theories with unitary gauge groups.\\

Given a Type IIB configuration of D3-D5-NS5-branes oriented as \eref{IIB-1}, consider an $\BZ_2$-orbifolding 
operation of the form $\CI_4\,(-1)^{F_L}$, where
\be
\CI_4\,:\, x^{3,4,5,6} \to - x^{3,4,5,6},
\ee
and $(-1)^{F_L}$ counts the number of left-moving worldsheet fermions modulo 2. The fixed point of the 
orbifolding operation is the 5-plane located at the point $x^{3,4,5,6}=0$ in the transverse direction, and 
we denote the 5-plane as ${\rm Orb}5$. The supersymmetry preserved on ${\rm Orb}5$ is the same as 
that preserved on the NS5-brane \cite{Gaiotto:2008ak}. The respective world-volumes of the D3, D5, NS5-branes 
and the ${\rm Orb}5$-plane can be summarized as:
\begin{align}\label{IIB-Orb}
& \text{D3:}\quad \BR^{2,1} \times L \times \{\vec X\}_{4,5,6} \times \{\vec Y\}_{7,8,9} \nn\\
& \text{D5:}\quad \BR^{2,1} \times \{X_3\}  \times \BR^{3}_{4,5,6} \times \{ \vec Y'\}_{7,8,9} \nn \\
& \text{NS5:}\quad \BR^{2,1} \times \{X'_3\}  \times  \{\vec X'\}_{4,5,6} \times \BR^3_{7,8,9},\nn \\
& \text{Orb5:}\quad \BR^{2,1}\times \{\vec{0} \}_{3,4,5,6} \times  \BR^3_{7,8,9}.
\end{align}

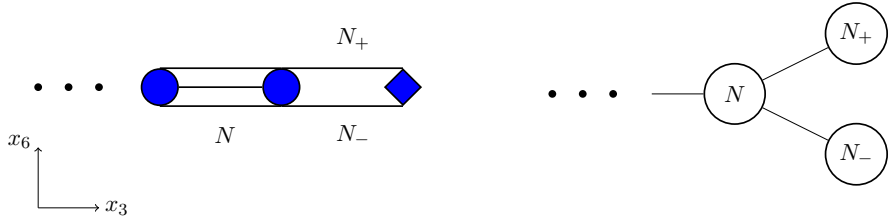
\begin{figure}[h]
\begin{center}
\scalebox{0.8}{\begin{tikzpicture}[
nnode/.style={circle,draw,thick, fill=blue,minimum size= 6mm},cnode/.style={circle,draw,thick,minimum size=4mm},snode/.style={rectangle,draw,thick,minimum size=6mm}]
\node[circle,draw,thick, fill, inner sep=1 pt] (1) at (0,0){} ;
\node[circle,draw,thick, fill, inner sep=1 pt] (2) at (0.5,0){} ;
\node[circle,draw,thick, fill, inner sep=1 pt] (3) at (1,0){} ;
\node[nnode] (4) at (2,0) {} ;
\node[nnode] (5) at (4,0) {};
\node[diamond, draw,thick, fill=blue,minimum size= 6mm] (6) at (6,0) {};
\draw[thick,-] (4.north) -- (5.north);
\draw[thick,-] (4.south) -- (5.south);
\draw[thick,-] (4.east) -- (5.west);
\draw[thick,-] (5.north) -- (6.north);
\draw[thick,-] (5.south) -- (6.south);
\draw[->] (0,-2) -- (1, -2);
\draw[->] (0,-2) -- (0,-1);
\node[text width=.2cm](9) at (1.2, -2){$x_3$};
\node[text width=1cm](10) at (0, -0.9){$x_6$};
\node[text width=.2cm](11) at (3, -0.8){$N$};
\node[text width=.2cm](12) at (5, -0.8){$N_-$};
\node[text width=.2cm](12) at (5, 0.8){$N_+$};
\end{tikzpicture}}
\qquad \qquad
\scalebox{0.8}{\begin{tikzpicture}[node distance=2cm, nnode/.style={circle,draw,thick, fill, inner sep=1 pt},cnode/.style={circle,draw,thick,minimum size=1.0 cm},snode/.style={rectangle,draw,thick,minimum size=1.0 cm}]
\node[nnode] (1) at (0,0){} ;
\node[nnode] (2) at (0.5,0){} ;
\node[nnode] (3) at (1,0){} ;
\node[](4) at (1.5,0){};
\node[cnode] (5) at (3,0) {$N$};
\node[cnode] (6) at (5, 1) {$N_+$};
\node[cnode] (7) at (5, -1) {$N_-$};
\node[] (8) at (5,-2){};
\draw[-] (4) -- (5);
\draw[-] (5) -- (6);
\draw[-] (5) -- (7);
\end{tikzpicture}}
\caption{\footnotesize{Type IIB realization of the bifurcated tail of a $D$-type quiver with unitary gauge groups. The blue circular nodes 
denote NS5-branes, while the blue rhombus denotes an Orb-5 plane. The D3-branes are denoted by black horizontal lines.}}
\label{Fig:HWOrb1}
\end{center}
\end{figure}

Consider a stack of $N$ semi-infinite D3-branes ending on the Orb5-plane. The orbifold action on the 
bosonic fields in the D3 world-volume (which form the bosonic part of an 4d $\CN=4$ vector multiplet) 
are given as follows:
\begin{align}
& A^\mu(\vec x, -x^3) = \gamma(g)\,A^\mu(\vec x, x^3)\, \gamma(g)^{-1}, \quad \mu=0,1,2.\\
& X^{A'}(\vec x, -x^3) = - \gamma(g)\,X^{A'}(\vec x, x^3)\, \gamma(g)^{-1}, \quad A'=3,4,5,6,\\
& Y^A(\vec x, -x^3) =\gamma(g)\,Y^A(\vec x, x^3)\, \gamma(g)^{-1}, \quad A=7,8,9,
\end{align}
where $g$ is the odd element of $\BZ_2$, and $\gamma(g)$ is a $N\times N$ matrix which acts on 
the Chan-Paton factors associated with the $N$ D3-branes. The most general form for $\gamma(g)$ 
is a diagonal matrix with entries
\be
\gamma_{ij}(g)={\rm diag}(1,\ldots,1,-1,\ldots,-1), 
\ee
where $i,j=1,\ldots, N$, with $N_{+}$ entries being $+1$, $N_{-}$ entries being -1, and $N= N_{+} + N_{-}$. 
Restricting these equations to $x^3=0$, we obtain 
a set of boundary conditions for the bosonic fields at the orbifold plane. The boundary conditions 
imply that gauge field $A^\mu$ and the scalars $Y^A$ combine to give the bosonic field content of a 3d
$\CN=4$ vector multiplet with a gauge group $U(N_{+}) \times U(N_{-})$. The scalars $X^{A'}$, on the 
other hand, combine to give the field content of a 3d $\CN=4$ hypermultiplet in the bifundamental 
representation of $U(N_{+}) \times U(N_{-})$. We will be mostly interested in D3-branes on a finite line 
segment stretched between an NS5-brane on the one side, and an orbifold plane on the other. 
The NS5-brane imposes Dirichlet boundary condition on the scalars $X^{A'}$, and Neumann boundary 
conditions on the remaining scalars. Therefore, the bifundamental hypermultiplets become massive 
on KK-reduction, and do not appear in the 3d Lagrangian. The brane configuration and 
the associated 3d quiver are shown in \figref{Fig:HWOrb1}.

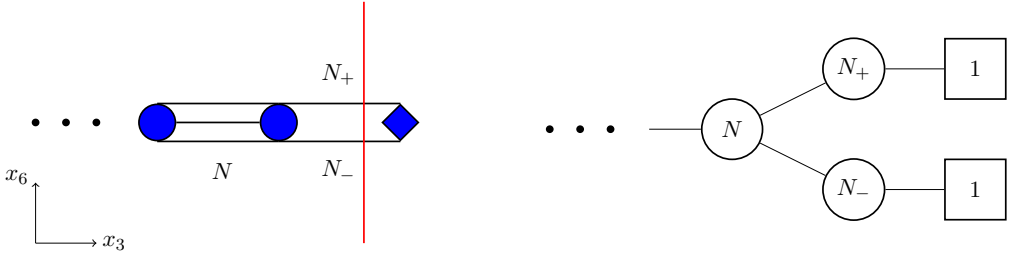
\begin{figure}[h]
\begin{center}
\scalebox{0.8}{\begin{tikzpicture}[
nnode/.style={circle,draw,thick, fill=blue,minimum size= 6mm},cnode/.style={circle,draw,thick,minimum size=4mm},snode/.style={rectangle,draw,thick,minimum size=6mm}]
\node[circle,draw,thick, fill, inner sep=1 pt] (1) at (0,0){} ;
\node[circle,draw,thick, fill, inner sep=1 pt] (2) at (0.5,0){} ;
\node[circle,draw,thick, fill, inner sep=1 pt] (3) at (1,0){} ;
\node[nnode] (4) at (2,0) {} ;
\node[nnode] (5) at (4,0) {};
\node[diamond, draw,thick, fill=blue,minimum size= 6mm] (6) at (6,0) {};
\draw[thick,-] (4.north) -- (5.north);
\draw[thick,-] (4.south) -- (5.south);
\draw[thick,-] (4.east) -- (5.west);
\draw[thick,-] (5.north) -- (6.north);
\draw[thick,-] (5.south) -- (6.south);
\draw[->] (0,-2) -- (1, -2);
\draw[->] (0,-2) -- (0,-1);
\draw[red, thick, -] (5.4,-2) -- (5.4,2); 
\node[text width=.2cm](9) at (1.2, -2){$x_3$};
\node[text width=1cm](10) at (0, -0.9){$x_6$};
\node[text width=.2cm](11) at (3, -0.8){$N$};
\node[text width=.2cm](12) at (4.8, -0.8){$N_-$};
\node[text width=.2cm](12) at (4.8, 0.8){$N_+$};
\end{tikzpicture}}
\qquad \qquad
\scalebox{0.8}{\begin{tikzpicture}[node distance=2cm, nnode/.style={circle,draw,thick, fill, inner sep=1 pt},cnode/.style={circle,draw,thick,minimum size=1.0 cm},snode/.style={rectangle,draw,thick,minimum size=1.0 cm}]
\node[nnode] (1) at (0,0){} ;
\node[nnode] (2) at (0.5,0){} ;
\node[nnode] (3) at (1,0){} ;
\node[](4) at (1.5,0){};
\node[cnode] (5) at (3,0) {$N$};
\node[cnode] (6) at (5, 1) {$N_+$};
\node[cnode] (7) at (5, -1) {$N_-$};
\node[] (8) at (5,-2){};
\node[snode] (9) at (7, 1) {$1$};
\node[snode] (10) at (7, -1) {$1$};
\draw[-] (4) -- (5);
\draw[-] (5) -- (6);
\draw[-] (5) -- (7);
\draw[-] (6) -- (9);
\draw[-] (7) -- (10);
\end{tikzpicture}}
\caption{\footnotesize{Type IIB configuration for a bifurcated tail of a $D$-type quiver with an additional D5-brane (denoted by a red vertical 
line). The associated 3d quiver is shown on the right.}}
\label{Fig:HWOrb2}
\end{center}
\end{figure}

Now, let us consider the scenario where one has also $M$ D5-branes coincident at the orbifold fixed point.
The orbifold action on the hypermultiplet scalars, arising from the $D3-D5$ massless open string spectrum,
is given as:
\be
h^a_{i I}(\vec x, 0)= \gamma_{ij}(g)\, h^a_{j J}(\vec x, 0)\, \gamma^{-1}_{JI}(g),
\ee 
where $a=1,2$ is the $SU(2)_H$ index, $i,j=1,\ldots,N$ are the CP indices for the D3-branes, and $I,J$ 
are the CP indices for the D5-branes. Here, one has to distinguish two cases -- one can either have
$M$ ``full" D5-branes which are free to move off the orbifold fixed point, or they can be stuck at the fixed point 
as fractional branes. In the first case, the matrix $\gamma_{IJ}(g)$ is given by the regular representation of a 
$\BZ_2$-orbifold, i.e.
\be
\gamma_{IJ}(g) ={\rm diag}\,(+1, \ldots, +1, -1, \ldots, -1),
\ee
where $I,J=1,\ldots, 2M$, and the number of $\pm1$ entries is $M$ in each case. In this case, the orbifold 
projection implies that the gauge groups $U(N_{+})$ and $U(N_{-})$ have $M$ fundamental hypers each.
For $M=1$, the brane configuration and the associated quiver is shown in \figref{Fig:HWOrb2}. 

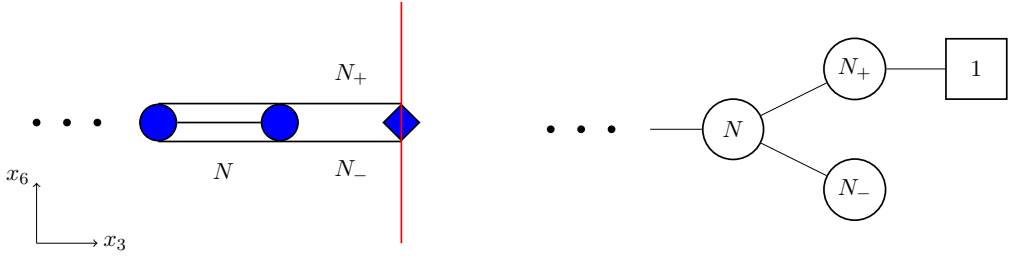
\begin{figure}[h]
\begin{center}
\scalebox{0.8}{\begin{tikzpicture}[
nnode/.style={circle,draw,thick, fill=blue,minimum size= 6mm},cnode/.style={circle,draw,thick,minimum size=4mm},snode/.style={rectangle,draw,thick,minimum size=6mm}]
\node[circle,draw,thick, fill, inner sep=1 pt] (1) at (0,0){} ;
\node[circle,draw,thick, fill, inner sep=1 pt] (2) at (0.5,0){} ;
\node[circle,draw,thick, fill, inner sep=1 pt] (3) at (1,0){} ;
\node[nnode] (4) at (2,0) {} ;
\node[nnode] (5) at (4,0) {};
\node[diamond, draw,thick, fill=blue,minimum size= 6mm] (6) at (6,0) {};
\draw[thick,-] (4.north) -- (5.north);
\draw[thick,-] (4.south) -- (5.south);
\draw[thick,-] (4.east) -- (5.west);
\draw[thick,-] (5.north) -- (6.north);
\draw[thick,-] (5.south) -- (6.south);
\draw[->] (0,-2) -- (1, -2);
\draw[->] (0,-2) -- (0,-1);
\draw[red, thick, -] (6,-2) -- (6,2);
\node[text width=.2cm](9) at (1.2, -2){$x_3$};
\node[text width=1cm](10) at (0, -0.9){$x_6$};
\node[text width=.2cm](11) at (3, -0.8){$N$};
\node[text width=.2cm](12) at (5, -0.8){$N_-$};
\node[text width=.2cm](12) at (5, 0.8){$N_+$};
\end{tikzpicture}}
\qquad \qquad
\scalebox{0.8}{\begin{tikzpicture}[node distance=2cm, nnode/.style={circle,draw,thick, fill, inner sep=1 pt},cnode/.style={circle,draw,thick,minimum size=1.0 cm},snode/.style={rectangle,draw,thick,minimum size=1.0 cm}]
\node[nnode] (1) at (0,0){} ;
\node[nnode] (2) at (0.5,0){} ;
\node[nnode] (3) at (1,0){} ;
\node[](4) at (1.5,0){};
\node[cnode] (5) at (3,0) {$N$};
\node[cnode] (6) at (5, 1) {$N_+$};
\node[cnode] (7) at (5, -1) {$N_-$};
\node[] (8) at (5,-2){};
\node[snode] (9) at (7, 1) {$1$};
\draw[-] (4) -- (5);
\draw[-] (5) -- (6);
\draw[-] (5) -- (7);
\draw[-] (6) -- (9);
\end{tikzpicture}}
\caption{\footnotesize{Type IIB configuration for a bifurcated tail of a $D$-type quiver with a fractional D5-brane (denoted by a red vertical 
line) stuck at the orbifold fixed plane. The associated 3d quiver is shown on the right.}}
\label{Fig:HWOrb3}
\end{center}
\end{figure}

For the fractional D5-branes, the matrix $\gamma_{IJ}(g)$ is 
given as
\be
\gamma_{IJ}(g) ={\rm diag}(1,\ldots,1,-1,\ldots,-1),
\ee
where $I,J=1,\ldots, M$, and with $M_+$ entries being $+1$, $M_-$ entries being -1, and $M= M_+ + M_-$. 
In this case, the orbifold projection implies that the gauge groups $U(N_{+})$ and $U(N_{-})$ will have 
$M_+$ and $M_-$ fundamental hypers respectively. The brane configuration for $M=1$, $M_+=1$, and $M_-=0$, 
and the associated quiver is shown in \figref{Fig:HWOrb3}.

\subsubsection{S-dual configurations: D3-branes ending on orientifolds}\label{D2-TypeIIB}
Let us now extend our discussion to 3d quivers for which the Type IIB configuration involves 
D3-branes ending on orientifold 5-planes, in addition to the NS5 and D5-branes. Generically, these 
boundary conditions lead to linear quiver gauge theories with unitary as well as symplectic gauge 
groups.\\

Given a Type IIB configuration of D3-D5-NS5-branes oriented as \eref{IIB-1}, consider an orientifolding operation 
$\CI_4\,\Omega$, where
\be
\CI_4\,:\, x^{3,7,8,9} \to - x^{3,7,8,9},
\ee
and $\Omega$ is the worldsheet orientation reversal operation. The fixed plane of the 
orientifolding operation is the 5-plane located at the point $x^{3,7,8,9}=0$ in the transverse direction. 
The gauge group for the world-volume theory on a stack of $M$ D5-branes, coincident with this fixed plane, 
can be $Sp(M)$ or $O(2M)$. In the former case, the fixed plane has a D5-brane charge $-1$, and is 
therefore referred to as the O5$^-$-plane. In latter case, the D5-brane charge $+1$, and the corresponding 
fixed plane is denoted as O5$^+$.
An orientifold 5-plane preserves the same supersymmetry as a parallel D5-brane. 
The respective world-volumes of the D3, D5, NS5-branes 
and the O5$^-$-plane can be summarized as:
\begin{align}\label{IIB-Orb}
& \text{D3:}\quad \BR^{2,1} \times L \times \{\vec X\}_{4,5,6} \times \{\vec Y\}_{7,8,9} \nn\\
& \text{D5:}\quad \BR^{2,1} \times \{X_3\}  \times \BR^{3}_{4,5,6} \times \{ \vec Y'\}_{7,8,9} \nn \\
& \text{NS5:}\quad \BR^{2,1} \times \{X'_3\}  \times  \{\vec X'\}_{4,5,6} \times \BR^3_{7,8,9},\nn \\
& \text{O5}^-: \quad \BR^{2,1}\times \{\vec{0} \}_{3,7,8,9} \times  \BR^3_{4,5,6}.
\end{align}
To begin with, consider a stack of $2N$ D3-branes ending on an O5$^-$-plane at one of the boundaries 
and an NS5-brane at the other boundary. The boundary conditions imply that gauge field $A^\mu$ and the scalars $Y^A$ combine to 
give the bosonic field content of a 3d $\CN=4$ vector multiplet with a gauge group $Sp(N)$. 
The NS5-brane imposes Dirichlet boundary condition on the scalars $X^3, X^{A'}$ at one end, 
so that they do not figure in the 3d Lagrangian. Additionally, if there are D5-branes in the D3-brane chamber, 
then the massless states from the D3-D5 strings give hypermultiplets in the fundamental representation of 
the gauge group $Sp(N)$.\\

In this work, we shall encounter two distinct boundary conditions involving orientifold 5-planes,  
which in turn arise on S-dualizing boundary conditions involving orbifold 5-planes. 
The first case is the S-dual of $2N$ D3-branes ending on an orbifold 5-plane with $N_+ =N_-=N$ \cite{Gaiotto:2008ak}. 
This is given by $2N$ D3-branes ending on an O5$^-$-plane with a single coincident D5-brane -- 
a combination we will refer to as an O5$^0$-plane, where the superscript indicates the zero total D5-brane charge.
For a D3-brane stack between an O5$^0$-plane and an NS5-brane, the associated 3d theory has an $Sp(N)$ 
vector multiplet and a single fundamental hypermultiplet.\\

The second case is the S-dual of $2N$ D3-branes ending on an orbifold 5-plane with $M$ fractional D5-branes, 
and $N_+ =N_-=N$ as before. 
This is given by $2N$ D3-branes ending on an O5$^0$-plane with $M$ NS5-branes stuck at the O5$^-$-plane.
Similar to the first case, we would be interested in D3-branes ending on an NS5 at the other boundary 
of $x^3$. In case of $M=1$, the associated 3d theory can be shown to be a $U(2N)$ gauge theory with a hypermultiplet 
in the antisymmetric representation and the fundamental representation of the gauge group respectively 
\cite{Hanany:1999sj, Hanany:1997gh}. Presence of additional D5-branes will contribute to fundamental 
hypermultiplets for the gauge group in the 3d theory. For $M > 1$, the associated 3d theory can again be read 
off from the brane configuration -- it turns out to be a linear quiver of unitary gauge nodes with a 
symplectic gauge node at the end.\\

Along with the standard D3-NS5-D5 system, the two boundary conditions with orbifold 5-planes (and their S-duals) 
are sufficient to engineer the D-type and the affine D-type quivers (and their mirror duals) we are interested in. 
The only exception will be the mirror pair discussed in \Secref{3d-bad} where the S-dual boundary condition 
involving orientifold 5-planes give a bad 3d quiver.

\subsection{Type IIB configurations with defects and S-duality}\label{D-TypeIIB-def}
In this section, we discuss the Type IIB brane configurations associated with vortex and Wilson defects in $D$-type 
and affine $D$-type quiver gauge theories with unitary gauge groups. Such configurations will generically involve 
introducing D1-branes and F1-strings respectively in a D3-D5-NS5 setting with one or two orbifold 5-planes. For 
each representative case, one can obtain the dual defect by $S$-dualizing this configuration, which will lead to a 
configuration with F1-strings and D1-branes respectively in a D3-D5-NS5 setting with one or two orientifold 
5-planes. We will restrict ourselves to Type IIB configuration of \figref{Fig:HWOrb1},
where there are no D5-branes between the orbifold 5-plane and the nearest NS5-brane.\\

A generic $D$-type quiver gauge theory, engineered by such a  Type IIB configuration, 
has a bifurcated tail, and the terminal gauge nodes at the tail do not have any fundamental 
matter. Away from the tail, the vortex/Wilson defects will have features similar to the linear quivers. 
Therefore, we will focus on defects for gauge groups around the bifurcated tail and study their S-dual counterparts.
For concreteness, we will choose the ranks of the gauge group and the matter content as follows.
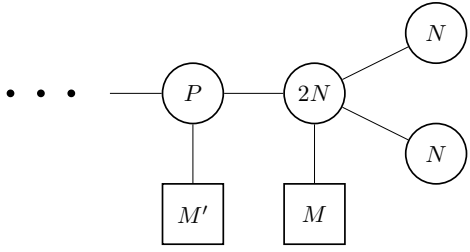
\begin{center}
\scalebox{0.8}{\begin{tikzpicture}[node distance=2cm, nnode/.style={circle,draw,thick, fill, inner sep=1 pt},cnode/.style={circle,draw,thick,minimum size=1.0 cm},snode/.style={rectangle,draw,thick,minimum size=1.0 cm}]
\node[nnode] (1) at (0,0){} ;
\node[nnode] (2) at (0.5,0){} ;
\node[nnode] (3) at (1,0){} ;
\node[](4) at (1.5,0){};
\node[cnode] (5) at (3,0) {$P$};
\node[cnode] (6) at (5,0) {$2N$};
\node[snode] (7) at (3,-2) {$M'$};
\node[snode] (8) at (5,-2) {$M$};
\node[cnode] (9) at (7, 1) {$N$};
\node[cnode] (10) at (7, -1) {$N$};
\draw[-] (4) -- (5);
\draw[-] (5) -- (6);
\draw[-] (5) -- (7);
\draw[-] (6) -- (8);
\draw[-] (6) -- (9);
\draw[-] (6) -- (10);
\end{tikzpicture}}
\end{center}
The integers $P,M,M',N$ are chosen such that every gauge node is either balanced or overbalanced. 
In addition, we will set $M'=0$ and $M=1$, for simplifying the presentation.

\subsubsection{Vortex defects in the presence of orbifold 5-planes}
Consider a vortex defect labelled by a representation $R=\otimes^l_{a=1} \CS_{k^{(a)}}\otimes^{l'}_{b=1} \CA_{k'^{(b)}} $ 
for the gauge node $U(2N)$ of the $D$-type quiver. Note that this is precisely the type of vortex defect we encountered 
in \Secref{V2W-D} for a $D_4$ quiver (see \figref{AbSU2}) and in \Secref{1-D-Sp(2)} for a $D_6$ quiver (see \figref{NAbEx1Sp}).
The defect can be introduced in a fashion similar to the linear quiver, reviewed in 
\Secref{LQ2BD}. We introduce stacks of  D1-branes stretching between the $2N$ D3-branes on one end and additional 
NS5 and/or NS5'-branes (displaced in the $x^6$-direction) on the other. The D1 stacks ending on the NS5-branes and the 
NS5'-branes correspond to the symmetric factors $ \CS_{k^{(a)}}$ and the antisymmetric factors $\CA_{k'^{(b)}}$ in the 
representation $R$ respectively. 

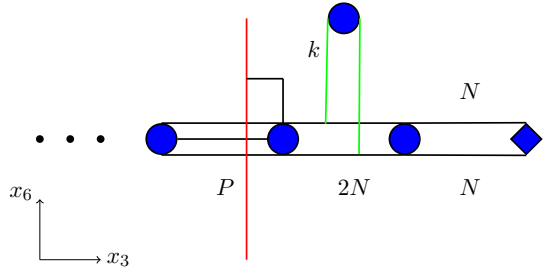
\begin{figure}[h]
\begin{center}
\scalebox{0.8}{\begin{tikzpicture}[
nnode/.style={circle,draw,thick, fill=blue,minimum size= 5mm},cnode/.style={circle,draw,thick,minimum size=4mm},snode/.style={rectangle,draw,thick,minimum size=6mm}]
\node[circle,draw,thick, fill, inner sep=1 pt] (1) at (0,0){} ;
\node[circle,draw,thick, fill, inner sep=1 pt] (2) at (0.5,0){} ;
\node[circle,draw,thick, fill, inner sep=1 pt] (3) at (1,0){} ;
\node[nnode] (4) at (2,0) {} ;
\node[nnode] (5) at (4,0) {};
\node[nnode] (6) at (6,0) {};
\node[diamond, draw,thick, fill=blue,minimum size= 5mm] (7) at (8,0) {};
\node[nnode] (8) at (5,2) {};
\draw[thick,-] (4.north) -- (5.north);
\draw[thick,-] (4.south) -- (5.south);
\draw[thick,-] (4.east) -- (5.west);
\draw[thick,-] (5.north) -- (6.north);
\draw[thick,-] (5.south) -- (6.south);
\draw[thick,-] (7.north) -- (6.north);
\draw[thick,-] (7.south) -- (6.south);
\draw[->] (0,-2) -- (1, -2);
\draw[->] (0,-2) -- (0,-1);
\draw[red, thick, -] (3.4,-2) -- (3.4,2);
\draw[thick,-] (3.4, 1) -- (4,1);
\draw[thick,-] (4, 1) -- (5.north);
\draw[green, thick] (8.west) -- (4.7,0.25);
\draw[green, thick] (8.east) -- (5.25,-0.25);
\node[text width=.2cm](9) at (1.2, -2){$x_3$};
\node[text width=1cm](10) at (0, -0.9){$x_6$};
\node[text width=.2cm](11) at (3, -0.8){$P$};
\node[text width=.2cm](12) at (5, -0.8){$2N$};
\node[text width=.2cm](13) at (7, -0.8){$N$};
\node[text width=.2cm](14) at (7, 0.8){$N$};
\node[text width=.2cm](15) at (4.5, 1.5){$k$};
\end{tikzpicture}}
\caption{\footnotesize{Type IIB configuration corresponding to a vortex defect for the $U(2N)$ gauge group in the 
generic $D$-type quiver discussed above. The green vertical lines denote $k_1,\ldots, k_N$ D1-branes respectively, where 
$k_1+\ldots+ k_N=k$. The configuration corresponds to a vortex defect in a representation $R=\CS_k$ of $U(2N)$.}}
\label{Fig:HWOrb4}
\end{center}
\end{figure}

The special case for $R=\CS_k$ is shown in \figref{Fig:HWOrb4}. In this case, $k_i$ D1-branes (denoted by vertical green lines) 
stretch between an NS5-brane and the $i$-th D3-brane such that $k_1 + \ldots +k_N= k$, with $\{k_i\}$ unrestricted. The set of integers $\{k_i \}$ are in 
one-to-one correspondence to the weights of the representation $R=\CS_k$ of $U(2N)$.
The configuration for a  more general representation can be written down in an analogous fashion following the rules in \Secref{LQ2BD}.\\

\begin{figure}[h]
\begin{center}
\begin{tabular}{ccc}
\scalebox{0.7}{\begin{tikzpicture}[
nnode/.style={circle,draw,thick, fill=blue,minimum size= 5mm},cnode/.style={circle,draw,thick,minimum size=4mm},snode/.style={rectangle,draw,thick,minimum size=6mm}]
\node[circle,draw,thick, fill, inner sep=1 pt] (1) at (0,0){} ;
\node[circle,draw,thick, fill, inner sep=1 pt] (2) at (0.5,0){} ;
\node[circle,draw,thick, fill, inner sep=1 pt] (3) at (1,0){} ;
\node[nnode] (4) at (2,0) {} ;
\node[nnode] (5) at (4,0) {};
\node[nnode] (6) at (6,0) {};
\node[diamond, draw,thick, fill=blue,minimum size= 5mm] (7) at (8,0) {};
\node[nnode] (8) at (6,2) {};
\draw[thick,-] (4.north) -- (5.north);
\draw[thick,-] (4.south) -- (5.south);
\draw[thick,-] (4.east) -- (5.west);
\draw[thick,-] (5.north) -- (6.north);
\draw[thick,-] (5.south) -- (6.south);
\draw[thick,-] (7.north) -- (6.north);
\draw[thick,-] (7.south) -- (6.south);
\draw[->] (0,-2) -- (1, -2);
\draw[->] (0,-2) -- (0,-1);
\draw[red, thick, -] (3.4,-2) -- (3.4,2);
\draw[thick,-] (3.4, 1) -- (4,1);
\draw[thick,-] (4, 1) -- (5.north);
\draw[green, thick] (8.south) -- (6.north);
\draw[green, thick] (8.south) -- (6.north);
\node[text width=.2cm](9) at (1.2, -2){$x_3$};
\node[text width=1cm](10) at (0, -0.9){$x_6$};
\node[text width=.2cm](11) at (3, -0.8){$P$};
\node[text width=.2cm](12) at (5, -0.8){$2N$};
\node[text width=.2cm](13) at (7, -0.8){$N$};
\node[text width=.2cm](14) at (7, 0.8){$N$};
\node[text width=.2cm](15) at (5.8, 1.5){$k$};
\end{tikzpicture}}
& \qquad &
\scalebox{0.7}{\begin{tikzpicture}[node distance=2cm, nnode/.style={circle,draw,thick, fill, inner sep=1 pt},cnode/.style={circle,draw,thick,minimum size=1.0 cm},snode/.style={rectangle,draw,thick,minimum size=1.0 cm},rnode/.style={red, circle,draw,thick,fill=red!30 ,minimum size=1.0cm}]
\node[nnode] (1) at (0,0){} ;
\node[nnode] (2) at (0.5,0){} ;
\node[nnode] (3) at (1,0){} ;
\node[](4) at (1.5,0){};
\node[cnode] (5) at (3,0) {$P$};
\node[cf-group] (6) at (5,0) {\rotatebox{-90}{$2N$}};
\node[rnode] (7) at (5,2) {$k$};
\node[snode] (8) at (5,-2) {$1$};
\node[cnode] (9) at (7, 1) {$N$};
\node[cnode] (10) at (7, -1) {$N$};
\draw[-] (4) -- (5);
\draw[-] (5) -- (6);
\draw[-] (6) -- (8);
\draw[-] (6) -- (9);
\draw[-] (6) -- (10);
\draw[red, thick,->] (6) -- (7);
\draw[red, thick,->] (7) -- (9);
\draw[red, thick,->] (7) -- (10);
\draw[red, thick, ->] (7) to [out=150,in=210, looseness=6] (7);
\end{tikzpicture}}\\
\scalebox{0.7}{\begin{tikzpicture}[
nnode/.style={circle,draw,thick, fill=blue,minimum size= 5mm},cnode/.style={circle,draw,thick,minimum size=4mm},snode/.style={rectangle,draw,thick,minimum size=6mm}]
\node[circle,draw,thick, fill, inner sep=1 pt] (1) at (0,0){} ;
\node[circle,draw,thick, fill, inner sep=1 pt] (2) at (0.5,0){} ;
\node[circle,draw,thick, fill, inner sep=1 pt] (3) at (1,0){} ;
\node[nnode] (4) at (2,0) {} ;
\node[nnode] (5) at (4,0) {};
\node[nnode] (6) at (6,0) {};
\node[diamond, draw,thick, fill=blue,minimum size= 5mm] (7) at (8,0) {};
\node[nnode] (8) at (4,2) {};
\draw[thick,-] (4.north) -- (5.north);
\draw[thick,-] (4.south) -- (5.south);
\draw[thick,-] (4.east) -- (5.west);
\draw[thick,-] (5.north) -- (6.north);
\draw[thick,-] (5.south) -- (6.south);
\draw[thick,-] (7.north) -- (6.north);
\draw[thick,-] (7.south) -- (6.south);
\draw[->] (0,-2) -- (1, -2);
\draw[->] (0,-2) -- (0,-1);
\draw[red, thick, -] (2.5,-2) -- (2.5,2);
\draw[thick,-] (2.5, 1) -- (3.4,1);
\draw[thick,-] (3.4, 1) -- (5.north);
\draw[green, thick] (8.south) -- (5.north);
\draw[green, thick] (8.south) -- (5.north);
\node[text width=.2cm](9) at (1.2, -2){$x_3$};
\node[text width=1cm](10) at (0, -0.9){$x_6$};
\node[text width=.2cm](11) at (3, -0.8){$P$};
\node[text width=.2cm](12) at (5, -0.8){$2N$};
\node[text width=.2cm](13) at (7, -0.8){$N$};
\node[text width=.2cm](14) at (7, 0.8){$N$};
\node[text width=.2cm](15) at (3.8, 1.5){$k$};
\end{tikzpicture}}
& \qquad &
\scalebox{0.7}{\begin{tikzpicture}[node distance=2cm, nnode/.style={circle,draw,thick, fill, inner sep=1 pt},cnode/.style={circle,draw,thick,minimum size=1.0 cm},snode/.style={rectangle,draw,thick,minimum size=1.0 cm},rnode/.style={red, circle,draw,thick,fill=red!30 ,minimum size=1.0cm}]
\node[nnode] (1) at (0,0){} ;
\node[nnode] (2) at (0.5,0){} ;
\node[nnode] (3) at (1,0){} ;
\node[](4) at (1.5,0){};
\node[cf-group] (5) at (3,0) {\rotatebox{-90}{$P$}};
\node[cf-group] (6) at (5,0) {\rotatebox{-90}{$2N$}};
\node[rnode] (7) at (4,2) {$k$};
\node[snode] (8) at (4,-2) {$1$};
\node[cnode] (9) at (7, 1) {$N$};
\node[cnode] (10) at (7, -1) {$N$};
\draw[-] (4) -- (5);
\draw[-] (5) -- (6);
\draw[-] (6) -- (8);
\draw[-] (6) -- (9);
\draw[-] (6) -- (10);
\draw[red, thick,->] (5) -- (7);
\draw[red, thick,->] (7) -- (6);
\draw[red, thick,->] (8) -- (7);
\draw[red, thick, ->] (7) to [out=60,in=120, looseness=6] (7);
\end{tikzpicture}}
\end{tabular}
\caption{\footnotesize{Vortex defects for the $U(2N)$ gauge group of the generic $D$-type quiver can be understood as the 
deformation of at least two 3d-1d coupled quivers. The two coupled quivers as well as the corresponding Type IIB realizations 
are shown.}}
\label{Fig:HWOrb5}
\end{center}
\end{figure}
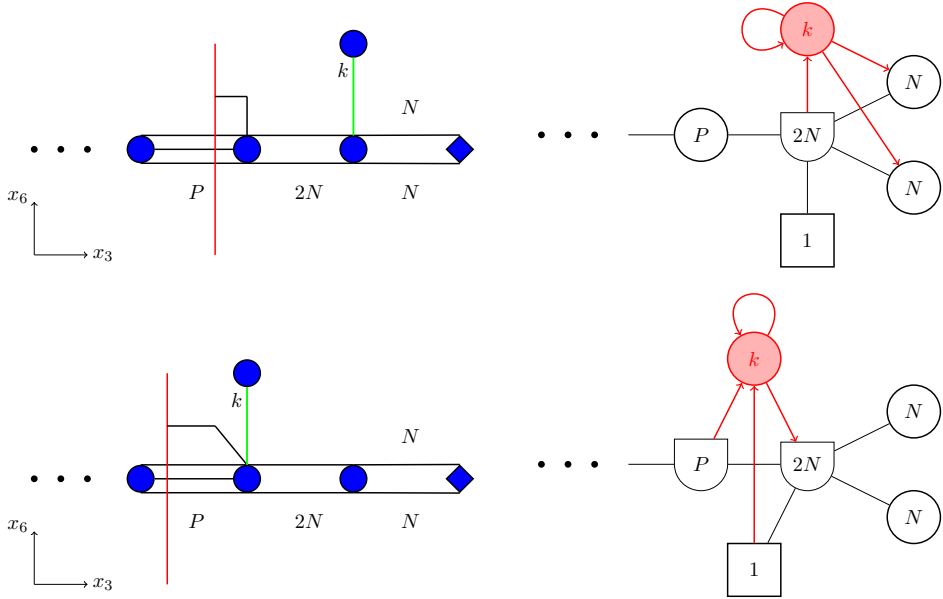

The vortex defect can be realized as deformations of two different coupled 3d-1d systems obtained by moving the 
stack of $k$ D1-branes to end either on the right NS5-brane or the left NS5-brane in \figref{Fig:HWOrb4}. 
The brane configurations and the associated coupled 3d-1d systems, which can be manifestly read off from the 
brane configurations, are shown in \figref{Fig:HWOrb5}. For $N=2$ and $P=3$, this is precisely the hopping duality 
we found in \figref{NAbEx1SpHop} of \Secref{1-D-Sp(2)} for the $D_6$ quiver gauge theory. 
The coupled systems $X'[V'^{(I)}_{2,R}]$ and $X[V'^{(II)}_{2,R}]$ 
in \figref{NAbEx1SpHop} were shown to describe the same vortex defect by arguing that the partition functions of the two 
systems are equal. The configurations in \figref{Fig:HWOrb5} give a Type IIB brane description of the hopping duality. 

For $N=1$ and $P=1$, the configurations of \figref{Fig:HWOrb5} reproduce the hopping duality found in 
\figref{AbEx1D4Hopping} of \Secref{1-D} for a $D_4$ quiver.


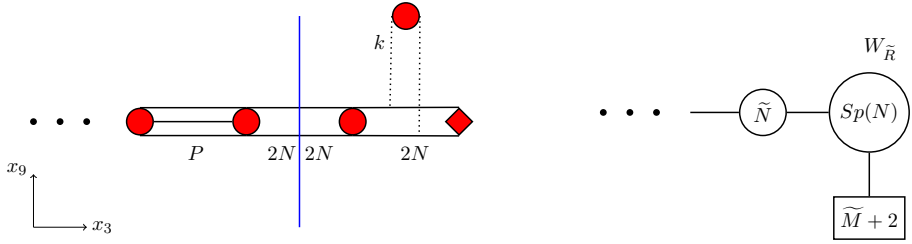
\begin{figure}[h]
\begin{center}
\scalebox{0.7}{\begin{tikzpicture}[
nnode/.style={circle,draw,thick, fill=red,minimum size= 5mm},cnode/.style={circle,draw,thick,minimum size=4mm},snode/.style={rectangle,draw,thick,minimum size=6mm}]
\node[circle,draw,thick, fill, inner sep=1 pt] (1) at (0,0){} ;
\node[circle,draw,thick, fill, inner sep=1 pt] (2) at (0.5,0){} ;
\node[circle,draw,thick, fill, inner sep=1 pt] (3) at (1,0){} ;
\node[nnode] (4) at (2,0) {} ;
\node[nnode] (5) at (4,0) {};
\node[nnode] (6) at (6,0) {};
\node[diamond, draw,thick, fill=red,minimum size= 5mm] (7) at (8,0) {};
\node[nnode] (8) at (7,2) {};
\draw[thick,-] (4.north) -- (5.north);
\draw[thick,-] (4.south) -- (5.south);
\draw[thick,-] (4.east) -- (5.west);
\draw[thick,-] (5.north) -- (6.north);
\draw[thick,-] (5.south) -- (6.south);
\draw[thick,-] (7.north) -- (6.north);
\draw[thick,-] (7.south) -- (6.south);
\draw[->] (0,-2) -- (1, -2);
\draw[->] (0,-2) -- (0,-1);
\draw[blue, thick, -] (5,-2) -- (5,2);
\draw[thick,dotted ] (8.west) -- (6.7,0.25);
\draw[thick,dotted] (8.east) -- (7.25,-0.25);
\node[text width=.2cm](9) at (1.2, -2){$x_3$};
\node[text width=1cm](10) at (0, -0.9){$x_9$};
\node[text width=.2cm](11) at (3, -0.6){$P$};
\node[text width=.2cm](12) at (5.2, -0.6){$2N$};
\node[text width=.2cm](12) at (4.5, -0.6){$2N$};
\node[text width=.2cm](13) at (7, -0.6){$2N$};
\node[text width=.2cm](15) at (6.5, 1.5){$k$};
\end{tikzpicture}}
\qquad  \qquad 
\scalebox{0.7}{\begin{tikzpicture}[
nnode/.style={circle,draw,thick, fill=red,minimum size= 4mm},cnode/.style={circle,draw,thick,minimum size=4mm},snode/.style={rectangle,draw,thick,minimum size=8mm}]
\node[circle,draw,thick, fill, inner sep=1 pt] (1) at (-1,0){} ;
\node[circle,draw,thick, fill, inner sep=1 pt] (2) at (-1.5,0){} ;
\node[circle,draw,thick, fill, inner sep=1 pt] (3) at (-2,0){} ;
\node[](4) at (-.5,0){};
\node[cnode] (5) at (3,0) {$Sp(N)$};
\node[cnode] (7) at (1,0) {$\wt{N}$};
\node[snode] (6) at (3,-2) {$\wt{M}+2$};
\node[text width=.2cm](10) at (3,1.2){$W_{\wt{R}}$};
\draw[thick] (7) -- (4);
\draw[thick] (5) -- (6);
\draw[thick] (5) -- (7);
\end{tikzpicture}}
\caption{\footnotesize{S-dual of the Type IIB configuration in \figref{Fig:HWOrb4}. The red rhombus denotes an O5$^0$-plane. 
The associated quiver with the Wilson defect is shown on the right.}}
\label{Fig:HWO7}
\end{center}
\end{figure}

To construct the dual defect and read off the mirror map, we consider the S-dual of the brane configuration presented above in \figref{Fig:HWOrb4}. 
As reviewed in \Secref{D2-TypeIIB}, D5 and NS5-branes are exchanged, D3-branes remain invariant, D1-branes become F1 strings, and the orbifold 5-plane 
becomes an O5$^0$-plane. One now needs to subject this brane configuration to a Hanany-Witten move where the NS5-brane is moved across the 
D5-brane to the immediate right. After the move, one can have $2N$ D3-branes in the rightmost chamber stretching between the NS5-brane and the 
O5$^0$-plane, while no D3-brane ends on the D5. The final configuration is given by \figref{Fig:HWO7}, from which the dual theory as well as the 
dual defect can be read off. 
As reviewed in \Secref{D2-TypeIIB}, the 3d theory associated with $2N$ D3-branes between the NS5-brane and the O5$^0$-plane is an $Sp(N)$ gauge theory.
The D5-branes in the chamber as well as the D3-branes stretching between the NS5 and D5-branes to the left will give fundamental hypermultiplets for $Sp(N)$. 
From the brane configuration, one can check that the total number of fundamental and bifundamental hypermultiplets of $Sp(N)$ is $2N+2$, i.e. 
$\wt{M} +\wt{N}=2N$ in the quiver. The precise integers $\wt{M},\wt{N}$ will depend on the details of the linear quiver tail. For the examples of the 
$D$-type quivers we have considered in this paper, $\wt{M}=2N$ and $\wt{N}=0$.

As shown in \figref{Fig:HWO7}, the defect is realized by a stack of $k$ F1-strings (denoted by vertical black dotted lines) stretched between a D5 and $2N$ D3-branes 
-- $k_i$ F1-strings end on the $i$-th D3-brane 
such that $k_1+\ldots +k_N=k$, with $0 \leq k_i \leq k$. Recall that the set of integers $\{k_i \}$ are in one-to-one correspondence to the weights of the 
representation $R=\CS_k$ of $U(2N)$. The configuration for a generic $R$ can be obtained in an analogous fashion. 
The configuration of F1-strings then generates a Wilson defect for the $Sp(N)$ gauge group, labelled 
by a representation $\wt{R}$, where $\wt{R}$ is the restriction of the representation $R$ of $U(2N)$ to the subgroup $Sp(N) \subset U(2N)$. 
Combined with the discussion of hopping duality, this leads to a mirror map of the form \eref{MM-D6-V2W1}, which was obtained for the 
mirror pair involving the $D_6$ quiver and the $Sp(2)$ gauge theory from sphere partition function analysis.


\subsubsection{Wilson defects in the presence of orbifold 5-planes}

Next, we consider a Wilson defect labelled by a representation $R=\otimes^l_{a=1} \CS_{k^{(a)}}\otimes^{l'}_{b=1} \CA_{k'^{(b)}} $ 
for the gauge node $U(2N)$ of the $D$-type quiver. Note that this is precisely the type of Wilson defect we encountered 
in \Secref{W2V-D} for a $D_4$ quiver (see \figref{AbEx2aD4}) and in \Secref{1-D-Sp(2)} for a $D_6$ quiver (see \figref{NAbEx2Sp}).
The defect can be introduced in a fashion similar to the linear quiver, reviewed in 
\Secref{LQ2BD}. We introduce stacks of  F1-strings stretching between the $2N$ D3-branes on one end and additional 
D5 and/or D5'-branes (displaced in the $x^6$-direction) on the other. The F1 stacks ending on the D5-branes and the 
D5'-branes correspond to the symmetric factors $ \CS_{k^{(a)}}$ and the antisymmetric factors $\CA_{k'^{(b)}}$ in the 
representation $R$ respectively. 

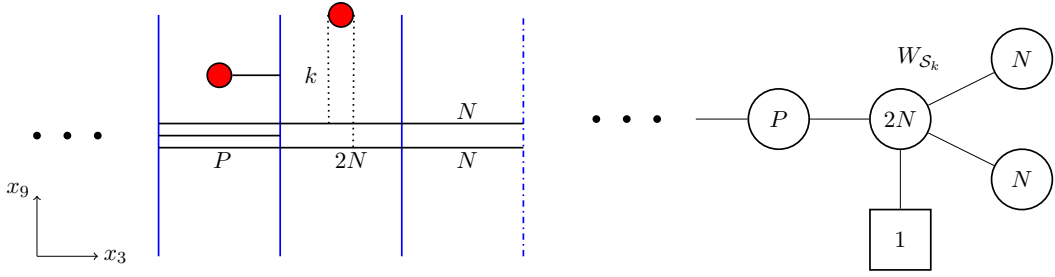
\begin{figure}[h]
\begin{center}
\scalebox{0.8}{\begin{tikzpicture}[node distance=2cm, nnode/.style={circle,draw,thick, fill, inner sep=1 pt},cnode/.style={circle,draw,thick,minimum size=1.0 cm},snode/.style={rectangle,draw,thick,minimum size=1.0 cm}]
\node[nnode] (1) at (0,0){} ;
\node[nnode] (2) at (0.5,0){} ;
\node[nnode] (3) at (1,0){} ;
\node[](4) at (1.5,0){};
\node[circle,draw,thick, fill=red,minimum size= 4mm] (5) at (3,1){};
\node[circle,draw,thick, fill=red,minimum size= 4mm] (6) at (5,2){};
\draw[blue, thick] (2,-2) -- (2,2);
\draw[blue, thick] (4,-2) -- (4,2);
\draw[blue, thick] (6,-2) -- (6,2);
\draw[blue, thick, dash dot] (8,-2) -- (8,2);
\draw[thick] (2,0) -- (4,0);
\draw[thick] (2,0.2) -- (4,0.2);
\draw[thick] (2,-0.2) -- (4,-0.2);
\draw[thick] (4,0.2) -- (6,0.2);
\draw[thick] (4,-0.2) -- (6,-0.2);
\draw[thick] (6,0.2) -- (8,0.2);
\draw[thick] (6,-0.2) -- (8,-0.2);
\draw[thick] (5) -- (4,1);
\draw[thick, dotted] (6.west) -- (4.8, 0.2);
\draw[thick, dotted] (6.east) -- (5.2, -0.2);
\node[text width=.2cm] (7) at (3,-0.4){$P$};
\node[text width=.2cm] (8) at (5,-0.4){$2N$};
\node[text width=.2cm] (9) at (7, 0.4){$N$};
\node[text width=.2cm] (10) at (7,-0.4){$N$};
\node[text width=.2cm] (11) at (4.5, 1){$k$};
\draw[->] (0,-2) -- (1, -2);
\draw[->] (0,-2) -- (0,-1);
\node[text width=.2cm](15) at (1.2, -2){$x_3$};
\node[text width=1cm](16) at (0, -0.9){$x_9$};
\end{tikzpicture}}
\qquad
\scalebox{0.8}{\begin{tikzpicture}[node distance=2cm, nnode/.style={circle,draw,thick, fill, inner sep=1 pt},cnode/.style={circle,draw,thick,minimum size=1.0 cm},snode/.style={rectangle,draw,thick,minimum size=1.0 cm}]
\node[nnode] (1) at (0,0){} ;
\node[nnode] (2) at (0.5,0){} ;
\node[nnode] (3) at (1,0){} ;
\node[](4) at (1.5,0){};
\node[cnode] (5) at (3,0) {$P$};
\node[cnode] (6) at (5,0) {$2N$};
\node[snode] (8) at (5,-2) {$1$};
\node[cnode] (9) at (7, 1) {$N$};
\node[cnode] (10) at (7, -1) {$N$};
\draw[-] (4) -- (5);
\draw[-] (5) -- (6);
\draw[-] (6) -- (8);
\draw[-] (6) -- (9);
\draw[-] (6) -- (10);
\node[text width=0.1cm] (11) at (5,1){$W_{\CS_k}$};
\end{tikzpicture}}
\caption{\footnotesize{Type IIB configuration corresponding to a Wilson defect for the $U(2N)$ gauge group in the 
generic $D$-type quiver discussed above. The blue dot-dashed line denotes the Orb5-plane. 
The black dotted vertical lines denote $k_1,\ldots, k_N$ F1-strings respectively, where $k_1+\dots +k_N=k$. 
The configuration corresponds to a Wilson defect in a representation $R=\CS_k$ of $U(2N)$.}}
\label{Fig:HWOrb7}
\end{center}
\end{figure}

The special case for $R=\CS_k$ is shown in \figref{Fig:HWOrb7}. In this case, $k_i$ F1-strings (denoted by vertical black dotted lines) 
stretch between a D5-brane and the $i$-th D3-brane such that $k_1 + \ldots +k_N= k$, with $0 \leq k_i \leq k$ for all $i$. 
The set of integers $\{k_i \}$ are in one-to-one correspondence to the weights of the representation $R=\CS_k$ of $U(2N)$.
The configuration for a  more general representation can be written down in an analogous fashion following the rules in \Secref{LQ2BD}.\\

\begin{figure}[h]
\begin{center}
\scalebox{0.8}{\begin{tikzpicture}[node distance=2cm, nnode/.style={circle,draw,thick, fill, inner sep=1 pt},cnode/.style={circle,draw,thick,minimum size=1.0 cm},snode/.style={rectangle,draw,thick,minimum size=1.0 cm}]
\node[nnode] (1) at (0,0){} ;
\node[nnode] (2) at (0.5,0){} ;
\node[nnode] (3) at (1,0){} ;
\node[](4) at (1.5,0){};
\node[circle,draw,thick, fill=blue,minimum size= 4mm] (5) at (4,0){};
\node[circle,draw,thick, fill=blue,minimum size= 4mm] (6) at (5,2){};
\draw[red, thick] (2,-2) -- (2,2);
\draw[red, thick] (3,-2) -- (3,2);
\draw[red, thick] (7,-2) -- (7,2);
\draw[red, thick, dash dot] (8,-2) -- (8,2);
\draw[thick] (2,0) -- (3,0);
\draw[thick] (2,0.2) -- (3,0.2);
\draw[thick] (2,-0.2) -- (3,-0.2);
\draw[thick] (3,0.2) -- (4,0.2);
\draw[thick] (3,-0.2) -- (4,-0.2);
\draw[thick] (4,0.2) -- (8,0.2);
\draw[thick] (4,-0.2) -- (8,-0.2);
\draw[thick, green] (6.west) -- (4.8, 0.2);
\draw[thick, green] (6.east) -- (5.2, -0.2);
\node[text width=.2cm] (7) at (2.5,-0.4){$P$};
\node[text width=.2cm] (8) at (5.5,-0.4){$2N$};
\node[text width=.2cm] (9) at (3.2,-0.4){$2N$};
\node[text width=.2cm] (11) at (4.5, 1){$k$};
\draw[->] (0,-2) -- (1, -2);
\draw[->] (0,-2) -- (0,-1);
\node[text width=.2cm](15) at (1.2, -2){$x_3$};
\node[text width=1cm](16) at (0, -0.9){$x_9$};
\end{tikzpicture}}
\qquad
\scalebox{0.8}{\begin{tikzpicture}[node distance=2cm, nnode/.style={circle,draw,thick, fill, inner sep=1 pt},cnode/.style={circle,draw,thick,minimum size=1.0 cm},snode/.style={rectangle,draw,thick,minimum size=1.0 cm}, rnode/.style={red, circle,draw,thick,fill=red!30 ,minimum size=1.0cm}]
\node[nnode] (1) at (0,0){} ;
\node[nnode] (2) at (0.5,0){} ;
\node[nnode] (3) at (1,0){} ;
\node[](4) at (1.5,0){};
\node[cf-group](5) at (3,0){\rotatebox{-90}{$\wt{N}$}};
\node[cf-group](6) at (5,0){\rotatebox{-90}{$Sp(N)$}};
\node[snode] (7) at (7,0){$2$} ;
\node[rnode](8) at (4,2) {$k$};
\node[snode] (10) at (4,-2){$\wt{M}$} ;
\draw[thick] (4) -- (5);
\draw[thick] (5) -- (6);
\draw[thick] (6) -- (7);
\draw[thick] (10) -- (6);
\draw[red, thick,->] (8) -- (6);
\draw[red, thick,->] (5) -- (8);
\draw[red, thick,->] (10) -- (8);
\draw[red, thick, ->] (8) to [out=60,in=120, looseness=6] (8);
\end{tikzpicture}}
\caption{\footnotesize{S-dual of the configuration in \figref{Fig:HWOrb7}. The red dot-dashed vertical line is an O5$^0$-plane. The 
Type IIB configuration realizes a vortex defect for the $Sp(N)$ gauge group. This vortex defect can be understood as a 
deformation of a 3d-1d coupled quiver, which is shown on the right.}}
\label{Fig:HWOrb8}
\end{center}
\end{figure}
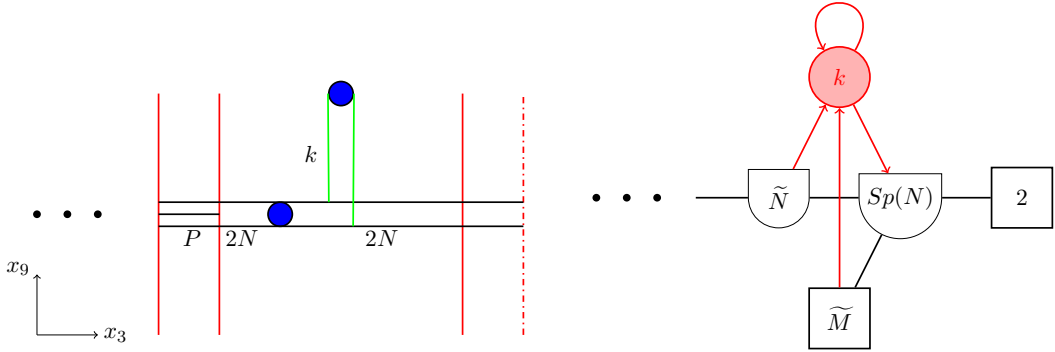

To construct the dual defect and read off the mirror map, we consider the S-dual of the brane configuration presented above in \figref{Fig:HWOrb7}. 
Under S-duality, D5 and NS5-branes are exchanged, D3-branes remain invariant, F1 strings become D1-branes, and the orbifold 5-plane 
becomes an O5$^0$-plane. One now needs to subject this brane configuration to a Hanany-Witten move where the NS5-brane is moved across the 
D5-brane to the immediate right (exactly what we did while S-dualizing the vortex configuration above). 
After the move, we can arrange the $2N$ D3-branes in the rightmost corner to stretch between the NS5-brane and the O5$^0$-plane. 
The final configuration is given in \figref{Fig:HWOrb8}, from which the dual 3d theory and the dual defect can be read off as before.

From \figref{Fig:HWOrb8}, one can see that the dual vortex defect is realized by a stack of $k$ D1-branes (denoted by vertical green lines) stretched between an NS5 and $2N$ D3-branes -- $k_i$ D1-branes end on the $i$-th D3-brane such that $k_1+\ldots +k_N=k$, with $0 \leq k_i \leq k$. The coupled 3d-1d system corresponding
to this vortex defect can be obtained by moving the stack of D1-branes to the left so that they all end on the NS5-brane. Then, using the rules reviewed 
in \Secref{LQ2BD}, the gauge group and the matter content of the coupled SQM can be read off -- the result is shown in \figref{Fig:HWOrb8}. 
The case of a generic representation can be handled in an analogous fashion. One therefore arrives at a mirror map between a Wilson defect 
in the bifurcated quiver and a vortex defect in the quiver containing an $Sp(N)$ gauge group. In the special case of 
$Sp(2)$ gauge theory with $N_f=6$ and its mirror dual, the mirror map reduces to \eref{MM-D6-W2V1}, which was obtained in 
\Secref{1-D-Sp(2)} using sphere partition functions. The coupled quiver in \figref{Fig:HWOrb8} clearly reduces to the coupled quiver 
in \figref{NAbEx2Sp} for the special case of $\wt{M}=2N$ and $\wt{N}=0$.

An interesting feature of the vortex defect realized in \figref{Fig:HWOrb8} is that it admits a single 3d-1d quiver, as opposed 
to two -- the left and the right -- for a linear quiver with unitary gauge groups. This is because there is only a single NS5-brane 
in \figref{Fig:HWOrb8} that the stack of D1-branes can approach without encountering a D5-brane. 
This is something that was found in \Secref{SwD-D4} and 
\Secref{1-D-Sp(2)} using the sphere partition function argument. 
The brane construction gives a more physical and intuitive understanding of this fact.

\section{Beyond Hanany-Witten :  A flavored $\wh{D}_4$ quiver}\label{3d-bad}
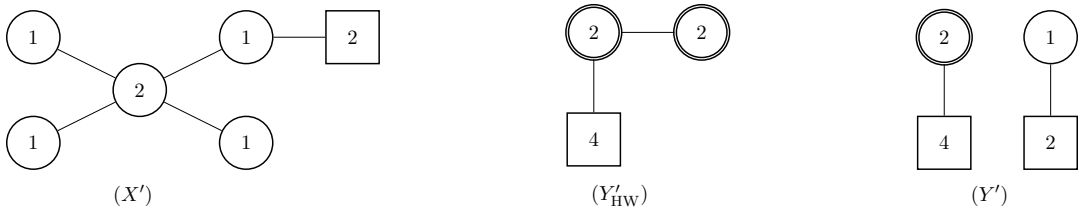
\begin{figure}[htbp]
\begin{center}
\scalebox{0.7}{\begin{tikzpicture}[node distance=2cm, nnode/.style={circle,draw,thick, red, fill=red!30, minimum size=2.0 cm},cnode/.style={circle,draw,thick,minimum size=1.0 cm},snode/.style={rectangle,draw,thick,minimum size=1.0 cm}]
\node[cnode] (1) at (0,1) {1} ;
\node[cnode] (2) at (2,0) {2};
\node[cnode] (3) at (0,-1) {1};
\node[snode] (4) at (6,1) {2};
\node[cnode] (5) at (4, 1) {1};
\node[cnode] (6) at (4, -1) {1};
\draw[-] (1) -- (2);
\draw[-] (2) -- (3);
\draw[-] (5) -- (4);
\draw[-] (2) -- (5);
\draw[-] (2) -- (6);
\node[text width=1cm](9) at (2, -2) {$(X')$};
\end{tikzpicture}}
\qquad \qquad \qquad
 \scalebox{0.7}{\begin{tikzpicture}[
cnode/.style={circle,draw,thick, minimum size=1.0cm},snode/.style={rectangle,draw,thick,minimum size=1cm},pnode/.style={circle,draw,double,thick, minimum size=1.0cm}]
\node[pnode] (9) at (10,0){2};
\node[snode] (10) [below=1cm of 9]{4};
\node[pnode] (11) [right=1cm of 9]{2};
\draw[-] (9) -- (10);
\draw[-] (9) -- (11);
\node[text width=0.1cm](21)[below=0.2 cm of 10]{$(Y'_{\rm HW})$};
\end{tikzpicture}}
\qquad \qquad \qquad
\scalebox{.7}{\begin{tikzpicture}[node distance=2cm,
cnode/.style={circle,draw,thick, minimum size=1.0cm},snode/.style={rectangle,draw,thick,minimum size=1cm}, pnode/.style={circle,draw,double,thick, minimum size=1.0cm}, lnode/.style = {shape = rounded rectangle, minimum size=1.0cm, rotate=90, rounded rectangle right arc = none, draw, double}]
\node[pnode] (1) at (0,0) {2} ;
\node[snode] (2) at (0,-2) {4};
\draw[-] (1) -- (2);
\node[text width=1cm](4) at (1, -3) {$(Y')$};
\node[cnode] (3) at (2,0) {1} ;
\node[snode] (4) at (2,-2) {2};
\draw[-] (3) -- (4);
\end{tikzpicture}}
\caption{\footnotesize{The Hanany-Witten dual of the theory $X'$ and the good dual. The former is a bad theory since the $SU(2)$ node 
without fundamental hypers is a bad node.}}
\label{Ex3dbad}
\end{center}
\end{figure}

In this section, we consider an example of mirror symmetry which is not directly realized by a Type IIB 
construction. Consider the $\wh{D}_4$ quiver gauge theory $X'$ in \figref{Ex3dbad}, which can be constructed 
by a Hanany-Witten configuration of the type discussed in \Secref{D-TypeIIB}. The set-up involves D3-D5-NS5-branes 
and a pair of orbifold 5-planes, one of which contains a pair of fractional D5-branes responsible for engineering 
the two fundamental hypermultiplets in $X'$.
The $S$-dual of the configuration leads to the quiver $Y'_{\rm HW}$ in \figref{Ex3dbad}. This quiver is evidently \textit{bad} 
in the Gaiotto-Witten sense, and therefore does not give the correct mirror dual of $X'$. A careful analysis of the 
IR physics of $Y'_{\rm HW}$, reveals a good quiver gauge theory dual to $X'$ \cite{Dey:2017fqs} -- which we denote as $Y'$ in 
\figref{Ex3dbad}. Alternatively, the theory $Y'$ can also be obtained by implementing an $S$-type operation on 
an appropriate linear quiver $X$ to engineer $X'$, and then reading off the dual from the partition 
function construction of \cite{Dey:2020hfe}. 

Note that the good quiver $Y'$ consists of two decoupled quiver gauge theories -- an $SU(2)$ gauge theory with $N_f=4$ flavors ($Y'_1$) 
and a $T(U(2))$ theory ($Y'_2$). We will study vortex and Wilson defects in the theory $X'$ and determine the mirror maps, 
demonstrating in particular how the dual defect ``factorizes" among the two decoupled theories.

\subsection{Vortex Defects in the $\wh{D}_4$ quiver}

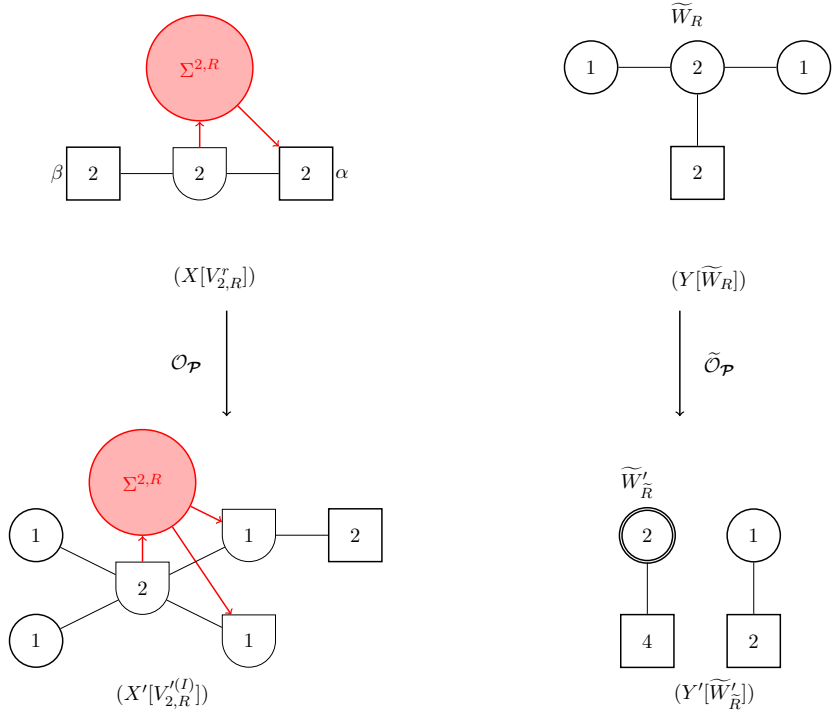
\begin{figure}[htbp]
\begin{center}
\begin{tabular}{ccc}
\scalebox{.7}{\begin{tikzpicture}[node distance=2cm,
cnode/.style={circle,draw,thick, minimum size=1.0cm},snode/.style={rectangle,draw,thick,minimum size=1.0cm}, nnode/.style={red, circle,draw,thick,fill=red!30 ,minimum size=2.0cm}]
\node[snode] (1) at (0,0) {2} ;
\node[cf-group] (2) at (2,0) {\rotatebox{-90}{2}};
\node[snode] (3) at (4,0) {2};
\node[nnode] (4) at (2,2) {$\Sigma^{2,R}$};
\node[text width=1cm](7) at (2, -2) {$(X[V^r_{2,R}])$};
\node[text width=0.1cm](8) at (4.6, 0) {$\alpha$};
\node[text width=0.1cm](9) at (-.75, 0) {$\beta$};
\draw[red, thick, ->] (2)--(4);
\draw[red, thick, ->] (4)--(3);
\draw[-] (1) -- (2);
\draw[-] (2) -- (3);
\end{tikzpicture}}
& \qquad \qquad \qquad
& \scalebox{.7}{\begin{tikzpicture}[node distance=2cm,
cnode/.style={circle,draw,thick, minimum size=1.0cm},snode/.style={rectangle,draw,thick,minimum size=1cm}, pnode/.style={red,rectangle,draw,thick, minimum size=1.0cm}]
\node[cnode] (1) at (0,0) {1} ;
\node[cnode] (2) at (2,0) {2} ;
\node[cnode] (3) at (4,0) {1} ;
\node[snode] (4) at (2,-2) {2};
\draw[-] (1) -- (2);
\draw[-] (2) -- (3);
\draw[-] (2) -- (4);
\node[text width=1cm](5) at (2, 1) {$\wt{W}_{R}$};
\node[text width=1cm](6) at (2, -4) {$(Y[\wt{W}_{R}])$};
\end{tikzpicture}}\\
 \scalebox{.7}{\begin{tikzpicture}
\draw[thick, ->] (15,-3) -- (15,-5);
\node[text width=0.1cm](20) at (14.0, -4) {$\CO_{\vec \CP}$};
\end{tikzpicture}}
&\qquad \qquad \qquad
& \scalebox{.7}{\begin{tikzpicture}
\draw[thick,->] (15,-3) -- (15,-5);
\node[text width=0.1cm](29) at (15.5, -4) {$\wt{\CO}_{\vec \CP}$};
\end{tikzpicture}}\\
\scalebox{.7}{\begin{tikzpicture}[node distance=2cm, nnode/.style={circle,draw,thick, red, fill=red!30, minimum size=2.0 cm},cnode/.style={circle,draw,thick,minimum size=1.0 cm},snode/.style={rectangle,draw,thick,minimum size=1.0 cm}]
\node[cnode] (1) at (0,1) {1} ;
\node[cf-group] (2) at (2,0) {\rotatebox{-90}{2}};
\node[cnode] (3) at (0,-1) {1};
\node[cf-group] (5) at (4, 1) {\rotatebox{-90}{1}};
\node[cf-group] (6) at (4, -1) {\rotatebox{-90}{1}};
\node[nnode] (7) at (2,2) {$\Sigma^{2,R}$};
\node[snode] (8) at (6,1){2};
\draw[red, thick, ->] (2)--(7);
\draw[red, thick, ->] (7)--(5);
\draw[red, thick, ->] (7)--(6);
\draw[-] (1) -- (2);
\draw[-] (2) -- (3);
\draw[-] (2) -- (5);
\draw[-] (2) -- (6);
\draw[-] (5) -- (8);
\node[text width=1cm](9) at (2, -2) {$(X'[V'^{(I)}_{2,R}])$};
\end{tikzpicture}}
&\qquad \qquad
& \scalebox{.7}{\begin{tikzpicture}[node distance=2cm,
cnode/.style={circle,draw,thick, minimum size=1.0cm},snode/.style={rectangle,draw,thick,minimum size=1cm}, pnode/.style={circle,draw,double,thick, minimum size=1.0cm}, lnode/.style = {shape = rounded rectangle, minimum size=1.0cm, rotate=90, rounded rectangle right arc = none, draw, double}]
\node[pnode] (1) at (0,0) {2} ;
\node[snode] (2) at (0,-2) {4};
\draw[-] (1) -- (2);
\node[text width=1cm](3) at (0,1) {$\wt{W}'_{\wt R}$};
\node[text width=1cm](4) at (1, -3) {$(Y'[\wt{W}'_{\wt R}])$};
\node[cnode] (3) at (2,0) {1} ;
\node[snode] (4) at (2,-2) {2};
\draw[-] (3) -- (4);
\end{tikzpicture}}
\end{tabular}
\caption{\footnotesize{Construction of a vortex defect in the central $U(2)$ gauge node of the theory $X'$ and its dual Wilson defect.}}
\label{Ex3dbad-V2W1}
\end{center}
\end{figure}

\begin{figure}[htbp]
\begin{center}
\scalebox{.7}{\begin{tikzpicture}[node distance=2cm, nnode/.style={circle,draw,thick, red, fill=red!30, minimum size=2.0 cm},cnode/.style={circle,draw,thick,minimum size=1.0 cm},snode/.style={rectangle,draw,thick,minimum size=1.0 cm}]
\node[cnode] (1) at (0,1) {1} ;
\node[cf-group] (2) at (2,0) {\rotatebox{-90}{2}};
\node[cnode] (3) at (0,-1) {1};
\node[cf-group] (5) at (4, 1) {\rotatebox{-90}{1}};
\node[cf-group] (6) at (4, -1) {\rotatebox{-90}{1}};
\node[nnode] (7) at (2,2) {$\Sigma^{2,R}$};
\node[snode] (8) at (6,1){2};
\draw[red, thick, ->] (2)--(7);
\draw[red, thick, ->] (7)--(5);
\draw[red, thick, ->] (7)--(6);
\draw[-] (1) -- (2);
\draw[-] (2) -- (3);
\draw[-] (2) -- (5);
\draw[-] (2) -- (6);
\draw[-] (5) -- (8);
\node[text width=1cm](9) at (2, -2) {$(X'[V'^{(I)}_{2,R}])$};
\end{tikzpicture}}
\qquad 
\scalebox{.7}{\begin{tikzpicture}[node distance=2cm, nnode/.style={circle,draw,thick, red, fill=red!30, minimum size=2.0 cm},cnode/.style={circle,draw,thick,minimum size=1.0 cm},snode/.style={rectangle,draw,thick,minimum size=1.0 cm}]
\node[cf-group] (1) at (0,1) {\rotatebox{-90}{1}} ;
\node[cf-group] (2) at (2,0) {\rotatebox{-90}{2}};
\node[cf-group] (3) at (0,-1) {\rotatebox{-90}{1}};
\node[cnode] (5) at (4, 1) {1};
\node[cnode] (6) at (4, -1) {1};
\node[nnode] (7) at (2,2) {$\Sigma^{2,R}$};
\node[snode] (8) at (6,1){2};
\draw[red, thick, ->] (7)--(2);
\draw[red, thick, ->] (1)--(7);
\draw[red, thick, ->] (3)--(7);
\draw[-] (1) -- (2);
\draw[-] (2) -- (3);
\draw[-] (2) -- (5);
\draw[-] (2) -- (6);
\draw[-] (5) -- (8);
\node[text width=1cm](9) at (2, -2) {$(X'[V'^{(II)}_{2,R}])$};
\end{tikzpicture}}
\caption{\footnotesize{Two different realizations of a vortex defect for the central $U(2)$ gauge node in the quiver $X'$.}}
\label{Ex3dbad-V2W1hop}
\end{center}
\end{figure}
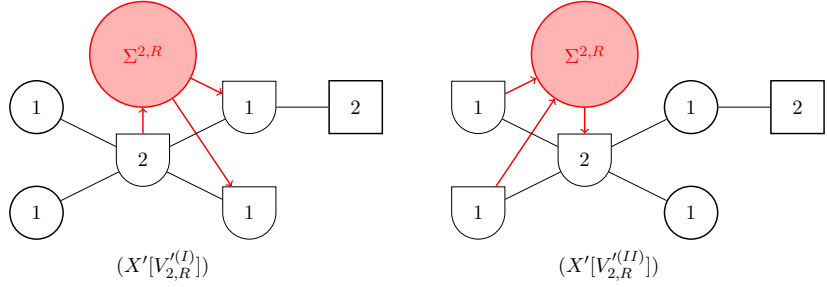

Let us consider vortex defects in the central $U(2)$ gauge node in the $\wh{D}_4$ quiver. 
The starting point is the dual pair of defect quivers -- $X[V^r_{2,R}]$ and $Y[\wt{W}_R]$, 
as shown in the first line of \figref{Ex3dbad-V2W1}. The fundamental masses in theory 
$X$ be labelled as $\{m_i| i=1,\ldots, 4\}$, such that the $U(2)_\beta$  and the $U(2)_\alpha$ flavor nodes  
in $X[V^r_{2,R}]$ are associated with the masses $(m_1,m_2)$ and $(m_3,m_4)$. 
We implement the following Abelian $S$-operation $\CO_{\vec \CP}$ on 
$X[V^r_{2,R}]$:
\be \label{AbS-Dhat2}
\CO_{\vec \CP} (X[V^r_{2,R}]) = G^{\beta_2}_{\CP_4} \circ G ^{\beta_1}_{\CP_3} \circ (G)^{\alpha_2}_{\CP_2} \circ (G\circ F)^{\alpha_1}_{\CP_1} (X[V^r_{2,R}]),
\ee
where $\alpha_1=\alpha$, $\beta_1=\beta$, $\alpha_2$ is the residual $U(1)$ flavor node from $U(2)_\alpha$ in the theory $G^{\alpha_1}_{\CP_1} (X)$, and $\beta_2$ is the residual $U(1)$ flavor node from $U(2)_\beta$. The flavoring operation in $(G\circ F)^{\alpha_1}$ adds two fundamental hypermultiplets. The mass parameters corresponding to the $U(1)^4$ global symmetry to be gauged are :
\be \label{u-choice8}
u_1=m_3, \quad u_2=m_4, \quad u_3=m_1, \quad u_4=m_2.
\ee 
The partition function $Z^{\CO_{\vec \CP}(X[V^r_{2,R}])}$ can be identified as the partition function of a 
coupled 3d-1d quiver $X'[V'^{(I)}_{2,R}]$, shown in \figref{Ex3dbad-V2W1}, where the 3d quiver is the 
$\wh{D}_4$ quiver gauge theory $X'$ and the SQM is $\Sigma^{2,R}$. The Witten index for the SQM is 
computed in the chamber $\vec \xi < 0$.
We redefine the partition function of the new 3d-1d quiver by a global Wilson defect factor, i.e.
\be
Z^{(X'[V'^{(I)}_{2,R}])}:= e^{-2\pi (t_1 + \frac{\eta_3+\eta_4}{2})|R|} \, Z^{\CO_{\vec \CP}(X[V^r_{2,R}])}(v, \vec m_{F}; \vec t, \vec \eta),
\ee
where $\vec{m}_F=(m_{1\,F}, m_{2\,F})$ are masses of the fundamental hypers in $X'$ for the $U(1)_1$ gauge node. 
The dual 3d-1d coupled system is then given by the quiver $Y'[\wt{W}'_{\wt R}]$ in \figref{Ex3dbad-V2W1}, where 
$Y'$ consists of two decoupled sectors as shown. The defect $\wt{W}'_{\wt R}$ is a gauge Wilson defect in a 
representation $\wt{R}$ for the gauge group $SU(2)$, where $\wt{R}$ is given by restricting the representation 
$R$ to a subgroup $SU(2) \subset U(2)$.
A second coupled 3d-1d quiver which realizes the same vortex defect 
can be obtained by implementing the $S$-type operation \eref{AbS-Dhat2} on the quiver  
$(X[V^{l}_{2,R}])$ -- the resultant quiver $(X'[V'^{(II)}_{2,R}])$ is shown in \figref{Ex3dbad-V2W1hop}. 
The Witten index for the SQM should be computed in the chamber $\vec \xi >0$.
Computing the dual partition function, one can again show that the dual defect is given 
by $Y'[\wt{W}'_{R}]$. This leads to the following mirror map:
\be \label{MM-D4hat-V2W}
\boxed{\langle V'^{(II)}_{2,R} \rangle_{X'}(\vec m_F; \vec t, \vec \eta)=\langle V'^{(I)}_{2,R} \rangle_{X'}(\vec m_F; \vec t, \vec \eta) = \langle \wt{W}'_{\wt R} \rangle_{Y'_1} \cdot \langle \BU \rangle_{Y'_2},}
\ee
where $\BU$ denotes a trivial Wilson defect in the theory $Y'_2$. 

\subsection{Wilson Defects in the $\wh{D}_4$ quiver}

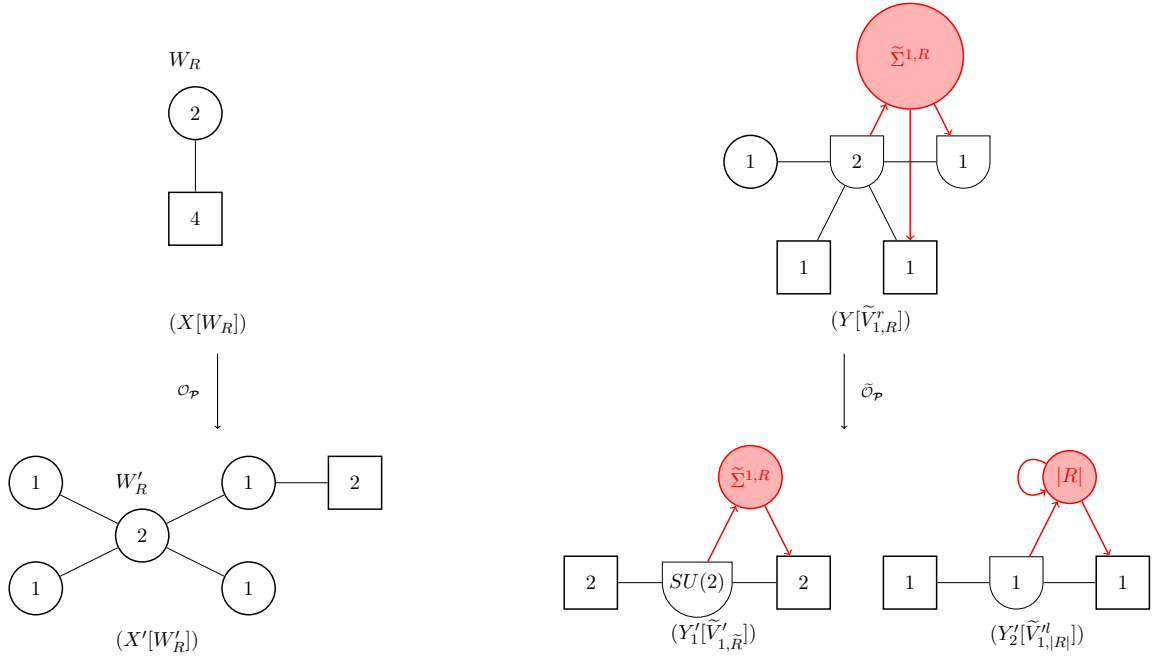
\begin{figure}[htbp]
\begin{center}
\begin{tabular}{ccc}
\scalebox{.7}{\begin{tikzpicture}[node distance=2cm,
cnode/.style={circle,draw,thick, minimum size=1.0cm},snode/.style={rectangle,draw,thick,minimum size=1cm}, pnode/.style={red,rectangle,draw,thick, minimum size=1.0cm}]
\node[cnode] (1) at (0,0) {2} ;
\node[snode] (2) at (0, -2) {4};
\draw[-] (1) -- (2);
\node[text width=1cm](5) at (0, 1) {$W_{R}$};
\node[text width=1cm](6) at (0, -4) {$(X[W_{R}])$};
\end{tikzpicture}}
& \qquad \qquad \qquad
& \scalebox{.7}{\begin{tikzpicture}[node distance=2cm,
cnode/.style={circle,draw,thick, minimum size=1.0cm},snode/.style={rectangle,draw,thick,minimum size=1cm}, nnode/.style={red, circle,draw,thick,fill=red!30 ,minimum size=2.0cm}]
\node[cnode] (1) at (0,0) {1} ;
\node[cf-group] (2) at (2,0) {\rotatebox{-90}{2}};
\node[cf-group] (3) at (4,0) {\rotatebox{-90}{1}};
\node[nnode] (4) at (3,2) {$\wt{\Sigma}^{1,R}$};
\node[snode] (5) at (3,-2) {1};
\node[snode] (6) at (1,-2) {1};
\node[text width=1cm](7) at (2, -3) {$(Y[\wt{V}^r_{1,R}])$};
\draw[red, thick, ->] (2)--(4);
\draw[red, thick, ->] (4)--(3);
\draw[red, thick, ->] (4)--(5);
\draw[-] (1) -- (2);
\draw[-] (2) -- (3);
\draw[-] (2) -- (5);
\draw[-] (2) -- (6);
\end{tikzpicture}}\\
\scalebox{.5}{\begin{tikzpicture}
\draw[thick, ->] (15,-3) -- (15,-5);
\node[text width=0.1cm](20) at (14.0, -4) {$\CO_{\vec \CP}$};
\end{tikzpicture}}
&\qquad \qquad \qquad
& \scalebox{.5}{\begin{tikzpicture}
\draw[thick,->] (15,-3) -- (15,-5);
\node[text width=0.1cm](29) at (15.5, -4) {$\wt{\CO}_{\vec \CP}$};
\end{tikzpicture}}\\
\scalebox{0.7}{\begin{tikzpicture}[node distance=2cm, nnode/.style={circle,draw,thick, red, fill=red!30, minimum size=2.0 cm},cnode/.style={circle,draw,thick,minimum size=1.0 cm},snode/.style={rectangle,draw,thick,minimum size=1.0 cm}]
\node[cnode] (1) at (0,1) {1} ;
\node[cnode] (2) at (2,0) {2};
\node[cnode] (3) at (0,-1) {1};
\node[snode] (4) at (6,1) {2};
\node[cnode] (5) at (4, 1) {1};
\node[cnode] (6) at (4, -1) {1};
\draw[-] (1) -- (2);
\draw[-] (2) -- (3);
\draw[-] (5) -- (4);
\draw[-] (2) -- (5);
\draw[-] (2) -- (6);
\node[text width=1cm](9) at (2, -2) {$(X'[W'_{R}])$};
\node[text width=1cm](15) at (2, 1) {$W'_{R}$};
\end{tikzpicture}}
& \qquad \qquad \qquad
& \scalebox{.7}{\begin{tikzpicture}[node distance=2cm,
cnode/.style={circle,draw,thick, minimum size=1.0cm},snode/.style={rectangle,draw,thick,minimum size=1cm}, pnode/.style={circle,draw,double,thick, minimum size=1.0cm}, nnode/.style={circle,draw,thick, red, fill=red!30, minimum size=1.0 cm},lnode/.style = {shape = rounded rectangle, minimum size=1.0cm, rotate=90, rounded rectangle right arc = none, draw, double}]
\node[cf-group] (1) at (0,0) {\rotatebox{-90}{$SU(2)$}} ;
\node[snode] (3) at (2,0) {2};
\node[nnode] (4) at (1,2) {$\wt{\Sigma}^{1,R}$};
\node[snode] (5) at (-2,0) {2};
\draw[-] (1) -- (2);
\draw[-] (5) -- (1);
\draw[red, thick, ->] (4)--(3);
\draw[red, thick, ->] (1)--(4);
\node[text width=1cm](6) at (0, -1) {$(Y'_1[\wt{V}'_{1, \wt{R}}])$};
\node[text width=1cm](10) at (6, -1) {$(Y'_2[\wt{V}'^l_{1, |R|}])$};
\node[snode] (7) at (4,0) {1} ;
\node[cf-group] (8) at (6,0) {\rotatebox{-90}{1}};
\node[snode] (9) at (8,0) {1};
\node[nnode] (10) at (7,2) {$|R|$};
\draw[red, thick, ->] (8)--(10);
\draw[red, thick, ->] (10)--(9);
\draw[-] (8) -- (7);
\draw[-] (8) -- (9);
\draw[red, thick, ->] (10) to [out=150,in=210,looseness=4] (10);
\end{tikzpicture}}
\end{tabular}
\caption{\footnotesize{Wilson defect in the central gauge node of the $\wh{D}_4$ quiver $X'$, and the dual vortex defect $Y'$.}}
\label{Ex3dbad-W2V1}
\end{center}
\end{figure}

Let us consider Wilson defects for the central $U(2)$ gauge node of the $\wh{D}_4$ quiver $X'$. 
The starting point is the dual pair of defects -- the Wilson defect $W_R$ in the theory $X$ and the 
vortex defect $\wt{V}_{1,R}$ in the theory $Y$, where the latter is realized by as a deformation of two 3d-1d 
quivers $\wt{V}^r_{1,R}$ and $\wt{V}^l_{1,R}$. The pair $({W}_R, \wt{V}^r_{1,R})$ is shown in the first line 
of \figref{Ex3dbad-W2V1}. 
We implement the Abelian $S$-type operation $\CO_{\vec \CP}$ on the system $X[{W}_R]$, where $\CO_{\vec \CP}$ is 
given as:
\begin{align}\label{AbS-Dhatb}
\CO_{\vec \CP} (X[{W}_R]) = G^{\alpha_4}_{\CP_4}\circ G^{\alpha_3}_{\CP_3} \circ G^{\alpha_2}_{\CP_2} \circ (G \circ F)^{\alpha_1}_{\CP_1} (X[{W}_R]),
\end{align}
where $\alpha_1=\alpha$, and the flavoring operation in $(G\circ F)^{\alpha_1}$ adds two fundamental hypermultiplets. 
The mass parameters corresponding to the following choice of the $U(1)^4$ global symmetry to be gauged:
\be \label{u-choice6b}
u_1=m_3, \quad u_2=m_4, \quad u_3=m_1, \quad u_4=m_2.
\ee

This $S$-operation on the quiver $X[W_{R}]$ leads to the defect quiver $X'[W'_{R}]$ in \figref{Ex3dbad-W2V1}. 
The dual vortex defect can be read off from the dual partition function and, after manipulations similar to previous 
examples, reduce to the following form:
\begin{align}
& Z^{(Y'[(W'_{R})^\vee])}= C\cdot Z^{(Y'_1[\wt{V}'_{1, \wt{R}}])}\cdot Z^{(Y'_2[\wt{V}'^r_{1, |R|}])}, \label{ZAbS5-D4hat}\\
& Z^{(Y'_1[\wt{V}'_{1, \wt{R}}])}= \lim_{z\to 1}\int\Big[d\vec \s\Big] \, \delta(\tr \vec\s)
Z^{U(2), N_f=4}_{\rm int} (\vec\s, \vec m'^{(1)}(\vec t, \vec\eta), \eta')\, \CI^{\wt{\Sigma}^{1,R}_r}(\vec \s, \vec m'^{(1)}, z| \vec \xi <0), \\
& Z^{(Y'_2[\wt{V}'^r_{1, |R|}])} = e^{2\pi i m_{2\,F}|R|} \cdot \lim_{z\to 1}\,\int d\tau\, Z^{T(U(2))}_{\rm int}(\tau, \vec m'^{(2)}(\vec t, \vec\eta),\vec m_{F})\, 
\CI^{\wt{\Sigma}^{1,|R|}_r}(\tau, \vec m'^{(2)}, z| \vec \xi <0),
\end{align}
where $C= C(\vec m_F, \vec \eta, \vec t)$ is a contact term and $\vec m'^{(1)}, \vec m'^{(2)}$ are mass parameters of the quivers $Y'_1$ and $Y'_2$ respectively. The FI parameters of $Y'_2$ are given by $\vec m_{F}= (m_{1F}, m_{2F})$, which are the fundamental masses in the $\wh{D}_4$ quiver 
gauge theory. Finally, the Witten indices are given as:
\begin{align}
& \CI^{\wt{\Sigma}^{1,R}_r}(\vec \s, \vec m'^{(1)}, z| \vec \xi <0)= \sum_{w \in R}\, \prod^2_{j=1} \prod^2_{i=1} \frac{\ch{(\s_j - m'^{(1)}_{i})}}{\ch{(\s_j + i w_j z - m'^{(1)}_{i})}},\\
& \CI^{\wt{\Sigma}^{1,|R|}_r}(\tau, \vec m'^{(2)}, z| \vec \xi <0)= \frac{\ch{(\tau - m'^{(2)}_1)}}{\ch{(\tau + i |R|z- m'^{(2)}_1)}}.
\end{align}
The RHS of \eref{ZAbS5-D4hat} shows that the dual defect factorizes into a vortex defect for the $SU(2)$ gauge 
group\footnote{Up to a global vortex defect.} in the 3d quiver $Y'_1$ (denoted as $Y'_1[\wt{V}'_{1, \wt{R}}]$), and a vortex defect the $U(1)$ gauge group of the 3d 
quiver $Y'_2$ (denoted as $Y'_2[\wt{V}'^r_{1, |R|}]$). The 3d-1d quiver is shown in the bottom right of \figref{Ex3dbad-W2V1}.  
We therefore have the following mirror map:
\be
\boxed{\langle W'_R \rangle_{X'} (\vec m_F; \vec t, \vec \eta) = \langle  \wt{V}'_{1, \wt{R}} \rangle_{Y'_1} \cdot \langle \wt{V}'^r_{1, |R|} \rangle_{Y'_2}.}
\ee
The dual of Wilson defects for the other gauge and flavor nodes can be similarly determined.

\section*{Acknowledgement}
The author would like to thank Ibrahima Bah for comments, and Vivek Saxena for collaboration on related issues.
The author is partially supported at the Johns Hopkins University by NSF grant PHY-1820784.

\appendix

\section{Line defects in a $U(2)$ gauge theory with $N_f=4$ flavors}\label{app:U(2)}

\subsection{Mirror symmetry and partition functions}\label{U24-PF}

The partition function of the quiver $X$ is
\begin{equation}
\begin{split}
& Z^{(X)}(\vec{m}; \vec{t})=\int \frac{d^2s}{2!} \, \frac{e^{2\pi i \tr{\vec{s}} ({t}_1- {t}_2)}  \sinh^2{\pi(s_1-s_2)}}{\prod^2_{i=1}\prod^4_{a=1} \cosh{\pi(s_i - {m}_a)}},
\end{split}
\label{eq:ZBex}
\end{equation}
where ${m}_1,{m}_2, {m}_3, {m}_4$ are the hypermultiplet masses, while the FI parameter of the gauge group is $\eta= {t}_1-{t}_2$.
Using Cauchy determinant identity:
\be \label{DI-1}
\frac{\sh{(x_1-x_2)}\,\sh{(y_1- y_2)}}{\prod_{i,j} \ch{(x_i -y_j)}} = \sum_{\rho \in \CS_2} (-1)^\rho \frac{1}{\prod_i \ch{(x_i - y_{\rho(i)})}},
\ee
the above integral can be decomposed into a sum of terms each involving Abelian integrals. Evaluating the Abelian integrals 
explicitly, we have the expression:
\begin{align}\label{PF-U24-1}
Z^{(X)}(\vec{m}; \vec{t})= & -\Big[ \frac{(e^{2\pi i m_1 \eta} - e^{2\pi i m_3 \eta})\,(e^{2\pi i m_2 \eta} - e^{2\pi i m_4 \eta})}{\sh{(m_1 -m_3)}\,\sh{(m_2-m_4)}}
 - \frac{(e^{2\pi i m_1 \eta} - e^{2\pi i m_4 \eta})\,(e^{2\pi i m_2 \eta} - e^{2\pi i m_3 \eta})}{\sh{(m_1 -m_4)}\,\sh{(m_2-m_3)}} \Big] \nn \\
 & \times \frac{1}{\sh{(m_1-m_2)}\, \sh{(m_3-m_4)}\, \sinh^2{\pi \eta}}.
\end{align}
\\

\noindent Similarly, the partition function of the quiver $Y$ is
\begin{equation}
\begin{split}
& Z^{(Y)}(\vec{t}; \vec{m})=\int d \s^{1}\,\Big[ d\vec{\s}^2 \Big]\,d \s^{3} \, \frac{e^{2\pi i \s^{1} (m_1- m_2)} e^{2\pi i \tr \vec{\s}^2 (m_2 -m_3)} e^{2\pi i \s^{3} (m_3- m_4)} \sinh^2{\pi(\s^2_1-\s^2_2)}}{\prod^2_{i=1}\cosh{\pi(\s^1-\s^2_i)}\prod^2_{a=1} \cosh{\pi(\s^2_i - t_a)}\cosh{\pi(\s^3 -\s^2_i)}},
\end{split}
\label{eq:ZAex}
\end{equation}
where $t_1$ and $t_2$ are the masses of the fundamental hypermultiplets in the middle node, and the FI parameters of the three gauge nodes are $\eta_1=m_1-m_2, \eta_2=m_2-m_3, \eta_3=m_3- m_4$. Performing the $\s_1$ and $\s_3$ integrals, the matrix integral can be 
reduced to a sum over terms each involving Abelian integrals, and after evaluating the latter, we have
\begin{align}
Z^{(Y)}(\vec{t}; \vec{m})=& \Big[ \frac{(e^{2\pi i t_1(m_1 -m_4)} - e^{2\pi i t_2(m_1-m_4)})\,(e^{2\pi i t_1 (m_2-m_3)} - e^{2\pi i t_2 (m_2-m_3)})}{\sh{(m_1 -m_4)}\,\sh{(m_2-m_3)}} 
\nn \\
& - \frac{(e^{2\pi i t_1 (m_1-m_3)} - e^{2\pi i t_2 (m_1-m_3)})\,(e^{2\pi i t_1 (m_2-m_4)} - e^{2\pi i t_2 (m_2-m_4)})}{\sh{(m_1 -m_3)}\,\sh{(m_2-m_4)}} \Big] \nn \\
 & \times \frac{1}{\sh{(m_1-m_2)}\, \sh{(m_3-m_4)}\, \sinh^2{\pi (t_1 -t_2)}}.
\end{align}
 \\
 
\noindent Using the explicit expressions for the two matrix integrals, one can check that
\begin{align}
& Z^{(X)}[\vec m ; \vec t] =C_{XY}(\vec m, \vec t) \, Z^{(Y)}[\vec t; - \vec m] = C_{XY}(\vec m, \vec t) \, Z^{(Y)}[-\vec t; \vec m], \label{MS-XY1a}\\
& C_{XY}(\vec m, \vec t)= e^{2\pi i t_1(m_1+ m_2)} e^{-2\pi i t_2(m_3+m_4)}.\label{MS-XY1b}
\end{align}
The expressions agree exactly (i.e. the contact term vanishes) when one imposes the constraints $t_1+t_2=0, \; m_1+m_2+m_3+m_4=0$.

\subsection{The vortex defects $V_{M, R}$ and the dual defects}\label{U24-V2W}
Using the ``right" SQM realization, the defect partition function is given as:
\begin{align}
Z^{(X[V^r_{M,R}])}(\vec m; \vec t) = & W_{\rm b.g.}(\vec t, R) \cdot \lim_{z\to 1} \int  \Big[d\vec s\Big] \, Z^{(X)}_{\rm int}(\vec s, \vec m, \vec t) \cdot \CI^{\Sigma^{M,R}_r} (\vec s, \vec m, z|\vec \xi <0), \\
=: & W_{\rm b.g.}(\vec t, R) \cdot \lim_{z\to 1} \int  \Big[d\vec s\Big] \, Z^{(X[V^r_{M,R}])}_{\rm int}(\vec s, \vec m, \vec t, z|\vec \xi <0),
\end{align} 
where the constituent functions are given as:
\begin{align}
& W_{\rm b.g.}(\vec t, R)= e^{2\pi t_2 |R|},\label{U24-V2W-main1}\\
& Z^{(X)}_{\rm int}(\vec s, \vec m, \vec t)= \frac{e^{2\pi i \tr{\vec{s}} ({t}_1- {t}_2)}  \sinh^2{\pi(s_1-s_2)}}{\prod^2_{j=1}\prod^4_{i=1} \cosh{\pi(s_j - {m}_i)}}, \label{U24-V2W-main2}\\
& \CI^{\Sigma^{M,R}_r}(\vec s, \vec m, z|\vec \xi <0)= \sum_{w \in R} \CF(s_j, z) \prod^2_{j=1} \prod^M_{i=(5-M)} \frac{\ch{(s_j - m_{i})}}{\ch{(s_j + i w_j z - m_{i})}},
\label{U24-V2W-main3}
\end{align}
and the explicit form of the Witten index $\CI^{\Sigma^{M,R}_r}$ depends on the value of $M$, with $0 \leq M \leq 4$. The function $\CF(s_j, z)$ is given as 
\be
\CF(s_j, z) = \prod_{\alpha, \kappa}\,\prod_{i \neq j}\, \frac{\sinh{(-\pi(s_i -s_j +\alpha\,z \pm z +\kappa/2))}}{\sh{(s_i -s_j+\alpha\,z +\kappa/2)}},
\ee
with $\alpha, \kappa$ taking real values which depend on the weights $w$. The poles of the function $\CF(s_j, z)$ have 
vanishing residues in the limit $z \to 1$, and therefore the limit can be taken trivially, which gives an overall sign. 
The defect partition function can then be written as
\begin{align}
& Z^{(X[V^r_{M,R}])}(\vec m; \vec t) =  W_{\rm b.g.}(\vec t, R) \cdot \sum_{w \in R} Z^{(X[V^r_{M,R}])}(\vec m; \vec t| w) \\
& Z^{(X[V^r_{M,R}])}(\vec m; \vec t| w)=  \lim_{z\to 1} \int  \Big[d\vec s\Big] \,\frac{e^{2\pi i \tr{\vec{s}} ({t}_1- {t}_2)}  \sinh^2{\pi(s_1-s_2)}}{\prod^2_{j=1}\prod^{4-M}_{i=1} \cosh{\pi(s_j - {m}_i)}\,\prod^{4}_{i=(5-M)} \ch{(s_j + i w_j z -m_i)}}. \label{ZVw}
\end{align} 
We now list the dual Wilson defects for different values of $M$ as follows.\\

\noindent \textbf{The case of $M=2$:} Using the Cauchy determinant \eref{DI-1}, we obtain the relation:
\be
Z^{(X[V^r_{2,R}])}(\vec m; \vec t | w) = \frac{1}{2}\,\sum_{\rho \in S_2} \,Z^{(X)}(m_1,m_2, m_3-iw_{\rho(1)}, m_4 - iw_{\rho(2)}; \vec t).
\ee
Using the fact that $S_2$ is the Weyl group of the $U(2)$ gauge group, we obtain
\begin{align}
Z^{(X[V^r_{2,R}])}(\vec m; \vec t) = & W_{\rm b.g.}(\vec t, R) \cdot \sum_{w} \,\frac{1}{2}\, \sum_{\rho \in S_2} \, Z^{(X)}(m_1,m_2, m_3-iw_{\rho(1)}, m_4 - iw_{\rho(2)}; \vec t) \nn \\
= &  W_{\rm b.g.}(\vec t, R) \cdot \sum_{w} \, Z^{(X)}(m_1,m_2, m_3-iw_{1}, m_4 - iw_{2}; \vec t). \label{XV2r1}
\end{align}

Using the mirror symmetry relation \eref{MS-XY1a}-\eref{MS-XY1b}, the above equation can be written as
\begin{align}
Z^{(X[V^r_{2,R}])}(\vec m; \vec t) = & W_{\rm b.g.}(\vec t, R) \cdot \sum_w \, Z^{(X)}(m_1,m_2, m_3-iw_{1}, m_4 - iw_{2}; \vec t) \nn \\
= & C_{XY}(\vec m, \vec t)\, \sum_w \, Z^{(Y)}(\vec t; -m_1, -m_2, -(m_3- iw_1), -(m_4 - iw_2)). \label{XV2r}
\end{align}

We now show that this vortex defect is mirror dual to the Wilson defect $\wt{W}_R$, associated with the central $U(2)$ gauge node of the quiver $Y$ in the same representation $R$. The latter defect has the following partition function:
\begin{align}
Z^{(Y[\wt{W}_{R}])}(\vec t; \vec m) =&  \int  d\s^1\,\Big[d\vec \s^2\Big] \, d\s^3\, Z^{(Y)}_{\rm int}(\{\vec \s^\gamma\}, \vec t, \vec m) 
\cdot Z_{\rm Wilson}(\vec \s^2, R), \\
= & \sum_{w \in R} \, Z^{(Y[\wt{W}_{R}])}(\vec t; \vec m | w),
\end{align} 
where the constituent functions are given as
\begin{align}
& Z^{(Y)}_{\rm int}(\{\vec \s^\gamma\}, \vec t, \vec m)=\frac{e^{2\pi i \s^{1} (m_1- m_2)} e^{2\pi i \tr \vec{\s}^2 (m_2 -m_3)} e^{2\pi i \s^{3} (m_3- m_4)} \sinh^2{\pi(\s^2_1-\s^2_2)}}{\prod^2_{i=1}\cosh{\pi(\s^1-\s^2_i)}\prod^2_{a=1} \cosh{\pi(\s^2_i - t_a)}\cosh{\pi(\s^3 -\s^2_i)}} ,\\
& Z_{\rm Wilson}(\vec \s^2, R)= \sum_{w \in R} \, \Big(e^{2\pi \sum_j w_j \s^2_j} \Big).
\end{align}
Integrating over $\s_3$ and manipulating the resultant expression, one can show that
\be
Z^{(Y[\wt{W}_{R}])}(\vec t; \vec m | w)= \frac{1}{2}\,\sum_{\rho' \in S_2}\, Z^{(Y)}(\vec t; m_1,m_2, m_3 + iw_{\rho(1)}, m_4 + iw_{\rho(2)}).
\ee
Again, recognizing that $S_2$ is the Weyl group of the $U(2)$ gauge group, we obtain
\begin{align} \label{YW}
Z^{(Y[\wt{W}_{R}])}(\vec t; \vec m) = \sum_{w \in R}\, Z^{(Y)}(\vec t; m_1,m_2, m_3 + iw_{1}, m_4 + iw_{2}). 
\end{align}
\\
Using \eref{YW} in the equation \eref{XV2r}, we obtain the following relation between the two defect partition functions, i.e.
\begin{align}
Z^{(X[V^r_{2,R}])}(\vec m; \vec t) = C_{XY}(\vec m, \vec t)\, Z^{(Y[\wt{W}_{R}])}(\vec t; -\vec m).
\end{align}
Normalizing by the respective partition functions, the expectation values of the two defects are related as:
\be
\boxed{ \langle V^r_{2,R} \rangle_X (\vec m; \vec t) = \langle \wt{W}_{R} \rangle_Y (\vec t; -\vec m).}
\ee
\\

\noindent \textbf{The case of $M=1,3$:} Consider the $M=1$ case first. The expression on the RHS of \eref{ZVw} can 
rewritten using the identity,
\begin{align}
\frac{\sh{(s_1-s_2 + i(w_1-w_2)z)}}{\prod^2_{i=1} \ch{(s_i + i w_i z -m_4)}} = - \sum_{\rho \in S_2}(-1)^\rho\,
\Big(\frac{e^{-\pi (s_{\rho(1)} + iw_{\rho(1)} -m_4) }}{\ch{(s_{\rho(1)} + iw_{\rho(1)} -m_4)}} \Big).
\end{align}
The resultant expression can be massaged into the following form:
\be
Z^{(X[V^r_{1,R}])}(\vec m; \vec t | w) = \frac{1}{2}\, (-1)^{w_1+w_2}\,\sum_{\rho \in S_2} \,Z^{(X)}(m_1,m_2, m_3, m_4 - iw_{\rho(1)}; \vec t).
\ee
Summing over all $w \in R$, and using the fact that $S_2$ is the Weyl group for the $U(2)$ gauge group, we get
\begin{align}
Z^{(X[V^r_{1,R}])}(\vec m; \vec t) = \sum_{w \in R} W_{\rm b.g.}(\vec t, R)\,(-1)^{w_1+w_2}\, Z^{(X)}(m_1,m_2, m_3, m_4 - iw_{1}; \vec t).
\end{align}
Using the mirror symmetry relation \eref{MS-XY1a}-\eref{MS-XY1b}, the above equation can be rewritten as
\begin{align}\label{XV1r}
Z^{(X[V^r_{1,R}])}(\vec m; \vec t) = C_{XY}(\vec m, \vec t)\, (-1)^{|R|} \,\sum_w \,e^{2\pi t_2 w_2} \, Z^{(Y)}(\vec t; -m_1, -m_2, -m_3, -(m_4 - iw_1)).
\end{align}
Mirror symmetry maps the vortex defect $V^r_{1,R}$ to a Wilson defect $\wt{W}_{\wt{R}}$ in the theory $Y$: 
\be \label{WY1}
\wt{W}_{\wt{R}} = \sum_{\kappa \in \Delta} \wt{W}^{\rm flavor} _{2, q^\kappa_2}\, \wt{W}^{(3)}_{q^\kappa_1},
\ee
where $\wt{W}^{\rm flavor} _{2, q^\kappa_2}$ is a flavor Wilson defect of charge $q^\kappa_2$ under a $U(1)_{t_2}$ subgroup of the $U(2)_f$ flavor symmetry of $Y$, embedded as $U(1)_{t_1} \times U(1)_{t_2} \subset U(2)_f$, and $\wt{W}^{(3)}_{q^\kappa_1}$ is a gauge Wilson defect 
of charge $q^\kappa_1$ for the rightmost $U(1)$ gauge node in the quiver $Y$. The charges $(q^\kappa_1, q^\kappa_2)$ are obtained 
from the decomposition of the representation $R$ under the subgroup $U(1) \times U(1) \subset U(2)$, and $\Delta$ is the set of such 
charge doublets counted with degeneracies. This mirror map can be directly read off from \eref{XV1r} as follows. Firstly, note that
\be
Z^{(Y)}(\vec t; -m_1, -m_2, -m_3, -(m_4 - iw_1)) = Z^{(Y[\wt{W}^{(3)}_{w_1}])}(\vec t; \vec m),
\ee
while the first factor in the summand on the RHS of \eref{XV1r} can be identified as the contribution of the flavor Wilson defect $\wt{W}^{\rm flavor} _{2, q^\kappa_2}$. Putting them together, we get
\begin{align}
Z^{(X[V^r_{1,R}])}(\vec m; \vec t) = & C_{XY}(\vec m, \vec t)\, \sum_{w \in R} \,Z_{\rm Wilson}(t_2, w_2) \, Z^{(Y[\wt{W}^{(3)}_{w_1}])}(\vec t; -\vec m) \\
= & C_{XY}(\vec m, \vec t)\,  \sum_{\kappa \in \Delta} \,Z_{\rm Wilson}(t_2, q^\kappa_2)\,Z^{(Y[\wt{W}^{(3)}_{q^\kappa_1}])}(\vec t; -\vec m) .
\end{align}
Normalizing by the respective partition functions, the expectation values of the two defects are related as:
\be
\boxed{ \langle V^r_{1,R} \rangle_X (\vec m; \vec t) = \langle  \sum_{\kappa \in \Delta} \wt{W}^{\rm flavor} _{2, q^\kappa_2}\, \wt{W}^{(3)}_{q^\kappa_1} \rangle_Y (\vec t; -\vec m).}
\ee
\\
The case of the vortex defect $V^r_{M,R}$ for $M=3$ can be handled in a similar fashion. Note that the RHS of \eref{ZVw} can 
be rewritten, after a change of variables $s_j \to s_j - iw_j z$, in the following fashion:
\begin{align}
Z^{(X[V^r_{3,R}])}(\vec m; \vec t | w) =  \frac{1}{2}\, (-1)^{w_1+w_2}\,e^{2\pi (w_1+w_2)(t_1 -t_2)}\, \sum_{\rho \in S_2} \, Z^{(X)}(m_1+ iw_{\rho(1)},m_2, m_3, m_4; \vec t).
\end{align}
Summing over the weights $w \in R$, as before, yields
\begin{align}
Z^{(X[V^r_{3,R}])}(\vec m; \vec t) = \sum_{w \in R} \,(-1)^{w_1+w_2}\, e^{2\pi t_1(w_1+w_2)}\,Z^{(X)}(m_1+ iw_1,m_2, m_3, m_4; \vec t).
\end{align}
Proceeding in the same fashion as before, we obtain 
\begin{align}
Z^{(X[V^r_{3,R}])}(\vec m; \vec t) = & C_{XY}(\vec m, \vec t)\, \sum_{w \in R} \,Z^{\rm flavor}_{\rm Wilson}(t_1, w_2) \, Z^{(Y[\wt{W}^{(1)}_{w_1}])}(\vec t; -\vec m) \\
= & C_{XY}(\vec m, \vec t)\,  \sum_{\kappa \in \Delta} \,Z^{\rm flavor}_{\rm Wilson}(t_1, q^\kappa_2)\,Z^{(Y[\wt{W}^{(1)}_{q^\kappa_1}])}(\vec t; -\vec m),
\end{align}
where $Z^{\rm flavor}_{\rm Wilson}(t_1, w_2)$ is the contribution of a flavor Wilson defect and $\wt{W}^{(1)}_{w_1}$ is a Wilson defect 
for the leftmost $U(1)$ gauge node in quiver $Y$. The charges $(q^\kappa_1, q^\kappa_2)$ and the set $\Delta$ are defined as in the 
$M=1$ case. This leads to the mirror symmetry map
\be
\boxed{ \langle V^r_{3,R} \rangle_X (\vec m; \vec t) = \langle  \sum_{\kappa \in \Delta} \wt{W}^{\rm flavor} _{1,q^\kappa_2}\, \wt{W}^{(1)}_{q^\kappa_1} \rangle_Y (\vec t; -\vec m).}
\ee
\\

\noindent \textbf{The case of $M=4,0$:} Similar to the $M=3$ case, we implement a change of variables -- $s_j \to s_j - iw_j z$ -- on the RHS of 
\eref{ZVw}, leading to the expression
\begin{align}
Z^{(X[V^r_{4,R}])}(\vec m; \vec t) =  & W_{\rm b.g.}(\vec t, R) \cdot \sum_{w \in R} (-1)^{w_1+w_2}\, e^{2\pi (w_1+w_2)(t_1-t_2)}\,Z^{(X)}(\vec m; \vec t)\\
= & \Big(\sum_{w \in R} (-1)^{w_1+w_2}\, e^{2\pi t_1 (w_1+w_2)}\,\Big)\,Z^{(X)}(\vec m; \vec t).
\end{align}
Following the procedure as above, this leads to the mirror map:
\be
\boxed{ \langle V^r_{4,R} \rangle_X (\vec m; \vec t) = \langle \wt{W}^{\rm flavor} _{1, |R|} \rangle_Y (\vec t; -\vec m),}
\ee
where $\wt{W}^{\rm flavor} _{|R|}$ is a flavor Wilson defect of charge $|R|$ in a $U(1)_{t_1}$ subgroup of the $U(2)_f$ flavor symmetry of the quiver $Y$,
embedded as $U(1)_{t_1} \times U(1)_{t_2} \subset U(2)_f$. \\
The $M=0$ defect also maps to a flavor Wilson defect in $Y$, which can be directly read off from \eref{U24-V2W-main1}-\eref{U24-V2W-main3}.
In this case, one can directly take the $z \to 1$ limit, since the residue of the $z$-dependent pole vanishes. The resultant mirror map is
\be
\boxed{ \langle V^r_{0,R} \rangle_X (\vec m; \vec t) = \langle \wt{W}^{\rm flavor} _{2, |R|} \rangle_Y (\vec t; -\vec m).}
\ee

\subsection{The Wilson defect $W_{R}$ and the dual defects}\label{U24-W2V}
Consider the Wilson defect $W_{R}$ for the $U(2)$ gauge node in the theory $X$ in the representation $R$. The defect partition function 
has the following form:
\begin{align}
Z^{(X[W_{R}])}(\vec m; \vec t) = &\int  \Big[d\vec s\Big] \, Z^{(X)}_{\rm int}(\vec s, \vec m, \vec t) \cdot Z_{\rm Wilson}(\vec s, R) \\
= & \sum_{w \in R} \, Z^{(X[W_{R}])}(\vec m; \vec t | w),
\end{align} 
where the constituent functions are given as
\begin{align}
& Z^{(X)}_{\rm int}(\vec s, \vec m, \vec t)= \frac{e^{2\pi i \tr{\vec{s}} ({t}_1- {t}_2)}  \sinh^2{\pi(s_1-s_2)}}{\prod^2_{j=1}\prod^4_{i=1} \cosh{\pi(s_j - {m}_i)}},
& Z_{\rm Wilson}(\vec s, R) = \sum_{w \in R} \, \Big(e^{2\pi \sum_j w_j s_j} \Big).
\end{align}
Using the Cauchy determinant identity twice to reduce $Z^{(X[W_{R}])}(\vec m; \vec t | w)$ in terms of Abelian integrals, and then performing those 
Abelian integrals, we obtain
\begin{align}
Z^{(X[W_{R}])}(\vec m; \vec t | w) = \sum_{\rho, \wt{\rho}}(-1)^{\rho + \wt{\rho}}\,
\frac{(-1)^{w_1+w_2}\,\prod^2_{i=1} \Big(e^{2\pi i m_{\rho(i)} (\eta - iw_i)} - e^{2\pi i m_{2+\wt{\rho}(i)} (\eta - iw_i)} \Big)}{\sh{m_{12}}\,\sh{m_{34}}\,\sinh^2{\pi \eta}\,\prod^2_{i=1}  \sh{(m_{\rho(i)} - m_{2+\wt{\rho}(i)})}},
\end{align}
where $m_{12}=m_1 -m_2$, $m_{34}=m_3 - m_4$, $\eta=t_1-t_2$, and $\rho, \wt{\rho}$ are elements of $S_2$.\\

The dual of the Wilson defect $W_{R}$ in the theory $X$, is the vortex defect $\wt{V}_{1,R}$ in the theory $Y$, where $R$ is a 
representation of the central $U(2)$ gauge group. The defect partition function for the latter, using a right SQM realization, 
is given by
\begin{align}
Z^{(Y[\wt{V}^r_{1,R}])}(\vec t; \vec m) = & W_{\rm b.g.}(\vec m, R) \cdot \lim_{z\to 1} \int  d\s^1\,\Big[d\vec \s^2\Big] \, d\s^3\, 
Z^{(Y)}_{\rm int}(\{\s^{\gamma'}\}, \vec t, \vec m) \cdot \CI^{\wt{\Sigma}^{1,R}_r}(\{\s^{\gamma'}\}, \vec t, z|\vec \xi <0), \\
=: & W_{\rm b.g.}(\vec m, R) \cdot \lim_{z\to 1} \int  d\s^1\,\Big[d\vec \s^2\Big] \, Z^{(Y[\wt{V}^r_{1,R}])}_{\rm int}(\{\s^{\gamma'}\}, \vec t, \vec m, z|\vec \xi <0).
\end{align} 

Using similar arguments as above, the defect partition function can be rewritten as
\begin{align}
Z^{(Y[\wt{V}^r_{1,R}])}(\vec t; \vec m) =  W_{\rm b.g.}(\vec m, R) \cdot \sum_{w \in R} Z^{(Y[\wt{V}^r_{1,R}])}(\vec t; \vec m | w),
\end{align}
where the individual functions can be written as
\begin{align}
& W_{\rm b.g.}(\vec t, R)= e^{2\pi m_3 |R| }, \\
& Z^{(Y[\wt{V}^r_{1,R}])}(\vec t; \vec m| w)=  \lim_{z\to 1}\,\int  d\s^1\,\Big[d\vec \s^2\Big] \, d\s^3\, 
\frac{Z^{(Y)}_{\rm int}(\{\s^{\gamma'}\}, \vec t, \vec m)\,\prod_i \ch{(\s^2_i -t_2)}\,\cosh{\pi(\s^3 -\s^2_i)}}{\prod_i \ch{(\s^2_i + iw_i z -t_2)}\,\cosh{\pi(\s^3 -\s^2_i - iw_i z)}}, \\
& Z^{(Y)}_{\rm int}(\{\s^{\gamma'}\}, \vec t, \vec m)=\frac{e^{2\pi i \s^{1} (m_1- m_2)} e^{2\pi i \tr \vec{\s}^2 (m_2 -m_3)} e^{2\pi i \s^{3} (m_3- m_4)} \sinh^2{\pi(\s^2_1-\s^2_2)}}{\prod^2_{i=1}\cosh{\pi(\s^1-\s^2_i)}\prod^2_{a=1} \cosh{\pi(\s^2_i - t_a)}\cosh{\pi(\s^3 -\s^2_i)}}.
\end{align}
Performing the Abelian integrals over $\s_1$ and $\s_3$, $Z^{(Y[\wt{V}^r_{1,R}])}(\vec m; \vec t| w)$ can be recast in the following form:
\begin{align}
& Z^{(Y[\wt{V}^r_{1,R}])}(\vec t; \vec m| w) \\
& = \lim_{z\to 1}\,\int  \Big[d\vec \s^2\Big]\,
\frac{(-1)^{w_1+w_2}\,e^{2\pi i m_{23} \tr\s^2} (e^{2\pi i \s^2_1 m_{12}} -e^{2\pi i \s^2_2 m_{12}})(e^{2\pi i (\s^2_1 + iw_1 z)m_{34}} -e^{2\pi i (\s^2_2 + iw_2 z)m_{34}})}{\sh{m_{12}}\,\sh{m_{34}}\,\prod_i \ch{(\s^2_i + iw_i z -t_2)}\, \cosh{\pi(\s^2_i - t_1)}}.
\end{align}
Finally, performing the Abelian integrals over $\s^2_1$ and $\s^2_2$, and manipulating the resultant expression, we obtain:
\begin{align}
Z^{(Y[\wt{V}^r_{1,R}])}(\vec t; \vec m| w) = e^{-2\pi m_3 (w_1+w_2)}\, e^{2\pi i t_1(m_1 +m_2)}\, e^{-2\pi i t_2(m_3 +m_4)}\,Z^{(X[W_{R}])}(\vec m; -\vec t | w).
\end{align}
The defect partition functions are therefore related as
\be \label{LQ-W2V-pf}
Z^{(X[W_{R}])}(\vec m; \vec t) = C_{XY}(\vec m, \vec t) \, Z^{(Y[\wt{V}^r_{1,R}])}(-\vec t; \vec m),
\ee
which leads to the following mirror map:
\be
\boxed{\langle W_{R} \rangle_{X}(\vec m; \vec t) = \langle  \wt{V}^r_{1,R} \rangle_{Y}(-\vec t; \vec m).}
\ee

\section{Abelian Gauging: Defects in the $D_4$ quiver gauge theory}\label{app:SU(2)}

\subsection{Vortex defects}\label{app:SU(2)-V2W}
Let us first consider the vortex defect, discussed in \Secref{V2W-D}, which can be constructed from the defect $V_{M,R}$ for $M=2$ in the quiver $X$ 
by an $S$-type operation. The $S$-type operation $\CO_{\vec \CP}$ is specified in \eref{AbS-D}-\eref{u-choice1}, and 
the corresponding partition function is given as\footnote{For the subsequent analysis in this section, we will drop the factors $\CF(s_j, z) $ for reasons noted 
in the main text of the paper.}:
\begin{align}
&Z^{\CO_{\vec \CP}(X[V^r_{2,R}])} = \lim_{z\to 1} \int \prod^3_{i=1} du_i \, \Big[d\vec s\Big] \, \prod^3_{i=1} Z_{\rm FI} (u_i, \eta_i)\,
Z^{(X)}_{\rm int}(\vec s, \vec u, v, \vec t)\, W_{\rm b.g.} (\vec t, R) \, \CI^{\Sigma^{2,R}_r}(\vec s, \vec u, z|\vec \xi <0), \label{ZAbS-D4-app}\\
& Z^{(X)}_{\rm int}(\vec s, \vec u, v, \vec t)= \Big\{Z^{(X)}_{\rm int}(\vec s, \vec m, \vec t)| m_3=u_1, m_4=u_2, m_1=u_3, m_2=v \Big\},\\
&W_{\rm b.g.} (\vec t, R)= e^{2\pi t_2 |R|}, \\
& \CI^{\Sigma^{2,R}_r}(\vec s, \vec u, z|\vec \xi <0)= \sum_{w \in R} \prod^2_{j=1} \prod^2_{i=1} \frac{\ch{(s_j -u_i)}}{\ch{(s_j + i w_j z - u_i)}}.
\end{align}
We will identify the resultant 3d-1d quiver as a vortex defect in the $D_4$ quiver gauge theory $X'$, and will denote it as $X'[V'^{(I)}_{2,R}]$.
\begin{align}
& Z^{(X'[V'^{(I)}_{2,R}])}:= e^{-2\pi (t_2 -\frac{\eta_1+\eta_2}{2})|R|} \, Z^{\CO_{\vec \CP}(X[V^r_{2,R}])}(v; \vec t, \vec \eta), \\
&= e^{2\pi (\frac{\eta_1+\eta_2}{2})|R|} \,\lim_{z\to 1} \int \prod^3_{i=1} du_i \, \Big[d\vec s\Big] \, \prod^3_{i=1} Z_{\rm FI} (u_i, \eta_i)\,
Z^{(X)}_{\rm int}(\vec s, \vec u, v, \vec t)\,  \CI^{\Sigma^{2,R}_r}(\vec s, \vec u, z|\vec \xi <0). \label{ZAbS-D4-app1}
\end{align}
\\
Mirror symmetry for the 3d defects $X[V_{2,R}]$ and $Y[\wt{W}_R]$ leads to the following identity:
\begin{align}
&\int\,\Big[d\vec s\Big]\,Z^{(X)}_{\rm int}(\vec s, \vec u, v, \vec t)\, W_{\rm b.g.} (\vec t, R) \, \CI^{\Sigma^{2,R}_r}(\vec s, \vec u, z|\vec \xi <0)\nn \\
= & \sum_{w \in R} W_{\rm b.g.} (\vec t, R) \,\frac{\sh{(u_1-u_2 - iw_1z + iw_2z)}}{\sh{(u_1 -u_2)}}\,Z^{(X)}(u_3, v, u_1-iw_{1}z, u_2 - iw_{2}z; \vec t) \nn \\
=& \sum_{w \in R} C_{XY}(\vec u, v,\vec t)\, \frac{\sh{(u_1-u_2 - iw_1z + iw_2z)}}{\sh{(u_1 -u_2)}}\,Z^{(Y)}(\vec t; -u_3, -v, -u_1+iw_1z, -u_2+iw_2z)\nn\\
=& C_{XY}(\vec u, v,\vec t)\,\int \,\prod^3_{\gamma'=1}\Big[d\s^{\gamma'}\Big] \,Z^{(Y)}_{\rm int}(\{\s^{\gamma'}\}, \vec t, -u_3,-v,-u_1,-u_2)\,\sum_{w \in R} \, e^{2\pi \sum_j w_j \s^2_j z}\\
=: & C_{XY}(\vec u, v,\vec t)\,\int \,\prod^3_{\gamma'=1}\Big[d\s^{\gamma'}\Big] \,Z^{(Y[\wt{W}_R])}_{\rm int}(\{\s^{\gamma'}\}, \vec t, -u_3,-v,-u_1,-u_2,z). \label{MS-Id1}
\end{align}
This is a concrete example of the identity \eref{MS-XYD-1} for the dual pair $(X, Y)$, $\CD=V_{2,R}$ is a vortex defect and 
$\CD^\vee = \wt{W}_R$ is a Wilson defect.

We can now compute the dual defect partition function, following the recipe given by \eref{PF-wtOPgenD-A2B}-\eref{CZ-wtOPD-A2B}. 
From \eref{CZ-wtOPD-A2B}, the function $\CZ_{\wt{\CO}_{\vec \CP}(Y)}$ is given as
\be
\CZ_{\wt{\CO}_{\vec \CP}(Y)}(\{ \s^{\gamma'} \}, \vec t, \vec \eta) = \delta(\eta_1+\tr \s^2 -\s^3 -t_2)\,\delta(\eta_2 -t_2 +\s^3)\,\delta(\eta_3 +t_1-\s^1).
\ee
The general expression in \eref{PF-wtOPgenD-A2B} for the dual defect partition function reduces to the following expression:
\begin{align}
Z^{\wt{\CO}_{\vec \CP}(Y[\wt{W}_R])}(\vec{t}, \vec{\eta}; v)
= \lim_{z\to1}\, \int  \prod^3_{\gamma'=1}\Big[d\s^{\gamma'}\Big]\,&\CZ_{\wt{\CO}_{\vec \CP}(Y)}(\{ \s^{\gamma'} \}, \vec t, \vec \eta)
\,C_{XY}(\vec u=0, v, \vec t)\nn \\
& \times Z^{(Y[\wt{W}_R])}_{\rm int}(\{\s^{\gamma'}\}, \vec t, \vec u=0,-v,z),
\end{align}
where the function $Z^{(Y[\wt{W}_R])}_{\rm int}(\ldots, z)$ can be read off from \eref{MS-Id1}:
\begin{align}
 Z^{(Y[\wt{W}_R])}_{\rm int}(\{\s^{\gamma'}\}, \vec t, \vec u=0,-v,z)= Z^{(Y)}_{\rm int}(\{\s^{\gamma'}\}, \vec t, \vec u=0,-v) \, \sum_{w \in R} \,e^{2\pi \sum_i w_i \s^2_i\,z} .
\end{align}
Since there are no $z$-dependent poles in the integrand, the $z \to 1$ limit can be taken at this stage.
Using the expression for $\CZ_{\wt{\CO}_{\vec \CP}(Y)}$, implementing the delta functions and after shifting integration variables appropriately, we obtain 
\begin{align}
Z^{\wt{\CO}_{\vec \CP}(Y[\wt{W}_R])}(\vec{t}, \vec{\eta}; v)=&C(v,\vec \eta, \vec t)\,\sum_{w \in R}\int\,\Big[d\vec \s^2\Big]\,\delta(\tr \vec\s^2)\,
Z^{\rm vec}_{\rm 1-loop}(\vec\s^2)\, Z^{\rm fund}_{\rm 1-loop}(\vec\s^2, \eta_3 + \eta +\frac{\eta_1+\eta_2}{2})\nn \\
& \times Z^{\rm fund}_{\rm 1-loop}(\vec\s^2, \frac{\eta_1-\eta_2}{2})\,\prod^2_{a=1} Z^{\rm fund}_{\rm 1-loop}(\vec\s^2,t_a + \frac{\eta_1+\eta_2}{2}-t_2)\,e^{2\pi \sum_i w_i \s^2_i},\\
=: & C(v, \vec \eta, \vec t)\,\int \,\Big[d\vec \s^2\Big] \, \delta(\tr \vec{\s}^2) \, Z^{U(2), N_f=4}_{\rm int} (\vec{\s^2}, \vec m'(\vec t, \vec\eta), \eta'=0)\, \sum_{w \in R} e^{2\pi \sum_i w_i \s^2_i}, \label{ZAbS-D4-app2}
\end{align}
where $\eta= t_1-t_2$ and $C(v,\vec \eta, \vec t)=e^{2\pi i v(\eta_1+\eta_2+\eta_3+2(t_1-t_2))}$. 
The masses $\vec m'$ of the theory $Y'$ are given in terms of the FI parameters of $X'$ as follows:
\begin{align}
& m'_1= \eta_3 + \eta +\frac{\eta_1+\eta_2}{2}, \label{mm-1}\\
& m'_2=\eta+ \frac{\eta_1+\eta_2}{2},\label{mm-2}\\
& m'_3=\frac{\eta_1+\eta_2}{2},\label{mm-3}\\
& m'_4=\frac{\eta_1-\eta_2}{2}.\label{mm-4}
\end{align}
The above expression \eref{ZAbS-D4-app2} reproduces the defect partition function in \eref{ZAbS-D4-d}.\\

The vortex defects which are constructed from the defect $V_{M, R}$ for $M \neq 2$ in quiver $X$ can be analyzed in a 
similar fashion as above.

\subsection{Wilson defects}\label{app:SU(2)-W2V}
Let us now consider a Wilson defect, discussed in \Secref{W2V-D}, which can be constructed from the 
defect $W_{R}$ in the quiver $X$ by an $S$-type operation $\CO_{\vec \CP}$ specified in 
\eref{AbS-Da}-\eref{u-choice1a}. The starting point is the dual pair of defects $X[W_{R}]$ and $Y[\wt{V}^r_{1,R}]$ 
discussed in \Secref{U24-W2V}.
The partition function $\CO_{\vec \CP}(X[W_{R}])$ is given as:
\begin{align}
& Z^{\CO_{\vec \CP}(X[W_{R}])} = \int \prod^3_{i=1} du_i \, \Big[d\vec s\Big] \, \prod^3_{i=1} Z_{\rm FI} (u_i, \eta_i)\,
Z^{(X)}_{\rm int}(\vec s, \vec u, v, \vec t)\, Z_{\rm Wilson}(\vec s, R), \\
& Z^{(X)}_{\rm int}(\vec s, \vec u, v, \vec t)= \Big\{Z^{(X)}_{\rm int}(\vec s, \vec m, \vec t)| m_1=u_1, m_2=u_2, m_4=u_3, m_3=v \Big\},\\
& Z_{\rm Wilson}(\vec s, R)= \sum_{w \in R} e^{\sum_j w_j\, s_j}.
\end{align}

Mirror symmetry of the defects $X[W_R]$ and $Y[\wt{V}^r_{1,R}]$ imply the following $z$-dependent identity:
\begin{align}
&\int  \Big[d\vec s\Big] \, Z^{(X[W_R])}_{\rm int}(\vec s, \vec u, v, \vec t, z)
=: \int  \Big[d\vec s\Big] \,Z^{(X)}_{\rm int}(\vec s, \vec u, v, \vec t)\, \sum_{w \in R} e^{\sum_j w_j\, s_j\,z}\nn \\
=& C_{XY}(\vec u, v, \vec t )\,\int  d\s^1\,\Big[d\vec \s^2\Big] \, d\s^3\,W_{\rm b.g.}(v, R)\,
Z^{(Y[\wt{V}^r_{1,R}])}_{\rm int}(\{\s^{\gamma'}\}, -\vec t, \vec u,v, z), \label{MS-Id2}
\end{align}
where the contact term $C_{XY}$ and the background Wilson defect $W_{\rm b.g.}$ are given as
\be
C_{XY}(\vec u, v, \vec t )= e^{2\pi i t_1(u_1 + u_2)}\,  e^{-2\pi i t_2(u_3 + v)}, \qquad W_{\rm b.g.}(v, R)= e^{2\pi v |R|}.
\ee
The integrand $Z^{(Y[\wt{V}^r_{1,R}])}_{\rm int}$ in \eref{MS-Id2} is given as
\begin{align}
& Z^{(Y[\wt{V}^r_{1,R}])}_{\rm int}(\{\s^{\gamma'}\}, \vec t, \vec u,v, z) = Z^{(Y)}_{\rm int}(\{\s^{\gamma'}\}, \vec t, \vec u,v) 
\cdot \CI^{\wt{\Sigma}^{1,R}_r} (\{\s^{\gamma'}\}, \vec t, z|\vec \xi <0),\\
& Z^{(Y)}_{\rm int}(\{\s^{\gamma'}\}, \vec t, \vec u,v)=\frac{e^{2\pi i \s^{1} (u_1- u_2)} e^{2\pi i \tr \vec{\s}^2 (u_2 - v)} e^{2\pi i \s^{3} (v- u_3)} \sinh^2{\pi(\s^2_1-\s^2_2)}}{\prod^2_{j=1}\cosh{\pi(\s^1-\s^2_j)}\prod^2_{a=1} \cosh{\pi(\s^2_j - t_a)}\cosh{\pi(\s^3 -\s^2_j)}}, \\
&\CI^{\wt{\Sigma}^{1,R}_r} (\{\s^{\gamma'}\}, \vec t, z|\vec \xi <0)= \sum_{w \in R}\frac{ \prod_i \ch{(\s^2_j -t_2)}\,\cosh{\pi(\s^2_j-\s^3 )}}{\prod_j \ch{(\s^2_j + iw_j z -t_2)}\,\cosh{\pi(\s^2_j + iw_j z-\s^3)}}.
\end{align}
The identity \eref{MS-Id2} is another concrete example of the general identity \eref{MS-XYD-1} for the dual pair $(X, Y)$, $\CD=W_{R}$ is a
Wilson defect and $\CD^\vee = \wt{V}^r_{1,R}$ is a vortex defect.\\

Following the general discussion around \eref{PF-wtOPgenD-B2A}-\eref{CZ-wtOPD-B2A}, the dual partition function 
$Z^{(Y'[(\wt{V}^r_{1,R})^\vee])}= Z^{\wt{\CO}_{\vec \CP}(Y[\wt{V}^r_{1,R}])}(\vec{t}, \vec{\eta}; v)$ can be written as:
\begin{align}\label{AbS-D4-V2W}
Z^{\wt{\CO}_{\vec \CP}(Y[\wt{V}^r_{1,R}])}(\vec{t}, \vec{\eta}; v)
= \lim_{z\to 1}\,\int  d\s^1\,\Big[d\vec \s^2\Big] \, d\s^3\, & \CZ_{\wt{\CO}_{\vec \CP}(Y)}(\{ \s^{\gamma'} \}, \vec t, \vec \eta)
\,C_{XY}(\vec u=0,v, \vec t )\, W_{\rm b.g.}(v, R)\nn \\
& \times Z^{(Y[\wt{V}^r_{1,R}])}_{\rm int}(\{\s^{\gamma'}\}, -\vec t, \vec u=0,v, z),
\end{align}

Finally, the function $\CZ_{\wt{\CO}_{\vec \CP}(Y)}$ can be computed from the general formula \eref{CZ-wtOPD-B2A}:
\be
\CZ_{\wt{\CO}_{\vec \CP}(Y)}(\{ \s^{\gamma'} \}, \vec t, \vec \eta) = \delta(\eta_1+\s^1 +t_1)\,\delta(\eta_2 -\s^1 + \tr \s^2 +t_1)\,\delta(\eta_3 -t_2-\s^3).
\ee
Putting together the different ingredients on the RHS of \eref{AbS-D4-V2W} and integrating over $\s^1$ and $\s^2$, we obtain the 
following expression for the dual partition function (after some trivial shifts in the integration variable):
\begin{align}
&  Z^{(Y'[(W'_{R})^\vee])} =C\cdot W_{\rm b.g.} \cdot \lim_{z\to 1}\int\Big[d\vec \s\Big] \, \delta(\tr \vec\s)
Z^{U(2), N_f=4}_{\rm int} (\vec\s, \vec m'(\vec t, \vec\eta), \eta') \CI^{\wt{\Sigma}^{1,R}_r}(\vec \s, \vec m', z| \vec \xi <0), \label{ZAbS2-D4}\\
&C := C(v, \vec \eta, \vec t)= e^{2\pi i v(\eta_1 +\eta_2+\eta_3 + 2(t_1-t_2))}, \qquad W_{\rm b.g.}:= W_{\rm b.g.} (v, R)= e^{2\pi v |R|}, \\
& \CI^{\wt{\Sigma}^{1,R}_r}(\vec \s, \vec m', z| \vec \xi <0)= \sum_{w \in R}\, \prod^2_{j=1} \prod^2_{i=1} \frac{\ch{(\s_j - m'_{i})}}{\ch{(\s_j + i w_j z - m'_{i})}},
\end{align}
where the masses $\{m'_i\}_{i=1,\ldots,4}$ are given in \eref{mm-1}-\eref{mm-4}. The 3d-1d quiver $Y'[\wt{V}'_{1,\wt{R}}]$ in \figref{AbEx2aD4} 
can be read off from the RHS of \eref{ZAbS2-D4}.\\

One can similarly implement the $S$-type operation \eref{AbS-Da}-\eref{u-choice1a} on the dual pair $X[W_{R}]$ and $Y[\wt{V}^l_{1,R}]$. 
Mirror symmetry then implies the following $z$-dependent identity:
\begin{align}
&\int  \Big[d\vec s\Big] \, Z^{(X[W_R])}_{\rm int}(\vec s, \vec u, v, \vec t, z)
=: \int  \Big[d\vec s\Big] \,Z^{(X)}_{\rm int}(\vec s, \vec u, v, \vec t)\, \sum_{w \in R} e^{\sum_j w_j\, s_j\,z}\nn \\
=& C_{XY}(\vec u, v, \vec t )\,\int  d\s^1\,\Big[d\vec \s^2\Big] \, d\s^3\,W_{\rm b.g.}(u_2, R)\,
Z^{(Y[\wt{V}^l_{1,R}])}_{\rm int}(\{\s^{\gamma'}\}, -\vec t, \vec u,v, z), \label{MS-Id3}
\end{align}
where the contact term $C_{XY}$ and the background Wilson defect $W_{\rm b.g.}$ are given as
\be
C_{XY}(\vec u, v, \vec t )= e^{2\pi i t_1(u_1 + u_2)}\,  e^{-2\pi i t_2(u_3 + v)}, \qquad W_{\rm b.g.}(u_2, R)= e^{2\pi u_2 |R|}.
\ee
The integrand $Z^{(Y[\wt{V}^l_{1,R}])}_{\rm int}$ in \eref{MS-Id2} is given as
\begin{align}
& Z^{(Y[\wt{V}^l_{1,R}])}_{\rm int}(\{\s^{\gamma'}\}, \vec t, \vec u,v, z) = Z^{(Y)}_{\rm int}(\{\s^{\gamma'}\}, \vec t, \vec u,v) 
\cdot \CI^{\wt{\Sigma}^{1,R}_l} (\{\s^{\gamma'}\}, \vec t, z|\vec \xi > 0),\\
&\CI^{\wt{\Sigma}^{1,R}_l} (\{\s^{\gamma'}\}, \vec t, z|\vec \xi > 0)= \sum_{w \in R}\frac{ \prod_i \ch{(\s^2_j -t_1)}\,\cosh{\pi(\s^2_j-\s^3 )}}{\prod_j \ch{(\s^2_j - iw_j z -t_1)}\,\cosh{\pi(\s^2_j - iw_j z-\s^3)}},
\end{align}
where $Z^{(Y)}_{\rm int}$ is given above. The dual partition function $Z^{(Y'[(\wt{V}^l_{1,R})^\vee])}= Z^{\wt{\CO}_{\vec \CP}(Y[\wt{V}^l_{1,R}])}(\vec{t}, \vec{\eta}; v)$ can be written as:
\begin{align}\label{AbS-D4-V2W2}
Z^{\wt{\CO}_{\vec \CP}(Y[\wt{V}^l_{1,R}])}(\vec{t}, \vec{\eta}; v)
= \lim_{z\to 1}\,\int  d\s^1\,\Big[d\vec \s^2\Big] \, d\s^3\, & \CZ_{\wt{\CO}_{\vec \CP}(Y)}(\{ \s^{\gamma'} \}, \vec t, \vec \eta)
\,C_{XY}(\vec u=0,v, \vec t )\nn \\
& \times Z^{(Y[\wt{V}^l_{1,R}])}_{\rm int}(\{\s^{\gamma'}\}, -\vec t, \vec u=0,v, z),
\end{align}
where the function $\CZ_{\wt{\CO}_{\vec \CP}(Y)}$ can be computed from the general formula \eref{CZ-wtOPD-B2A}:
\be
\CZ_{\wt{\CO}_{\vec \CP}(Y)}(\{ \s^{\gamma'} \}, \vec t, \vec \eta) = \delta(\eta_1+\s^1 +t_1)\,\delta(\eta_2 -\s^1 + \tr \s^2 +t_1-i|R|z)\,\delta(\eta_3 -t_2-\s^3).
\ee
Note that the form of $\CZ_{\wt{\CO}_{\vec \CP}(Y)}$ is different from what we got in the case of the right SQM. Now, performing 
a change of variables $\s^2_j \to \s^2_j + iw_j\,z$ on the RHS of \eref{AbS-D4-V2W2}, and then 
integrating over $\s^1$ and $\s^2$, we can rewrite the dual partition function (after some trivial shifts in the integration variable) 
in the following form:
\begin{align}
& Z^{\wt{\CO}_{\vec \CP}(Y[\wt{V}^l_{1,R}])} =C\cdot W_{\rm b.g.} \cdot \lim_{z\to 1}\int\Big[d\vec \s\Big] \, \delta(\tr \vec\s)
Z^{U(2), N_f=4}_{\rm int} (\vec\s, \vec m'(\vec t, \vec\eta), \eta') \CI^{\wt{\Sigma}^{1,R}_r}(\vec \s, \vec m', z| \vec \xi <0), \label{ZAbS2-D4-2}
\end{align}
where $C$, $W_{\rm b.g.}$, and $\CI^{\wt{\Sigma}^{1,R}_r}$ are functions given above. The final 3d-1d quiver is therefore identical to the 
one that we obtained in the earlier case.

\section{Non-Abelian Gauging : Defects in the $D_6$ quiver}\label{app: NAG-Ex}
In this section, we explicitly construct the mirror map in \figref{NAbEx1Sp}, which involves a vortex defect in the $D_6$ quiver 
on the one hand and a Wilson defect in the $Sp(2)$ gauge theory on the other. The starting point is the pair of linear quivers 
with defects in \figref{NAEx-LQdef}, where we implement the $S$-type operation \eref{NAbS-D} on the 3d-1d system $(X[V^l_{2,R}])$.

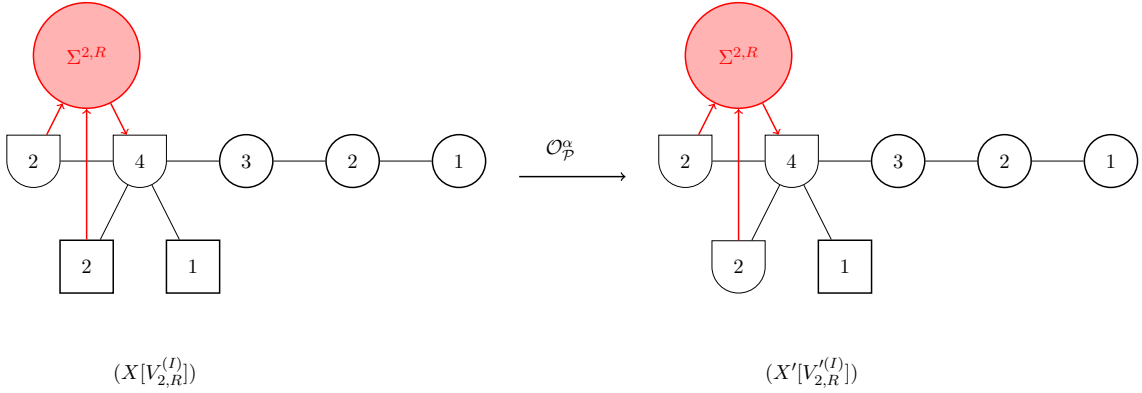
\begin{center}
\begin{tabular}{ccc}
\scalebox{.7}{\begin{tikzpicture}[node distance=2cm,
cnode/.style={circle,draw,thick, minimum size=1.0cm},snode/.style={rectangle,draw,thick,minimum size=1cm}, pnode/.style={red,rectangle,draw,thick, minimum size=1.0cm},nnode/.style={red, circle,draw,thick,fill=red!30 ,minimum size=2.0cm}]
\node[cf-group] (1) at (0,0) {\rotatebox{-90}{2}} ;
\node[cf-group] (2) at (2,0) {\rotatebox{-90}{4}};
\node[snode] (3) at (3,-2) {1};
\node[snode] (11) at (1,-2) {2};
\node[cnode] (4) at (4,0) {3};
\node[cnode] (5) at (6,0) {2};
\node[cnode] (6) at (8,0) {1};
\node[nnode] (10) at (1,2) {$\Sigma^{2,R}$};
\draw[-] (1) -- (2);
\draw[-] (2) -- (3);
\draw[-] (2) -- (4);
\draw[-] (4) -- (5);
\draw[-] (5) -- (6);
\draw[-] (2) -- (11);
\draw[red, thick, ->] (1)--(10);
\draw[red, thick, ->] (11)--(10);
\draw[red, thick, ->] (10)--(2);
\node[text width=1cm](8) at (2, -4) {$(X[V^{(I)}_{2,R}])$};
\end{tikzpicture}}
& \scalebox{.7}{\begin{tikzpicture} \draw[thick, ->] (0,0) -- (2,0); 
\node[] at (1,-4) {};
\node[text width=1cm](1) at (1,0.5) {$\CO^\alpha_{\CP}$};
\end{tikzpicture}} &
\scalebox{.7}{\begin{tikzpicture}[node distance=2cm,
cnode/.style={circle,draw,thick, minimum size=1.0cm},snode/.style={rectangle,draw,thick,minimum size=1cm}, pnode/.style={red,rectangle,draw,thick, minimum size=1.0cm},nnode/.style={red, circle,draw,thick,fill=red!30 ,minimum size=2.0cm}]
\node[cf-group] (1) at (0,0) {\rotatebox{-90}{2}} ;
\node[cf-group] (2) at (2,0) {\rotatebox{-90}{4}};
\node[snode] (3) at (3,-2) {1};
\node[cf-group] (11) at (1,-2) {\rotatebox{-90}{2}};
\node[cnode] (4) at (4,0) {3};
\node[cnode] (5) at (6,0) {2};
\node[cnode] (6) at (8,0) {1};
\node[nnode] (10) at (1,2) {$\Sigma^{2,R}$};
\draw[-] (1) -- (2);
\draw[-] (2) -- (3);
\draw[-] (2) -- (4);
\draw[-] (4) -- (5);
\draw[-] (5) -- (6);
\draw[-] (2) -- (11);
\draw[red, thick, ->] (1)--(10);
\draw[red, thick, ->] (11)--(10);
\draw[red, thick, ->] (10)--(2);
\node[text width=1cm](8) at (2, -4) {$(X'[V'^{(I)}_{2,R}])$};
\end{tikzpicture}}
\end{tabular}
\end{center}

\begin{align}
Z^{G^\alpha_{\CP}(X[V^l_{2,R}])} = \lim_{z\to 1}\, \int \Big[d \vec u\Big] \, \prod^5_{\gamma=1}\,\Big[d\vec s^\gamma\Big] \, & \CZ_{G^\alpha_{\CP}} (\vec u, \eta^\alpha)\,
Z^{(X,\CP)}_{\rm int}(\{\vec s^\gamma \}, \vec u, v, \vec t)\, W_{\rm b.g.} (\vec t, R) \nn \\
& \times \CI^{\Sigma^{2,R}_l}(\vec s^1, \vec s^2, \vec u, z|\vec \xi >0), \label{ZNAbS-D6-app}
\end{align}
where the label $\gamma$ of the gauge nodes of $X$ increases from the left to the right. 
The constituent functions appearing in the matrix integral above -- $\CZ_{G^\alpha_{\CP}}$, $Z^{(X)}_{\rm int}$, $W_{\rm b.g.} $ and $\CI^{\Sigma^{2,R}_l}$ 
are given as 
\begin{align}
&\CZ_{G^\alpha_{\CP}} (\vec u, \eta^\alpha)= e^{2\pi i \eta_\alpha (u_1+u_2)}\, \sinh^2{\pi(u_1 -u_2)},  \\
& Z^{(X,\CP)}_{\rm int}(\{\vec s^\gamma \}, \vec u, v, \vec t)= Z^{(X)}_{\rm int}(\{\vec s^\gamma \}, m_1=u_1, m_2=u_2, m_3=v, \vec t),\\
&W_{\rm b.g.} (\vec t, R)= e^{2\pi t_2 |R|}, \\
& \CI^{\Sigma^{2,R}_l}(\vec s^1, \vec s^2, \vec u, z|\vec \xi > 0)= \sum_{w \in R} \prod^4_{j=1} \Big(\prod^2_{k=1} \frac{\ch{(s^2_j -s^1_k)}}{\ch{(s^2_j - i w_j z - s^1_k)}}\Big)
\cdot \Big( \prod^2_{i=1} \frac{\ch{(s^2_j -u_i)}}{\ch{(s^2_j - i w_j z - u_i)}}\Big).
\end{align}

Mirror symmetry implies that one can write down an identity of the form \eref{MS-XYD-1} involving the $z$-dependent partition functions of $X$ and $Y$.
In the present case, this identity takes the form:
\begin{align}
Z^{(X[V^l_{2,R}],\CP)}(\vec u, v, \vec t |z ) = & \int \prod^5_{\gamma=1}\,\Big[d\vec s^\gamma\Big] \,
Z^{(X,\CP)}_{\rm int}(\{\vec s^\gamma \}, \vec u, v, \vec t)\, W_{\rm b.g.} (\vec t, R) \, \CI^{\Sigma^{2,R}_l}(\vec s^1, \vec s^2, \vec u, z|\vec \xi >0) \nn \\
=& \int \prod^2_{\gamma'=1}\,\Big[d\vec \s^{\gamma'}\Big] \, Z^{(Y)}_{\rm int} (\{ \vec \s^{\gamma'} \}, \vec t, -\vec u, -v) \, \sum_{w \in R}\, e^{2\pi \sum_j w_j \s^2_j z}.
\end{align}

\subsection{Non-Abelian Gauging via abelianization of the partition function} \label{app: NAG-Ex-1}

The non-Abelian gauging operation can then be Abelianized following the procedure of \eref{NAG-AG}:
\begin{align}
Z^{G^\alpha_{\CP}(X[V^l_{2,R}])} =& \lim_{z\to 1}\, \int \prod^2_{i=1}\,du_i\, e^{2\pi i \eta_\alpha (u_1+u_2)}\, (e^{2\pi u_1}\, e^{-2\pi u_2} +e^{-2\pi u_1}\, e^{2\pi u_2} -2)\,Z^{(X[V^l_{2,R}],\CP)}(\vec u, v, \vec t |z ) \\
=: &\,  Z^{\overline{\CO}^{2, \alpha}_{\CP}(X[V^l_{2,R}])}(v, \vec t | 1,-1) + Z^{\overline{\CO}^{2, \alpha}_{\CP}(X[V^l_{2,R}])}(v, \vec t | -1,1) - 2 Z^{\overline{\CO}^{2, \alpha}_{\CP}(X[V^l_{2,R}])}(v, \vec t | 0, 0),
\end{align}
where we have used the notation of \eref{Ab-defg}. The function $Z^{\overline{\CO}^{2, \beta}_{\CP}(X[V^l_{2,R}])} (v, \vec t| q_1, q_2)$ corresponds to 
the partition function of a theory obtained from the quiver $X[V^l_{2,R}]$ by a sequence of two Abelian defect-gauging $S$-type operations of the 
Wilson type, and is given as
\begin{align}
Z^{\overline{\CO}^{2, \alpha}_{\CP}(X[V^l_{2,R}])} (v, \vec t| q_1, q_2) = \lim_{z\to 1}\, \int \prod^2_{i=1}\,du_i\, e^{2\pi i \eta_\alpha (u_1+u_2)}\,e^{2\pi q_1 u_1}\, e^{2\pi q_2 u_2}\, Z^{(X[V^l_{2,R}],\CP)}(\vec u, v, \vec t |z ).
\end{align}
To evaluate a partition function of the above form, we will use an analytic continuation trick of the following form:
\begin{align}
Z^{\overline{\CO}^{2, \alpha}_{\CP}(X[V^l_{2,R}])} (v, \vec t| q_1, q_2) = \lim_{\substack{z\to 1 \\ z' \to 1}}\, \int \prod^2_{i=1}\,du_i\, e^{2\pi i \eta_\alpha (u_1+u_2)}\,e^{2\pi q_1 u_1 z'}\, e^{2\pi q_2 u_2z'}\, Z^{(X[V^l_{2,R}],\CP)}(\vec u, v, \vec t |z ),
\end{align}
where we compute the integral assuming $z' \in i\BR$ and then take the $z' \to 1$ limit.\\

The partition function of the dual theory $\wt{G}^\alpha_{\CP}(Y[\wt{W}_{R}])$ can then be written as
\be \label{D6-pf-dual-Ab}
Z^{\wt{G}^\alpha_{\CP}(Y[\wt{W}_{R}])} =  Z^{\wt{\overline{\CO}}^{2, \alpha}_{\CP}(Y[\wt{W}_{R}])}(v, \vec t | 1,-1) + Z^{\wt{\overline{\CO}}^{2, \alpha}_{\CP}(Y[\wt{W}_{R}])}(v, \vec t | -1,1) - 2 Z^{\wt{\overline{\CO}}^{2, \alpha}_{\CP}(Y[\wt{W}_{R}])}(v, \vec t | 0, 0),
\ee
where the dual partition function for the Abelian $S$-type operation can be computed following the general prescription of 
\eref{PF-wtOPgenD-A2B}-\eref{CZ-wtOPD-A2B}. Explicitly, we have
\begin{align}
 \CZ_{\overline{\CO}^{2, \alpha}_{\CP}(X)} = & e^{2\pi i \eta_\alpha (u_1+u_2)}\,e^{2\pi q_1 u_1 z'}\, e^{2\pi q_2 u_2z'}, \\
\CZ_{\wt{\overline{\CO}}^{2, \alpha}_{\CP}(Y)}= & \int \prod^2_{i=1}\,du_i\, \CZ_{\overline{\CO}^{2, \alpha}_{\CP}(X)}\, \prod^2_{i=1}\, e^{2\pi i \, g_i(\{\vec \s^{\gamma'} \}, \CP)\, u_i} \nn\\
= & \delta \Big(-\tr \vec \s^1 + \eta_\alpha -i q_1 z'\Big)\,  \delta \Big( \tr \vec \s^1- \tr \vec \s^2 +  \eta_\alpha - i q_2 z'\Big) ,
\end{align}
where the functions $g_i(\{\vec \s^{\gamma'} \}, \CP) $ are given by
\be
g_1(\{\vec \s^{\gamma'} \}, \CP) = - \tr \vec \s^1, \qquad g_2(\{\vec \s^{\gamma'} \}, \CP) = \tr \vec \s^1 - \tr \vec \s^2.
\ee
In writing the above formula for $\CZ_{\wt{\overline{\CO}}^{2, \alpha}_{\CP}(Y)}$, we have ignored the $\vec u$-dependent contact terms, 
for the sake of simplifying the computation. These can be easily reinstated and will lead to contact terms and some additional background 
Wilson defect. However, the latter can be absorbed by an appropriate redefinition of the vortex defect.\\

From the general expression \eref{PF-wtOPgenD-A2B}, one can write down the dual partition function 
$Z^{\wt{\overline{\CO}}^{2, \alpha}_{\CP}(Y[\wt{W}_{R}])}(v, \vec t| q_1, q_2)$ as follows:
\begin{align}
Z^{\wt{\overline{\CO}}^{2, \alpha}_{\CP}(Y[\wt{W}_{R}])} = \lim_{\substack{z\to 1 \\ z' \to 1}}\, \int \prod^2_{\gamma'=1}\,\Big[d\vec \s^{\gamma'}\Big] \, 
& \delta \Big(-\tr \vec \s^1 + \eta_\alpha -i q_1 z'\Big)\,  \delta \Big( \tr \vec \s^1- \tr \vec \s^2 +  \eta_\alpha - i q_2 z'\Big) \nn \\
&\times \,Z^{(Y)}_{\rm int} (\{ \vec \s^{\gamma'} \}, \vec t, -\vec u=0, -v) \, \sum_{w \in R}\, e^{2\pi \sum_j w_j \s^2_j z}.
\end{align}
The limit $z \to 1$ can be trivially implemented at this stage. After a change of integration variables $\s^1_i \to \s^1_i + \eta_\alpha/2 - iq_1 z'/2$, 
one can rewrite the above partition function as follows:
\begin{align}
 Z^{\wt{\overline{\CO}}^{2, \alpha}_{\CP}(Y[\wt{W}_{R}])} =& \lim_{z' \to 1}\, \int \prod^2_{\gamma'=1}\,\Big[d\vec \s^{\gamma'}\Big] \, 
 \delta \Big(\tr \vec \s^1\Big)\,  \delta \Big( \tr \vec \s^1- \tr \vec \s^2 + 2\eta_\alpha - i (q_1 +q_2) z'\Big)\,e^{2\pi i v \tr \vec \s^2} \nn \\
&\times \, \prod^2_{\gamma'=1}\, Z^{\rm vec}_{\rm 1-loop} (\vec \s^{\gamma'})\, Z^{\rm bif}_{\rm 1-loop}(\vec \s^1, \vec \s^2, -\eta_\alpha/2 + iq_1 z'/2)\,
Z^{\rm fund}_{\rm 1-loop} (\vec \s^{2}, \vec t)\, \sum_{w \in R}\, e^{2\pi \sum_j w_j \s^2_j }.
\end{align}
Since $q_1+q_2=0$ in the present case, $z'$-dependent term drops off from the delta function. 
In addition, performing a change of variables $\s^2_j \to \s^2_j + \eta_\alpha/2 $, the above integral can be rewritten as (up to contact terms)
\begin{align}
 Z^{\wt{\overline{\CO}}^{2, \alpha}_{\CP}(Y[\wt{W}_{R}])} & = \lim_{z' \to 1}\, \int \prod^2_{\gamma'=1}\,\Big[d\vec \s^{\gamma'}\Big] \, 
 \delta \Big(\tr \vec \s^1\Big)\,  \delta \Big(  \tr \vec \s^2 \Big)\,e^{2\pi |R|(\eta_\alpha/2)}\,\prod^2_{\gamma'=1}\, Z^{\rm vec}_{\rm 1-loop} (\vec \s^{\gamma'}) \nn \\
&\times \, Z^{\rm bif}_{\rm 1-loop}(\vec \s^1, \vec \s^2,  iq_1 z'/2)\,
Z^{\rm fund}_{\rm 1-loop} (\vec \s^{2}, \vec t -\eta_\alpha/2)\, \sum_{w \in R}\, e^{2\pi \sum_j w_j \s^2_j }.
\end{align}

Now, let us focus the $\vec \s^1$-dependent part of the integral, which can be written in the following form:
\begin{align}
& I := \int \, \Big[d\vec \s^{1}\Big] \,\delta \Big(\tr \vec \s^1\Big)\, Z^{\rm vec}_{\rm 1-loop} (\vec \s^{1})\,Z^{\rm bif}_{\rm 1-loop}(\vec \s^1, \vec \s^2,  iq_1 z'/2) \nn \\
&= \int \, d\xi \, \Big[d\vec \s^{1}\Big] \, e^{2\pi i \xi \tr \vec \s^1}\,  Z^{\rm vec}_{\rm 1-loop} (\vec \s^{1})\,Z^{\rm bif}_{\rm 1-loop}(\vec \s^1, \vec \s^2,  iq_1 z'/2) \nn \\
&= \int \, d\xi \, Z^{(U(2),N_f=4)}(\vec\s^2 +  iq_1 z'/2, \xi).
\end{align}
Using the result for the partition function of a $U(2)$ gauge theory with $N_f=4$ flavors in \eref{PF-U24-1}, we get 
\begin{align}
 I = & \int \, d\xi \,\Big[ \frac{(e^{2\pi i \s^2_1 \xi} - e^{2\pi i \s^2_3 \xi})\,(e^{2\pi i \s^2_2 \xi} - e^{2\pi i \s^2_4 \xi})}{\sh{(\s^2_1 - \s^2_3)}\,\sh{(\s^2_2-\s^2_4)}}
 - \frac{(e^{2\pi i \s^2_1 \xi} - e^{2\pi i \s^2_4 \xi})\,(e^{2\pi i \s^2_2 \xi} - e^{2\pi i \s^2_3 \xi})}{\sh{(\s^2_1 - \s^2_4)}\,\sh{(\s^2_2- \s^2_3)}} \Big] \nn \\
 & \times \frac{e^{-2\pi q_1 z' \xi }}{\sh{(\s^2_1- \s^2_2)}\, \sh{(\s^2_3-\s^2_4)}\, \sinh^2{\pi \xi}} \nn \\
 = &  \int \, d\xi \,\Big[ (e^{2\pi i \xi (\s^2_1 +\s^2_2)} + e^{2\pi i \xi (\s^2_3 +\s^2_4)} )\,\Big( \frac{1}{\sh{\s^2_{13}} \, \sh{\s^2_{24}}} -  \frac{1}{\sh{\s^2_{14}} \, \sh{\s^2_{23}}}\Big) \nn \\
& - \frac{e^{2\pi i \xi (\s^2_1 +\s^2_4)}}{\sh{\s^2_{13}} \, \sh{\s^2_{24}}} - \frac{e^{2\pi i \xi (\s^2_2 +\s^2_3)}}{\sh{\s^2_{13}} \, \sh{\s^2_{24}}} 
 + \frac{e^{2\pi i \xi (\s^2_1 +\s^2_3)}}{\sh{\s^2_{14}} \, \sh{\s^2_{23}}} + \frac{e^{2\pi i \xi (\s^2_2 +\s^2_4)}}{\sh{\s^2_{14}} \, \sh{\s^2_{23}}} \Big] \nn \\
 & \times \frac{e^{-2\pi q_1 z' \xi }}{\sh{(\s^2_{12})}\, \sh{(\s^2_{34})}\, \sinh^2{\pi \xi}}\nn \\
 = &  \int \, d\xi \,\Big[ - (e^{2\pi i \xi (\s^2_1 +\s^2_2)} + e^{2\pi i \xi (\s^2_3 +\s^2_4)} )\,\Big( \frac{\sh{\s^2_{12}}\, \sh{\s^2_{34}}}{\sh{\s^2_{13}} \, \sh{\s^2_{24}}\,\sh{\s^2_{14}} \, \sh{\s^2_{23}}}\Big) \nn \\
& - \frac{e^{2\pi i \xi (\s^2_1 +\s^2_4)}}{\sh{\s^2_{13}} \, \sh{\s^2_{24}}} - \frac{e^{2\pi i \xi (\s^2_2 +\s^2_3)}}{\sh{\s^2_{13}} \, \sh{\s^2_{24}}} 
 + \frac{e^{2\pi i \xi (\s^2_1 +\s^2_3)}}{\sh{\s^2_{14}} \, \sh{\s^2_{23}}} + \frac{e^{2\pi i \xi (\s^2_2 +\s^2_4)}}{\sh{\s^2_{14}} \, \sh{\s^2_{23}}} \Big] \nn \\
 & \times \frac{e^{-2\pi q_1 z' \xi }}{\sh{(\s^2_{12})}\, \sh{(\s^2_{34})}\, \sinh^2{\pi \xi}}.
\end{align}
One can now take the $z' \to 1$ limit trivially, since there are no $z'$-dependent poles in the integral. Performing the sum over the terms in \eref{D6-pf-dual-Ab} 
gives a $\sinh^2{\pi \xi}$ factor in the numerator which cancels against the $\sinh^2{\pi \xi}$ factor in the denominator. The integration over $\xi$ then gives 
a delta function and the dual partition function assumes the form:
\begin{align}\label{DualPF-fin0}
& Z^{(Y'[(V'^{(I)}_{2,R})^\vee])} = e^{-2\pi |R|(\eta_\alpha/2)}\,Z^{\wt{G}^\alpha_{\CP}(Y[\wt{W}_{R}])} \nn \\
& =\int \,\Big[d\vec \s^{2}\Big] \, \delta \Big(  \tr \vec \s^2 \Big)\, Z^{\rm vec}_{\rm 1-loop} (\vec \s^{2})\,Z^{\rm fund}_{\rm 1-loop} (\vec \s^{2}, \vec t -\eta_\alpha/2)\,\, \sum_{w \in R}\, e^{2\pi \sum_j w_j \s^2_j } \nn \\
& \times \Big[ - (\delta(\s^2_1 +\s^2_2) + \delta(\s^2_3 +\s^2_4) )\,\Big( \frac{\sh{\s^2_{12}}\, \sh{\s^2_{34}}}{\sh{\s^2_{13}} \, \sh{\s^2_{24}}\,\sh{\s^2_{14}} \, \sh{\s^2_{23}}}\Big) \nn \\
& - \frac{\delta(\s^2_1 +\s^2_4)}{\sh{\s^2_{13}} \, \sh{\s^2_{24}}} - \frac{\delta(\s^2_2 +\s^2_3)}{\sh{\s^2_{13}} \, \sh{\s^2_{24}}} 
 + \frac{ \delta(\s^2_1 +\s^2_3)}{\sh{\s^2_{14}} \, \sh{\s^2_{23}}} + \frac{\delta(\s^2_2 +\s^2_4)}{\sh{\s^2_{14}} \, \sh{\s^2_{23}}} \Big] \nn \\
 & \times \frac{1}{\sh{(\s^2_{12})}\, \sh{(\s^2_{34})} }.
\end{align}
Note that the terms inside the $\Big[ \,\,\Big]$ brackets have different delta function coefficients. We now make a change of variables such that 
each term has the same delta coefficient $\delta(\s^2_1 +\s^2_2)$. For example, in the second term this is achieved by the change of variables 
$\s^2_3 \leftrightarrow \s^2_1$ and $\s^2_4 \leftrightarrow \s^2_2$, and so on. Note that the integrand in the first line on the RHS of \eref{DualPF-fin0}, 
including the Wilson defect factor $\sum_{w \in R}\, e^{2\pi \sum_j w_j \s^2_j }$, is invariant under these change of variables. 
Factoring out the common delta function coefficient, the RHS of the matrix integral can then be simplified as follows:
\begin{align}
Z^{(Y'[(V'^{(I)}_{2,R})^\vee])} =& \int \,\Big[d\vec \s^{2}\Big] \, \delta \Big(  \tr \vec \s^2 \Big)\, Z^{\rm vec}_{\rm 1-loop} (\vec \s^{2})
\,Z^{\rm fund}_{\rm 1-loop} (\vec \s^{2}, \vec t -\eta_\alpha/2)\,\, \sum_{w \in R}\, e^{2\pi \sum_j w_j \s^2_j } \nn \\
& \times \Big[ \delta(\s^2_1 +\s^2_2)\,\Big( \frac{1}{\sh{\s^2_{13}} \, \sh{\s^2_{24}}\,\sh{\s^2_{14}} \, \sh{\s^2_{23}}}\Big) \Big] \nn \\
= & \int \,\Big[d\vec \s^{2}\Big] \, \delta(\s^2_1 +\s^2_2)\,\delta(\s^2_3 +\s^2_4)\,Z^{\rm fund}_{\rm 1-loop} (\vec \s^{2}, \vec t -\eta_\alpha/2)\, 
\sum_{w \in R}\, e^{2\pi \sum_j w_j \s^2_j }\nn \\
& \times \Big[ \sinh^2{\pi\,\s^2_{12}}\,\sh{\s^2_{13}} \, \sh{\s^2_{24}}\,\sh{\s^2_{14}} \, \sh{\s^2_{23}}
\, \sinh^2{\pi \, \s^2_{34}}\Big].
\end{align}
Choosing $x= \s^2_1$ and $y= \s^2_3$, the term in the parenthesis (after implementing the delta functions) can be identified as the 
1-loop contribution of an $Sp(2)$ vector multiplet, i.e.
\begin{align}
& \Big[ \sinh^2{\pi\,\s^2_{12}}\,\,\sh{\s^2_{13}} \, \sh{\s^2_{24}}\,\sh{\s^2_{14}} \, \sh{\s^2_{23}}
\, \sinh^2{\pi\,\s^2_{34}}\,\Big] \nn \\
& \longrightarrow  \sinh^2{\pi\, 2x} \, \sinh^2{\pi\,2y} \,  \sinh^2{\pi \,(x+y)} \,  \sinh^2{\pi \,(x-y)} =: Z^{{\rm vec}, Sp(2)}_{\rm 1-loop} (\vec \s^{2}).
\end{align}
Therefore, the final form of the matrix integral is given as:
\begin{align} \label{DualPF-fin}
Z^{(Y'[(V'^{(I)}_{2,R})^\vee])} = \int \,\Big[d\vec \s^{2}\Big] \, \delta(\s^2_1 +\s^2_2)\,\delta(\s^2_3 +\s^2_4)\,Z^{{\rm vec}, Sp(2)}_{\rm 1-loop} (\vec \s^{2})\,
Z^{\rm fund}_{\rm 1-loop} (\vec \s^{2}, \vec t -\eta_\alpha/2)\, \sum_{w \in R}\, e^{2\pi \sum_j w_j \s^2_j },
\end{align}
which reproduces \eref{ZNAbS-d} in the main text \footnote{In performing the computation above, we haven't carried over the overall 
numerical/combinatorial factors in the intermediate steps to make the presentation less cumbersome. They have, however, been reinstated 
in the final answer and one can readily check that the matrix integrals have the correct Weyl factors.}.\\

The mirror map between the Wilson defect in the $D_6$ quiver and vortex defect in the $Sp(2)$ quiver can be similarly worked out by starting 
from the pair in \figref{NAEx-LQdef2} and implementing an $S$-type operation on $X[W_R]$. As before, the procedure involves abelianizing the 
$S$-type operation and then computing the dual partition function following \eref{PF-wtOPgenD-B2A}-\eref{CZ-wtOPD-B2A}.

\subsection{Non-Abelian Gauging without using abelianization}  \label{app: NAG-Ex-2}

Let us now perform the non-Abelian gauging operation without resorting to the abelianization procedure. 
The gauging operation on the defect partition function of $X[V^{(I)}_{2,R}]$ is given by \eref{ZNAbS-D6-app}, 
which manifestly gives the defect partition function of $X'[V'^{(I)}_{2,R}]$. 

The dual partition function can be constituted from the general expression \eref{PF-wtOPgenD-A2B}, as follows:
\begin{align}\label{DualPF-woAb}
Z^{\wt{G}^{\alpha}_{\CP}(Y[\wt{W}_{R}])} = \lim_{z\to 1}\, \int \prod^2_{\gamma'=1}\,\Big[d\vec \s^{\gamma'}\Big] \, 
& \CZ_{\wt{G}^{\alpha}_{\CP}(Y)}( \{ \vec \s^{\gamma'} \}, \eta_\alpha) \,Z^{(Y)}_{\rm int} (\{ \vec \s^{\gamma'} \}, \vec t, \vec u=0, -v) \, 
\sum_{w \in R}\, e^{2\pi \sum_j w_j \s^2_j z},
\end{align}
where the limit $z \to 1$ can be trivially implemented at this stage. The function $\CZ_{\wt{G}^{\alpha}_{\CP}(Y)}$ is given as:
\begin{align}
\CZ_{\wt{G}^{\alpha}_{\CP}(Y)}( \{ \vec \s^{\gamma'} \}, \eta_\alpha) =& \int\, \frac{d^2u}{2!}\, \CZ_{{G}^{\alpha}_{\CP}(X)}(\vec u, \eta_\alpha)\,\prod^2_{i=1}
\,e^{2\pi i g_i (\{\vec \s^{\gamma'} \}, \CP)\,u_i} \nn \\
=& \int\, \frac{d^2u}{2!}\, \CZ_{{G}^{\alpha}_{\CP}(X)}(\vec u, \eta_\alpha)\, e^{-2\pi i (u_1 -u_2)\,\tr \vec \s^1} \, e^{-2\pi i u_2\,\tr \vec \s^2},
\end{align}
with the functions $ \CZ_{{G}^{\alpha}_{\CP}(X)}, g_i$ and $Z^{(Y)}_{\rm int}$ being
\begin{align}
& \CZ_{{G}^{\alpha}_{\CP}(X)}(\vec u, \eta_\alpha) = e^{2\pi i \eta_\alpha (u_1+u_2)}\, \sinh^2{\pi(u_1 -u_2)}, \\
& g_1(\{\vec \s^{\gamma'} \}, \CP) = - \tr \vec \s^1, \qquad g_2(\{\vec \s^{\gamma'} \}, \CP) = \tr \vec \s^1 - \tr \vec \s^2, \\
& Z^{(Y)}_{\rm int} (\{ \vec \s^{\gamma'} \}, \vec t, \vec u=0, -v)= e^{2\pi i v \tr \vec \s^2} \, \prod^2_{\gamma'=1}\, Z^{\rm vec}_{\rm 1-loop} (\vec \s^{\gamma'})\, Z^{\rm bif}_{\rm 1-loop}(\vec \s^1, \vec \s^2, 0)\,Z^{\rm fund}_{\rm 1-loop} (\vec \s^{2}, \vec t).
\end{align}

Given the expression in \eref{DualPF-woAb}, we will first perform the integration over $\vec \s^1$. Isolating 
the $\vec \s^1$-dependent part of the matrix integral, and using the result \eref{PF-U24-1}, we have
\begin{align} \label{sigma1-int}
& \int \Big[d\vec \s^{1}\Big] \,Z^{\rm vec}_{\rm 1-loop} (\vec \s^{1})\,Z^{\rm bif}_{\rm 1-loop}(\vec \s^1, \vec \s^2, 0)\, e^{-2\pi i (u_1 -u_2)\,\tr \vec \s^1} \nn \\
& = \Big[ \frac{(e^{2\pi i \s^2_1 (u_2-u_1)} - e^{2\pi i \s^2_3 (u_2-u_1)})\,(e^{2\pi i \s^2_2 (u_2-u_1)} - e^{2\pi i \s^2_4 (u_2-u_1)})}{\sh{(\s^2_{13})}\,\sh{(\s^2_{24})}}  \nn \\
& - \frac{(e^{2\pi i \s^2_1 (u_2-u_1)} - e^{2\pi i \s^2_4 (u_2-u_1)})\,(e^{2\pi i \s^2_2 (u_2-u_1)} - e^{2\pi i \s^2_3 (u_2-u_1)})}{\sh{(\s^2_{14})}\,\sh{(\s^2_{23})}} \Big] \times \frac{1}{\sh{(\s^2_{12})}\, \sh{(\s^2_{34})}\, \sinh^2{\pi (u_2-u_1)}} \nn \\
 & = \Big[ - (e^{2\pi i (u_2-u_1) (\s^2_1 +\s^2_2)} + e^{2\pi i (u_2-u_1)(\s^2_3 +\s^2_4)} )\,\Big( \frac{\sh{\s^2_{12}}\, \sh{\s^2_{34}}}{\sh{\s^2_{13}} \, \sh{\s^2_{24}}\,\sh{\s^2_{14}} \, \sh{\s^2_{23}}}\Big) \nn \\
& - \frac{e^{2\pi i (u_2-u_1) (\s^2_1 +\s^2_4)}}{\sh{\s^2_{13}} \, \sh{\s^2_{24}}} - \frac{e^{2\pi i (u_2-u_1) (\s^2_2 +\s^2_3)}}{\sh{\s^2_{13}} \, \sh{\s^2_{24}}} 
 + \frac{e^{2\pi i (u_2-u_1) (\s^2_1 +\s^2_3)}}{\sh{\s^2_{14}} \, \sh{\s^2_{23}}} + \frac{e^{2\pi i (u_2-u_1) (\s^2_2 +\s^2_4)}}{\sh{\s^2_{14}} \, \sh{\s^2_{23}}} \Big] \nn \\
 & \times \frac{1}{\sh{(\s^2_{12})}\, \sh{(\s^2_{34})}\, \sinh^2{\pi (u_2-u_1)}},
\end{align}
where for the second equality, we have simply rearranged the terms within the parenthesis. The next step is to perform the integral over the variables $\{u_i\}$. 
First, note that the factor $1/\sinh^2{\pi (u_2-u_1)}$ in \eref{sigma1-int} exactly cancels the $\sinh^2$ term in $\CZ_{{G}^{\alpha}_{\CP}(X)}(\vec u, \eta_\alpha)$, 
and therefore $u_1,u_2$ only appear in the exponential terms of the integrand. 
It is convenient to redefine the integration variables as $u'_1= u_1 -u_2$, $u'_2=u_2$ (the resultant determinant of the Jacobian matrix is unity).
Performing the integration over $u'_1$ and $u'_2$, and shifting the $\vec \s^2$ variables as $\vec \s^2_j \to \vec \s^2_j + \eta_\alpha/2$ 
($j=1,\ldots,4$), one arrives at the following expression for the dual partition function:
\begin{align}
& Z^{(Y'[(V'^{(I)}_{2,R})^\vee])}= e^{-2\pi |R|(\eta_\alpha/2)}\,Z^{\wt{G}^\alpha_{\CP}(Y[\wt{W}_{R}])} \nn \\
& =\int \,\Big[d\vec \s^{2}\Big] \, \delta \Big(  \tr \vec \s^2 \Big)\, Z^{\rm vec}_{\rm 1-loop} (\vec \s^{2})
\,Z^{\rm fund}_{\rm 1-loop} (\vec \s^{2}, \vec t -\eta_\alpha/2)\,\, \sum_{w \in R}\, e^{2\pi \sum_j w_j \s^2_j } \nn \\
& \times \Big[ - (\delta(\s^2_1 +\s^2_2) + \delta(\s^2_3 +\s^2_4) )\,\Big( \frac{\sh{\s^2_{12}}\, \sh{\s^2_{34}}}{\sh{\s^2_{13}} \, \sh{\s^2_{24}}\,\sh{\s^2_{14}} \, \sh{\s^2_{23}}}\Big) \nn \\
& - \frac{\delta(\s^2_1 +\s^2_4)}{\sh{\s^2_{13}} \, \sh{\s^2_{24}}} - \frac{\delta(\s^2_2 +\s^2_3)}{\sh{\s^2_{13}} \, \sh{\s^2_{24}}} 
 + \frac{ \delta(\s^2_1 +\s^2_3)}{\sh{\s^2_{14}} \, \sh{\s^2_{23}}} + \frac{\delta(\s^2_2 +\s^2_4)}{\sh{\s^2_{14}} \, \sh{\s^2_{23}}} \Big] \nn \\
 & \times \frac{1}{\sh{(\s^2_{12})}\, \sh{(\s^2_{34})} },
\end{align}
which is identical to \eref{DualPF-fin0}. Then, after performing an identical sequence of manipulations of the matrix model integral, one arrives at the final answer 
\eref{DualPF-fin}, which reproduces the result \eref{ZNAbS-d} in the main text.

\bibliography{cpn1-1}
\bibliographystyle{JHEP}

\end{document}